# Applications of Kac-Moody Algebras to Gravity and String Theory

Keith Glennon

*Department of Mathematics, King's College London,*

*The Strand, London WC2R 2LS, United Kingdom*

Thesis Supervisor: Professor Peter West FRS

Thesis submitted in partial fulfilment of the requirements

of the Degree of Doctor of Philosophy

Date of Submission: 08/08/22

## Abstract

It has previously been proposed that the the theory of strings and branes possesses a large symmetry group generated by the Kac-Moody algebra $E_{11}$. It has also previously been proposed that the the theory of gravitation in four dimensions possesses a large symmetry group generated by the Kac-Moody algebra $A_1^{+++}$. These symmetry groups predict the existence of an infinite collection of fields in each theory, including a field that can be interpreted as a dual graviton. In this thesis we will discuss developments related to these approaches to string theory and gravity.

We begin by reviewing the method of nonlinear realizations, then introducing the Kac-Moody algebra $A_1^{+++}$, along with its first fundamental representation representation which introduces space-time, to low levels. We will then consider a nonlinear realization involving their semi-direct product, and use the associated Cartan forms to construct dynamics for the low level fields of this nonlinear realization. We will then propose duality relations between fields of the theory at different levels, and then propose a nonlinear equation of motion for the dual graviton arising from $A_1^{+++}$.

We then consider the analogous results for $E_{11}$. After introducing $E_{11}$ along with its first fundamental representation, we will review its nonlinear realization and the construction of the dynamics and duality relations in terms of the resulting Cartan forms. We will then describe a proposed nonlinear dual graviton equation in the context of string theory.

We next discuss irreducible representations of $E_{11}$. We will construct isotropy groups associated to massive and massless particles, and branes such as the M2 and M5 branes. We will show how the irreducible representations based on $E_{11}$ when applied to a massless particle result in the known degrees of freedom of a massless graviton in eleven-dimensional supergravity. We will trace the reduced number of degrees of freedom to the existence of an ideal annihilating the fields of the representation. We will then apply these ideas to the IIA string, constructing an isotropy group and propose a structure for its generators to all orders.

Finally we present an approach to introducing spinorial representations of the Kac-Moody algebra $E_{11}$, generalizing the well-known way fermions are usually introduced into relativistic theories.

**Acknowledgements**

I would like to thank my advisor Peter West for offering me an opportunity to work on the material contained in this thesis. I am extremely grateful to him for giving his time, support, guidance, insight and encouragement throughout my studies. I would also like to thank Paul Cook for his time, support and insight into $E_{11}$.

In addition I would like to thank the Mathematics Department at King's College London for their support, in particular for supporting my PhD studies with an 'NMS Faculty Studentship', and the Irish Government/people for their 'Back to Education Allowance' scheme support during my undergraduate studies.

I would also like to thank my family: my mother Mary and father Michael, and (what have become throughout my studies) my *musician*/philosopher/computer-scientist brother Ronan and *professional wrestler* sister Karen, for all their support. I would also like to thank all my cousins for their support, such as Brendan, Catherine, Irene, Greg and Dermot Glennon and their families, and my aunt Anne Murray as well as my cousins Conor and Susan Bracken, Patricia and Kate Murray. In particular I would like dedicate this thesis to the memories of my grandmother Beatrice Glennon, my uncle Colm Murray, aunt Cathy Bracken and uncle Aidan Bracken.

I would also like to thank my friends, in particular Steven Horan and his family, Kevin and Joseph Twomey and their family, and everyone else, for all their support. Lastly I would like to thank Aman Thind for her constant support throughout my studies.

# Contents

*"String theory is a very promising (and fertile) framework for a consistent theory of quantum gravity. However, we still appear to be at a rather preliminary stage of our understanding of this theory.*

*In particular a non-perturbative formulation of the theory and uncovering its symmetries are important open issues.*

*Matrix models and AdS/CFT correspondence mark recent success on the former issue, while hyperbolic Kac–Moody algebras might be the right language for the latter."*

— STRING THEORY AS A THEORY OF QUANTUM GRAVITY: A STATUS REPORT [5]

# Chapter 1

# Introduction

It is currently a prevailing belief in the string theory community that the five different superstring theories in ten dimensions are simply different manifestations of a single eleven-dimensional theory, unifying the disparate string theories, referred to as 'M-theory' [22]. There is, however, no dynamical description of M-theory, so it is worth reviewing some of the arguments in favor of the existence of this Missing theory with no dynamical description especially given the years of, what were considered to be, apparently strong arguments against eleven-dimensional theories [22], [40, Sec. 11.5].

It is known that M-theory in the low energy limit is described by a unique eleven-dimensional supergravity theory [14], thus we begin from this theory. The Nahm classification of supersymmetries [44] describes the unique nature of this supergravity theory, and this very uniqueness, along with its utility in constructing the $N = 8$ four-dimensional supergravity via dimensional reduction [12], were strong arguments for its importance in physics, before the discovery of M-theory. However, the superstring programme of the 1980's was criticized for neglecting the existence of this eleven dimensional theory [22]. Performing Kaluza-Klein dimensional reduction of eleven-dimensional supergravity on a circle one finds the unique (non-chiral) ten-dimensional IIA supergravity theory [7], which is the low energy limit of Type IIA superstring theory. This however was in fact taken as a *criticism* of the eleven-dimensional theory [40], since it implies the existence of extra Kaluza-Klein states which were not seen in the ten-dimensional superstring theory. This apparent defect later turned out to be a *virtue* of the eleven-dimensional theory, when it was recognized that the additional Kaluza-Klein states in fact do arise in the ten-dimensional superstring theories in the form of 'soliton-like' solutions associated to (mem)branes of the theory. In other words, the ten-dimensional supergravity theory with the additional soliton-like states appeared to be an eleven-dimensional theory in disguise [40].

An important aspect of supergravity theories is the occurence of coset space symmetries that determine the way the scalars enter these theories [37], [11], [85, Sec. 13.6]. In particular, the scalars in a supergravity multiplet are known to be associated to a nonlinear realization, with the full supergravity possessing the 'exceptional symmetry group' of that nonlinear realization. For example, in [13] it was shown that the 70 scalars in the above-mentioned $N = 8$ maximal supergravity theory in four dimensions could be associated to the nonlinear realization of the coset $E_7/\mathrm{SU}(8)$, where SU(8) is the maximal compact subgroup of $E_7$. Similar properties were found of the other maximal supergravity theories obtainable from the eleven-dimensional supergravity theory by dimensional reduction. The Type IIB supergravity theory, though not straightforwardly related to the other cases [34], can also be formulated as a coset model given the above arguments. These results are listed in the following table:

| Dimension | Exceptional Symmetry Group | Coset Space |
|---|---|---|
| 10 (IIA) | $\mathrm{O}(1,1)$ | - |
| 10 (IIB) | $\mathrm{SL}(2)$ | $\mathrm{SL}(2)/\mathrm{SO}(2)$ |
| 9 | $\mathrm{GL}(2)$ | $\mathrm{GL}(2)/\mathrm{SO}(2)$ |
| 8 | $E_3 \sim \mathrm{SL}(3) \times \mathrm{SL}(2)$ | $\mathrm{SL}(3) \times \mathrm{SL}(2)/\mathrm{SO}(3) \times \mathrm{SO}(2)$ |
| 7 | $E_4 \sim \mathrm{SL}(5)$ | $\mathrm{SL}(5)/\mathrm{SO}(5)$ |
| 6 | $E_5 \sim \mathrm{SO}(5,5)$ | $\mathrm{SO}(5,5)/\mathrm{SO}(5) \times \mathrm{SO}(5)$ |
| 5 | $E_6$ | $E_6/\mathrm{Sp}(8)$ |
| 4 | $E_7$ | $E_7/\mathrm{SU}(8)$ |
| 3 | $E_8$ | $E_8/\mathrm{SO}(16)$ |

Continuing below dimension 3 directly in terms of the eleven-dimensional supergravity action is complicated by the fact that the Einstein-Hilbert action reduces to a total derivative [37]. However, the above table suggests some 'empirical rules of dimensional reduction and group oxidation' [37]: we can decrease/increase the dimension in a given theory from $D$ to $D \mp 1$ by adding/removing a node in the $E_{(11-D)}$ Dynkin diagram obtaining $E_{11-(D\mp1)}$ and considering the nonlinear realization in terms of its associated maximal compact subgroup[1] of $E_{11-(D\mp1)}$. Thus by examining the above table one might expect $E_9$ and $E_{10}$ to arise in dimension 2 and 1 respectively. For example, as far back as 1982 it was conjectured that the dimensional reduction of eleven-dimensional supergravity to one (time) dimension would possess an $E_{10}$ symmetry [38]. In general relativity it is well known cosmological models near singularities can be taken to reduce to effectively one-dimensional models with oscillatory behaviour as part of the BKL programme [3], [42]. In [16] this behaviour was shown to extend to the five superstring theories, and in [17] this behaviour was related to the properties of the $E_{10}$ Kac-Moody algebra. Similarly the relevance of $E_9$ was discussed e.g. in [45] and [47] where it is noted that the above 'empirical rules' naturally lead to the conjecture that $E_9$ is a symmetry of eleven-dimensional supergravity when dimensionally reduced down to two dimensions.

A natural question arises from the above procedure as to whether the above procedure lifts to eleven dimensions and whether $E_{11}$ was involved [45]. However, it was nearly universally thought for some time that these symmetries did not lift to eleven-dimensions [68]. In [78] it was proposed that in fact the symmetries do lift, resulting in a proposal that the low energy effective action of the theory of strings and branes possesses a symmetry described the Kac-Moody algebra $E_{11}$. The Kac-Moody algebra $E_{11}$ possesses the Dynkin diagram

$$\begin{array}{ccccccccccccccccccc} & & & & & & & & & & & & & & \otimes\ 11 & & & & \\ & & & & & & & & & & & & & & | & & & & \\ \bullet & - & \bullet & - & \bullet & - & \bullet & - & \bullet & - & \bullet & - & \bullet & - & \bullet & - & \bullet & - & \bullet \\ 1 & & 2 & & 3 & & 4 & & 5 & & 6 & & 7 & & 8 & & 9 & & 10 \end{array} \tag{1.1}$$

It can be seen from the diagram that, on deleting the $D$'th node along the $A_{10}$ line in this diagram, that one finds all of the previous $E_{11-D}$ exceptional symmetry groups, along with a residual $A_{D-1}$ Dynkin diagram, that in fact can directly be associated to gravity and can be referred to as the 'gravity line', as we will review in Chapter 2. In addition, the $E_{11}$ approach naturally incorporates both the IIA and IIB theories in a unified manner [81].

One of the consequences of working with $E_{11}$ is that one is forced to work with an infinite collection of fields, the first few of which are $h_a{}^b$, $A_{a_1a_2a_3}$, $A_{a_1\ldots a_6}$, $h_{a_1\ldots a_8,b}$. The first field can be used to describe the gravitational field, the second and third describing gauge fields arising in eleven-dimensional supergravity, where the latter can be interpreted as a 'dual graviton', while an interpretation of the higher fields is an open question. One might compare the presence of an infinite number of fields to the discussion of soliton-like states in the IIA string as being indicative of a very complicated structure underlying M-theory.

[1]The maximal compact subgroups can be described in terms of a Cartan involution operator on the Lie algebra [85, Ch. 13], we will see this explicitly for $A_1^{+++}$ and $E_{11}$ in later chapters.

Similarly, there had been indications in the literature that an extension beyond the usual space-time was required in discussions of M-theory. For example, in [20] it was suggested that an extension of eleven-dimensional supergravity to 'BPS-extended supergravities' based on a 'hidden exceptional geometry' may be needed, in which the central charges would be associated with extra dimensions, thus incorporating BPS multiplets and Kaluza-Klein multiplets into this larger theory. From the perspective of $E_{11}$ it was independently proposed [79] that space-time could be incorporated into the above $E_{11}$ proposal in an intrinsic manner by considering the first fundamental representation of $E_{11}$ as generating a 'generalized space-time'. This generalized space-time incorporated not only the familiar momentum $P_a$ of point particles, but treated the eleven-dimensional supergravity central charges $Z^{a_1a_2}$ and $Z^{a_1\cdots a_5}$, as the first of an infinite collection of (momentum-space) coordinates, on which the above fields depend. Indeed, not only does $E_{11}$ lead to a generalized notion of space-time and 'exceptional geometry' [84], but it also leads to a generalized notion of BPS conditions [83].

It was proposed that a nonlinear realization of the semi-direct product of $E_{11}$ with its first fundamental representation $l_1$, $E_{11} \otimes_s l_1$, with respect to the maximal compact subgroup of $E_{11}$ called the 'Cartan Involution Invariant Subgroup' $I_c(E_{11})$, would result in the dynamics of eleven dimensional supergravity. This conjecture has essentially been proved in [70] and [69], and is reviewed in [87]. Given the recent interest in questioning the foundations of space-time [1], the fact that '$E$ Theory' [87] leads to a new generalized space-time, with its own associated generalized geometry [84], and more general applications[2] of Kac-Moody algebras such as those of $A_1^{+++}$ (with its own notion of generalized space-time [57]) to gravity (as discussed in this thesis) may play an increasing role in these discussions.

We have mentioned above that the fields $h_{ab}$ and $h_{a_1\ldots a_8,b}$ can be interpreted as a graviton and dual-graviton field respectively. The existence of an analogue of electric-magnetic duality in a gravitational context had been considered [27], and in [15] an independent 'dual graviton field' was first introduced. Similarly, possible S-duality relations in a gravitational context had also been explored [48]. The dual graviton field arising independently from and $E_{11}$ perspective has the properties of the dual graviton introduced in [15], thus a formulation of gravity in which the dual graviton field is incorporated is naturally suggested by the $E_{11}$ perspective.

[2]A recent discussion of some applications of Kac-Moody algebras to physics is given in [24].

The question naturally arises as to whether similar results can be formulated without the additional string-theoretic fields $A_{a_1a_2a_3}$, $A_{a_1\ldots a_6}$ (and higher level analogues) arising from the Kac-Moody algebra, rather than artificially setting them to zero. It was first proposed in [41] that the Kac-Moody algebra $A_{D-3}^{+++}$ in $D$ dimensions is a symmetry of Einstein's general relativity. In $D = 11$ one finds $h_{ab}$ and $h_{a_1\ldots a_8,b}$ as the first fields arising in the theory. Some solutions based on this model have been discussed in [10]. The Kac-Moody formulation naturally encodes duality relations between these fields as fundamental as we will see in the ensuing chapters. We can rediscover the above Kac-Moody discussion of gravity, without necessarily demanding the existence of a dual graviton field in advance, instead arguing in terms of the empirical rules of group oxidation, based on the following historical results.

In [37], [38] it was shown[3] that dimensional reduction of the Einstein-Hilbert action of four-dimensional gravity to three dimensions can be reformulated as a nonlinear realization on the coset space $\mathrm{SL}(2,\mathbb{R})/\mathrm{SO}(2)$, and possessed $A_1 = \mathrm{SL}(2,\mathbb{R})$ as a symmetry group. This 'exceptional symmetry group' was called the 'Ehler's symmetry group'. The origin of this symmetry at the time was mysterious, but utilizing the results of E theory [78], an explanation can be given. From this perspective we can say that since, as mentioned previously, gravity in 3 dimensions can be associated to the 'gravity line' Dynkin diagram $A_2$, the following Dynkin diagrams can be associated to this theory:

$$\bullet - \bullet \qquad \bullet \tag{1.2}$$

Similarly, dimensional reduction to two dimensions (with an associated 'gravity line' $A_1$) resulted in an 'exceptional symmetry group' at the time called the 'Geroch group', the Kac-Moody algebra $A_1^+$, with Dynkin diagram $\bullet = \bullet$. From the perspective of E theory we can say that the following Dynkin diagrams are associated to the theory:

$$\bullet \qquad \bullet = \bullet \tag{1.3}$$

[3] [55] provides a useful review.

Furthermore, in [46], building on the above results, a discussion of dimensionally reducing gravity down to one dimension was discussed, and evidence for the Kac-Moody algebra $A_1^{++}$ being a symmetry was given, where its Dynkin diagram is given by

$$\bullet \; - \; \bullet \; = \; \bullet \tag{1.4}$$

In this reference, however, an explicit dependence on supergravity was utilised. Recent studies of the one-dimensional case, without any reliance on supergravity, have been conducted [54], and although they do not utilize the known results of [41] partial evidence of an underlying Kac-Moody structure was still discovered. Given the above-mentioned success of $E_{11}$ in lifting the empirical rules of dimensional reduction up to the eleven-dimension, it is natural to propose that four-dimensional general relativity is described by a nonlinear realization of the semi-direct product of the Kac-Moody algebra $A_1^{+++}$ and its $l_1$ representation, $A_1^{+++} \otimes_s l_1$, with respect to its maximal compact subgroup, the Cartan involution invariant subgroup denoted $I_c(A_1^{+++})$. The Dynkin diagram is given by

$$\begin{array}{ccccccc} \bullet & - & \bullet & - & \bullet & = & \bullet \\ 1 & & 2 & & 3 & & 4 \end{array} \tag{1.5}$$

Here we have explicitly numbered the nodes 1 to 4. The above theories in $d < 4$ dimensions then arise from this diagram by deleting the $(4-d)$'th node, treating the diagram to the left of the deleted node as the gravity line of the $d$-dimensional theory, and the diagram to the right of this deleted node as the internal symmetry group of the $d$-dimensional theory.

In this thesis we begin by reviewing the notion of a nonlinear realization, and formulate Einstein's theory of gravity as a nonlinear realization to set the stage. We will then introduce the Kac-Moody algebra $A_1^{+++}$ along with its vector representation, and construct the algebra explicitly, to low levels,. We then present the material of [29] in which we considered a nonlinear realization involving their semi-direct product $A_1^{+++} \otimes_s l_1$ with respect to the maximal compact subgroup of $A_1^{+++}$, where the resulting Cartan forms were used to construct dynamics associated to the low level fields of this nonlinear realization. Results will include a proposed nonlinear dual graviton equation of motion, and duality relations between low level fields of the theory. The study of duality relations in the context of the theory of gravitation in particular is a novel feature of the Kac-Moody perspective.

We then consider the analogous results for $E_{11}$. After introducing $E_{11}$ along with its first fundamental representation, we will review its nonlinear realization and the construction of the dynamics and duality relations in terms of the resulting Cartan forms. We will then describe a proposed nonlinear dual graviton equation in the context of string theory [30].

We next discuss irreducible representations of $E_{11}$. We construct isotropy groups associated to massive and massless particles, and branes such as the M2 and M5 branes [89]. We will show how the irreducible representations based on $E_{11}$ when applied to a massless particle result in the known degrees of freedom of a massless graviton in eleven-dimensional supergravity [31]. We will trace the reduced number of degrees of freedom to the existence of an ideal annihilating the fields of the representation. We will then apply these ideas to the IIA string [32], constructing an isotropy group and then propose a structure for its generators to all levels.

Finally, we present an approach to introducing spinorial representations of the Kac-Moody algebra $E_{11}$ [32], generalizing the well-known way fermions are usually introduced into relativistic theories.

# Chapter 2

# The Method of Nonlinear Realizations

## 2.1 Motivation from Spontaneous Symmetry Breaking

Spontaneous symmetry breaking of local symmetry groups in the standard model of particle physics is a well known method for introducing gauge fields into a physical theory as Goldstone bosons. In this chapter we will show that the pure Yang-Mills gauge field, and indeed even the gravitational field of general relativity, can be formulated as Goldstone bosons. To do this we first formalise the familiar method of spontaneous symmetry breaking into representation-independent group-theoretical language, and is more generally referred to as the 'method of nonlinear realizations'. The notion of a nonlinear realization is discussed for example in [8], [61], [73], [74], [75], [51], [85], and [87], which we refer to for further information.

Consider a theory invariant under a local symmetry group $G$, containing a local subgroup $H$. A useful example to keep in mind is the spontaneous symmetry breaking of a Lagrangian invariant under $\mathrm{SO}(D)$ whose vacuum is invariant under $\mathrm{SO}(D-1)$ [75], [39]. However, it is very important to note that the method is independent of a Lagrangian formulation. In terms of the Lie algebra we can set $H = \{H_i\}$ and $G/H = \{T_a\}$, where $i$ and $a$ are the indices of the subgroup $H$ and quotient $G/H$ generators respectively, and consider group

elements of the form

$$g(x) \in G/H \quad \rightarrow \quad g(x) = e^{i\phi^a(x)T_a} \quad , \quad h(x) \in H \quad \rightarrow \quad h(x) = e^{iu^i(x)H_i} \tag{2.1}$$

A general element of $G$ (near the identity) can always be written as a product in the form

$$g(x)h(x) = e^{i\phi^a(x)T_a} e^{iu^i(x)H_i}. \tag{2.2}$$

If $G$ is a compact group, it means that there exists a basis in which the structure constants are anti-symmetric [75, Sec. 19.6], thus the Lie algebra can be written as

$$[H_i, H_j] = f_{ij}{}^k H_k \quad , \quad [H_i, T_a] = f_{ia}{}^b T_b \quad , \quad [T_a, T_b] = f_{ab}{}^c T_c + f_{ab}{}^i H_i. \tag{2.3}$$

A matter representation $\Psi(x)$ of $G$ transforms as

$$\Psi(x) \rightarrow \Psi'(x) = D[g(x)]\Psi(x) \ . \tag{2.4}$$

Suppose we single out a vacuum $\langle\Psi\rangle_{Vac}$ of $\Psi(x)$ and find that it is preserved not by $G$ but rather only by a subgroup $H \subset G$,

$$h(x)\langle\Psi\rangle_{Vac} = \langle\Psi\rangle_{Vac}. \tag{2.5}$$

To obtain a particle interpretation we must parametrize the field $\Psi(x)$ around the vacuum. To do this it is natural to begin from fields $\psi(x)$ which form a representation under $H$ so that, on taking the vacuum expectation value of any such field, it satisfies $h(x)\langle\psi\rangle_{Vac} = \langle\psi\rangle_{Vac}$. Here $\psi(x)$ can be interpreted as a linear realization of the subgroup $H$, where we use the word realization to denote an element of a representation, and the representation is a linear representation since $\psi(x)$ is a vector in a vector space acted on by this representation.

In general we can act on $\psi(x)$ by elements from all of $G$, not just those from $H$. Acting on $\psi(x)$ by a representation $D[g(x)]$ of an element $g(x)$ from the quotient $G/H$ results in a new element in the representation of $G$, and so we can define

$$\Psi(x) = D[g(x)]\psi(x). \tag{2.6}$$

Thus we have an association between elements of the quotient $G/H$ and fields $\Psi(x)$ in the representation.

In order to perform $G$ transformations on a general field $\Psi(x)$, it is sufficient to study transformations in the coset $G/H$. We first focus on performing rigid $g_0$ transformations in the coset (i.e. $g_0$ is independent of $x$). The action of $g_0 \in G$ on $g(x) \in G/H$ can always be written as

$$g_0 g(x) = g'(x, g_0) h(x, g_0) \tag{2.7}$$

where the element $g_0 g(x)$ is always an element of some new coset whose coset representative is denoted $g'(x, g_0)$, and the precise element in this coset is $g'(x, g_0)h(x, g_0)$. We can thus study transformations

$$g(x) \quad \to \quad g'(x, g_0) = g_0 g(x) h^{-1}(x, g_0). \tag{2.8}$$

Under a rigid $g_0$ transformation, $\Psi(x)$ transforms as $\Psi'(x) = D[g_0]\psi(x)$. However, written in the form (2.6) we find

$$\Psi(x) \quad \to \quad \Psi'(x) = D[g'(x, g_0)]\psi'(x) = D[g_0]D[g(x)]D[h^{-1}(x, g_0)]\psi'(x), \tag{2.9}$$

which tells us that $\psi(x)$ must transform as

$$\psi'(x) = D[h(x, g_0)]\psi(x). \tag{2.10}$$

Indeed, isolating

$$\psi(x) = D[g^{-1}(x)]\Psi(x) \tag{2.11}$$

we find

$$\psi'(x) = D[g'^{-1}(x)]\Psi'(x) = D[h(x, g_0)]D[g^{-1}(x)]D[g_0^{-1}]D[g_0]\Psi(x) = D[h(x, g_0)]\psi(x). \tag{2.12}$$

Notice that for $g_0 = h_0 \in H$ we have

$$h_0 g(x) = h_0 g(x) h_0^{-1} h_0 = e^{i\phi^a h_0 T_a h_0^{-1}} h_0, \tag{2.13}$$

where $g'(x, h_0) = e^{i\phi^a h_0 T_a h_0^{-1}} \in G/H$ is a pure coset representative, so that $\psi(x)$ transforms as

$$\psi'(x) = D[g'^{-1}(x)]\Psi'(x) = D[h_0 g(x) h_0^{-1}]^{-1} D[h_0]\psi(x) = D[h_0]\psi(x). \tag{2.14}$$

This shows that, under the action of rigid elements $h_0$ of $H$, $\psi(x)$ transforms under a linear representation of $H$ due to (2.14), i.e. as a linear realization of $H$, yet under a rigid element $g_0 \in G/H$ $\psi(x)$ transforms nonlinearly under an element of $H$ due to (2.12), and so $\psi(x)$ is called a nonlinear realization of $G$.

We now consider the derivative of $\Psi(x)$. Expressed in terms of (2.6) we have

$$\partial_\mu \Psi(x) = D[g(x)][\partial_\mu + D[g^{-1}(x)\partial_\mu g(x)]\psi(x) \quad . \tag{2.15}$$

Thus we can define a derivative on the nonlinear realization $\psi(x)$ as

$$\begin{aligned} \Delta_\mu \psi(x) &:= D[g^{-1}(x)]\partial_\mu \Psi(x) \\ &= \{\partial_\mu + D[g^{-1}(x)\partial_\mu g(x)]\}\psi(x). \end{aligned} \tag{2.16}$$

Here the quantity

$$\Omega_\mu(x) = g^{-1}(x)\partial_\mu g(x) \tag{2.17}$$

lives in the tangent space to $G$, and so it can be expanded in terms of the Lie algebra generators

$$\Omega_\mu := f_\mu{}^a T_a + w_\mu{}^i H_i := f_\mu + w_\mu \quad . \tag{2.18}$$

Similarly we can consider this in the representation as given by

$$D(\Omega_\mu) = F_\mu{}^a T_a + W_\mu{}^i H_i \quad . \tag{2.19}$$

The partial derivative of $\Psi(x)$ transforms under a *rigid* $g_0$ transformation as

$$\partial_\mu \Psi(x) \quad \rightarrow \quad \partial_\mu \Psi'(x) = D(g_0)\partial_\mu \Psi(x) \quad . \tag{2.20}$$

In terms of the decomposition $\partial_\mu \Psi = D[g]\Delta_\mu \psi$ this transformation reads as

$$\begin{aligned} \partial_\mu \Psi' = D[g'][\partial_\mu + D(\Omega'_\mu)]\psi' &= D[g_0 g h^{-1}]\{\partial_\mu + D[g_0 g h^{-1}]^{-1}\partial_\mu D[g_0 g h^{-1}]\}D[h]\psi \\ &= D[g_0]D[g(x)]\Delta_\mu \psi(x) \\ &= D[g_0]\partial_\mu \Psi(x) \quad , \end{aligned} \tag{2.21}$$

where $\Delta_\mu \psi$ transforms as

$$\begin{aligned} \Delta_\mu \psi \rightarrow \Delta'_\mu \psi' &= [\partial_\mu + g'^{-1}\partial_\mu g']D(h)\psi \\ &= D(h)\Delta_\mu \psi \quad . \end{aligned} \tag{2.22}$$

In studying the transformation of $\partial_\mu \Psi(x)$, we considered a rigid element $g_0$. The above procedure can be generalized to consider local $g_0$ transformations, which would necessitate the introduction of a new gauge field [61]. However the transformation (2.21) involves a gauge-like cancellation on the $D[h^{-1}]\partial_\mu D[h]$ terms. Indeed $\Omega_\mu$ itself transforms under a rigid $g_0$ as

$$\Omega_\mu \quad \to \quad \Omega'_\mu = h\Omega_\mu h^{-1} - h^{-1}\partial_\mu h \quad , \tag{2.23}$$

The quantity $h^{-1}\partial_\mu h$ lives entirely in the subgroup $H$. In terms of the generators of equation (2.18) the transformation law of equation (2.23) reads as

$$f'_\mu = hf_\mu h^{-1} \quad , \quad w'_\mu = w_\mu - h^{-1}\partial_\mu h. \tag{2.24}$$

Thus ${f_\mu}^a$ can be interpreted as a vielbein on $G/H$, while $w_\mu$ is the connection associated to local $H$ transformations. Similarly we can construct other covariant quantities such as curvatures via $[\Delta_\mu, \Delta_\nu]\psi = H_i {R^i}_{\mu\nu}\psi$, etc... Thus [75, Sec. 19.6], any Lagrangian invariant under $H$ and constructed $\psi$, $\Delta_\mu\psi$ and from ${f_\mu}^a$, will also be invariant under $G$. However the possible presence of a Lagrangian, or even a particular matter representation of $G$, are additional concepts.

In the above discussion, the space-time coordinates $x$ have so far been treated as dummy variables on which the group elements $g$ are assumed to depend via $g = g(x)$, so that space-time is not intrinsically associated to the nonlinear realization. Applying the above procedure to the Poincaré group $P$ in $D$ dimensions [87], where $P$ is defined as the semi-direct product of the Lorentz group $\mathrm{SO}(1, D-1) = \{J_{ab}\}$ and its vector representation denoted $T^D = \{P_a\}$, $a, b = 0, 1, \ldots, D-1$,

$$P = \mathrm{SO}(1, D-1) \otimes_s T^D \quad , \tag{2.25}$$

and choosing the local subgroup to be $H = \mathrm{SO}(1, D-1)$, results in a group element

$$g = e^{x^a P_a} \quad , \tag{2.26}$$

where the $x^a$ are fields that can be interpreted as Minkowski space coordinates.

In the next section we will abstract away from any particular matter representation, and, generalizing the space-time example just given, doing so in a manner that allows one to introduce a notion of space-time that is intrinsic to the group of the nonlinear realization.

## 2.2 Nonlinear Realizations

We take our definition of a nonlinear realization of a group $G$ with local subgroup $H$ to be a set of dynamical equations constructed out of a group element $g \in G/H$, where a given $g \in G$ thus satisfies

$$g \simeq gh^{-1} \quad , \tag{2.27}$$

and transforms under a rigid $g_0 \in G$ as

$$g \;\; \rightarrow \;\; g' = g_0 g h^{-1} \simeq g_0 g \quad . \tag{2.28}$$

Thus we are constructing the nonlinear realization out of elements of the quotient $G/H$ which transform under rigid elements of $G$. We have used the inverse $h^{-1}$, however this is purely a convention that was natural in the previous section. From now on we will instead write this element without the inverse notation, and so work with the nonlinear realisation as being constructed out of group elements $g$ satisfying $g \simeq gh$, which we write as $g \rightarrow gh$, subjected to $g \rightarrow g_0 gh \simeq g_0 g$, which we can write as $g \rightarrow g_0 g$, that is:

$$g \;\; \rightarrow \;\; g_0 g \quad \text{as well as} \quad g \;\; \rightarrow \;\; gh \quad . \tag{2.29}$$

This is how it is commonly written in the literature (for example, [87]). The dynamics associated to this nonlinear realization are then constructed in terms of the Cartan form which we write as

$$\mathcal{V} = g^{-1} dg \quad , \tag{2.30}$$

and which transforms under (2.29) as

$$\mathcal{V} \;\; \rightarrow \;\; h^{-1} \mathcal{V} h + h^{-1} dh. \tag{2.31}$$

In the remainder of this thesis we will take the group $G$ to be of a special form, given as the semi-direct product of a group $\hat{G}$ and the vector representation of $\hat{G}$ denoted $l_1$, which is given by

$$G = \hat{G} \otimes_s l_1, \tag{2.32}$$

where we assume that the local subgroup $H$ is a subgroup of $\hat{G}$, and that the generators of $l_1$ commute. The group element of $G$ can now be taken in the form

$$g = e^{x^A l_A} e^{A_\alpha R^\alpha} = g_l g_R. \tag{2.33}$$

In the case of the Poincaré group, the group element is $g = e^{x^a l_a} e^{A_\alpha(x) R^\alpha}$, where $A_\alpha(x)$ is a field depending on the space-time coordinates $x^a$. In general we can interpret the $A_\alpha$ as fields which depend on 'coordinates' $x^A$, associated to the vector representation of $G$, $A_\alpha = A_\alpha(x^A)$. When applied to the groups $\hat{G} = E_{11}$ or $\hat{G} = A_1^{+++}$ the associated vector representation will not only contain Minkowski space-time, but additional 'space-time coordinates' that are natural and unavoidable when one demands that $\hat{G}$ be a symmetry.

Setting $\hat{G} = \{R^\alpha\}$ and $l = \{l_A\}$ the algebra is then given by

$$[R^\alpha, R^\beta] = f^{\alpha\beta}{}_\gamma R^\gamma \quad , \quad [R^\alpha, l_A] = -(D^\alpha)_A{}^B l_B \quad , \quad [l_A, l_B] = 0. \tag{2.34}$$

The minus sign in the second commutator in equation (2.34) ensures that the matrices $(D^\alpha)_A{}^B$ form a representation of $\hat{G}$; indeed commuting with $R^\beta$ and using the Jacobi identity one finds

$$[D^\alpha, D^\beta]_A{}^B l_B = f^{\alpha\beta}{}_\gamma (D^\gamma)_A{}^B l_B. \tag{2.35}$$

We now consider the Cartan form associated to the group element (2.33) [84, Sec.13.2]. This can be written as

$$\begin{aligned}
\Omega &= g^{-1} dg \\
&= (g_l g_R)^{-1} d(g_l g_R) \\
&= g_R^{-1}(g_l^{-1} dg_l) g_R + g_R^{-1} dg_R \\
&= g_R^{-1}(dx^A l_A) g_R + dx^\Pi G_{\Pi,\alpha} R^\alpha \\
&= dx^\Pi E_\Pi{}^A l_A + dx^\Pi G_{\Pi,\alpha} R^\alpha \\
&= dx^\Pi E_\Pi{}^A (l_A + (E^{-1})_A{}^{\Pi'} G_{\Pi',\alpha} R^\alpha) \\
&= dx^\Pi E_\Pi{}^A (l_A + G_{A,\alpha} R^\alpha) \quad ,
\end{aligned} \tag{2.36}$$

where we set

$$\begin{aligned}
dx^A = dx^\Pi \delta_\Pi{}^A \quad &, \quad l_A = l_\Pi \delta_A{}^\Pi \quad , \\
g_R^{-1} l_\Pi g_R &= E_\Pi{}^A l_A \ , \\
G_{A,\alpha} &= (E^{-1})_A{}^\Pi G_{\Pi,\alpha} \ .
\end{aligned} \tag{2.37}$$

We have re-written $\Omega$ as above for the following reasons. Since the $dx^\Pi$ transform under $g \to g_0 g$, due to $g_0 g = (g_0 g_l g_0^{-1})(g_0 g_R)$, we take $dx^\Pi$ to transform in the $\Pi$ index under $g_0$. Since $\Omega$ is invariant under $g \to g_0 g$, the $(l_A + G_{A,\alpha} R^\alpha)$ are left invariant under $g \to g_0 g$, so that the coefficients $dx^\Pi E_\Pi{}^A$ and $dx^\Pi G_{\Pi,\alpha}$ must also be left invariant, and so the $dx^\Pi$ transformation under $g_0$ must be counteracted by transformations of the $\Pi$ index in $E_\Pi{}^A$, while the $A$ index is left invariant. Therefore it is natural to interpret $E_\Pi{}^A$ as a 'generalized vielbein' on $G$, and to distinguish the indices of $dx^\Pi$ from those of $dx^A$ via $dx^A = dx^\Pi \delta_\Pi{}^A$. We can similarly consider 'generalized derivatives'

$$\partial_\Pi = \frac{\partial}{\partial x^\Pi} \tag{2.38}$$

with respect to the generalized coordinates $x^\Pi$. For example, in the case of $E_{11}$ we will find generalized coordinates which read as $x^a, x_{b_1 b_2}, x_{b_1 \dots b_5}, \dots$ and so have the generalized derivatives $\partial_a, \partial_{a_1 a_2}, \partial_{a_1 \dots a_5}, \dots$.

Setting $g_0^{-1} l_\Pi g_0 = D(g_0)_\Pi{}^\Lambda l_\Lambda$ for $D$ any representation of $G$ containing the $l_\Pi$ generators, we see $x^\Pi \partial_\Pi$ is invariant under

$$dx^\Pi \to dx'^\Pi = dx^\Lambda D(g_0^{-1})_\Lambda{}^\Pi \quad \text{and} \quad \partial_\Pi \to \partial'_\Pi = D(g_0)_\Pi{}^\Lambda \partial_\Lambda \ . \tag{2.39}$$

Similarly note that $E_\Pi{}^A$ transforms under $g_0$ as

$$E'_\Pi{}^A = E_\Pi{}^\Lambda D(g_0)_\Lambda{}^A \quad , \tag{2.40}$$

which follows by considering

$$g_0^{-1} g_R^{-1} l_\Pi g_R g_0 = g_0^{-1} E_\Pi{}^\Lambda l_\Lambda g_0 = E_\Pi{}^\Lambda D(g_0)_\Lambda{}^A l_A. \tag{2.41}$$

We have discussed the group element (2.33) as it related to $G$. If we restrict to a group element of $G/H$, thus only including those $R^\alpha$ which generate $G/H$ in (2.33), then the Cartan form coefficients $G_{A,\alpha}$ associated to the generators $R^\alpha$ in equation (2.36) can be interpreted as 'covariant derivatives' of the fields $A_\alpha$ associated to the generators $R^\alpha$.

In practice, the Cartan form $g^{-1}dg$ is explicitly computed using the formulae [85, Sec.17.5]

$$\begin{aligned} e^{-A} B e^{A} &= e^{-A} \wedge B = B - [A,B] + \frac{1}{2}[A,[A,B]] + \dots \quad , \\ e^{-A} de^{A} &= \frac{1 - e^{-A}}{A} \wedge dA = dA - \frac{1}{2}[A,dA] + \frac{1}{3!}[A,[A,dA]] - \dots \quad , \end{aligned} \tag{2.42}$$

where $1 \wedge B = B$, $A \wedge B = [A,B]$, $A^2 \wedge B = [A,[A,B]]$, etc...

## 2.3 Einstein's Gravity as a Nonlinear Realization

We now show how Einstein's gravity can be formulated in terms of a nonlinear realization of $G = \mathrm{IGL}(4,\mathbb{R})$, generated by $P_a$ and $K^a{}_b$ satisfying

$$[K^a{}_b, K^c{}_d] = \delta_b{}^c K^a{}_d - \delta_d{}^a K^b{}_c \quad , \quad [K^a{}_b, P_c] = -\delta_c{}^a P_b \quad , \tag{2.43}$$

with respect to the local subgroup $H = \mathrm{SO}(1,3)$ generated by $J_{ab} = \eta_{[a|e}K^e{}_{|b]}$. The local group element of $G/H$ can be taken as

$$g(x) = e^{x^a P_a} e^{h_a{}^b(x) K^a{}_b}, \tag{2.44}$$

where we do not make use of the local subgroup in this parametrization so that $h_a{}^b$ is not symmetric. This model was considered in [85, Sec. 13.2] (and is also discussed in [66, Sec. 1.3.2] and [56, Sec. 1.5]). To compute the Cartan form $g^{-1}dg$ it is useful to note the following calculation involving $dx^b e^{-h_{a_1}{}^{a_2} K^{a_1}{}_{a_2}} P_b e^{h_{a_1}{}^{a_2} K^{a_1}{}_{a_2}} = dx^b e^{-h_{a_1}{}^{a_2} K^{a_1}{}_{a_2}} \wedge P_b$:

$$\begin{aligned} &dx^b e^{-h_{a_1}{}^{a_2} K^{a_1}{}_{a_2}} \wedge P_b \\ &\quad = dx^b (P_b - h_{a_1}{}^{a_2} [K^{a_1}{}_{a_2}, P_b] + \frac{1}{2} h_{a_1}{}^{a_2} h_{a_3}{}^{a_4} [K^{a_3}{}_{a_4}, [K^{a_1}{}_{a_2}, P_b]] + ..) \\ &\quad = dx^b (P_b + h_b{}^{a_2} P_{a_2} + \frac{1}{2} h_b{}^{a_1} h_{a_1}{}^{a_2} P_{a_2} + \ldots) \\ &\quad = dx^b (e^h)_b{}^{a_1} P_{a_1} = dx^b \delta_b{}^\mu (e^h)_\mu{}^{a_1} P_{a_1} = dx^\mu e_\mu{}^{a_1} P_{a_1} \quad , \end{aligned} \tag{2.45}$$

where the change to 'space-time' $\mu$ indices is a simple relabelling. We can thus treat the conjugation of any operator $A_{a_1..}{}^{b_1\cdots}$ by $e^{-h_a{}^b K^a{}_b}$ as a tensor transformation in its indices using the vielbein and/or its inverse, for example,

$$\Phi^b_{a_1 a_2} e^{-h_c{}^d(x) K^c{}_d} K^{a_1 a_2}{}_b e^{h_c{}^d K^c{}_d} = \Phi_{\mu_1 \mu_2}{}^\nu (e^{-1})_{a_1}{}^{\mu_1} (e^{-1})_{a_2}{}^{\mu_2} e_\nu{}^b K^{a_1 a_2}{}_b. \tag{2.46}$$

Similarly, we can derive

$$e^{-h_a{}^b(x) K^a{}_b} d e^{h_a{}^b(x) K^a{}_b} = dx^\mu (e^{-1}\partial_\mu e)_a{}^b K^a{}_b \ . \tag{2.47}$$

Using these results, the Cartan form is given by

$$\begin{aligned} g^{-1}dg &= e^{-h_a{}^b(x) K^a{}_b} e^{-x^a P_a} d[e^{x^a P_a} e^{h_a{}^b(x) K^a{}_b}] \\ &= e^{-h_a{}^b(x) K^a{}_b} dx^a P_a e^{h_a{}^b(x) K^a{}_b} + e^{-h_a{}^b(x) K^a{}_b} d e^{h_a{}^b(x) K^a{}_b} \\ &= dx^\mu e_\mu{}^a P_a + dx^\mu (e^{-1}\partial_\mu e)_a{}^b K^a{}_b \\ &= dx^\mu e_\mu{}^a [P_a + \Omega_{a,b}{}^c K^b{}_c] \\ &= dx^\mu e_\mu{}^a (P_a + S_{a,b}{}^c T^b{}_c + Q_{a,b}{}^c J^b{}_c) \quad , \end{aligned} \tag{2.48}$$

where we used $K^a{}_b = K^{(a}{}_{b)} + K^{[a}{}_{b]} := T^a{}_b + J^a{}_b$, and (using $g_{\mu\nu} = e_\mu{}^a e_\nu{}^b \eta_{ab}$)

$$S_{a,b}{}^c = e_a{}^\lambda \Omega_{\lambda,(b}{}^{c)} = e_a{}^\lambda (e^{-1}\partial_\lambda e)_{(b}{}^{c)} = \frac{1}{2}[e_a{}^\lambda \partial_\lambda g_{\mu\nu}] e_b{}^\mu e_d{}^\nu \eta^{dc} \quad , \tag{2.49}$$

is the 'covariant derivative' of the Goldstone boson $h_{(a}{}^{b)}$ associated to $T^a{}_b$, while the connection $Q_{a,b}{}^c$ associated to the local $H = \mathrm{SO}(1,3)$ transformations is defined as

$$Q_{a,b}{}^c = \Omega_{a,[c}{}^{b]} = e_a{}^\mu (e^{-1}\partial_\mu e)_{[b}{}^{c]} \ . \tag{2.50}$$

Clearly the connection $Q_{a,b}{}^c$ is *not* the connection of general relativity. This connection can be used to define covariant derivatives $D_a = \partial_a + Q_a$ (ignoring the $b, c$ indices in $Q_{a,b}{}^c$), which, for example, act on tensors like $S_{b,c}{}^d$ as

$$D_a S_{b,c}{}^d = \partial_a S_{b,c}{}^d + Q_{a,b}{}^e S_{e,c}{}^d + Q_{a,c}{}^e S_{b,e}{}^d - Q_{a,e}{}^d S_{b,c}{}^e \quad , \tag{2.51}$$

but again these are *not* the covariant derivatives of general relativity, since $D_a = \partial_a + Q_a$ does not use the connection of general relativity. Instead, if we want to have any hope of using the above nonlinear realization to describe general relativity, we must somehow use the above information to construct the covariant derivatives of general relativity directly. One way to do this is to recall the fact discussed at the end of Section 2.1, that any action which is linearly invariant under the local subgroup $H$, and constructed out of the 'covariant derivative' $S_{a,b}{}^c$, or covariant quantities constructed from this such as $D_a S_{b,c}{}^d$, will also be invariant (nonlinearly) under $G$. We can thus use the above covariant derivatives to construct the most general minimal action possible:

$$\begin{aligned} S = \int d^D x \det e[ & d_1 \eta^{ab} D_a S_{b,c}{}^c + d_2 D_a S_{b,}{}^{ba} + d_3 S_{a,}{}^{ab} S_{c,b}{}^c + d_4 S_{a,}{}^{bc} S_{b,c}{}^a \\ & + d_5 \eta^{ab} S_{a,c}{}^d S_{b,d}{}^c + d_6 S_{a,}{}^{ab} S_{b,c}{}^c + d_7 \eta^{ab} S_{a,c}{}^c S_{b,d}{}^d] \end{aligned} \tag{2.52}$$

Our hope is that this action reduces to the Einstein-Hilbert action for some choice of constants. However the action is invariant under local $\mathrm{GL}(4,\mathbb{R})$ transformations for arbitrary choices of the constants $d_i$, $i = 1, \ldots, 7$, thus we can generate symmetry groups other than the diffeomorphism group in this approach. Here the constants are *not* determined by the nonlinear realization, thus even if we do arrive at the Einstein-Hilbert action we are *not* uniquely fixing the diffeomorphism group of general relativity simply by making $\mathrm{GL}(4,\mathbb{R})$ local over $\mathrm{SO}(1,3)$.

There are now two ways to proceed. One way is to now *arbitrarily* impose the *additional* assumption of diffeomorphism invariance of $S$, i.e. invariance under general coordinate transformations. Requiring diffeomorphism invariance fixes

$$(d_1, d_2, d_3, d_4, d_5, d_6, d_7) = \left(-\frac{1}{2}, \frac{1}{2}, -\frac{1}{2}, \frac{1}{2}, -\frac{1}{4}, \frac{1}{2}, -\frac{1}{4}\right). \tag{2.53}$$

and so the above action reduces to the Einstein-Hilbert action

$$S = \int d^D x (\det e) R \tag{2.54}$$

as a *special case* [85, Sec. 13.2]. This does not change the fact that the connection ${S_{a,b}}^c$ associated to the nonlinear realization is not the connection of general relativity. Instead, the connection of general relativity arises *implicitly* in the action (2.52) for a special choice of constants $d_i$ from having added various combinations of the 'covariant derivative' ${S_{a,b}}^c$ and its 'covariant derivatives' $D_a {S_{b,c}}^d$ under $D_a$.

Equivalently, instead of working with (2.51), we can add linear combinations of ${S_{b,c}}^d$ to this expression without changing the transformation properties under local $\mathrm{SO}(1,3)$ transformations. Thus, instead of the covariant derivative $D_a = \partial_a + Q_a$, we can consider

$$\tilde{D}_a = \partial_a + Q_a + \sum_i \lambda_i S_i \quad . \tag{2.55}$$

See equations (25) to (29) of reference [6] to see the correct index structure and terms in equation (2.55), presented in a matter field representation[1]. The coefficients $\lambda_i$ in this approach are in general arbitrary, and must be fixed by the additional imposition of diffeomorphism invariance in order to reproduce the covariant derivatives of general relativity.

In the following two chapters we will consider a nonlinear realisation based on the Kac-Moody algebras $A_1^{+++}$ and $E_{11}$. These two algebras will contain $\mathrm{GL}(D,\mathbb{R})$ as subalgebras for $D = 4$ and $D = 11$ respectively. We will see that the additional symmetries in $A_1^{+++}$ and $E_{11}$ will however essentially uniquely determine the covariant derivatives of general relativity, in contrast to the situation of working with $\mathrm{GL}(4,\mathbb{R})$ alone, and result in essentially unique dynamics associated to new fields such as the dual gravity field as we will discuss, indicating the powerful nature of these Kac-Moody algebras.

[1] A review, in the context of supergravity, is given in [2] and [9], based on the work of [77].

# Chapter 3

# The Kac-Moody Algebra $A_1^{+++}$, Gravity and Dual Gravity

In this chapter we will construct the Kac-Moody algebra $A_1^{+++}$ that was motivated in Chapter 1, along with its vector representation which introduces generalized space-time, at low levels. We will then consider its applications to the theory of gravity, and the predicted dual gravity field arising from the theory. A very short review of Kac-Moody algebras is given in Appendix B, and we refer to [85, Ch. 16] for a more detailed introduction.

On first reading, it may be sufficient to skip the formal nature of Appendix B, and take it on trust that: the theory of Kac-Moody algebras which we will discuss do indeed result in the generators listed in the tables in each chapter; the tables of generators can be found by inserting the associated Dynkin diagram into the program [49]; and that they can be organized in terms of a notion of discrete positive and negative 'levels' ..., $-2, -1, 0, 1, 2, \ldots$. Once one obtains these generators, much of the remaining discussion should feel familiar to those acquainted with Lie algebras.

A short review on Lie algebras has been added in Appendix A. In particular, we note the discussion of fundamental representations of $A_{D-1} = \mathrm{SL}(D)$ discussed in Appendix A.6, on describing $A_4$ as a level decomposition in terms of representations of $A_3$ in Appendix A.7, and the discussion on constructing new Lie Algebras from a given Lie algebra along with its fundamental representations in Appendix A.8, may provide enough of the analogous arguments, in a Lie algebra context, to follow the text without Appendix B.

## 3.1 The Kac-Moody Algebra $A_1^{+++}$

The Dynkin diagram for the Kac-Moody algebra $A_1^{+++}$ is given by

$$\begin{array}{ccccccc} \bullet & - & \bullet & - & \bullet & = & \otimes \\ 1 & & 2 & & 3 & & 4 \end{array} \tag{3.1}$$

This corresponds to the following generalized Cartan matrix

$$A = \begin{bmatrix} 2 & -1 & 0 & 0 \\ -1 & 2 & -1 & 0 \\ 0 & -1 & 2 & -2 \\ 0 & 0 & -2 & 2 \end{bmatrix} \tag{3.2}$$

We now decompose the algebra $A_1^{+++}$ in terms of its $A_3$ subalgebra by deleting node four, which is indicated by the $\otimes$ in the above diagram, and considering the level decomposition with respect to this node[1] . To do this we can use the direct methods of Appendix B, or simply insert the above Dynkin diagram into the program [49] which has automated this procedure to low levels, to find the output from level zero to level plus two and level minus two as

Table 3.1: $A_3$ representations in $A_1^{+++}$

| $l$ | $A_3$ Irrep | $A_1^{+++}$ Root | $\alpha^2$ | $d_r$ | $\mu$ | $R^\alpha$ |
|---|---|---|---|---|---|---|
| -2 | 0 1 2 | 0 0 -1 -2 | 2 | 45 | 1 | $R_{a_1a_2,(b_1b_2)}$ |
| -1 | 0 0 2 | 0 0 0 -1 | 2 | 10 | 1 | $R_{(ab)}$ |
| +0 | 0 0 0 | 0 0 0 0 | 0 | 1 | 1 | $D$ |
| +0 | 1 0 1 | 1 1 1 0 | 2 | 15 | 1 | $\hat{K}^a{}_b$ |
| +1 | 2 0 0 | 2 2 2 1 | 2 | 10 | 1 | $R^{(ab)}$ |
| +2 | 2 1 0 | 3 4 4 2 | 2 | 45 | 1 | $R^{a_1a_2,(b_1b_2)}$ |

The first column represents the level in $A_1^{+++}$. The second column represents the highest weight representations of $A_3$ appearing at a given level, where $(a, b, c) = a\mu_1 + b\mu_2 + c\mu_3$ holds, for $\mu_i$ the $i$'th fundamental representation of $A_3$. The third column represents the actual $A_1^{+++}$ root, the fourth column the value that this root squares to. The fifth column represents the dimension $d_r$ of the irreducible representation, the sixth represents the 'outer multiplicity' $\mu$ of that representation which is how many copies appear of this representation, and the final column denotes these generators explicitly. We can define the level of a generator directly from the table as the number of up minus down indices, all divided by two. Higher level generators are listed in [29].

[1]As motivation for this step: in Appendix A.7 we establish the analogous situation of studying the Lie algebra $A_4$, $\bullet - \bullet - \bullet - \bullet$. There we will temporarily ignore its fourth node, resulting in $A_3$, $\bullet - \bullet - \bullet$, and then describe $A_4$ in terms of a level decomposition with respect to $A_3$, denoting this as $\bullet - \bullet - \bullet - \otimes$.

At level $l = 0$, the $A_3$ highest weight irreducible representation $(1,0,1) = \mu_1 + \mu_3$ arises. This is the adjoint representation of $A_3$, generated by $\hat{K}^a{}_b$ (satisfying $\sum_{c=1}^4 \hat{K}^c{}_c = 0$), along with a generator denoted $D$. We can interpret $D$ as the trace of a GL(4) generated by $K^a{}_b$, $D = \sum_{c=1}^4 K^c{}_c$, and then combine the level zero generators into the GL(4) generators

$$K^a{}_b = \hat{K}^a{}_b + \frac{1}{4}\delta^a{}_b D \ . \tag{3.3}$$

Similarly, at level one we find $2\mu_1$, two copies of the vector representation symmetrized i.e. a rank two symmetric tensor $R^{ab}$ satisfying $R^{(ab)} = R^{ab}$, and at level two we find $2\mu_1 + \mu_2$, described by a tensor denoted $R^{a_1a_2,b_1b_2}$ and satisfying $R^{[a_1a_2],(b_1b_2)} = R^{a_1a_2,b_1b_2}$. Thus the positive level generators of $A_1^{+++}$ to level two are thus given by

$$K^a{}_b \ ; \quad R^{(ab)} \ ; \ R^{a_1a_2,(b_1b_2)} \ , \ \ldots \tag{3.4}$$

where the generators at levels zero, one, two, ... are all separated by ; semi-colons, blocks of indices are separated by , commas, and the indices in each block are assumed to be anti-symmetric, except when indices are contained in ( ) brackets, indicating that such indices are symmetric. The negative level generators to level minus two are similarly given by

$$R_{(ab)} \ ; \ R_{a_1a_2,(b_1b_2)} \ ; . \tag{3.5}$$

The above listed generators possess the following $A_3$ irreducibility properties[2]

$$R^{[a_1a_2,b_1]b_2} = 0 \quad , \quad R_{[a_1a_2,b_1]b_2} = 0 \quad . \tag{3.6}$$

We now consider the algebra of these generators [67]. This algebra can be constructed directly, from first principles, by postulating the most general relations consistent with the index structure of a given commutator, and then requiring that the Jacobi identities be satisfied with all generators listed up to that level, level by level, up to normalization coefficients that have been chosen below. This is analogous to the procedure in the Lie Algebra case in Appendix A.8. More details on this, and the relationship of this procedure to the generators listed in the Serre presentation definition of a Kac-Moody algebra, are discussed in [85, Ch. 16].

[2]Without such irreducibility conditions, we could construct new tensors with different index symmetries by contracting against $\varepsilon_{ca_1a_2b_1}$, and so $R^{a_1a_2,b_1b_2}$ would not be an irreducible representation of $A_3$.

Since the above generators belong to representations of GL(4), the commutators of ${K^a}_b$ with the positive generators are

$$\begin{aligned} [{K^a}_b,\, {K^c}_d] &= {\delta^c}_b {K^a}_d - \delta^a_d {K^c}_b \ , \\ [{K^a}_b,\, R^{c_1c_2}] &= 2\,\delta_b^{(c_1}\, R^{|a|c_2)} \ , \ [{K^a}_b, R_{c_1c_2}] = -2\delta^a_{(c_1}\, R_{|b|c_2)} \ , \\ [{K^a}_b, R^{cd,ef}] &= \delta^c_b\, R^{ad,ef} + \delta^d_b\, R^{ca,ef} + \delta^e_b\, R^{cd,af} + \delta^f_b\, R^{cd,ea} \ , \\ [{K^a}_b,\, R_{cd,ef}] &= -\delta^a_c R_{bd,ef} - \delta^a_d R_{cb,ef} - \delta^a_e\, R_{cd,bf} - \delta^a_f R_{cd,eb} \ . \end{aligned} \tag{3.7}$$

The commutators of two level one generators must give on the right-hand side the unique level 2 generators, and similarly for level minus one generators, thus we can postulate a relation of the form

$$[R^{ab}, R^{cd}] = \lambda R^{(a|(c,d)|b)} \quad , \tag{3.8}$$

where $\lambda$ is a normalization constant, and similarly for $[R^{ab}, R^{cd}]$. On checking the Jacobi identities of this relation with all generators to this level, using the above commutators and irreducibility conditions, we find it is sufficient to take the commutators in the form

$$[R^{ab}, R^{cd}] = R^{ac,bd} + R^{bd,ac} \ , \ [R_{ab}, R_{cd}] = R_{ac,bd} + R_{bd,ac}. \tag{3.9}$$

The normalisation of the level 2 ($-2$) generators are fixed by these relations. The commutators between the positive and negative level generators are given by

$$\begin{aligned} [R^{ab}, R_{cd}] &= 2\delta^{(a}_{(c} {K^{b)}}_{d)} - \delta^{(ab)}_{cd} \sum_e {K^e}_e \ , \\ [R^{ab,cd}, R_{ef}] &= \delta^{(bd)}_{ef}\, R^{ac} + \delta^{(bc)}_{ef} R^{ad} - \delta^{(ac)}_{ef} R^{bd} - \delta^{(ad)}_{ef} R^{bc} \ , \\ [R_{ab,cd}, R^{ef}] &= \delta^{ef}_{bd} R_{ac} + \delta^{(ef)}_{bc} R_{ad} - \delta^{(ef)}_{ac} R_{bd} - \delta^{ef}_{ad}\, R_{bc} \ , \end{aligned} \tag{3.10}$$

where we used $\delta^{(ab)}_{cd} = \delta^{(a}_c \delta^{b)}_d$.

As discussed in Chapter 1, the local subgroup associated to the nonlinear realization in each dimension $D$ in the dimensional reduction procedure can be interpreted as the Cartan involution invariant subgroup of the associated $E_{11-D}$ symmetry group, and similarly for $A_1^{+++}$. We thus next introduce a Cartan involution operator $I_c$ on $A_1^{+++}$. This is an operator $I_c$ on the Lie algebra that acts as an involution, $I_c^2(R^\alpha) = R^\alpha$ for $R^\alpha \in A_1^{+++}$. Thus the Cartan involution $I_c$ can be defined to act on the above-listed generators of $A_1^{+++}$ as

$$I_c({K^a}_b) = -\eta^{ad}\eta_{bc}{K^c}_d \ , \ I_c(R_{ab}) = -\eta_{ac}\eta_{bd}R^{cd} \ , \ I_c(R^{ab,cd}) = \eta^{ae}\eta^{bf}\eta^{cg}\eta^{dh}R_{ef,gh} \ , \tag{3.11}$$

where $\eta_{ab}$ the Minkowski metric, and it satisfies $I_c^2(R^\alpha) = R^\alpha$ for any $A_1^{+++}$ generator $R^\alpha$. Cartan-involution invariant generators are then given by

$$\begin{aligned} J_{ab} &= \eta_{ac}{K^c}_b - \eta_{bc}{K^c}_a \quad , \quad S_{ab} = R^{cd}\eta_{ca}\eta_{db} - R_{ab} \ , \\ S_{a_1a_2,b_1b_2} &= R^{c_1c_2,d_1d_2}\eta_{c_1a_1}\eta_{c_2a_2}\eta_{d_1b_1}\eta_{d_2b_2} + R_{a_1a_2,b_1b_2} \ . \end{aligned} \tag{3.12}$$

These generators result in the Cartan involution-invariant subalgebra denoted by $I_c(A_1^{+++})$. Those constructed from the lowest level generators result in the local Lorentz group $\mathrm{SO}(1,3)$. The commutators of this subalgebra are given to low levels as

$$\begin{aligned} [J_{a_1a_2}, J_{b_1b_2}] &= \eta_{a_2b_1}J_{a_1b_2} - \eta_{a_2b_2}J_{a_1b_1} - \eta_{a_1b_1}J_{a_2b_2} + \eta_{a_1b_2}J_{a_2b_1} \\ [J_{a_1a_2}, S_{b_1b_2}] &= \eta_{a_2b_1}S_{a_1b_2} + \eta_{a_2b_2}S_{a_1b_1} - \eta_{a_1b_1}S_{a_2b_2} - \eta_{a_1b_2}S_{a_2b_1} \\ [S_{a_1a_2}, S_{b_1b_2}] &= 2S_{(a_1|(b_1,b_2)|a_2)} - 2\eta_{(b_1|(a_1}J_{a_2)|b_2)}. \end{aligned} \tag{3.13}$$

The first fundamental representation of $A_1^{+++}$ is denoted by $l_1$ and is referred to as the vector representation. This representation can be found by the following trick, discussed also in Appendix A.7. Consider first a simpler case: the algebra $A_{D-1}$, generated by ${\hat{K}^a}_b$, with $a = 1,..,D$. By adding an extra node to the first node of the Dynkin diagram of $A_{D-1}$ we generalize to $A_D$ generated by ${\hat{K}^0}_0$, ${\hat{K}^a}_0$, ${\hat{K}^0}_b$, ${\hat{K}^a}_b$. The added generators ${\hat{K}^0}_b$ and ${\hat{K}^a}_0$ transform as vectors with respect to the $A_{D-1}$ subalgebra ${\hat{K}^a}_b$. We can interpret the 'vector' ${\hat{K}^a}_0$ as the first fundamental representation of $A_{D-1}$, i.e. the 'level one' representation of $A_D$ is the vector representation of $A_{D-1}$. Note that the commutation relations between a level zero $A_{D-1}$ generator of $A_D$, and a level one generator only gives level one generators. Similarly for a Kac-Moody algebra such as $A_1^{+++}$, on adding an additional node connected to the first node, we find $A_1^{++++}$ with the Dynkin diagram

$$\begin{array}{ccccccccc} \otimes & - & \bullet & - & \bullet & - & \bullet & = & \otimes \\ 0 & & 1 & & 2 & & 3 & & 4 \end{array} \tag{3.14}$$

Considering a level decomposition with respect to the zero'th node as denoted in the diagram, we note that at level one the resulting generators form the vector representation of $A_1^{+++}$, while generators at other levels are irrelevant. We have expressed these generators in terms of the $A_3$ subalgebra by also deleting node 4 as denoted in the above diagram. Using the program [49] one finds the table:

Table 3.2: $A_3$ representations in $A_1^{++++}$

| $l$ | $A_3$ Irrep | $A_1^{++++}$ Root | $a^2$ | $d_r$ | $\mu$ | $Z^\alpha$ |
|---|---|---|---|---|---|---|
| 0 0 | 0 0 0 | 0 0 0 0 0 | 0 | 1 | 2 | |
| 0 0 | 1 0 1 | 0 1 1 1 0 | 2 | 15 | 1 | |
| 1 0 | 0 0 1 | 1 1 1 1 0 | 2 | 4 | 1 | $P_a$ |
| 0 1 | 2 0 0 | 0 2 2 2 1 | 2 | 10 | 1 | |
| 1 1 | 1 0 0 | 1 2 2 2 1 | 0 | 4 | 1 | $Z^a$ |
| 2 1 | 0 0 0 | 2 2 2 2 1 | 2 | 1 | 1 | |
| 0 2 | 2 1 0 | 0 3 4 4 2 | 2 | 45 | 1 | |
| 1 2 | 1 1 0 | 1 3 4 4 2 | -2 | 20 | 1 | $Z^{a_1a_2,b}$ |
| 1 2 | 3 0 0 | 1 4 4 4 2 | 2 | 20 | 1 | $Z^{(a_1a_2a_3)}$ |

Here the levels are now given by $l = (l_1, l_2)$, where $l_1$ is the level of the zero'th node, $l_2$ is the level of the fourth node four. For the fixed level $l_1 = 1$, we can interpret the resulting generators, classified in terms of different levels $l_2$, as the generators of the vector representation of $A_1^{+++}$, and ignore all $l_1 \neq 1$ contributions. At level $l_2 = 1$, from $(0, 0, 1)$, we find the tensor $P^{bcd}$, and from this we can construct $P_a = \frac{1}{3!}\varepsilon_{abcd}P^{bcd}$, the usual four-dimensional space-time translation generator. This representation thus has, to $l_2$ level two, the generators

$$P_a \ ; \quad Z^a \ ; \quad Z^{(a_1a_2a_3)} = Z^{a_1a_2a_3}, \quad Z^{a_1a_2,b} \quad . \tag{3.15}$$

where the upper indices are anti-symmetric except when symmetric ( ) brackets are present, and, at higher levels, subscripts label generators with multiplicity greater than one. These generators also satisfy irreducibility conditions, now given by[3]

$$Z^{[a_1a_2,b]} = 0 \ . \tag{3.16}$$

The semi-direct product of $A_1^{+++}$ with the $l_1$ representation is denoted $A_1^{+++} \otimes_s l_1$. The commutators of the non-negative level $A_1^{+++}$ generators with those of the vector representation are given to low levels as [67], [79]

$$\begin{aligned}
[K^a{}_b,\, P_c] &= -\delta^a_c P_b + \frac{1}{2}\delta^a_b P_c \ , \ [K^a{}_b,\, Z^c] = \delta^c_b\, Z^a + \frac{1}{2}\,\delta^a_b\, Z^c \ , \\
[K^a{}_b,\, Z^{cde}] &= \delta^c_b Z^{ade} + \delta^d_b Z^{cae} + \delta^e_b Z^{cda} + \frac{1}{2}\delta^a_b Z^{cde} \ , \\
[K^a{}_b,\, Z^{cd,e}] &= \delta^c_b Z^{ad,e} + \delta^d_b Z^{ca,e} + \delta^e_b Z^{cd,a} + \frac{1}{2}\delta^a_b Z^{cd,e} \ , \\
[R^{ab},\, P_c] &= \delta^{(a}_c Z^{b)} \ , \ [R^{ab},\, Z^c] = Z^{abc} + Z^{c(a,b)} \ , \\
[R^{ab,cd},\, P_e] &= -\delta^{[a}_e Z^{b]cd} + \frac{1}{4}(\delta^a_e Z^{b(c,d)} - \delta^b_e Z^{a(c,d)}) - \frac{3}{8}(\delta^c_e Z^{ab,d} + \delta^d_e Z^{ab,c}) \ .
\end{aligned} \tag{3.17}$$

The commutators with the negative level $A_1^{+++}$ generators are given by

$$\begin{aligned}
[R_{ab},\, P_c] &= 0 \ , \ [R_{ab}, Z^c] = 2\delta_{(a}{}^c P_{b)} \ , \\
[R_{ab}, Z^{cde}] &= \frac{2}{3}(\delta^{cd}_{(ab)}\, Z^e + \delta^{de}_{(ab)}\, Z^c + \delta^{ec}_{(ab)}\, Z^d) \ , \\
[R_{ab},\, Z^{cd,e}] &= \frac{4}{3}(\delta^{de}_{(ab)} Z^c - \delta^{ce}_{(ab)} Z^d).
\end{aligned} \tag{3.18}$$

Finally, the generators of $I_c(A_1^{+++})$ have commutation relations with the generators of the $l_1$ representation as follows

$$\begin{aligned}
[J_{a_1a_2}, P_b] &= 2P_{[a_1}\eta_{a_2]b} \ , \ [S^{ab}, P_c] = \delta^{(a}{}_c Z^{b)} \ , \\
[J^{a_1a_2}, Z^b] &= -2\eta^{b[a_1} Z^{a_2]} \ , \\
[S^{a_1a_2}, Z^b] &= (Z^{a_1a_2b} + Z^{b(a_1,a_2)}) - 2\eta^{b(a_1} P^{a_2)} \ .
\end{aligned} \tag{3.19}$$

## 3.2 Nonlinear Realization of $A_1^{+++}$

We now consider the nonlinear realization of $A_1^{+++} \otimes_s l_1$ with respect to its Cartan involution invariant subgroup $I_c(A_1^{+++})$. We begin with the group element $g \in A_1^{+++} \otimes_s l_1$ which is written as

$$g = g_l g_A \ . \tag{3.20}$$

[3]Without such an irreducibility condition, we could construct a new tensor with different index symmetries by contracting against $\varepsilon_{ca_1a_2b}$, and so $Z^{a_1a_2,b}$ would not be an irreducible representation of $A_3$.

In this equation $g_A$ is a group element of $A_1^{+++}$, and the local $I_c(A_1^{+++})$ subgroup can be used to write $g_A$ in terms of only non-negative level generators $R^{\underline{\alpha}}$ of $A_1^{+++}$, i.e. $g_A = \Pi_{\underline{\alpha}} e^{A_{\underline{\alpha}} R^{\underline{\alpha}}}$. The group element $g_l$ is formed from the generators of the $l_1$ representation and so has the form $\Pi_A e^{z^A L_A}$ where $z^A$ are the coordinates of the generalised space-time. The fields $A_{\underline{\alpha}}$ depend on the coordinates $z^A$.

The above group elements can thus be written in the explicit form

$$g_A = \ldots e^{A_{a_1 a_2, b_1 b_2} R^{a_1 a_2, b_1 b_2}} e^{A_{a_1 a_2} R^{a_1 a_2}} e^{h_a{}^b K^a{}_b} \ . \tag{3.21}$$

Similarly, the group element $g_l$ can be taken in the form

$$g_l = e^{x^a P_a} e^{y_a Z^a} e^{x_{abc} Z^{abc}} e^{x_{ab,c} Z^{ab,c}} \ldots \tag{3.22}$$

Here we have introduced generalized coordinates of the $A_1^{+++}$ generalized space-time

$$x^a \ ; \ y_a \ ; \ x_{abc}, \ x_{ab,c} \ ; \ \ldots \tag{3.23}$$

which possess the same symmetries as the associated generators of the vector representation, for example $x_{abc} = x_{(a_1 a_2 a_3)}$.

In the above group elements we have introduced the fields

$$h_a{}^b \ ; \ A_{(a_1 a_2)} \ ; \ A_{a_1 a_2, (b_1 b_2)} \ ; \ \ldots \tag{3.24}$$

which similarly possess the same index symmetries as the associated $A_1^{+++}$ generator, and obey irreducibility conditions of the same form, which to the level considered read as

$$A_{[a_1 a_2, b_1] b_2} = 0 \ . \tag{3.25}$$

These fields depend on all of the generalized coordinates $x^a, y_a, \ldots$.

The field $h_a{}^b$ is defined as the graviton, the field $A_{ab}$ is defined as the dual graviton, the field $A_{ab,cd}$ is defined as the dual dual-graviton etc.... The coordinates $x^a$ are the usual coordinates of space-time, while the coordinates $y_a, x_{abc}, \ldots$ are additional coordinates in the associated generalized space-time.

The dynamics of the nonlinear realization is just a set of equations of motion, that are invariant under the transformations of equation (2.29). We will construct the dynamics of the $A_1^{+++} \otimes_s l_1$ nonlinear realization from the Cartan forms which are given by

$$\mathcal{V} \equiv g^{-1} dg = \mathcal{V}_A + \mathcal{V}_l \ \ , \tag{3.26}$$

where

$$\mathcal{V}_A = g_A^{-1} dg_A \equiv dz^\Pi G_{\Pi,\underline{\alpha}} R^{\underline{\alpha}} \ , \text{ and } \mathcal{V}_l = g_A^{-1}(g_l^{-1} dg_l) g_A = g_A^{-1} dz \cdot l g_A \equiv dz^\Pi E_\Pi{}^A l_A \ . \quad (3.27)$$

Here $\mathcal{V}_A$ belongs to the $A_1^{+++}$ algebra, it is the Cartan form of $A_1^{+++}$, while $\mathcal{V}_l$ is in the space of generators of the $l_1$ representation. Both $\mathcal{V}_A$ and $\mathcal{V}_l$ are invariant under rigid transformations, and under local $I_c(A_1^{+++})$ transformations they change as

$$\mathcal{V}_A \to h^{-1} \mathcal{V}_A h + h^{-1} dh \quad \text{and} \quad \mathcal{V}_l \to h^{-1} \mathcal{V}_l h \quad (3.28)$$

Inserting the group element of equation (3.21) into $\mathcal{V}_A$, the Cartan forms are given by

$$\mathcal{V}_A = G_a{}^b K^a{}_b + \overline{G}_{a_1 a_2} R^{a_1 a_2} + G_{a_1 a_2, b_1 b_2} R^{a_1 a_2, b_1 b_2} + \dots \quad (3.29)$$

where, with $e_a{}^b = (e^h)_a{}^b$,

$$\begin{gathered} G_a{}^b = (e^{-1} de)_a{}^b \quad , \quad \overline{G}_{a_1 a_2} = e_{a_1}{}^{\mu_1} e_{a_1}{}^{\mu_2} dA_{\mu_1 \mu_2}, \\ G_{a_1 a_2, b_1 b_2} = e_{a_1}{}^{\mu_1} e_{a_2}{}^{\mu_2} e_{b_1}{}^{\nu_1} e_{b_2}{}^{\nu_2} (dA_{\mu_1 \mu_2, \nu_1 \nu_2} - A_{[\mu_1 | (\nu_1} dA_{\nu_2) | \mu_2]}) \ . \end{gathered} \quad (3.30)$$

Since $A_{\mu_1 \mu_2}$ is symmetric, $G_{a_1 a_2, b_1 b_2}$ satisfies the irreducibility condition $G_{[a_1 a_2, b_1] b_2} = 0$.

The generalised vielbein $E_\Pi{}^A$ and its inverse can be derived [67], [66] from its definition $g_R^{-1} l_\Pi g_R = E_\Pi{}^A l_A$ in equation (2.37) or equivalently from $\mathcal{V}_l$ in equation (3.27). To level one it suffices to consider (here we write $l_\Pi = \{P_\mu, Z^\mu, \dots\}$ where $P_\mu = \delta_\mu{}^a P_a$ etc...), also only keeping terms to first order in $h_a{}^b$,

$$\begin{aligned} & e^{-h_a{}^b K^a{}_b} e^{-A_{a_1 a_2} R^{a_1 a_2}} \{P_\mu, Z^\mu\} e^{A_{a_1 a_2} R^{a_1 a_2}} e^{h_a{}^b K^a{}_b} \\ &= e^{-h_a{}^b K^a{}_b} \{P_\mu - A_{\mu\nu} Z^\nu, Z^\mu - A_{\nu\rho} Z^{\mu\nu\rho}\} e^{h_a{}^b K^a{}_b} \\ &= \{e^{-h_a{}^b K^a{}_b} P_\mu e^{h_a{}^b K^a{}_b} - A_{\mu\nu} e^{-h_a{}^b K^a{}_b} Z^\nu e^{h_a{}^b K^a{}_b} \ , \ e^{-h_a{}^b K^a{}_b} Z^\mu e^{h_a{}^b K^a{}_b}\} \\ &= \{(P_\mu - h_a{}^b [K^a{}_b, P_\mu]) - A_{\mu\nu}(Z^\nu - h_a{}^b [K^a{}_b, Z^\nu]) \ , \ (Z^\mu - h_a{}^b [K^a{}_b, Z^\mu])\} \\ &= \{(P_\mu + h_a{}^b \delta_\mu{}^a P_b - \frac{1}{2} h_a{}^b \delta^a{}_b P_\mu) - A_{\mu\nu}(Z^\nu - h_a{}^b \delta^\nu{}_b Z^a - \frac{1}{2} h_a{}^b \delta^a{}_b Z^\nu) \ , \\ &\qquad (Z^\mu - h_a{}^b \delta^\mu{}_b Z^a - \frac{1}{2} h_a{}^b \delta^a{}_b Z^\mu)\} \\ &= \{P_\mu + h_\mu{}^b P_b - \frac{1}{2} h_c{}^c P_\mu - A_{\mu\nu} Z^\nu + A_{\mu\nu} h_a{}^\nu Z^a + \frac{1}{2} h_c{}^c A_{\mu\nu} Z^\nu \ , \\ &\qquad Z^\mu - h_a{}^\mu Z^a - \frac{1}{2} h_c{}^c Z^\mu\}. \end{aligned} \quad (3.31)$$

Thus, on exponentiating the $h_a{}^b$ terms to all orders, the generalized vielbein and its inverse are given to level one by

$$E_\Pi{}^A = (\det e)^{-\frac{1}{2}} \begin{pmatrix} e_\mu{}^a & -e_\mu{}^b A_{ba} \\ 0 & e_a{}^\mu \end{pmatrix} \ , \ \left(E^{-1}\right)_A{}^\Pi = (\det e)^{\frac{1}{2}} \begin{pmatrix} e_a{}^\mu & A_{ab} e_\mu{}^b \\ 0 & e_\mu{}^a \end{pmatrix} \ , \quad (3.32)$$

where the factor $(\det e)^{-\frac{1}{2}}$ arises in $E_\Pi{}^A$ due to the $\frac{1}{2}\delta^a{}_b l_\Pi$ terms in the $[K^a{}_b, l_\Pi]$ commutators of equation (3.17).

The Cartan involution invariant subalgebra at level zero is SO(1, 3) and the Cartan forms transform under this as Lorentz tensors. Under a local $I_c(A_1^{+++})$ transform to level one, i.e. under the group element $h = I - \Lambda_{a_1a_2}S^{a_1a_2} \in I_c(A_1^{+++})$, the Cartan forms transform as

$$\delta\mathcal{V}_A = [\Lambda_{a_1a_2}S^{a_1a_2}, \mathcal{V}_A] - S^{a_1a_2}d\Lambda_{a_1a_2} \quad , \tag{3.33}$$

which in terms of the Cartan form coefficients read as

$$\begin{aligned} \delta G_a{}^b &= 2\Lambda^{cb}\overline{G}_{ca} - \delta_a{}^b\Lambda^{c_1c_2}\overline{G}_{c_1c_2}, \\ \delta\overline{G}_{a_1a_2} &= -2\Lambda_{(a_1}{}^b G_{a_2)b} - 4G_{(a_1|b_1|,a_2)b_2}\Lambda^{b_1b_2} - d\Lambda_{a_1a_2} \quad , \\ \delta G_{a_1a_2,b_1b_2} &= 2\Lambda_{[a_1|(b_1|}G_{|a_2]|b_2)} \quad . \end{aligned} \tag{3.34}$$

If one were to only consider the local $I_c(A_1^{+++})$ transformation $\delta\mathcal{V}_A = [\Lambda_{a_1a_2}S^{a_1a_2}, \mathcal{V}_A]$, the result would involve both positive and negative level $A_1^{+++}$ generators. However, we must stipulate that the local transformations preserve the gauge choice of having no non-positive level generators present. We can however use the freedom of local $I_c(A_1^{+++})$ invariance to perform transformations on these conditions. Demanding that the transformed Cartan form has no negative level contributions from the algebra requires introducing the $-d\Lambda_{a_1a_2}S^{a_1a_2}$ transformation that we can think of as a local gauge transformation. Examining the resulting coefficient of $R_{a_1a_2}$, this then restricts $d\Lambda^{a_1a_2}$ to satisfy

$$d\Lambda_{a_1a_2} - 2\Lambda_{(a_1}{}^b G_{|b|a_2)} = 0. \tag{3.35}$$

Using this condition (3.35) in the variations of equation (3.34) we find that we can re-express $\delta\overline{G}_{a_1a_2}$ as

$$\delta\overline{G}_{a_1a_2} = -4\Lambda_{(a_1}{}^b G_{(a_2)b)} - 4G_{(a_1|b_1|,a_2)b_2}\Lambda^{b_1b_2}. \tag{3.36}$$

The Cartan forms of equation (3.30), written as forms involving $d = dx^\Pi\partial_\Pi$, are invariant under the above transformations. However, once we consider them in terms of components and remove the $dx^\Pi$ contributions (which transform as in equation (2.39)), they are no longer invariant under the rigid transformation $g_0 \in A_1^{+++} \otimes_s l_1$, as discussed above equation (2.39). To find objects that are invariant under rigid transformations we must, as discussed in and below equation (2.37), consider

$$G_{A,\bullet} = (E^{-1})_A{}^\Pi G_{\Pi,\bullet} \tag{3.37}$$

where $\bullet$ is any $A_1^{+++}$ index. The $A$ indices must therefore also transform under the local $h \in I_c(A_1^{+++})$ transformations, and so

$$\delta G_{A,\bullet} = [\delta(E^{-1})_A{}^\Pi]G_{\Pi,\bullet} + (E^{-1})_A{}^\Pi \delta G_{\Pi,\bullet} \tag{3.38}$$

where $\delta G_{\Pi,\bullet}$ transforms as $\delta G_\bullet$. We thus need to determine $\delta(E^{-1})_A{}^\Pi$. To do this we can follow the derivation of equation (2.40), for simplicity now conjugating the quantity $dx^\Pi g_R^{-1} l_\Pi g_R = dx^\Pi E_\Pi{}^A l_A$ by $h$ to find

$$h^{-1}dx^\Pi g_R^{-1} l_\Pi g_R h = h^{-1} dx^\Pi E_\Pi{}^A l_A h = dx^\Pi E_\Pi{}^\Lambda D(h)_\Lambda{}^A l_A = dx^\Pi E'_\Pi{}^A l_A. \tag{3.39}$$

For $h = I - \Lambda_{ab} S^{ab}$ this reduces to computing

$$dx^\Pi \delta E_\Pi{}^A l_A = \Lambda_{ab} dx^\Pi E_\Pi{}^A [S^{ab}, l_A] \tag{3.40}$$

and using the result we can find $\delta(E^{-1})_A{}^\Pi$ from

$$\delta(E^{-1})_A{}^\Pi = -(E^{-1})_A{}^{\Pi'} \delta E_{\Pi'}{}^B (E^{-1})_B{}^\Pi. \tag{3.41}$$

Explicitly, we have

$$\begin{aligned} dx^\Pi \delta E_\Pi{}^a P_a + dx^\Pi \delta E_{\Pi,a} Z^a + &= \Lambda_{ab} dx^\Pi [S^{ab}, E_\Pi{}^c P_c + E_{\Pi,c} Z^c + \ldots] \\ &= dx^\Pi(-2E_{\Pi,b}\Lambda^{ba})P_a + dx^\Pi E_\Pi{}^b \Lambda_{ba} Z^a + \ldots \end{aligned} \tag{3.42}$$

that is

$$\delta E_\Pi{}^a = -2E_{\Pi,b}\Lambda^{ba} \quad , \quad \delta E_{\Pi,a} = E_\Pi{}^b \Lambda_{ba} \quad . \tag{3.43}$$

Thus we have

$$\begin{aligned} \delta(E^{-1})_a{}^\Pi &= -(E^{-1})_a{}^{\Pi'} \delta E_{\Pi'}{}^b (E^{-1})_b{}^\Pi - (E^{-1})_a{}^{\Pi'} \delta E_{\Pi',b} (E^{-1})^{b,\Pi} \\ &= -(E^{-1})_a{}^{\Pi'}(-2E_{\Pi',c}\Lambda^{cb})(E^{-1})_b{}^\Pi - (E^{-1})_a{}^{\Pi'}(E_{\Pi'}{}^c \Lambda_{cb})(E^{-1})^{b,\Pi} \\ &= 0 - \Lambda_{ab}(E^{-1})^{b,\Pi} = -\Lambda_{ab}(E^{-1})^{b,\Pi} \quad , \end{aligned} \tag{3.44}$$

and similarly

$$\delta(E^{-1})^{a,\Pi} = -(E^{-1})^{a,\Pi'} \delta E_{\Pi'}{}^b (E^{-1})_b{}^\Pi - (E^{-1})^{a,\Pi'} \delta E_{\Pi',b} (E^{-1})^{b,\Pi} = 2\Lambda^{ab}(E^{-1})_b{}^\Pi \quad . \tag{3.45}$$

As a result, on the $l_1$ index, the Cartan forms transform as

$$\delta G_{a,\bullet} = -\Lambda_{ab} \hat{G}^b{}_{,\bullet} \quad , \quad \delta \hat{G}^{a,}{}_\bullet = 2\Lambda^{ab} G_{b,\bullet} \quad , \tag{3.46}$$

where the hat is included to indicate a derivative with respect to a level one coordinate $y_a$.

## 3.3 Derivation of First Order Duality Relations

### 3.3.1 General Dynamical Principles

We now seek to construct dynamics from the Cartan forms associated to the nonlinear realization. It is important to note that this formulation of dynamical equations of motion does not necessarily involve a Lagrangian formulation, yet still promises a method of constructing dynamical equations of motion involving the various fields of the nonlinear realization. Given the difficulties of studying Lagrangian formulations, for example in string theory and higher spin fields [58], the nonlinear realization approach may yet have much to offer.

We will use symmetry principles in order to construct these equations: our equations should be preserved under rigid $g_0 \in A_1^{+++} \otimes_s l_1$ transformations, and also under local $h \in I_c(A_1^{+++})$ transformations. Since these Cartan forms are invariant under the rigid $g_0 \in A_1^{+++}$ transformations, we do not need to take further account of these transformations, and so our focus is on the $h$ transformations. Since $I_c(A_1^{+++})$ can be generated from the level one transformations, we focus on level one transformations $h = I - \Lambda^{a_1 a_2} S_{a_1 a_2}$. If we denote a general equation of motion by $E_i = 0$, then under $h$ an equation of motion must vary into other equations of motion in order to be consistent:

$$\delta E_i \doteq \Lambda_i{}^j E_j \quad . \tag{3.47}$$

This allows us to propose $E_i = 0$ as the equations of motion, and such equations are then preserved by local $I_c(A_1^{+++})$ variations. More generally, however, we must also allow for extra terms arising from 'gauge transformations', and so in general write:

$$\delta E_i = \Lambda_i{}^j E_j + \partial \tilde{\Lambda}_i \quad . \tag{3.48}$$

These terms have been referred to as 'modulo terms' [71], and the dot notation $\doteq$ signifies that such terms had been suppressed. As discussed in [71], these modulo contributions can be removed by taking certain derivatives in a manner similar to that in other gauge theories.

We begin by postulating equations that are first order in derivatives, from the Cartan forms of equation (3.30). The Cartan forms transform in non-trivial manner under the local $h \in I_c(A_1^{+++})$ transformations, as can be seen from equations (3.34), and so we seek first order equations that remain invariant under these transformations. At level zero the $I_c(A_1^{+++})$ transformations are local Lorentz transformations $\mathrm{SO}(1,3)$, thus all first order equations must transform in a covariant manner. Transformations involving $I_c(A_1^{+++})$ at level one, as given in equation (3.34), mix Cartan forms at a given level into Cartan forms whose level is increased or decreased by one. Thus, in order to ensure invariance under local $h \in I_c(A_1^{+++})$ transformations up to level one, we require covariant first order equations constructed from Cartan forms relating fields at different levels, which we refer to as duality relations, and they must vary into other duality relations under these local $h$ transformations. Since $A_1^{+++} \otimes_s l_1$ involves an infinite number of levels, we can expect these duality relations may also involve additional terms involving higher level derivatives, and indeed we will find this behaviour below.

### 3.3.2 Spin Connection in Terms of Cartan Forms

In the next section we will need the spin connection of general relativity written in terms of Cartan forms as follows

$$
\begin{aligned}
\omega_{a,b_1b_2} &= e_a{}^{\mu}\omega_{\mu,b_1b_2} \\
&= e_a{}^{\mu}\frac{1}{2}\left[e_{b_1}{}^{\rho}(\partial_\mu e_{\rho b_2} - \partial_\rho e_{\mu b_2}) - e_{b_2}{}^{\rho}(\partial_\mu e_{\rho b_1} - \partial_\rho e_{\mu b_1}) - e_{b_1}{}^{\lambda}e_{b_2}{}^{\rho}(\partial_\lambda e_{\rho c} - \partial_\rho e_{\lambda c})e_\mu{}^{c}\right] \\
&= \left[\frac{1}{2}e_a{}^{\mu}(e_{b_1}{}^{\rho}\partial_\mu e_{\rho b_2} - e_{b_2}{}^{\rho}\partial_\mu e_{\rho b_1}) - \frac{1}{2}e_{b_1}{}^{\rho}(e_a{}^{\lambda}\partial_\rho e_{\lambda b_2} + e_{b_2}{}^{\lambda}\partial_\rho e_{\lambda a})\right. \\
&\qquad\qquad\qquad\qquad \left. + \frac{1}{2}e_{b_2}{}^{\rho}(e_a{}^{\lambda}\partial_\rho e_{\lambda b_1} + e_{b_1}{}^{\lambda}\partial_\rho e_{\lambda a})\right] \\
&= (-e_{b_1}{}^{\rho}G_{\rho,(b_2a)} + e_{b_2}{}^{\rho}G_{\rho,(b_1a)} + e_a{}^{\rho}G_{\mu,[b_1b_2]}) \\
&= (\det e)^{-\frac{1}{2}}(-G_{b_1,(b_2a)} + G_{b_2,(b_1a)} + G_{a,[b_1b_2]}) \quad ,
\end{aligned}
\tag{3.49}
$$

where we have used the generalised spin connection of equation (3.32) in the form

$$
G_{a,b}{}^{c} = E_a{}^{\mu}G_{\mu,b}{}^{c} = (\det e)^{\frac{1}{2}}e_a{}^{\mu}(e_b{}^{\nu}\partial_\mu e_\nu{}^{c}). \tag{3.50}
$$

### 3.3.3 First Order Duality Relations

We now use the spin connection, expressed in terms of the Cartan forms, to construct a first order equation involving the level one Cartan form $\overline{G}_{a,b_1b_2}$, and possessing the correct level zero Lorentz covariance properties. We postulate the following equation:

$$
D_{a,b_1b_2} = (\det e)^{\frac{1}{2}}\omega_{a,b_1b_2} + \frac{\tilde{e}_1}{2}\varepsilon_{b_1b_2}{}^{c_1c_2}\overline{G}_{c_1,c_2a} = 0, \tag{3.51}
$$

where $\tilde{e}_1$ is a constant. Note that in order to postulate a duality relation involving only level zero derivatives and without introducing any additional extraneous fields into the theory we can only use the 'invariant (pseudo-)tensor' $\varepsilon_{a_1\ldots a_4}$ to relate fields at different levels, and the only possible way to contract it against $\overline{G}_{c_1,c_2a}$ is as written above. This then fixes the level zero combination to be an anti-symmetric quantity constructed from the Cartan form coefficients $G_{a,b}{}^c$. The computation below will imply that this can only be the spin connection, thus we simply take it as the spin connection from the beginning. The duality relation is still only determined up to the constant $\tilde{e}_1$ which we will see is in fact uniquely fixed by local $I_c(A_1^{+++})$ covariance. This is in marked contrast to the situation for the nonlinear realization of gravity based solely on local GL(4) discussed earlier.

Equation (3.51) possesses level zero Lorentz covariance. The constant will now be uniquely determined by ensuring the variations under level one $I_c(A_1^{+++})$ transformations lead to consistent results.

Equation (3.51) is technically only the beginning of an equation which also involves derivatives at higher levels, as we will see below. When we include terms involving derivatives at level one we refer to this more general equation as the corresponding $l_1$-extended equation. The $l_1$ extension of the first term in $D_{a,b_1b_2}$, the spin connection term, will be determined as part of the process of applying the level one $I_c(A_1^{+++})$ transformations to $D_{a,b_1b_2}$.

Applying the level one $I_c(A_1^{+++})$ transformations (3.34) to the first term in equation (3.51) we find

$$\begin{aligned}
\delta[(\det e)^{\frac{1}{2}}\omega_{a,b_1b_2}] &= \frac{1}{2}\delta(-G_{b_1,b_2a} - G_{b_1,ab_2} + G_{b_2,b_1a} + G_{b_2,ab_1} + G_{a,b_1b_2} - G_{a,b_2b_1}) \\
&= \frac{1}{2}\Big[-(2\overline{G}_{b_1,b_2e}\Lambda^e{}_a - \eta_{b_2a}\overline{G}_{b_1,e_1e_2}\Lambda^{e_1e_2}) - (2\overline{G}_{b_1,ae}\Lambda^e{}_{b_2} - \eta_{ab_2}\overline{G}_{b_1,e_1e_2}\Lambda^{e_1e_2}) \\
&\quad + (2\overline{G}_{b_2,b_1e}\Lambda^e{}_a - \eta_{b_1a}\overline{G}_{b_2,e_1e_2}\Lambda^{e_1e_2}) + (2\overline{G}_{b_2,ae}\Lambda^e{}_{b_1} - \eta_{ab_1}\overline{G}_{b_2,e_1e_2}\Lambda^{e_1e_2}) \\
&\quad +(2\overline{G}_{a,b_1e}\Lambda^e{}_{b_2} - \eta_{b_1b_2}\overline{G}_{a,e_1e_2}\Lambda^{e_1e_2}) - (2\overline{G}_{a,b_2e}\Lambda^e{}_{b_1} - \eta_{b_1b_2}\overline{G}_{a,e_1e_2}\Lambda^{e_1e_2})\Big] \\
&= 2\Lambda^e{}_a\overline{G}_{[b_2,b_1]e} + 2\Lambda^e{}_{b_1}\overline{G}_{[b_2,a]e} + 2\Lambda^e{}_{b_2}\overline{G}_{[a,b_1]e} + 2\eta_{ab_2}\overline{G}_{[b_1,e_1]e_2}\Lambda^{e_1e_2} \\
&\quad + 2\eta_{ab_1}\overline{G}_{[e_1,b_2]e_2}\Lambda^{e_1e_2} + \eta_{ab_2}\overline{G}_{e_1,b_1e_2}\Lambda^{e_1e_2} - \eta_{ab_1}\overline{G}_{e_1,b_2e_2}\Lambda^{e_1e_2} \\
&= 2\Lambda^e{}_a\overline{G}_{[b_2,b_1]e} + 2\Lambda^e{}_{b_1}\overline{G}_{[b_2,a]e} + 2\Lambda^e{}_{b_2}\overline{G}_{[a,b_1]e} + 2\eta_{ab_2}\overline{G}_{[b_1,e_1]e_2}\Lambda^{e_1e_2} \\
&\quad + 2\eta_{ab_1}\overline{G}_{[e_1,b_2]e_2}\Lambda^{e_1e_2} - \delta\left[-\frac{1}{2}\eta_{ab_2}\hat{\overline{G}}{}^{e,}{}_{b_1e} + \frac{1}{2}\eta_{ab_1}\hat{\overline{G}}{}^{e,}{}_{b_2e}\right]. \qquad (3.52)
\end{aligned}$$

This suggests that the $l_1$ extended spin connection should be defined as

$$(\det e)^{\frac{1}{2}}\Omega_{a,b_1b_2} = (\det e)^{\frac{1}{2}}\omega_{a,b_1b_2} - \frac{1}{2}\eta_{b_2a}\hat{\overline{G}}^{e_2,}{}_{b_1e_2} + \frac{1}{2}\eta_{b_1a}\hat{\overline{G}}^{e_2,}{}_{b_2e_2}, \tag{3.53}$$

and that its variation is given by

$$\begin{aligned}\delta[(\det e)^{\frac{1}{2}}\Omega_{a,b_1b_2}] &= 2\Lambda^e{}_a\overline{G}_{[b_2,b_1]e} + 2\Lambda^e{}_{b_1}\overline{G}_{[b_2,a]e} + 2\Lambda^e{}_{b_2}\overline{G}_{[a,b_1]e}\\ &\quad + 2\eta_{ab_2}\Lambda^{e_1e_2}\overline{G}_{[b_1,e_1]e_2} + 2\eta_{ab_1}\Lambda^{e_1e_2}\overline{G}_{[e_1,b_2]e_2}.\end{aligned} \tag{3.54}$$

To process the variation of the second term in equation (3.51) is more involved. To keep track of things, at times we will not simplify obvious coefficients, leaving coefficients like $2\frac{1}{2}$ in this form rather than simplifying to 1. To begin we apply the variation (3.36) and find

$$\begin{aligned}\frac{\tilde{e}_1}{2}\varepsilon_{b_1b_2}{}^{c_1c_2}\delta\overline{G}_{c_1,c_2a} &= \frac{\tilde{e}_1}{2}\varepsilon_{b_1b_2}{}^{c_1c_2}\,[-\Lambda_{c_2}{}^eG_{c_1,ae} - \Lambda_a{}^eG_{c_1,c_2e} - \Lambda_{c_2}{}^eG_{c_1,ea} - \Lambda_a{}^eG_{c_1,ec_2}\\ &\qquad -2G_{c_1,c_2e_1,ae_2}\Lambda^{e_1e_2} - 2G_{c_1,ae_1,c_2e_2}\Lambda^{e_1e_2}]\\ &= \frac{\tilde{e}_1}{2}\varepsilon_{b_1b_2}{}^{c_1c_2}\Big[2\Lambda_{c_2}{}^e(\det e)^{1/2}\omega_{e,c_1a} + \Lambda_a{}^e(\det e)^{1/2}\omega_{e,c_1c_2}\\ &\qquad - 3G_{[c_1,c_2e_1],ae_2}\Lambda^{e_1e_2} - 6G_{[c_1,ae_1],c_2e_2}\Lambda^{e_1e_2}\\ &\qquad + 2G_{e_1,c_1a,c_2e_2}\Lambda^{e_1e_2} + G_{e_1,c_1c_2,ae_2}\Lambda^{e_1e_2} - \Lambda_a{}^eG_{e,[c_1c_2]}\\ &\qquad -2\Lambda_{c_2}{}^eG_{e,[c_1a]} + 2G_{a,e_1c_1,c_2e_2}\Lambda^{e_1e_2} - 2\Lambda_{c_2}{}^eG_{a,(c_1e)}\Big]\quad,\end{aligned} \tag{3.55}$$

which can be written as

$$\begin{aligned}\frac{\tilde{e}_1}{2}\varepsilon_{b_1b_2}{}^{c_1c_2}\delta\overline{G}_{c_1,c_2a} &= 2\frac{\tilde{e}_1}{2}\varepsilon_{b_1b_2}{}^{c_1c_2}\Lambda_{c_2}{}^e(\det e)^{1/2}\omega_{e,c_1a} + \frac{\tilde{e}_1}{2}\varepsilon_{b_1b_2}{}^{c_1c_2}\Lambda_a{}^e(\det e)^{1/2}\omega_{e,c_1c_2}\\ &\quad - 3\frac{\tilde{e}_1}{2}\varepsilon_{b_1b_2}{}^{c_1c_2}G_{[c_1,c_2e_1],ae_2}\Lambda^{e_1e_2} - 6\frac{\tilde{e}_1}{2}\varepsilon_{b_1b_2}{}^{c_1c_2}G_{[c_1,ae_1],c_2e_2}\Lambda^{e_1e_2}\\ &\quad - \delta[\frac{\tilde{e}_1}{2}\varepsilon_{b_1b_2}{}^{c_1c_2}(-\hat{G}^{e,}{}_{c_1a,c_2e} - \frac{1}{2}\hat{G}^{e,}{}_{c_1c_2,ae} + \frac{1}{2}\hat{G}_{a,[c_1c_2]} + \hat{G}_{c_2,[c_1a]})]\\ &\quad + 2\frac{\tilde{e}_1}{2}\varepsilon_{b_1b_2}{}^{c_1c_2}G_{a,e_1c_1,c_2e_2}\Lambda^{e_1e_2} - 2\frac{\tilde{e}_1}{2}\varepsilon_{b_1b_2}{}^{c_1c_2}\Lambda_{c_2}{}^eG_{a,(c_1e)}.\end{aligned} \tag{3.56}$$

We now express the first two terms in terms of the duality relation $D_{a,b_1b_2}$ of equation (3.51), and simplify the remaining terms. The first term satisfies

$$\begin{aligned}2\frac{\tilde{e}_1}{2}\varepsilon_{b_1b_2}{}^{c_1c_2}\Lambda_{c_2}{}^e(\det e)^{\frac{1}{2}}\omega_{e,c_1a} &= 2\frac{\tilde{e}_1}{2}\varepsilon_{b_1b_2}{}^{c_1c_2}\Lambda_{c_2}{}^e(D_{e,c_1a} - \frac{\tilde{e}_1}{2}\varepsilon_{c_1a}{}^{d_1d_2}\overline{G}_{d_1,d_2e})\\ &= 2\frac{\tilde{e}_1}{2}\varepsilon_{b_1b_2}{}^{c_1c_2}\Lambda_{c_2}{}^eD_{e,c_1a} + 4\frac{\tilde{e}_1}{2}\frac{\tilde{e}_1}{2}\Lambda_a{}^e\overline{G}_{[b_1,b_2]e}\\ &\quad - 4\frac{\tilde{e}_1}{2}\frac{\tilde{e}_1}{2}\eta_{ab_1}\Lambda^{c_2e}\overline{G}_{[c_2,b_2]e} + 4\frac{\tilde{e}_1}{2}\frac{\tilde{e}_1}{2}\eta_{ab_2}\Lambda^{c_2e}\overline{G}_{[c_2,b_1]e}\quad.\end{aligned} \tag{3.57}$$

The second term reduces similarly

$$\begin{aligned}\frac{\tilde{e}_1}{2}\varepsilon_{b_1b_2}{}^{c_1c_2}\Lambda_a{}^e(\det e)^{\frac{1}{2}}\omega_{e,c_1c_2} &= \frac{\tilde{e}_1}{2}\varepsilon_{b_1b_2}{}^{c_1c_2}\Lambda_a{}^e(D_{e,c_1c_2} - \frac{\tilde{e}_1}{2}\varepsilon_{c_1c_2}{}^{d_1d_2}\overline{G}_{d_1,d_2e})\\ &= \frac{\tilde{e}_1}{2}\varepsilon_{b_1b_2}{}^{c_1c_2}\Lambda_a{}^eD_{e,c_1c_2} + 4\frac{\tilde{e}_1}{2}\frac{\tilde{e}_1}{2}\Lambda_a{}^e\overline{G}_{[b_1,b_2]e}\quad.\end{aligned} \tag{3.58}$$

The third term involves no apparent duality relation so far in our derivation, but it can be simplified as

$$\begin{aligned}
-3\frac{\tilde{e}_1}{2}\varepsilon_{b_1b_2}{}^{c_1c_2}G_{[c_1,c_2e_1],ae_2}\Lambda^{e_1e_2} &= -3\frac{\tilde{e}_1}{2}\varepsilon_{b_1b_2}{}^{c_1c_2}\delta^{p_1p_2p_3}_{c_1c_2e_1}G_{p_1,p_2p_3,ae_2}\Lambda^{e_1e_2} \\
&= -3\frac{\tilde{e}_1}{2}\varepsilon^{c_1c_2}{}_{b_1b_2}(-\frac{1}{3!}\varepsilon_{c_1c_2e_1d}\varepsilon^{p_1p_2p_3d})G_{p_1,p_2p_3,ae_2}\Lambda^{e_1e_2} \qquad (3.59)\\
&= \frac{\tilde{e}_1}{2}\varepsilon_{b_2}{}^{c_1c_2c_3}G_{c_1,c_2c_3,ae}\Lambda_{b_1}{}^{e} - \frac{\tilde{e}_1}{2}\varepsilon_{b_1}{}^{c_1c_2c_3}G_{c_1,c_2c_3,ae}\Lambda_{b_2}{}^{e}
\end{aligned}$$

The fourth term simplifies similarly

$$\begin{aligned}
&-6\frac{\tilde{e}_1}{2}\varepsilon_{b_1b_2}{}^{c_1c_2}G_{[c_1,ae_1],c_2e_2}\Lambda^{e_1e_2} \\
&= \frac{\tilde{e}_1}{2}\varepsilon_{b_2}{}^{c_1c_2c_3}G_{c_1,c_2c_3,ae}\Lambda_{b_1}{}^{e} - \frac{\tilde{e}_1}{2}\varepsilon_{b_1}{}^{c_1c_2c_3}G_{c_1,c_2c_3,ae}\Lambda_{b_2}{}^{e} \\
&- \frac{\tilde{e}_1}{2}\eta_{ab_1}\varepsilon_{b_2}{}^{c_1c_2c_3}G_{c_1,c_2c_3,e_1e_2}\Lambda^{e_1e_2} - \frac{\tilde{e}_1}{2}\eta_{ab_1}\varepsilon^{p_1p_2p_3p_4}G_{p_1,p_2p_3,p_4e}\Lambda_{b_2}{}^{e} \\
&+ \frac{\tilde{e}_1}{2}\eta_{ab_2}\varepsilon_{b_1}{}^{p_1p_2p_3}G_{p_1,p_2p_3,e_1e_2}\Lambda^{e_1e_2} + \frac{\tilde{e}_1}{2}\eta_{ab_2}\varepsilon^{p_1p_2p_3p_4}G_{p_1,p_2p_3,p_4e}\Lambda_{b_1}{}^{e} \qquad (3.60)\\
&= \frac{\tilde{e}_1}{2}\varepsilon_{b_2}{}^{c_1c_2c_3}G_{c_1,c_2c_3,ae}\Lambda_{b_1}{}^{e} - \frac{\tilde{e}_1}{2}\varepsilon_{b_1}{}^{c_1c_2c_3}G_{c_1,c_2c_3,ae}\Lambda_{b_2}{}^{e} \\
&- \frac{\tilde{e}_1}{2}\eta_{ab_1}\varepsilon_{b_2}{}^{c_1c_2c_3}G_{c_1,c_2c_3,e_1e_2}\Lambda^{e_1e_2} + \frac{\tilde{e}_1}{2}\eta_{ab_2}\varepsilon_{b_1}{}^{p_1p_2p_3}G_{p_1,p_2p_3,e_1e_2}\Lambda^{e_1e_2}.
\end{aligned}$$

Here we have used the irreducibility condition $G_{p_1,[p_2p_3,p_4]e} = 0$ to go from the second last line to the last line. Collecting these results, we find that the second term of equation (3.51) varies into

$$\begin{aligned}
&\delta[\frac{\tilde{e}_1}{2}\varepsilon_{b_1b_2}{}^{c_1c_2}\overline{G}_{c_1,c_2a}] \\
&= 2\frac{\tilde{e}_1}{2}\varepsilon_{b_1b_2}{}^{c_1c_2}\Lambda_{c_2}{}^{e}D_{e,c_1a} + 4\frac{\tilde{e}_1}{2}\frac{\tilde{e}_1}{2}\Lambda_a{}^{e}\overline{G}_{[b_1,b_2]e} \\
&- 4\frac{\tilde{e}_1}{2}\frac{\tilde{e}_1}{2}\eta_{ab_1}\Lambda^{c_2e}\overline{G}_{[c_2,b_2]e} + 4\frac{\tilde{e}_1}{2}\frac{\tilde{e}_1}{2}\eta_{ab_2}\Lambda^{c_2e}\overline{G}_{[c_2,b_1]e} \\
&+ \frac{\tilde{e}_1}{2}\varepsilon_{b_1b_2}{}^{c_1c_2}\Lambda_a{}^{e}D_{e,c_1c_2} + 4\frac{\tilde{e}_1}{2}\frac{\tilde{e}_1}{2}\Lambda_a{}^{e}\overline{G}_{[b_1,b_2]e} \\
&+ \frac{\tilde{e}_1}{2}\varepsilon_{b_2}{}^{c_1c_2c_3}G_{c_1,c_2c_3,ae}\Lambda_{b_1}{}^{e} - \frac{\tilde{e}_1}{2}\varepsilon_{b_1}{}^{c_1c_2c_3}G_{c_1,c_2c_3,ae}\Lambda_{b_2}{}^{e} \qquad (3.61)\\
&+ \frac{\tilde{e}_1}{2}\varepsilon_{b_2}{}^{c_1c_2c_3}G_{c_1,c_2c_3,ae}\Lambda_{b_1}{}^{e} - \frac{\tilde{e}_1}{2}\varepsilon_{b_1}{}^{c_1c_2c_3}G_{c_1,c_2c_3,ae}\Lambda_{b_2}{}^{e} \\
&- \frac{\tilde{e}_1}{2}\eta_{ab_1}\varepsilon_{b_2}{}^{c_1c_2c_3}G_{c_1,c_2c_3,e_1e_2}\Lambda^{e_1e_2} + \frac{\tilde{e}_1}{2}\eta_{ab_2}\varepsilon_{b_1}{}^{p_1p_2p_3}G_{p_1,p_2p_3,e_1e_2}\Lambda^{e_1e_2} \\
&\quad - \delta\left[\frac{\tilde{e}_1}{2}\varepsilon_{b_1b_2}{}^{c_1c_2}(-\hat{G}^{e,}{}_{c_1a,c_2e} - \frac{1}{2}\hat{G}^{e,}{}_{c_1c_2,ae} + \frac{1}{2}\hat{G}_{a,[c_1c_2]} + \hat{G}_{c_2,[c_1a]})\right] \\
&\quad + 2\frac{\tilde{e}_1}{2}\varepsilon_{b_1b_2}{}^{c_1c_2}G_{a,e_1c_1,c_2e_2}\Lambda^{e_1e_2} - 2\frac{\tilde{e}_1}{2}\varepsilon_{b_1b_2}{}^{c_1c_2}\Lambda_{c_2}{}^{e}G_{a,(c_1e)} \quad ,
\end{aligned}$$

which we write as

$$\begin{aligned}
&\delta[\frac{\tilde{e}_1}{2}\varepsilon_{b_1b_2}{}^{c_1c_2}\overline{G}_{c_1,c_2a}] \\
&= \frac{\tilde{e}_1}{2}\varepsilon_{b_1b_2}{}^{c_1c_2}\Lambda_a{}^eD_{e,c_1c_2} + 2\frac{\tilde{e}_1}{2}\varepsilon_{b_1b_2}{}^{c_1c_2}\Lambda_{c_2}{}^eD_{e,c_1a} + 8\frac{\tilde{e}_1}{2}\frac{\tilde{e}_1}{2}\Lambda_a{}^e\overline{G}_{[b_1,b_2]e} \\
&+ \eta_{ab_1}\Lambda^{e_1e_2}(4\frac{\tilde{e}_1}{2}\frac{\tilde{e}_1}{2}\overline{G}_{[b_2,e_1]e_2} - \frac{\tilde{e}_1}{2}\varepsilon_{b_2}{}^{c_1c_2c_3}G_{c_1,c_2c_3,e_1e_2}) \\
&- \eta_{ab_2}\Lambda^{e_1e_2}(4\frac{\tilde{e}_1}{2}\frac{\tilde{e}_1}{2}\overline{G}_{[b_1,e_1]e_2} - \frac{\tilde{e}_1}{2}\varepsilon_{b_1}{}^{p_1p_2p_3}G_{p_1,p_2p_3,e_1e_2}) \\
&+ \frac{\tilde{e}_1}{2}\Lambda_{b_1}{}^e2\varepsilon_{b_2}{}^{c_1c_2c_3}G_{c_1,c_2c_3,ae} - \frac{\tilde{e}_1}{2}\Lambda_{b_2}{}^e2\varepsilon_{b_1}{}^{c_1c_2c_3}G_{c_1,c_2c_3,ae} \\
&\quad - \delta[\frac{\tilde{e}_1}{2}\varepsilon_{b_1b_2}{}^{c_1c_2}(-\hat{G}^{e,}{}_{c_1a,c_2e} - \frac{1}{2}\hat{G}^{e,}{}_{c_1c_2,ae} + \frac{1}{2}\hat{G}_{a,[c_1c_2]} + \hat{G}_{c_2,[c_1a]})] \\
&\quad + 2\frac{\tilde{e}_1}{2}\varepsilon_{b_1b_2}{}^{c_1c_2}G_{a,e_1c_1,c_2e_2}\Lambda^{e_1e_2} - 2\frac{\tilde{e}_1}{2}\varepsilon_{b_1b_2}{}^{c_1c_2}\Lambda_{c_2}{}^eG_{a,(c_1e)} \quad .
\end{aligned} \tag{3.62}$$

We thus see the variation of the $l_1$-extended equation of motion

$$\begin{aligned}
\tilde{\mathcal{D}}_{a,b_1b_2} &= (\det e)^{\frac{1}{2}}\omega_{a,b_1b_2} - \frac{1}{2}\eta_{ab_2}\hat{\overline{G}}^{e_2,}{}_{b_1e_2} + \frac{1}{2}\eta_{ab_1}\hat{\overline{G}}^{e_2,}{}_{b_2e_2} + \frac{\tilde{e}_1}{2}\varepsilon_{b_1b_2}{}^{c_1c_2}\overline{G}_{c_1,c_2a} \\
&+ \frac{\tilde{e}_1}{2}\varepsilon_{b_1b_2}{}^{c_1c_2}(-\hat{G}^{e,}{}_{c_1a,c_2e} - \frac{1}{2}\hat{G}^{e,}{}_{c_1c_2,ae} + \frac{1}{2}\hat{G}_{a,[c_1c_2]} + \hat{G}_{c_2,[c_1a]})
\end{aligned} \tag{3.63}$$

is given by

$$\begin{aligned}
\delta\tilde{\mathcal{D}}_{a,b_1b_2} &= 2\Lambda^e{}_a\overline{G}_{[b_2,b_1]e} + 2\Lambda^e{}_{b_2}\overline{G}_{[a,b_1]e} + 2\Lambda^e{}_{b_1}\overline{G}_{[b_2,a]e} \\
&+ 2\eta_{b_2a}\Lambda^{e_1e_2}\overline{G}_{[b_1,e_1]e_2} - 2\eta_{b_1a}\Lambda^{e_1e_2}\overline{G}_{[b_2,e_1]e_2} \\
&+ \frac{\tilde{e}_1}{2}\varepsilon_{b_1b_2}{}^{c_1c_2}\Lambda_a{}^eD_{e,c_1c_2} + 2\frac{\tilde{e}_1}{2}\varepsilon_{b_1b_2}{}^{c_1c_2}\Lambda_{c_2}{}^eD_{e,c_1a} + 8\frac{\tilde{e}_1}{2}\frac{\tilde{e}_1}{2}\Lambda_a{}^e\overline{G}_{[b_1,b_2]e} \\
&+ \eta_{ab_1}\Lambda^{e_1e_2}(4\frac{\tilde{e}_1}{2}\frac{\tilde{e}_1}{2}\overline{G}_{[b_2,e_1]e_2} - \frac{\tilde{e}_1}{2}\varepsilon_{b_2}{}^{c_1c_2c_3}G_{c_1,c_2c_3,e_1e_2}) \\
&- \eta_{ab_2}\Lambda^{e_1e_2}(4\frac{\tilde{e}_1}{2}\frac{\tilde{e}_1}{2}\overline{G}_{[b_1,e_1]e_2} - \frac{\tilde{e}_1}{2}\varepsilon_{b_1}{}^{p_1p_2p_3}G_{p_1,p_2p_3,e_1e_2}) \\
&+ \frac{\tilde{e}_1}{2}\Lambda_{b_1}{}^e2\varepsilon_{b_2}{}^{c_1c_2c_3}G_{c_1,c_2c_3,ae} - \frac{\tilde{e}_1}{2}\Lambda_{b_2}{}^e2\varepsilon_{b_1}{}^{c_1c_2c_3}G_{c_1,c_2c_3,ae} \\
&\quad + 2\frac{\tilde{e}_1}{2}\varepsilon_{b_1b_2}{}^{c_1c_2}G_{a,e_1c_1,c_2e_2}\Lambda^{e_1e_2} - 2\frac{\tilde{e}_1}{2}\varepsilon_{b_1b_2}{}^{c_1c_2}\Lambda_{c_2}{}^eG_{a,(c_1e)} \\
&= \frac{\tilde{e}_1}{2}\varepsilon_{b_1b_2}{}^{c_1c_2}\Lambda_a{}^eD_{e,c_1c_2} + 2\frac{\tilde{e}_1}{2}\varepsilon_{b_1b_2}{}^{c_1c_2}\Lambda_{c_2}{}^eD_{e,c_1a} \\
&+ 8\frac{\tilde{e}_1}{2}\frac{\tilde{e}_1}{2}\Lambda_a{}^e\overline{G}_{[b_1,b_2]e} + 2\Lambda^e{}_a\overline{G}_{[b_2,b_1]e} \\
&- 2\eta_{b_1a}\Lambda^{e_1e_2}\overline{G}_{[b_2,e_1]e_2} + \eta_{ab_1}\Lambda^{e_1e_2}(4\frac{\tilde{e}_1}{2}\frac{\tilde{e}_1}{2}\overline{G}_{[b_2,e_1]e_2} - \frac{\tilde{e}_1}{2}\varepsilon_{b_2}{}^{c_1c_2c_3}G_{c_1,c_2c_3,e_1e_2}) \\
&+ 2\eta_{b_2a}\Lambda^{e_1e_2}\overline{G}_{[b_1,e_1]e_2} - \eta_{ab_2}\Lambda^{e_1e_2}(4\frac{\tilde{e}_1}{2}\frac{\tilde{e}_1}{2}\overline{G}_{[b_1,e_1]e_2} - \frac{\tilde{e}_1}{2}\varepsilon_{b_1}{}^{p_1p_2p_3}G_{p_1,p_2p_3,e_1e_2}) \\
&+ 2\Lambda_{b_1}{}^e\overline{G}_{[b_2,a]e} + \frac{\tilde{e}_1}{2}\Lambda_{b_1}{}^e2\varepsilon_{b_2}{}^{c_1c_2c_3}G_{c_1,c_2c_3,ae} \\
&- 2\Lambda_{b_2}{}^e\overline{G}_{[b_1,a]e} - \frac{\tilde{e}_1}{2}\Lambda_{b_2}{}^e2\varepsilon_{b_1}{}^{c_1c_2c_3}G_{c_1,c_2c_3,ae} \\
&\quad + 2\frac{\tilde{e}_1}{2}\varepsilon_{b_1b_2}{}^{c_1c_2}G_{a,e_1c_1,c_2e_2}\Lambda^{e_1e_2} - 2\frac{\tilde{e}_1}{2}\varepsilon_{b_1b_2}{}^{c_1c_2}\Lambda_{c_2}{}^eG_{a,(c_1e)} \quad ,
\end{aligned} \tag{3.64}$$

which simplifies to

$$\begin{aligned}
\delta\tilde{\mathcal{D}}_{a,b_1b_2} &= \frac{\tilde{e}_1}{2}\varepsilon_{b_1b_2}{}^{c_1c_2}\Lambda_a{}^eD_{e,c_1c_2} + 2\frac{\tilde{e}_1}{2}\varepsilon_{b_1b_2}{}^{c_1c_2}\Lambda_{c_2}{}^eD_{e,c_1a} \\
&+ (2\tilde{e}_1^2-2)\Lambda_a{}^e\overline{G}_{[b_1,b_2]e} + \eta_{ab_1}\Lambda^{e_1e_2}[(\tilde{e}_1^2-2)\overline{G}_{[b_2,e_1]e_2} - \frac{\tilde{e}_1}{2}\varepsilon_{b_2}{}^{c_1c_2c_3}G_{c_1,c_2c_3,e_1e_2}] \\
&- \eta_{ab_2}\Lambda^{e_1e_2}[(\tilde{e}_1^2-2)\overline{G}_{[b_1,e_1]e_2} - \frac{\tilde{e}_1}{2}\varepsilon_{b_1}{}^{p_1p_2p_3}G_{p_1,p_2p_3,e_1e_2}] \\
&+ 2\Lambda_{b_1}{}^e[\overline{G}_{[b_2,a]e} + \frac{\tilde{e}_1}{2}\varepsilon_{b_2}{}^{c_1c_2c_3}G_{c_1,c_2c_3,ae}] - 2\Lambda_{b_2}{}^e[\overline{G}_{[b_1,a]e} + \frac{\tilde{e}_1}{2}\varepsilon_{b_1}{}^{c_1c_2c_3}G_{c_1,c_2c_3,ae}] \\
&\qquad + \tilde{e}_1\varepsilon_{b_1b_2}{}^{c_1c_2}G_{a,e_1c_1,c_2e_2}\Lambda^{e_1e_2} - \tilde{e}_1\varepsilon_{b_1b_2}{}^{c_1c_2}\Lambda_{c_2}{}^eG_{a,(c_1e)} \quad ,
\end{aligned} \tag{3.65}$$

that is,

$$\begin{aligned}
\delta\tilde{\mathcal{D}}_{a,b_1b_2} &= \frac{\tilde{e}_1}{2}\varepsilon_{b_1b_2}{}^{c_1c_2}\Lambda_a{}^eD_{e,c_1c_2} + 2\frac{\tilde{e}_1}{2}\varepsilon_{b_1b_2}{}^{c_1c_2}\Lambda_{c_2}{}^eD_{e,c_1a} + (2\tilde{e}_1^2-2)\Lambda_a{}^e\overline{G}_{[b_1,b_2]e} \\
&+ \eta_{ab_1}\Lambda^{e_1e_2}[(\tilde{e}_1^2-2)\overline{G}_{[b_2,e_1]e_2} - \frac{\tilde{e}_1}{2}\varepsilon_{b_2}{}^{c_1c_2c_3}G_{c_1,c_2c_3,e_1e_2}] \\
&- \eta_{ab_2}\Lambda^{e_1e_2}[(\tilde{e}_1^2-2)\overline{G}_{[b_1,e_1]e_2} - \frac{\tilde{e}_1}{2}\varepsilon_{b_1}{}^{p_1p_2p_3}G_{p_1,p_2p_3,e_1e_2}] \\
&+ 2\Lambda_{b_1}{}^e[\overline{G}_{[b_2,a]e} + \frac{\tilde{e}_1}{2}\varepsilon_{b_2}{}^{c_1c_2c_3}G_{c_1,c_2c_3,ae}] - 2\Lambda_{b_2}{}^e[\overline{G}_{[b_1,a]e} + \frac{\tilde{e}_1}{2}\varepsilon_{b_1}{}^{c_1c_2c_3}G_{c_1,c_2c_3,ae}] \\
&\qquad + \tilde{e}_1\varepsilon_{b_1b_2}{}^{c_1c_2}G_{a,e_1c_1,c_2e_2}\Lambda^{e_1e_2} - \tilde{e}_1\varepsilon_{b_1b_2}{}^{c_1c_2}\Lambda_{c_2}{}^eG_{a,(c_1e)} \quad .
\end{aligned} \tag{3.66}$$

In order to eliminate the third term on the first line of equation (3.66), which is not a modulo term, not an $l_1$ term, and is part of no duality relation, it is now clear that we have no other choice but to set $\tilde{e}_1 = \pm 1$, and we choose

$$\tilde{e}_1 = 1. \tag{3.67}$$

On doing this, new duality relations between the dual graviton and dual dual graviton start to become apparent:

$$\begin{aligned}
\delta\tilde{\mathcal{D}}_{a,b_1b_2} &= \frac{1}{2}\varepsilon_{b_1b_2}{}^{c_1c_2}\Lambda_a{}^eD_{e,c_1c_2} + \varepsilon_{b_1b_2}{}^{c_1c_2}\Lambda_{c_2}{}^eD_{e,c_1a} \\
&- \eta_{ab_1}\Lambda^{e_1e_2}[\overline{G}_{[b_2,e_1]e_2} + \frac{1}{2}\varepsilon_{b_2}{}^{c_1c_2c_3}G_{c_1,c_2c_3,e_1e_2}] \\
&+ \eta_{ab_2}\Lambda^{e_1e_2}[\overline{G}_{[b_1,e_1]e_2} + \frac{1}{2}\varepsilon_{b_1}{}^{p_1p_2p_3}G_{p_1,p_2p_3,e_1e_2}] \\
&+ 2\Lambda_{b_1}{}^e[\overline{G}_{[b_2,a]e} + \frac{1}{2}\varepsilon_{b_2}{}^{c_1c_2c_3}G_{c_1,c_2c_3,ae}] \\
&- 2\Lambda_{b_2}{}^e[\overline{G}_{[b_1,a]e} + \frac{1}{2}\varepsilon_{b_1}{}^{c_1c_2c_3}G_{c_1,c_2c_3,ae}] \\
&\qquad + \varepsilon_{b_1b_2}{}^{c_1c_2}G_{a,e_1c_1,c_2e_2}\Lambda^{e_1e_2} - \varepsilon_{b_1b_2}{}^{c_1c_2}\Lambda_{c_2}{}^eG_{a,(c_1e)} \quad .
\end{aligned} \tag{3.68}$$

However, the $\Lambda^{e_1e_2}$ coefficients are symmetric in the $e_1e_2$ indices, while e.g. the coefficient of $\Lambda_{b_1}{}^e$ is not, and so we introduce this symmetry by re-writing the expression as

$$\begin{aligned}\delta\tilde{\mathcal{D}}_{a,b_1b_2} = &\frac{1}{2}\varepsilon_{b_1b_2}{}^{c_1c_2}\Lambda_a{}^e D_{e,c_1c_2} + \varepsilon_{b_1b_2}{}^{c_1c_2}\Lambda_{c_2}{}^e D_{e,c_1a} \\ &- \frac{1}{2}\eta_{ab_1}\Lambda^{e_1e_2}[\overline{G}_{b_2,e_1e_2} + \varepsilon_{b_2}{}^{c_1c_2c_3}G_{c_1,c_2c_3,e_1e_2}] \\ &+ \frac{1}{2}\eta_{ab_2}\Lambda^{e_1e_2}[\overline{G}_{b_1,e_1e_2} + \varepsilon_{b_1}{}^{p_1p_2p_3}G_{p_1,p_2p_3,e_1e_2}] \\ &+ \Lambda_{b_1}{}^e[\overline{G}_{b_2,ae} + \varepsilon_{b_2}{}^{c_1c_2c_3}G_{c_1,c_2c_3,ae}] \\ &- \Lambda_{b_2}{}^e[\overline{G}_{b_1,ae} + \varepsilon_{b_1}{}^{c_1c_2c_3}G_{c_1,c_2c_3,ae}] \\ &\quad + \eta_{ab_1}\Lambda^{e_1e_2}\frac{1}{2}\overline{G}_{e_1,b_2e_2} - \eta_{ab_2}\Lambda^{e_1e_2}\frac{1}{2}\overline{G}_{e_1,b_1e_2} \\ &\quad - \Lambda_{b_1}{}^e\overline{G}_{a,b_2e} + \Lambda_{b_2}{}^e\overline{G}_{a,b_1e} + \varepsilon_{b_1b_2}{}^{c_1c_2}G_{a,e_1c_1,c_2e_2}\Lambda^{e_1e_2} - \varepsilon_{b_1b_2}{}^{c_1c_2}\Lambda_{c_2}{}^e G_{a,(c_1e)} \quad .\end{aligned} \tag{3.69}$$

The sixth line of this equation can be written as $l_1$ terms, while the last line takes the form of a 'modulo' transformation:

$$\begin{aligned}\partial_a\tilde{\Lambda}_{b_1b_2} &:= -\Lambda_{b_1}{}^e\overline{G}_{a,b_2e} + \Lambda_{b_2}{}^e\overline{G}_{a,b_1e} + \varepsilon_{b_1b_2}{}^{c_1c_2}G_{a,e_1c_1,c_2e_2}\Lambda^{e_1e_2} - \varepsilon_{b_1b_2}{}^{c_1c_2}\Lambda_{c_2}{}^e G_{a,(c_1e)} \\ &= -\varepsilon_{b_1b_2}{}^{c_1c_2}[\Lambda_{c_2}{}^e G_{a,(c_1e)} - \frac{1}{2}\varepsilon_{c_1c_2}{}^{d_1d_2}\Lambda_{d_1}{}^e\overline{G}_{a,d_2e} - G_{a,e_1c_1,c_2e_2}\Lambda^{e_1e_2}].\end{aligned} \tag{3.70}$$

This results in the dual-graviton dual-dual-graviton duality relation

$$\overline{D}_{a,b_1b_2} = \overline{G}_{a,b_1b_2} + \varepsilon_a{}^{e_1e_2e_3}G_{e_1,e_2e_3,b_1b_2} = 0 \quad , \tag{3.71}$$

and the graviton dual-graviton equation of motion (now with the coefficient fixed) is thus given by

$$D_{a,b_1b_2} = (\det e)^{\frac{1}{2}}\omega_{a,b_1b_2} + \frac{1}{2}\varepsilon_{b_1b_2}{}^{c_1c_2}\overline{G}_{c_1,c_2a} = 0 \quad . \tag{3.72}$$

In summary, we have found that the $l_1$ extension of the graviton dual-graviton equation given by

$$\begin{aligned}\mathcal{D}_{a,b_1b_2} = &(\det e)^{\frac{1}{2}}\omega_{a,b_1b_2} - \frac{1}{2}\eta_{ab_2}\hat{\overline{G}}^{e_2,}{}_{b_1e_2} + \frac{1}{2}\eta_{ab_1}\hat{\overline{G}}^{e_2,}{}_{b_2e_2} \\ &+ \frac{\tilde{e}_1}{2}\varepsilon_{b_1b_2}{}^{c_1c_2}\overline{G}_{c_1,c_2a} - \frac{1}{4}\eta_{ab_1}\overline{G}^{e,}{}_{b_2e} + \frac{1}{4}\eta_{ab_2}\overline{G}^{e,}{}_{b_1e_2} \\ &+ \frac{\tilde{e}_1}{2}\varepsilon_{b_1b_2}{}^{c_1c_2}(-\hat{G}^{e,}{}_{c_1a,c_2e} - \frac{1}{2}\hat{G}^{e,}{}_{c_1c_2,ae} + \frac{1}{2}\hat{G}_{a,[c_1c_2]} + \hat{G}_{c_2,[c_1a]}) \quad ,\end{aligned} \tag{3.73}$$

varies into

$$\begin{aligned}\delta\mathcal{D}_{a,b_1b_2} = &\frac{1}{2}\varepsilon_{b_1b_2}{}^{c_1c_2}\Lambda_a{}^e D_{e,c_1c_2} + \varepsilon_{b_1b_2}{}^{c_1c_2}\Lambda_{c_2}{}^e D_{e,c_1a} - \frac{1}{2}\eta_{ab_1}\Lambda^{e_1e_2}\overline{D}_{b_2,e_1e_2} \\ &+ \frac{1}{2}\eta_{ab_2}\Lambda^{e_1e_2}\overline{D}_{b_1,e_1e_2} + \Lambda_{b_1}{}^e\overline{D}_{b_2,ae} - \Lambda_{b_2}{}^e\overline{D}_{b_1,ae} + \partial_a\tilde{\Lambda}_{b_1b_2}.\end{aligned} \tag{3.74}$$

In principle we could repeat this procedure on $\overline{D}_{a,b_1b_2}$, and repeat this again with the resulting duality relations etc..., to discover an infinite collection of first order duality relations relating fields at different levels, all uniquely determined by the nonlinear realization.

## 3.4 Second Order Gravity and Dual Gravity Equations of Motion

In this section we now use the symmetries of the nonlinear realization to consider second order equations of motion for the graviton and the dual graviton. It is natural to begin from the Riemann tensor for the graviton field. Since a local $h = I - \Lambda_{a_1 a_2} S^{a_1 a_2} \in I_c(A_1^{+++})$ transformation sends a Cartan form coefficient at level $n$ into others whose levels are at $n \pm 1$ (this is displayed in equation (3.34) to level $n = 2$), the variation of the gravity equation, in principle, must lead to the dual gravity equation.

### 3.4.1 Ricci Tensor in Terms of Cartan Forms

We begin with the the Ricci tensor in standard form

$$(\det e) R_a{}^b = (\det e)\{e_a{}^\mu \partial_\mu(\omega_{\nu,}{}^{bd}) e_d{}^\nu - \partial_\nu(\omega_\mu{}^{bd}) e_d{}^\nu e_a{}^\mu + \omega_{a,}{}^b{}_c \omega_{d,}{}^{cd} - \omega_{d,}{}^b{}_c \omega_{a,}{}^{cd}\}. \tag{3.75}$$

We take this as our starting equation of motion $E_a{}^b = 0$, where

$$E_a{}^b := (\det e) R_a{}^b. \tag{3.76}$$

We now re-express this in terms of the Cartan forms so that we are able to perform local $I_c(A_1^{+++})$ transformations on this equation. We begin by labelling the four terms in equation (3.75) as $I$, $II$, $III$ and $IV$, i.e. setting

$$E_a{}^b = I + II + III + IV \quad . \tag{3.76'}$$

The first term, $I$, simplifies as

$$\begin{aligned} I &= (\det e)\{e_a{}^\mu \partial_\mu(\omega_{\nu,}{}^{bd}) e_d{}^\nu\} \\ &= (\det e)\{e_a{}^\mu \partial_\mu(\omega_{\nu,}{}^{bd} e_d{}^\nu) - \omega_{\nu,}{}^{bd} e_a{}^\mu \partial_\mu e_d{}^\nu\} \\ &= (\det e)^{1/2} e_a{}^\mu \partial_\mu(\omega_{d,}{}^{bd})(\det e)^{1/2} - (\det e)^{1/2} \omega_{c,}{}^{bd} (\det e)^{1/2} (e_a{}^\mu \partial_\mu e_d{}^\nu) e^c{}_\nu \\ &= (\det e)^{1/2} e_a{}^\mu \partial_\mu[(\det e)^{1/2} \omega_{d,}{}^{bd}] - (\det e)^{1/2} \omega_{d,}{}^{bd} \{e_a{}^\mu \partial_\mu (\det e)^{1/2}\} \\ &\qquad\qquad + (\det e)^{1/2} \omega_{c,}{}^{bd} (\det e)^{1/2} (e_a{}^\mu \partial_\mu e_\nu{}^c) e_d{}^\nu \\ &= (\det e)^{1/2} e_a{}^\mu \partial_\mu[(\det e)^{1/2} \omega_{d,}{}^{bd}] - \frac{1}{2} G_{a,c}{}^c (\det e)^{1/2} \omega_{d,}{}^{bd} + G_{a,d}{}^c (\det e)^{1/2} \omega_{c,}{}^{bd} \quad . \end{aligned} \tag{3.77}$$

The second term, $II$, simplifies as

$$\begin{aligned}
II &= (\det e)\{-\partial_\nu(\omega_{\mu,}{}^{bd})e_d{}^\nu e_a{}^\mu\} \\
&= -(\det e)^{1/2} e_d{}^\nu \partial_\nu(\omega_{\mu,}{}^{bd} e_a{}^\mu)(\det e)^{1/2} + (\det e)^{1/2}(\det e)^{1/2} e_d{}^\nu e^c{}_\mu(\partial_\nu e_a{}^\mu)\omega_{c,}{}^{bd} \\
&= -(\det e)^{1/2} e_d{}^\nu \partial_\nu((\det e)^{1/2}\omega_{a,}{}^{bd}) + (\det e)^{1/2}\omega_{a,}{}^{bd}\{e_d{}^\nu \partial_\nu(\det e)^{1/2}\} \\
&\qquad - (\det e)^{1/2}(\det e)^{1/2} e_d{}^\nu(\partial_\nu e_\mu{}^c)e_a{}^\mu \omega_{c,}{}^{bd} \\
&= -(\det e)^{1/2} e_d{}^\nu \partial_\nu((\det e)^{1/2}\omega_{a,}{}^{bd}) + \frac{1}{2}G_{d,c}{}^c(\det e)^{1/2}\omega_{a,}{}^{bd} - (\det e)^{1/2}G_{d,a}{}^c\omega_{c,}{}^{bd} \quad .
\end{aligned} \tag{3.78}$$

The third term, $III$, simplifies as

$$\begin{aligned}
III &= (\det e)\omega_{a,}{}^b{}_c\omega_{d,}{}^{cd} \\
&= (\det e)^{1/2}\omega_{a,}{}^b{}_c(-G^{c,(d}{}_{d)} + G^{d,(c}{}_{d)} + G_{d,}{}^{[cd]}) \\
&= \frac{1}{2}(\det e)^{1/2}\omega_{a,}{}^b{}_c(-2G^{c,d}{}_d + G^{d,c}{}_d + G^{d,}{}_d{}^c + G_{d,}{}^{cd} - G_{d,}{}^{dc}) \\
&= \frac{1}{2}(\det e)^{1/2}\omega_{a,}{}^b{}_c(-2G^{c,d}{}_d + 2G_{d,}{}^{cd}) \\
&= -G_{d,c}{}^c(\det e)^{1/2}\omega_{a,}{}^{bd} + G_{c,d}{}^c(\det e)^{1/2}\omega_{a,}{}^{bd} \quad .
\end{aligned} \tag{3.79}$$

We now have

$$\begin{aligned}
E_a{}^b &= (\det e)^{1/2} e_a{}^\mu \partial_\mu[(\det e)^{1/2}\omega_{d,}{}^{bd}] - (\det e)^{1/2} e_d{}^\nu \partial_\nu((\det e)^{1/2}\omega_{a,}{}^{bd}) \\
&+ G_{a,d}{}^c(\det e)^{1/2}\omega_{c,}{}^{bd} - (\det e)^{1/2}G_{d,a}{}^c\omega_{c,}{}^{bd} - (\det e)\omega_{d,}{}^b{}_c\omega_{a,}{}^{cd} \\
&- \frac{1}{2}G_{a,c}{}^c(\det e)^{1/2}\omega_{d,}{}^{bd} + \frac{1}{2}G_{d,c}{}^c(\det e)^{1/2}\omega_{a,}{}^{bd} \\
&- G_{d,c}{}^c(\det e)^{1/2}\omega_{a,}{}^{bd} + G_{c,d}{}^c(\det e)^{1/2}\omega_{a,}{}^{bd}.
\end{aligned} \tag{3.80}$$

The seventh and eighth terms of equation (3.80) combine into $-\frac{1}{2}G_{d,c}{}^c(\det e)^{1/2}\omega_{a,}{}^{bd}$. The second line of equation (3.80) simplifies as

$$\begin{aligned}
&G_{a,d}{}^c(\det e)^{1/2}\omega_{c,}{}^{bd} - (\det e)^{1/2}G_{d,a}{}^c\omega_{c,}{}^{bd} - (\det e)\omega_{d,}{}^b{}_c\omega_{a,}{}^{cd} \\
&= -(\det e)^{1/2}\omega^{c,bd}[-G_{a,dc} + G_{d,ac}] - (\det e)\omega_{d,}{}^b{}_c\omega_{a,}{}^{cd} \\
&= -(\det e)^{1/2}\omega^{c,bd}[-G_{a,(dc)} + G_{d,(ac)} - G_{a,[dc]} - G_{d,[ca]}] - (\det e)\omega_{c,}{}^b{}_d\omega_{a,}{}^{dc} \\
&= -(\det e)^{1/2}\omega^{c,bd}[-G_{a,(dc)} + G_{d,(ac)} + G_{c,[ad]} - G_{c,[ad]} - G_{a,[dc]} - G_{d,[ca]}] - (\det e)\omega^{c,bd}\omega_{a,dc} \\
&= (\det e)^{1/2}\omega^{c,bd}[G_{a,(dc)} - G_{d,(ac)} - G_{c,[ad]} + G_{c,[ad]} + G_{a,[dc]} + G_{d,[ca]} - \omega_{a,dc}] \\
&= (\det e)^{1/2}\omega^{c,bd}[G_{a,(dc)} - G_{d,(ac)} - G_{c,[ad]} + G_{c,[ad]} + G_{a,[dc]} + G_{d,[ca]} + G_{d,(ca)} - G_{c,(da)} - G_{a,[dc]}] \\
&= (\det e)^{1/2}\omega^{c,bd}[G_{a,(dc)} + G_{d,[ca]} - G_{c,(da)}] \\
&= (\det e)\omega^{c,bd}\omega_{d,ca}.
\end{aligned} \tag{3.81}$$

This gives $E_a{}^b$ to be

$$\begin{aligned} E_a{}^b &= (\det e)^{1/2} e_a{}^\mu \partial_\mu [(\det e)^{1/2} \omega_{d,}{}^{bd}] - (\det e)^{1/2} e_d{}^\nu \partial_\nu ((\det e)^{1/2} \omega_{a,}{}^{bd}) \\ &+ (\det e) \omega^{c,bd} \omega_{d,ca} + G_{c,d}{}^c (\det e)^{1/2} \omega_{a,}{}^{bd} \\ &- \frac{1}{2} G_{a,c}{}^c (\det e)^{1/2} \omega_{d,}{}^{bd} - \frac{1}{2} G_{d,c}{}^c (\det e)^{1/2} \omega_{a,}{}^{bd}. \end{aligned} \tag{3.82}$$

### 3.4.2 Variation of the Ricci Tensor

We now determine the correct $l_1$ extension of $E_{ab}$, now given by equation (3.82), and its variation under $I_c(A_1^{+++})$. This calculation is quite involved, however it simply uses the same principles as those in the calculation of the previous section. Due to its importance we will give the calculation in explicit detail. Due to its size we will often label terms in the equations to keep track of things, using color-coded symbols and superscripts, such as red circles ①, ②, blue circles ①, ②, blue boxes $\square_1$, $\square_2$, etc... to make collecting terms easier.

We begin by replacing the spin connection terms $\omega_{a,bc}$ in equation (3.82) by the corresponding $l_1$-extended spin connection terms $\Omega_{a,bc}$, where $\Omega_{a,bc}$ is defined in equation (3.53). Denoting the result as $\tilde{\mathcal{E}}_{ab}$, this is:

$$\begin{aligned} \tilde{\mathcal{E}}_{ab} &= (\det e)^{1/2} e_a{}^\mu \partial_\mu [(\det e)^{1/2} \Omega_{d,b}{}^d] - (\det e)^{1/2} e^{c\nu} \partial_\nu [(\det e)^{1/2} \Omega_{a,bc}] \\ &+ (\det e) \Omega_{c,b}{}^d \Omega_{d,ca} + (\det e)^{1/2} \Omega_{a,b}{}^d G_{c,d}{}^c \\ &- \frac{1}{2} G_{a,c}{}^c (\det e)^{1/2} \Omega_{d,}{}^{bd} - \frac{1}{2} G_{d,c}{}^c (\det e)^{1/2} \Omega_{a,b}{}^d \\ &= ① + \ldots + ⑥ \end{aligned} \tag{3.83}$$

Here we have numbered the six terms ①, ..., ⑥, as they appear.

#### Varying ①, ..., ⑥

Using the variations

$$\begin{aligned} \delta G_{c,a}{}^b &= 2\Lambda^{eb} \overline{G}_{c,ea} - \delta_a{}^b \Lambda^{e_1 e_2} \overline{G}_{c,e_1 e_2} \\ \delta[(\det e)^{1/2} \Omega_{a,bc}] &= -2\Lambda^e{}_a \overline{G}_{[b,c]e} + 2\Lambda^e{}_c \overline{G}_{[a,b]e} - 2\Lambda^e{}_b \overline{G}_{[a,c]e} \\ &\qquad + 2\eta_{ac} \Lambda^{e_1 e_2} \overline{G}_{[b,e_1]e_2} - 2\eta_{ab} \Lambda^{e_1 e_2} \overline{G}_{[c,e_1]e_2} \quad , \\ \delta[(\det e)^{1/2} \Omega_{a,b}{}^a] &= 2\Lambda^{e_1 e_2} \overline{G}_{[b,e_1]e_2} \quad , \end{aligned} \tag{3.84}$$

the first term, $\textcircled{1}$, varies into

$$
\begin{aligned}
\delta\textcircled{1} &= \delta\{(\det e)^{\frac{1}{2}} e_a{}^{\mu}\partial_\mu[(\det e)^{1/2}\Omega_{d,b}{}^{d}]\} \\
&= (\det e)^{\frac{1}{2}} e_a{}^{\mu}\partial_\mu[2\Lambda^{e_1e_2}\overline{G}_{[b,e_1]e_2}]
\end{aligned}
\tag{3.85}
$$

and the second term, $\textcircled{2}$, varies into

$$
\begin{aligned}
\delta\textcircled{2} &= \delta\{-(\det e)^{\frac{1}{2}} e^{c\nu}\partial_\nu[(\det e)^{1/2}\Omega_{a,bc}]\} \\
&= (\det e)^{\frac{1}{2}} e^{c\nu}\partial_\nu[2\Lambda^{e}{}_{a}\overline{G}_{[b,c]e} - 2\Lambda^{e}{}_{c}\overline{G}_{[a,b]e} + 2\Lambda^{e}{}_{b}\overline{G}_{[a,c]e} - 2\eta_{ac}\Lambda^{e_1e_2}\overline{G}_{[b,e_1]e_2} + 2\eta_{ab}\Lambda^{e_1e_2}\overline{G}_{[c,e_1]e_2}] \\
&= (2.1) + (2.2) + (2.3) + (2.4)^{Cancels\ \textcircled{1}} + (2.5)\ .
\end{aligned}
\tag{3.86}
$$

Their combination thus varies into

$$
\begin{aligned}
\delta[\textcircled{1}+\textcircled{2}] &= (\det e)^{\frac{1}{2}} e^{c\nu}\partial_\nu[2\Lambda^{e}{}_{a}\overline{G}_{[b,c]e} - 2\Lambda^{e}{}_{c}\overline{G}_{[a,b]e} + 2\Lambda^{e}{}_{b}\overline{G}_{[a,c]e} + 2\eta_{ab}\Lambda^{e_1e_2}\overline{G}_{[c,e_1]e_2}] \\
&= 2(\det e)^{\frac{1}{2}} e^{c\nu}\partial_\nu[\Lambda^{e}{}_{a}\overline{G}_{[b,c]e}] - 2(\det e)^{1/2} e^{c\nu}\partial_\nu[\overline{G}_{[a,b]e}\Lambda^{e}{}_{c}] \\
&+ 2(\det e)^{\frac{1}{2}} e^{c\nu}\partial_\nu[\overline{G}_{[a,c]e}\Lambda^{e}{}_{b}] + 2\eta_{ab}(\det e)^{\frac{1}{2}} e^{c\nu}\partial_\nu[\overline{G}_{[c,e_1]e_2} e_{\tau_1}{}^{e_1} e_{\tau_2}{}^{e_2}\Lambda^{\tau_1\tau_2}] \\
&= (2.1) + (2.2) + (2.3) + (2.5)\quad ,
\end{aligned}
\tag{3.87}
$$

where

$$
\begin{aligned}
(2.1) &= 2(\det e)^{\frac{1}{2}} e^{c\nu}\partial_\nu[\Lambda^{ep}\overline{G}_{[b,c]e}\eta_{pa}] \\
&= -\Lambda^{e}{}_{a}\{2(\det e)^{\frac{1}{2}} e^{c\nu}[\partial_\nu\overline{G}_{[c,b]e}] + 2G^{c,}{}_{e}{}^{d}\overline{G}_{[c,b]d}\} - 2\Lambda^{e_1e_2}G^{c,}{}_{e_1a}\overline{G}_{[c,b]e_2}\ ,
\end{aligned}
\tag{3.88}
$$

and

$$
\begin{aligned}
(2.2) &= -2(\det e)^{1/2} e^{c\nu}\partial_\nu[\overline{G}_{[a,b]e}\Lambda^{e}{}_{c}] = -2(\det e)^{1/2} e_c{}^{\nu}\partial_\nu[\overline{G}_{[a,b]e}\Lambda^{e\tau} e_\tau{}^{c}] \\
&= -2(\det e)^{1/2} e_c{}^{\nu}\partial_\nu[\overline{G}_{[a,b]e}\Lambda^{e\tau}] e_\tau{}^{c} - 2(\det e)^{1/2}\overline{G}_{[a,b]e}\Lambda^{e\tau} e_c{}^{\nu}\partial_\nu[e_\tau{}^{c}] \\
&= -2(\det e)^{1/2} e_c{}^{\nu}\partial_\nu[\overline{G}_{[a,b]e}\Lambda^{e\tau}] e_\tau{}^{c\,(2.2.2\ l_1)} - 2\Lambda^{e_1e_2}G_{c,e_1}{}^{c}\ \overline{G}_{[a,b]e_2} \\
&= (2.2.1) - 2(\det e)^{1/2} e_c{}^{\nu}\partial_\nu[\overline{G}_{[a,b]e}\Lambda^{e\tau}] e_\tau{}^{c\,(2.2.2\ l_1)}\quad ,
\end{aligned}
\tag{3.89}
$$

and

$$
\begin{aligned}
(2.3) &= 2(\det e)^{1/2} e^{c\nu}\partial_\nu[\overline{G}_{[a,c]e}\Lambda^{e}{}_{b}] \\
&= -\Lambda^{e}{}_{b}\{2(\det e)^{1/2}[e^{c\nu}\partial_\nu\overline{G}_{[c,a]e}] + 2G^{c,}{}_{e}{}^{d}\overline{G}_{[c,a]d}\} - 2\Lambda^{e_1e_2}G^{c,}{}_{e_1b}\overline{G}_{[c,a]e_2}\quad ,
\end{aligned}
\tag{3.90}
$$

and

$$
\begin{aligned}
(2.5) &= 2\eta_{ab}(\det e)^{1/2} e^{c\nu}\partial_\nu[\Lambda^{e_1e_2}\overline{G}_{[c,e_1]e_2}] \\
&= \eta_{ab}\Lambda^{e_1e_2}\{2(\det e)^{1/2}[e^{c\nu}\partial_\nu\overline{G}_{[c,e_1]e_2}] + 2G^{c,}{}_{e_1}{}^{d}\overline{G}_{[c,d]e_2} + 2G^{c,}{}_{e_2}{}^{d}\overline{G}_{[c,e_1]d}\}\quad .
\end{aligned}
\tag{3.91}
$$

The total variation of ① + ② now reads as

$$
\begin{aligned}
&\delta[①+②] \\
&\quad = (2.1) + (2.2) + (2.3) + (2.5) \\
&\quad = -\Lambda^{e}{}_{a}\{2(\det e)^{\frac{1}{2}} e^{c\nu}[\partial_\nu \overline{G}_{[c,b]e}] + 2G^{c,}{}_{e}{}^{d}\overline{G}_{[c,b]d}\} - 2\Lambda^{e_1e_2}G^{c,}{}_{e_1a}\overline{G}_{[c,b]e_2} \\
&\quad - 2\Lambda^{e_1e_2}G_{c,e_1}{}^{c}\overline{G}_{[a,b]e_2} - 2(\det e)^{1/2}e_c{}^{\nu}\partial_\nu[\overline{G}_{[a,b]e}\Lambda^{e\tau}]e_\tau{}^{c(2.2.2\ l_1)} \\
&\quad - \Lambda^{e}{}_{b}\{2(\det e)^{1/2}[e^{c\nu}\partial_\nu\overline{G}_{[c,a]e}] + 2G^{c,}{}_{e}{}^{d}\overline{G}_{[c,a]d}\} - 2\Lambda^{e_1e_2}G^{c,}{}_{e_1b}\overline{G}_{[c,a]e_2} \\
&\quad + \eta_{ab}\Lambda^{e_1e_2}\{2(\det e)^{1/2}[e^{c\nu}\partial_\nu\overline{G}_{[c,e_1]e_2}] + 2G^{c,}{}_{e_1}{}^{d}\overline{G}_{[c,d]e_2} + 2G^{c,}{}_{e_2}{}^{d}\overline{G}_{[c,e_1]d}\} \\
&\quad = -\Lambda_{ea}\{4(\det e)^{\frac{1}{2}} e^{\nu[c}[\partial_\nu\overline{G}_{[c,b]}{}^{e]}] + 2G^{c,e}{}_{d}\overline{G}_{[c,b]}{}^{d}\} \quad (eom) \\
&\qquad - \Lambda_{eb}\{4(\det e)^{1/2}[e^{\nu[c}\partial_\nu\overline{G}_{[c,a]}{}^{e]}] + 2G^{c,e}{}_{d}\overline{G}_{[c,a]}{}^{d}\} \quad (eom) \qquad (3.92) \\
&\qquad + \eta_{ab}\Lambda^{e_1e_2}\{4(\det e)^{1/2}[e^{\nu[c}\partial_\nu\overline{G}_{[c,e_1]}{}^{e_2]}] + 2G^{c,}{}_{e_1}{}^{d}\overline{G}_{[c,d]e_2}\ \square_1 + 2G^{c,}{}_{e_2}{}^{d}\overline{G}_{[c,e_1]d}\ \square_2\} \quad (eom) \\
&\qquad - 2\Lambda^{e_1e_2}G_{c,e_1}{}^{c}\overline{G}_{[a,b]e_2} \quad \triangle \\
&\qquad - 2\Lambda^{e_1e_2}G^{c,}{}_{e_1a}\overline{G}_{[c,b]e_2} \quad \square \\
&\qquad - 2\Lambda^{e_1e_2}G^{c,}{}_{e_1b}\overline{G}_{[c,a]e_2} \quad \bigcirc \\
&\qquad - 2(\det e)^{1/2}e_c{}^{\nu}\partial_\nu[\overline{G}_{[a,b]e}\Lambda^{e\tau}]e_\tau{}^{c(2.2.2\ l_1)} \\
&\qquad - 2\Lambda_{ae}(\det e)^{1/2}e^{\nu e}[\partial_\nu\overline{G}_{[c,b]}{}^{c}] - 2\Lambda_{be}(\det e)^{1/2}[e^{\nu e}\partial_\nu\overline{G}_{[c,a]}{}^{c}] + 2\eta_{ab}\Lambda^{e_1}{}_{e_2}(\det e)^{1/2}[e^{\nu e_2}\partial_\nu\overline{G}_{[c,e_1]}{}^{c}] \ (eom\ l_1)
\end{aligned}
$$

We now vary the third term, ③, to find

$$
\begin{aligned}
\delta③ = &-2(\det e)^{1/2}\Omega^{d,c}{}_{a}\Lambda^{e}{}_{c}\overline{G}_{[b,d]e}\boxed{1} + 2(\det e)^{1/2}\Omega_{d,}{}^{c}{}_{a}\Lambda^{ed}\overline{G}_{[c,b]e}\boxed{2} - 2(\det e)^{1/2}\Omega^{d,c}{}_{a}\Lambda^{e}{}_{b}\overline{G}_{[c,d]e}{}^{\triangle_1} \\
&+ 2(\det e)^{1/2}\Omega_{c,}{}^{c}{}_{a}\Lambda^{e_1e_2}\overline{G}_{[b,e_1]e_2}\boxed{1} - 2(\det e)^{1/2}\Omega^{d,}{}_{ba}\Lambda^{e_1e_2}\overline{G}_{[d,e_1]e_2}{}^{②} \\
&- 2(\det e)^{1/2}\Omega^{c,}{}_{b}{}^{d}\Lambda^{e}{}_{d}\overline{G}_{[c,a]e}{}^{①} + 2(\det e)^{1/2}\Omega^{c,}{}_{b}{}^{d}\Lambda^{e}{}_{a}\overline{G}_{[d,c]e}{}^{①} - 2(\det e)^{1/2}\Omega^{c,}{}_{b}{}^{d}\Lambda^{e}{}_{c}\overline{G}_{[d,a]e}{}^{②} \\
&+ 2(\det e)^{1/2}\Omega^{c,}{}_{ba}\Lambda^{e_1e_2}\overline{G}_{[c,e_1]e_2}{}^{①} - 2(\det e)^{1/2}\Omega^{c,}{}_{bc}\Lambda^{e_1e_2}\overline{G}_{[a,e_1]e_2}{}^{①} \quad . \qquad (3.93)
\end{aligned}
$$

The variation of the fourth term, ④, is given by

$$
\begin{aligned}
\delta④ = &-2\Lambda^{e}{}_{a}G_{c,}{}^{dc}\overline{G}_{[b,d]e}{}^{②} + 2\Lambda^{ed}G_{c,d}{}^{c}\overline{G}_{[a,b]e}{}^{\triangle_1} - 2\Lambda^{e}{}_{b}G_{c,}{}^{dc}\overline{G}_{[a,d]e}{}^{\triangle_2} \\
&+ 2\Lambda^{e_1e_2}G_{c,a}{}^{c}\overline{G}_{[b,e_1]e_2}\boxed{2} - 2\eta_{ab}\Lambda^{e_1e_2}G_{c,}{}^{dc}\overline{G}_{[d,e_1]e_2}{}^{\triangle_1} \qquad (3.94) \\
&+ 2\Lambda^{e_1e_2}(\det e)^{1/2}\Omega_{a,b}{}^{d}\overline{G}_{e_1,e_2d}\boxed{l_1\ from\ \delta④} - \Lambda^{e_1e_2}(\det e)^{1/2}\Omega_{a,b}{}^{d}\overline{G}_{d,e_1e_2}{}^{③}.
\end{aligned}
$$

The variation of the fifth term, ⑤, is given by

$$
\delta⑤ = \Lambda^{e_1e_2}\overline{G}_{a,e_1e_2}(\det e)^{1/2}\Omega_{d,b}{}^{d②} - G_{a,c}{}^{c}\Lambda^{e_1e_2}\overline{G}_{[b,e_1]e_2}\boxed{3} \quad , \qquad (3.95)
$$

and the variation of the sixth term, $\textcircled{6}$, is given by

$$\begin{aligned}\delta\textcircled{6} &= \Lambda^{e_1e_2}(\det e)^{1/2}\Omega_{a,b}{}^{d}\overline{G}_{d,e_1e_2}{}^{\textcircled{4}} \\ &\quad + \Lambda^{e}{}_{a}G^{d,c}{}_{c}\overline{G}_{[b,d]e}{}^{\textcircled{3}} - \Lambda^{ed}G_{d,c}{}^{c}\overline{G}_{[a,b]e}{}^{\triangle_2} + \Lambda^{e}{}_{b}G^{d,c}{}_{c}\overline{G}_{[a,d]e}{}^{\triangle_3} \\ &\quad - \Lambda^{e_1e_2}G_{a,c}{}^{c}\overline{G}_{[b,e_1]e_2}{}^{\boxed{4}} + \eta_{ab}\Lambda^{e_1e_2}G^{d,c}{}_{c}\overline{G}_{[d,e_1]e_2}{}^{\triangle_2} \ .\end{aligned} \tag{3.96}$$

### Collecting $\Lambda^{e}{}_{a}$ Terms

We now collect all $\Lambda^{e}{}_{a}$ terms. The $\Lambda^{e}{}_{a}$ terms from $\delta[\textcircled{1}+\textcircled{2}]$ are

$$\begin{aligned}&-\Lambda_{ea}\{4(\det e)^{\frac{1}{2}}e^{\nu[c}[\partial_\nu\overline{G}_{[c,b]}{}^{e]}] + 2G^{c,e}{}_{d}\overline{G}_{[c,b]}{}^{d}\} \quad \textit{(eom)} \\ &-2\Lambda_{ae}(\det e)^{1/2}e^{\nu e}\partial_\nu\overline{G}_{[c,b]}{}^{c} \quad (\textit{eom } l_1^a)\end{aligned} \tag{3.97}$$

and, from $\delta\textcircled{3}$ to $\delta\textcircled{6}$, are

$$\begin{aligned}&+2(\det e)^{1/2}\Omega^{c,}{}_{b}{}^{d}\Lambda^{e}{}_{a}\overline{G}_{[d,c]e}{}^{\textcircled{1}} - 2\Lambda^{e}{}_{a}G_{c,}{}^{dc}\overline{G}_{[b,d]e}{}^{\textcircled{2}} + \Lambda^{e}{}_{a}G^{d,c}{}_{c}\overline{G}_{[b,d]e}{}^{\textcircled{3}} \\ &\quad = \Lambda^{e}{}_{a}\{2(\det e)^{1/2}\Omega^{c,}{}_{b}{}^{d}\overline{G}_{[d,c]e}{}^{\textcircled{1}} - 2G_{c,}{}^{dc}\overline{G}_{[b,d]e}{}^{\textcircled{2}} + G^{d,c}{}_{c}\overline{G}_{[b,d]e}{}^{\textcircled{3}}\} \\ &\quad = \Lambda_{ea}\{2G^{c,d}{}_{b}\overline{G}_{[c,d]}{}^{e\textcircled{1}} - 2G_{c,}{}^{dc}\overline{G}_{[b,d]}{}^{e\textcircled{2}} + G^{d,c}{}_{c}\overline{G}_{[b,d]}{}^{e\textcircled{3}}\} \ .\end{aligned} \tag{3.98}$$

These add to give

$$\begin{aligned}\Lambda^{e}{}_{a}\text{ terms} &= -\Lambda_{ea}\{4(\det e)^{\frac{1}{2}}e^{\nu[c}[\partial_\nu\overline{G}_{[c,b]}{}^{e]}] + 2G^{c,e}{}_{d}\overline{G}_{[c,b]}{}^{d}\} \quad \textit{(eom)} \\ &\quad + \Lambda_{ea}\{2G^{c,d}{}_{b}\overline{G}_{[c,d]}{}^{e\textcircled{1}} - 2G_{c,}{}^{dc}\overline{G}_{[b,d]}{}^{e\textcircled{2}} + G^{d,c}{}_{c}\overline{G}_{[b,d]}{}^{e\textcircled{3}}\} \\ &\quad - 2\Lambda_{ae}(\det e)^{1/2}e^{\nu e}\partial_\nu\overline{G}_{[c,b]}{}^{c} \quad (\textit{eom } l_1^a) \\ &= -\Lambda_{ea}\{4(\det e)^{\frac{1}{2}}e^{\nu[c}\partial_\nu\overline{G}_{[c,b]}{}^{e]} + 4G^{[d,e]c}\overline{G}_{[d,b]c}\} \quad \textit{(eom)} \\ &\quad + \Lambda_{ea}\{4G^{[c,|d|}{}_{b}\overline{G}_{[c,d]}{}^{e]\textcircled{1}} - 4G^{[c,|d|}{}_{c}\overline{G}_{[b,d]}{}^{e]\textcircled{2}} + 2G^{[d,|c|}{}_{c}\overline{G}_{[b,d]}{}^{e]\textcircled{3}}\} \\ &\quad - 2\Lambda_{ae}(\det e)^{1/2}e^{\nu e}\partial_\nu\overline{G}_{[c,b]}{}^{c} - 2\Lambda_{ea}G^{e,c}{}_{d}\overline{G}_{[c,b]}{}^{d} \quad (\textit{eom } l_1^a) \\ &\quad + 2\Lambda_{ea}G^{e,d}{}_{b}\overline{G}_{[c,d]}{}^{c\textcircled{1}\ l_1} - 2\Lambda_{ea}G^{e,d}{}_{c}\overline{G}_{[b,d]}{}^{c\textcircled{2}\ l_1} + \Lambda_{ea}G^{e,c}{}_{c}\overline{G}_{[b,d]}{}^{d\textcircled{3}\ l_1} \\ &= -4\Lambda_{ea}\{(\det e)^{\frac{1}{2}}e^{\nu[c}\partial_\nu\overline{G}_{[c,b]}{}^{e]} + G^{[d,e]c}\overline{G}_{[d,b]c} - G^{[c,|d|}{}_{b}\overline{G}_{[c,d]}{}^{e]\textcircled{1}} \\ &\qquad\qquad - G^{[c,|d|}{}_{c}\overline{G}_{[d,b]}{}^{e]\textcircled{2}} + \frac{1}{2}G^{[d,|c|}{}_{c}\overline{G}_{[d,b]}{}^{e]\textcircled{3}}\} \\ &\quad - 2\Lambda_{ae}(\det e)^{1/2}e^{\nu e}\partial_\nu\overline{G}_{[c,b]}{}^{c} - 2\Lambda_{ea}G^{e,c}{}_{d}\overline{G}_{[c,b]}{}^{d} \quad (\textit{eom } l_1^a) \\ &\quad + 2\Lambda_{ea}G^{e,d}{}_{b}\overline{G}_{[c,d]}{}^{c\textcircled{1}\ l_1} - 2\Lambda_{ea}G^{e,d}{}_{c}\overline{G}_{[b,d]}{}^{c\textcircled{2}\ l_1} + \Lambda_{ea}G^{e,c}{}_{c}\overline{G}_{[b,d]}{}^{d\textcircled{3}\ l_1}\end{aligned} \tag{3.99}$$

The $-4\Lambda_{ae}$ coefficient $\{\ldots\}$ has the appearance of an equation of motion (though it has not yet been fully processed), and the additional terms are all $l_1$ or modulo terms.

**Collecting $\Lambda^e{}_b$ Terms**

We now collect all $\Lambda^e{}_b$ terms. The $\Lambda^e{}_b$ terms from $\delta[①+②]$ are

$$\begin{aligned} &-\Lambda_{eb}\{4(\det e)^{1/2}[e^{\nu[c}\partial_\nu \overline{G}_{[c,a]}{}^{e]}] + 2G^{c,e}{}_d\overline{G}_{[c,a]}{}^d\} \quad \textit{(eom)} \\ &-2\Lambda_{be}(\det e)^{1/2}[e^{\nu e}\partial_\nu \overline{G}_{[c,a]}{}^c](\textit{eom } l_1^b) \end{aligned} \tag{3.100}$$

and, from $\delta③$ to $\delta⑥$, labelled with blue triangles, we have:

$$\begin{aligned} &-2(\det e)^{1/2}\Omega^{d,c}{}_a\Lambda^e{}_b\overline{G}_{[c,d]e}{}^{\triangle_1} - 2\Lambda^e{}_bG_{c,}{}^{dc}\overline{G}_{[a,d]e}{}^{\triangle_2} + \Lambda^e{}_bG^{d,c}{}_c\overline{G}_{[a,d]e}{}^{\triangle_3} \\ &= \Lambda^e{}_b\{-2(\det e)^{1/2}\Omega^{d,c}{}_a\overline{G}_{[c,d]e}{}^{\triangle_1} + 2G_{c,}{}^{dc}\overline{G}_{[d,a]e}{}^{\triangle_2} - G^{d,c}{}_c\overline{G}_{[d,a]e}{}^{\triangle_3}\} \\ &= \Lambda^e{}_b\{2(\det e)^{1/2}\Omega^{d,c}{}_a\overline{G}_{[d,c]e}{}^{\triangle_1} + 2G_{c,}{}^{dc}\overline{G}_{[d,a]e}{}^{\triangle_2} - G^{d,c}{}_c\overline{G}_{[d,a]e}{}^{\triangle_3}\} \\ &= \Lambda^e{}_b\{2G_{d,ca}\overline{G}^{[d,c]}{}_e{}^{\triangle_1} - 2G_{c,}{}^{dc}\overline{G}_{[a,d]e}{}^{\triangle_2} + G^{d,c}{}_c\overline{G}_{[a,d]e}{}^{\triangle_3}\} \quad , \end{aligned} \tag{3.101}$$

(where we used:

$$\begin{aligned} 2\Lambda_{be}(\det e)^{1/2}\omega_{d,ca}\overline{G}^{[d,c]e} &= 2\Lambda_{be}(\det e)^{1/2}\omega_{d,ac}\overline{G}^{[c,d]e} \\ &= 2\Lambda_{be}G_{d,ca}\overline{G}^{[d,c]e}) \quad . \end{aligned} \tag{3.102}$$

These add to give

$$\begin{aligned} \Lambda^e{}_b \text{ terms} &= -\Lambda_{eb}\{4(\det e)^{1/2}[e^{\nu[c}\partial_\nu\overline{G}_{[c,a]}{}^{e]}] + 2G^{c,e}{}_d\overline{G}_{[c,a]}{}^d\} \quad \textit{(eom)} \\ &\quad + \Lambda^e{}_b\{2G_{d,ca}\overline{G}^{[d,c]e\triangle_1} + 2G_{c,}{}^{dc}\overline{G}_{[d,a]e}{}^{\triangle_2} - G^{d,c}{}_c\overline{G}_{[d,a]e}{}^{\triangle_3}\} \\ &\quad - 2\Lambda_{be}(\det e)^{1/2}[e^{\nu e}\partial_\nu\overline{G}_{[c,a]}{}^c](\textit{eom } l_1^b) \\ &= -\Lambda_{eb}\{4(\det e)^{1/2}e^{\nu[c}\partial_\nu\overline{G}_{[c,a]}{}^{e]} + 4G^{[c,e]}{}_d\overline{G}_{[c,a]}{}^d\} \quad \textit{(eom)} \\ &\quad + \Lambda_{eb}\{4G^{[d,|c|}{}_a\overline{G}_{[d,c]}{}^{e]\triangle_1} + 4G^{[c,|d|}{}_c\overline{G}_{[d,a]}{}^{e]\triangle_2} - 2G^{[d,|c|}{}_c\overline{G}_{[d,a]}{}^{e]\triangle_3}\} \\ &\quad - 2\Lambda_{be}(\det e)^{1/2}e^{\nu e}\partial_\nu\overline{G}_{[c,a]}{}^c - 2\Lambda_{eb}G^{e,c}{}_d\overline{G}_{[c,a]}{}^d \quad (\textit{eom } l_1^b) \\ &\quad + 2\Lambda_{eb}G^{e,c}{}_a\overline{G}_{[d,c]}{}^{d\triangle_1\ l_1} + 2\Lambda_{eb}G^{e,d}{}_c\overline{G}_{[d,a]}{}^{c\triangle_2\ l_1} - \Lambda_{eb}G^{e,c}{}_c\overline{G}_{[d,a]}{}^{d\triangle_3\ l_1} \\ &= -4\Lambda_{eb}\{(\det e)^{1/2}e^{\nu[c}\partial_\nu\overline{G}_{[c,a]}{}^{e]} + G^{[c,e]}{}_d\overline{G}_{[c,a]}{}^d - G^{[d,|c|}{}_a\overline{G}_{[d,c]}{}^{e]\triangle_1} \\ &\qquad\qquad - G^{[c,|d|}{}_c\overline{G}_{[d,a]}{}^{e]\triangle_2} + \frac{1}{2}G^{[d,|c|}{}_c\overline{G}_{[d,a]}{}^{e]\triangle_3}\} \\ &\quad - 2\Lambda_{be}(\det e)^{1/2}e^{\nu e}\partial_\nu\overline{G}_{[c,a]}{}^c - 2\Lambda_{eb}G^{e,c}{}_d\overline{G}_{[c,a]}{}^d \quad (\textit{eom } l_1^b) \\ &\quad + 2\Lambda_{eb}G^{e,d}{}_a\overline{G}_{[c,d]}{}^{c\triangle_1\ l_1} - 2\Lambda_{eb}G^{e,d}{}_c\overline{G}_{[a,d]}{}^{c\triangle_2\ l_1} + \Lambda_{eb}G^{e,c}{}_c\overline{G}_{[a,d]}{}^{d\triangle_3\ l_1} \end{aligned} \tag{3.103}$$

The $-4\Lambda_{eb}$ coefficient $\{...\}$ has the appearance of an equation of motion (though it has not yet been fully evaluated), and is similar in structure to the $-4\Lambda_{eb}$ coefficient, while the additional terms are all $l_1$ or modulo terms.

### Collecting $\Lambda^{e_1e_2}$ Terms

We now collect all $\Lambda^{e_1e_2}$ terms. This step is more involved, however our goal is to end up a potential equation of motion which is the same as what we have found above. We begin by first simplifying terms from $\delta\textcircled{3}$ to $\delta\textcircled{6}$. Collecting green circles, we find

$$\begin{aligned}\textcircled{c} &= 2(\det e)^{1/2}\Omega^{c,}{}_{ba}\Lambda^{e_1e_2}\overline{G}_{[c,e_1]e_2}{}^{\textcircled{1}} - 2(\det e)^{1/2}\Omega^{d,}{}_{ba}\Lambda^{e_1e_2}\overline{G}_{[d,e_1]e_2}{}^{\textcircled{2}} \\ &\quad - \Lambda^{e_1e_2}(\det e)^{1/2}\Omega_{a,b}{}^{d}\overline{G}_{d,e_1e_2}{}^{\textcircled{3}} + \Lambda^{e_1e_2}(\det e)^{1/2}\Omega_{a,b}{}^{d}\overline{G}_{d,e_1e_2}{}^{\textcircled{4}} \\ &= \Lambda^{e_1e_2}(\det e)^{\frac{1}{2}}\{2\Omega^{c,}{}_{ba}\overline{G}_{[c,e_1]e_2}{}^{\textcircled{1}} - 2\Omega^{d,}{}_{ba}\overline{G}_{[d,e_1]e_2}{}^{\textcircled{2}} - \Omega_{a,b}{}^{d}\overline{G}_{d,e_1e_2}{}^{\textcircled{3}} + \Omega_{a,b}{}^{d}\overline{G}_{d,e_1e_2}{}^{\textcircled{4}}\} \\ &= 0.\end{aligned} \tag{3.104}$$

Collecting blue boxes, we find:

$$\begin{aligned}\boxed{f} &= 2(\det e)^{1/2}\Omega_{c,}{}^{c}{}_{a}\Lambda^{e_1e_2}\overline{G}_{[b,e_1]e_2}{}^{\boxed{1}} + 2\Lambda^{e_1e_2}G_{c,a}{}^{c}\overline{G}_{[b,e_1]e_2}{}^{\boxed{2}} \\ &\quad - G_{a,c}{}^{c}\Lambda^{e_1e_2}\overline{G}_{[b,e_1]e_2}{}^{\boxed{3}} - \Lambda^{e_1e_2}G_{a,c}{}^{c}\overline{G}_{[b,e_1]e_2}{}^{\boxed{4}} \\ &= \Lambda^{e_1e_2}\{2(\det e)^{1/2}\Omega_{c,}{}^{c}{}_{a}\overline{G}_{[b,e_1]e_2}{}^{\boxed{1}} + 2G_{c,a}{}^{c}\overline{G}_{[b,e_1]e_2}{}^{\boxed{2}} - G_{a,c}{}^{c}\overline{G}_{[b,e_1]e_2}{}^{\boxed{3}} - G_{a,c}{}^{c}\overline{G}_{[b,e_1]e_2}{}^{\boxed{4}}\} \\ &= 0 \quad .\end{aligned} \tag{3.105}$$

We now collect blue circles

$$\begin{aligned}\textcircled{b} &= -2(\det e)^{1/2}\Omega^{c,}{}_{bc}\Lambda^{e_1e_2}\overline{G}_{[a,e_1]e_2}{}^{\textcircled{1}} + \Lambda^{e_1e_2}\overline{G}_{a,e_1e_2}(\det e)^{1/2}\Omega_{d,b}{}^{d\textcircled{2}} \\ &= -2(\det e)^{1/2}\Omega^{c,}{}_{bc}\Lambda^{e_1e_2}\overline{G}_{[a,e_1]e_2}{}^{\textcircled{1}} + 2\Lambda^{e_1e_2}\overline{G}_{a,e_1e_2}(\det e)^{1/2}\Omega_{d,b}{}^{d\textcircled{2}} \\ &\quad + \Lambda^{e_1e_2}\overline{G}_{e_1,ae_2}(\det e)^{1/2}\Omega_{d,b}{}^{d\textcircled{2}l_1^{(3)}} \\ &= \Lambda^{e_1e_2}\overline{G}_{e_1,ae_2}(\det e)^{1/2}\Omega_{d,b}{}^{d\textcircled{2}l_1^{(3)}} \quad ,\end{aligned} \tag{3.106}$$

and black circles

$$\begin{aligned}\textcircled{d} &= -2(\det e)^{1/2}\Omega^{c,}{}_{b}{}^{d}\Lambda^{e}{}_{d}\overline{G}_{[c,a]e}{}^{\textcircled{1}} - 2(\det e)^{1/2}\Omega^{c,}{}_{b}{}^{d}\Lambda^{e}{}_{c}\overline{G}_{[d,a]e}{}^{\textcircled{2}} \\ &= 2\Lambda^{e}{}_{d}\overline{G}_{[a,c]e}\{(\det e)^{1/2}\Omega^{c,}{}_{b}{}^{d\textcircled{1}} + (\det e)^{1/2}\Omega^{d,}{}_{b}{}^{c\textcircled{2}}\} \\ &= 2\Lambda^{e_1}{}_{e_2}\overline{G}_{[a,c]e_1}\{(\det e)^{1/2}\Omega^{c,}{}_{b}{}^{e_2\textcircled{1}} + (\det e)^{1/2}\Omega^{e_2,}{}_{b}{}^{c\textcircled{2}}\} \quad ,\end{aligned} \tag{3.107}$$

and black boxes:

$$\begin{aligned}\boxed{e} &= -2(\det e)^{1/2}\Omega^{d,c}{}_{a}\Lambda^{e}{}_{c}\overline{G}_{[b,d]e}{}^{\boxed{1}} + 2(\det e)^{1/2}\Omega_{d,}{}^{c}{}_{a}\Lambda^{ed}\overline{G}_{[c,b]e}{}^{\boxed{2}} \\ &= -2(\det e)^{1/2}\Omega^{c,d}{}_{a}\Lambda^{e}{}_{d}\overline{G}_{[b,c]e}{}^{\boxed{1}} - 2(\det e)^{1/2}\Omega_{d,}{}^{c}{}_{a}\Lambda^{ed}\overline{G}_{[b,c]e}{}^{\boxed{2}} \\ &= -2\Lambda^{e}{}_{d}\overline{G}_{[b,c]e}\{(\det e)^{1/2}\Omega^{c,d}{}_{a}{}^{\boxed{1}} + (\det e)^{1/2}\Omega^{d,c}{}_{a}\Lambda^{ed\boxed{2}}\} \quad ,\end{aligned} \tag{3.108}$$

and red triangles:

$$\begin{aligned} \triangle_g &= +2\Lambda^{ed} G_{c,d}{}^{c}\overline{G}_{[a,b]e}{}^{\triangle_1} - \Lambda^{ed} G_{d,c}{}^{c}\overline{G}_{[a,b]e}{}^{\triangle_2} \\ &= \Lambda^{e_1e_2}\overline{G}_{[a,b]e_1}\{2G_{c,e_2}{}^{c\triangle_1} - G_{e_2,c}{}^{c\triangle_2}\} \quad , \end{aligned} \tag{3.109}$$

and green triangles

$$\begin{aligned} \triangle_i &= -2\eta_{ab}\Lambda^{e_1e_2} G_{c,}{}^{dc}\overline{G}_{[d,e_1]e_2}{}^{\triangle_1} + \eta_{ab}\Lambda^{e_1e_2} G^{d,c}{}_{c}\overline{G}_{[d,e_1]e_2}{}^{\triangle_2} \\ &= \Lambda^{e_1e_2}\{-2\eta_{ab} G_{c,}{}^{dc}\overline{G}_{[d,e_1]e_2}{}^{\triangle_1} + \eta_{ab} G^{d,c}{}_{c}\overline{G}_{[d,e_1]e_2}{}^{\triangle_2}\} \quad . \end{aligned} \tag{3.110}$$

These terms add up to give

$$\begin{aligned} &+ \Lambda^{e_1e_2}\overline{G}_{e_1,ae_2}(\det e)^{1/2}\Omega_{d,b}{}^{d ② l_1^{(3)}} + 2\Lambda^{e_1}{}_{e_2}\overline{G}_{[a,c]e_1}\{(\det e)^{1/2}\Omega^{c,}{}_{b}{}^{e_2 ①} + (\det e)^{1/2}\Omega^{e_2,}{}_{b}{}^{c ②}\} \\ &\quad + 2\Lambda^{e_1}{}_{e_2}\overline{G}_{[b,c]e_1}\{(\det e)^{1/2}\Omega^{c,}{}_{a}{}^{e_2 \boxed{1}} + (\det e)^{1/2}\Omega^{e_2,}{}_{a}{}^{c \boxed{2}}\} + \Lambda^{e_1e_2}\overline{G}_{[a,b]e_1}\{2G_{c,e_2}{}^{c\triangle_1} - G_{e_2,c}{}^{c\triangle_2}\} \\ &\quad + \Lambda^{e_1e_2}\{-2\eta_{ab} G_{c,}{}^{dc}\overline{G}_{[d,e_1]e_2}{}^{\triangle_1} + \eta_{ab} G^{d,c}{}_{c}\overline{G}_{[d,e_1]e_2}{}^{\triangle_2}\} \quad . \end{aligned} \tag{3.111}$$

All that's left are the $\Lambda^{e_1e_2}$ terms coming from $\delta[① + ②]$ and the remaining ones from $\delta④$

$$\begin{aligned} &\eta_{ab}\Lambda^{e_1e_2}\{4(\det e)^{1/2}[e^{\nu[c}\partial_\nu\overline{G}_{[c,e_1]}{}^{e_2]}] + 2G^{c,}{}_{e_1}{}^{d}\overline{G}_{[c,d]e_2}\,\square_1 + 2G^{c,}{}_{e_2}{}^{d}\overline{G}_{[c,e_1]d}\,\square_2\} \quad (eom) \\ &\quad - 2\Lambda^{e_1e_2} G_{c,e_1}{}^{c}\overline{G}_{[a,b]e_2} \quad \triangle \quad - 2\Lambda^{e_1e_2} G^{c,}{}_{e_1a}\overline{G}_{[c,b]e_2} \\ &\quad - 2\Lambda^{e_1e_2} G^{c,}{}_{e_1b}\overline{G}_{[c,a]e_2} \quad \bigcirc \quad - 2(\det e)^{1/2} e_c{}^{\nu}\partial_\nu[\overline{G}_{[a,b]e}\Lambda^{e\tau}]e_\tau{}^{c (2.2.2\ l_1)} \\ &\quad + 2\eta_{ab}\Lambda^{e_1}{}_{e_2}(\det e)^{1/2} e^{\nu e_2}\partial_\nu\overline{G}_{[c,e_1]}{}^{c} \quad (eom\ l_1^{\eta_{ab}}) \\ &\quad + 2\Lambda^{e_1e_2}(\det e)^{1/2}\Omega_{a,b}{}^{d}\overline{G}_{e_1,e_2d}\boxed{l_1\ from\ \delta④} \quad . \end{aligned} \tag{3.112}$$

On adding equation (3.111) and (3.112), we have thus collected all $\Lambda^{e_1e_2}$ terms:

$\Lambda^{e_1e_2}$ terms

$$\begin{aligned} &= \eta_{ab}\Lambda^{e_1e_2}\{4(\det e)^{1/2}[e^{\nu[c}\partial_\nu\overline{G}_{[c,e_1]}{}^{e_2]}] + 2G^{c,}{}_{e_1}{}^{d}\overline{G}_{[c,d]e_2}\,\square_1 + 2G^{c,}{}_{e_2}{}^{d}\overline{G}_{[c,e_1]d}\,\square_2\} \quad (eom) \\ &\quad + \eta_{ab}\Lambda^{e_1e_2}\{-2G_{c,}{}^{dc}\overline{G}_{[d,e_1]e_2}{}^{\triangle_1} + G^{d,c}{}_{c}\overline{G}_{[d,e_1]e_2}{}^{\triangle_2}\} \\ &\quad - 2\Lambda^{e_1e_2} G_{c,e_1}{}^{c}\overline{G}_{[a,b]e_2} \quad \triangle \quad + \Lambda^{e_1e_2}\overline{G}_{[a,b]e_2}\{2G_{c,e_1}{}^{c\triangle_1} - G_{e_1,c}{}^{c\triangle_2}\} \\ &\quad + 2\Lambda^{e_1e_2} G^{c,}{}_{e_1a}\overline{G}_{[b,c]e_2} \quad \square \quad + 2\Lambda^{e_1}{}_{e_2}\overline{G}_{[b,c]e_1}\{(\det e)^{1/2}\Omega^{c,}{}_{a}{}^{e_2 \boxed{1}} + (\det e)^{1/2}\Omega^{e_2,}{}_{a}{}^{c \boxed{2}}\} \\ &\quad + 2\Lambda^{e_1e_2} G^{c,}{}_{e_1b}\overline{G}_{[a,c]e_2} \quad \bigcirc \quad + 2\Lambda^{e_1}{}_{e_2}\overline{G}_{[a,c]e_1}\{(\det e)^{1/2}\Omega^{c,}{}_{b}{}^{e_2 ①} + (\det e)^{1/2}\Omega^{e_2,}{}_{b}{}^{c ②}\} \\ &\quad - 2(\det e)^{1/2} e_c{}^{\nu}\partial_\nu[\overline{G}_{[a,b]e}\Lambda^{e\tau}]e_\tau{}^{c (2.2.2\ l_1)} \\ &\quad + 2\eta_{ab}\Lambda^{e_1}{}_{e_2}(\det e)^{1/2} e^{\nu e_2}\partial_\nu\overline{G}_{[c,e_1]}{}^{c} \quad (eom\ l_1^{\eta_{ab}}) \quad + 2\Lambda^{e_1e_2}(\det e)^{1/2}\Omega_{a,b}{}^{d}\overline{G}_{e_1,e_2d}\boxed{l_1\ from\ \delta④} \\ &\quad + \Lambda^{e_1e_2}\overline{G}_{e_1,ae_2}(\det e)^{1/2}\Omega_{d,b}{}^{d ② l_1^{(3)}} \end{aligned} \tag{3.113}$$

### Simplifying $\Lambda^{e_1e_2}$ Terms Further

We still need to process this result. We first evaluate the black boxes in equation (3.113):

$$\begin{aligned}
\square &= 2\Lambda^{e_1}{}_{e_2}\overline{G}_{[b,c]e_1}\{(\det e)^{1/2}\Omega^{c,}{}_{a}{}^{e_2}\boxed{1} + (\det e)^{1/2}\Omega^{e_2,}{}_{a}{}^{c}\boxed{2}\} \\
&= 2\Lambda^{e_1}{}_{e_2}\overline{G}_{[b,c]e_1}\{-G_{a,}{}^{(e_2c)} + G^{e_2,}{}_{(a}{}^{c)} + G^{c,}{}_{[a}{}^{e_2]} - G_{a,}{}^{(ce_2)} + G^{c,}{}_{(a}{}^{e_2)} + G^{e_2,}{}_{[a}{}^{c]}\} \\
&= 2\Lambda^{e_1}{}_{e_2}\overline{G}_{[b,c]e_1}\{-2G_{a,}{}^{(e_2c)} + G^{e_2,}{}_{(a}{}^{c)} + G^{e_2,}{}_{[a}{}^{c]} + G^{c,}{}_{[a}{}^{e_2]} + G^{c,}{}_{(a}{}^{e_2)}\} \\
&= 2\Lambda^{e_1}{}_{e_2}\overline{G}_{[b,c]e_1}\{-2G_{a,}{}^{(e_2c)} + G^{e_2,}{}_{a}{}^{c} + G^{c,}{}_{a}{}^{e_2}\} \quad .
\end{aligned} \tag{3.114}$$

The black circles in equation (3.113) with $a \leftrightarrow b$ reduce in the same way. Thus equation (3.113) reduces to

$\Lambda^{e_1e_2}$ terms

$$\begin{aligned}
&= \eta_{ab}\Lambda^{e_1e_2}\{4(\det e)^{1/2}[e^{\nu[c}\partial_\nu\overline{G}_{[c,e_1]}{}^{e_2]}] + 2G^{c,}{}_{e_1}{}^{d}\overline{G}_{[c,d]e_2}\,\square_1 + 2G^{c,}{}_{e_2}{}^{d}\overline{G}_{[c,e_1]d}\,\square_2\} \quad \textit{(eom)} \\
&\quad + \eta_{ab}\Lambda^{e_1e_2}\{-2G_{c,}{}^{dc}\overline{G}_{[d,e_1]e_2}{}^{\triangle_1} + G^{d,c}{}_{c}\overline{G}_{[d,e_1]e_2}{}^{\triangle_2}\} \\
&\quad - \Lambda^{e_1e_2}\overline{G}_{[a,b]e_2}G_{e_1,c}{}^{c\triangle_2\ l_1} \\
&\quad + 2\Lambda^{e_1e_2}\overline{G}_{[b,c]e_1}\{G^{c,}{}_{e_2a} + (-2G_{a,}{}^{(e_2c)} + G^{e_2,}{}_{a}{}^{c} + G^{c,}{}_{a}{}^{e_2})\} \quad (= I_1 + .. + I_4) \\
&\quad + 2\Lambda^{e_1e_2}\overline{G}_{[a,c]e_1}\{G^{c,}{}_{e_2b} + (-2G_{b,}{}^{(e_2c)} + G^{e_2,}{}_{b}{}^{c} + G^{c,}{}_{b}{}^{e_2})\} \quad (= i_1 + .. + i_4) \\
&\quad - 2(\det e)^{1/2}e_c{}^{\nu}\partial_\nu[\overline{G}_{[a,b]e}\Lambda^{e\tau}]e_\tau{}^{c(2.2.2\ l_1)} \\
&\quad + 2\eta_{ab}\Lambda^{e_1}{}_{e_2}(\det e)^{1/2}e^{\nu e_2}\partial_\nu\overline{G}_{[c,e_1]}{}^{c} \quad (\textit{eom } l_1^{\eta_{ab}}) \quad + 2\Lambda^{e_1e_2}(\det e)^{1/2}\Omega_{a,b}{}^{d}\overline{G}_{e_1,e_2d}\boxed{l_1 \textit{ from } \delta④} \\
&\quad + \Lambda^{e_1e_2}\overline{G}_{e_1,ae_2}(\det e)^{1/2}\Omega_{d,b}{}^{d②l_1^{(3)}} \quad .
\end{aligned} \tag{3.115}$$

Now using

$$\begin{aligned}
E_{a,b_1b_2} &= \omega_{a,b_1b_2} + \frac{1}{2}\varepsilon_{b_1b_2}{}^{c_1c_2}\overline{G}_{[c_1,c_2]a} \\
\varepsilon^{b_1b_2}{}_{d_1d_2}E_{a,b_1b_2} &= \varepsilon^{b_1b_2}{}_{d_1d_2}\omega_{a,b_1b_2} + \frac{1}{2}\varepsilon^{b_1b_2}{}_{d_1d_2}\varepsilon_{b_1b_2}{}^{c_1c_2}\overline{G}_{[c_1,c_2]a} \\
&= \varepsilon^{b_1b_2}{}_{d_1d_2}\omega_{a,b_1b_2} - 2\overline{G}_{[d_1,d_2]a} \\
\overline{G}_{[d_1,d_2]a} &= \frac{1}{2}\varepsilon^{b_1b_2}{}_{d_1d_2}\omega_{a,b_1b_2} - \frac{1}{2}\varepsilon^{b_1b_2}{}_{d_1d_2}E_{a,b_1b_2}
\end{aligned} \tag{3.116}$$

we evaluate the $I_1 + .. + I_4$ line

$$
\begin{aligned}
I_{1..} &= 2\Lambda^{e_1e_2}\overline{G}_{[b,c]e_1}\{G^{c,}{}_{e_2a} + (-2G_{a,}{}^{(e_2c)} + G^{e_2,}{}_{a}{}^{c} + G^{c,}{}_{a}{}^{e_2})\} \\
&= 2\Lambda^{e_1e_2}\overline{G}_{[b,c]e_1}\{2G^{c,}{}_{(e_2a)} - 2G_{a,}{}^{(e_2c)} + G^{e_2,}{}_{a}{}^{c}\} \\
&= 2\Lambda^{e_1e_2}\overline{G}_{[b,c]e_1}\{-2G_{a,}{}^{(e_2c)} + 2G^{c,}{}_{(e_2a)} + 2G^{e_2,}{}_{[a}{}^{c]}\} + 2\Lambda^{e_1e_2}\overline{G}_{[b,c]e_1}G^{e_2,c}{}_{a}{}^{I_1+..+I_4\ l_1} \\
&= 4\Lambda^{e_1e_2}\overline{G}_{[b,c]e_1}(\det e)^{1/2}\omega^{e_2,}{}_{a}{}^{c} + 2\Lambda^{e_1e_2}\overline{G}_{[b,c]e_1}G^{e_2,c}{}_{a}{}^{I_1+..+I_4\ l_1} \\
&= 4\Lambda^{e_1e_2}[\frac{1}{2}\varepsilon_{bc}{}^{f_1f_2}(\det e)^{1/2}\omega_{e_1,f_1f_2} - \frac{1}{2}\varepsilon_{bc}{}^{f_1f_2}E_{e_1,f_1f_2}](\det e)^{1/2}\omega^{e_2,}{}_{a}{}^{c} + 2\Lambda^{e_1e_2}\overline{G}_{[b,c]e_1}G^{e_2,c}{}_{a}{}^{I_1+..+I_4\ l_1} \\
&= 2\Lambda^{e_1e_2}\varepsilon_{bc}{}^{f_1f_2}(\det e)^{1/2}\omega_{e_1,f_1f_2}(\det e)^{1/2}\omega^{e_2,}{}_{a}{}^{c} - 2\Lambda^{e_1e_2}\varepsilon_{bc}{}^{f_1f_2}E_{e_1,f_1f_2}(\det e)^{1/2}\omega^{e_2,}{}_{a}{}^{c} \\
&\quad + 2\Lambda^{e_1e_2}\overline{G}_{[b,c]e_1}G^{e_2,c}{}_{a}{}^{I_1+..+I_4\ l_1} \qquad (3.117) \\
&= 2\Lambda^{e_1e_2}\varepsilon_{bc}{}^{f_1f_2}(\det e)^{1/2}\omega_{e_1,f_1f_2}[E^{e_2,}{}_{a}{}^{e} - \frac{1}{2}\varepsilon_{a}{}^{cg_1g_2}\overline{G}_{[g_1,g_2]}{}^{e_2}] \\
&\quad - 2\Lambda^{e_1e_2}\varepsilon_{bc}{}^{f_1f_2}E_{e_1,f_1f_2}(\det e)^{1/2}\omega^{e_2,}{}_{a}{}^{c} + 2\Lambda^{e_1e_2}\overline{G}_{[b,c]e_1}G^{e_2,c}{}_{a}{}^{I_1+..+I_4\ l_1} \\
&= +3!\delta^{bf_1f_2}_{ag_1g_2}\Lambda^{e_1e_2}(\det e)^{1/2}\omega_{e_1,f_1f_2}\overline{G}^{[g_1,g_2]e_2} \\
&+ 2\Lambda^{e_1e_2}\varepsilon_{bc}{}^{f_1f_2}(\det e)^{1/2}\omega_{e_1,f_1f_2}E^{e_2,}{}_{a}{}^{e} - 2\Lambda^{e_1e_2}\varepsilon_{bc}{}^{f_1f_2}E_{e_1,f_1f_2}(\det e)^{1/2}\omega^{e_2,}{}_{a}{}^{c} \\
&\quad + 2\Lambda^{e_1e_2}\overline{G}_{[b,c]e_1}G^{e_2,c}{}_{a}{}^{I_1+..+I_4\ l_1} \\
&= 2\eta_{ab}\Lambda^{e_1e_2}(\det e)^{1/2}\omega_{e_1,f_1f_2}\overline{G}^{[f_1,f_2]}{}_{e_2} - 2\Lambda^{e_1e_2}(\det e)^{1/2}\omega_{e_1,ag}\overline{G}^{[b,g]}{}_{e_2} \\
&\quad + 2\Lambda^{e_1e_2}(\det e)^{1/2}\omega_{e_1,ag}\overline{G}^{[g,b]}{}_{e_2} + 2\Lambda^{e_1e_2}\varepsilon_{bc}{}^{f_1f_2}(\det e)^{1/2}\omega_{e_1,f_1f_2}E^{e_2,}{}_{a}{}^{e} \\
&\quad - 2\Lambda^{e_1e_2}\varepsilon_{bc}{}^{f_1f_2}E_{e_1,f_1f_2}(\det e)^{1/2}\omega^{e_2,}{}_{a}{}^{c} + 2\Lambda^{e_1e_2}\overline{G}_{[b,c]e_1}G^{e_2,c}{}_{a}{}^{I_1+..+I_4\ l_1} \\
&= 2\eta_{ab}\Lambda^{e_1e_2}(\det e)^{1/2}\omega_{e_1,f_1f_2}\overline{G}^{[f_1,f_2]}{}_{e_2} - 4\Lambda^{e_1e_2}(\det e)^{1/2}\omega_{e_1,ag}\overline{G}^{[b,g]}{}_{e_2} \\
&+ 2\Lambda^{e_1e_2}\varepsilon_{bc}{}^{f_1f_2}(\det e)^{1/2}\omega_{e_1,f_1f_2}E^{e_2,}{}_{a}{}^{e} - 2\Lambda^{e_1e_2}\varepsilon_{bc}{}^{f_1f_2}E_{e_1,f_1f_2}(\det e)^{1/2}\omega^{e_2,}{}_{a}{}^{c} \\
&+ 2\Lambda^{e_1e_2}\overline{G}_{[b,c]e_1}G^{e_2,c}{}_{a}{}^{I_1+..+I_4\ l_1}
\end{aligned}
$$

which tells us that

$$
\begin{aligned}
I_{1...} &= 4\Lambda^{e_1e_2}(\det e)^{1/2}\omega^{e_2,}{}_{a}{}^{c}\overline{G}_{[b,c]e_1} + 2\Lambda^{e_1e_2}\overline{G}_{[b,c]e_1}G^{e_2,c}{}_{a}{}^{I_1+..+I_4\ l_1} \\
&= 2\eta_{ab}\Lambda^{e_1e_2}(\det e)^{1/2}\omega_{e_1,f_1f_2}\overline{G}^{[f_1,f_2]}{}_{e_2} - 4\Lambda^{e_1e_2}(\det e)^{1/2}\omega_{e_2,a}{}^{c}\overline{G}_{[b,c]e_1} \qquad (3.118) \\
&+ 2\Lambda^{e_1e_2}\varepsilon_{bc}{}^{f_1f_2}(\det e)^{1/2}\omega_{e_1,f_1f_2}E^{e_2,}{}_{a}{}^{e} - 2\Lambda^{e_1e_2}\varepsilon_{bc}{}^{f_1f_2}E_{e_1,f_1f_2}(\det e)^{1/2}\omega^{e_2,}{}_{a}{}^{c} \qquad (3.119) \\
&+ 2\Lambda^{e_1e_2}\overline{G}_{[b,c]e_1}G^{e_2,c}{}_{a}{}^{I_1+..+I_4\ l_1}
\end{aligned}
$$

reduces to the equality

$$
\begin{aligned}
8\Lambda^{e_1e_2}(\det e)^{1/2}\omega^{e_2,}{}_{a}{}^{c}\overline{G}_{[b,c]e_1} &= 2\eta_{ab}\Lambda^{e_1e_2}(\det e)^{1/2}\omega_{e_1,f_1f_2}\overline{G}^{[f_1,f_2]}{}_{e_2} \qquad (3.120) \\
&+ 2\Lambda^{e_1e_2}\varepsilon_{bc}{}^{f_1f_2}(\det e)^{1/2}\omega_{e_1,f_1f_2}E^{e_2,}{}_{a}{}^{e} \qquad (3.121) \\
&- 2\Lambda^{e_1e_2}\varepsilon_{bc}{}^{f_1f_2}E_{e_1,f_1f_2}(\det e)^{1/2}\omega^{e_2,}{}_{a}{}^{c}
\end{aligned}
$$

or rather

$$\begin{aligned}4\Lambda^{e_1e_2}(\det e)^{1/2}\omega^{e_2,}{}_a{}^c\overline{G}_{[b,c]e_1} &= \eta_{ab}\Lambda^{e_1e_2}(\det e)^{1/2}\omega_{e_1,f_1f_2}\overline{G}^{[f_1,f_2]}{}_{e_2}\\ &+\Lambda^{e_1e_2}\varepsilon_{bc}{}^{f_1f_2}(\det e)^{1/2}\omega_{e_1,f_1f_2}E^{e_2,}{}_a{}^e\\ &-\Lambda^{e_1e_2}\varepsilon_{bc}{}^{f_1f_2}E_{e_1,f_1f_2}(\det e)^{1/2}\omega^{e_2,}{}_a{}^c\end{aligned}\tag{3.122}$$

which we plug back into the first line of (3.118)

$$\begin{aligned}I_1+.. &= 4\Lambda^{e_1e_2}(\det e)^{1/2}\omega^{e_2,}{}_a{}^c\overline{G}_{[b,c]e_1}+2\Lambda^{e_1e_2}\overline{G}_{[b,c]e_1}G^{e_2,c}{}_a{}^{I_1+..+I_4\ l_1}\\ &= \eta_{ab}\Lambda^{e_1e_2}(\det e)^{1/2}\omega_{e_1,f_1f_2}\overline{G}^{[f_1,f_2]}{}_{e_2}\\ &\quad+\Lambda^{e_1e_2}\varepsilon_{bc}{}^{f_1f_2}(\det e)^{1/2}\omega_{e_1,f_1f_2}E^{e_2,}{}_a{}^e-\Lambda^{e_1e_2}\varepsilon_{bc}{}^{f_1f_2}E_{e_1,f_1f_2}(\det e)^{1/2}\omega^{e_2,}{}_a{}^c\\ &\quad+2\Lambda^{e_1e_2}\overline{G}_{[b,c]e_1}G^{e_2,c}{}_a{}^{I_1+..+I_4\ l_1}\end{aligned}\tag{3.123}$$

We now use this in our expression for the $\Lambda^{e_1e_2}$ terms (noting $i_1+\ldots$ with $a\leftrightarrow b$ is the same as this last result)

$\Lambda^{e_1e_2}$ terms

$$\begin{aligned}&=\eta_{ab}\Lambda^{e_1e_2}\{4(\det e)^{1/2}[e^{\nu[c}\partial_\nu\overline{G}_{[c,e_1]}{}^{e_2]}]+2G^{c,}{}_{e_1}{}^d\overline{G}_{[c,d]e_2}\ \square_1+2G^{c,}{}_{e_2}{}^d\overline{G}_{[c,e_1]d}\ \square_2\}\quad \textit{(eom)}\\ &+\eta_{ab}\Lambda^{e_1e_2}\{-2G_{c,}{}^{dc}\overline{G}_{[d,e_1]e_2}{}^{\triangle_1}+G^{d,c}{}_c\overline{G}_{[d,e_1]e_2}{}^{\triangle_2}\}\\ &-\Lambda^{e_1e_2}\overline{G}_{[a,b]e_2}G_{e_1,c}{}^{c\triangle_2\ l_1}\\ &+\eta_{ab}\Lambda^{e_1e_2}(\det e)^{1/2}\omega_{e_1,f_1f_2}\overline{G}^{[f_1,f_2]}{}_{e_2}\\ &\quad+\Lambda^{e_1e_2}\varepsilon_{bc}{}^{f_1f_2}(\det e)^{1/2}\omega_{e_1,f_1f_2}E^{e_2,}{}_a{}^e-\Lambda^{e_1e_2}\varepsilon_{bc}{}^{f_1f_2}E_{e_1,f_1f_2}(\det e)^{1/2}\omega^{e_2,}{}_a{}^c\\ &\quad+2\Lambda^{e_1e_2}\overline{G}_{[b,c]e_1}G^{e_2,c}{}_a{}^{I_1+..+I_4\ l_1}\quad \text{(all three lines } = I_1+..+I_4 \text{ tersm)}\\ &+\eta_{ab}\Lambda^{e_1e_2}(\det e)^{1/2}\omega_{e_1,f_1f_2}\overline{G}^{[f_1,f_2]}{}_{e_2}\\ &\quad+\Lambda^{e_1e_2}\varepsilon_{ac}{}^{f_1f_2}(\det e)^{1/2}\omega_{e_1,f_1f_2}E^{e_2,}{}_b{}^e-\Lambda^{e_1e_2}\varepsilon_{ac}{}^{f_1f_2}E_{e_1,f_1f_2}(\det e)^{1/2}\omega^{e_2,}{}_b{}^c\\ &\quad+2\Lambda^{e_1e_2}\overline{G}_{[a,c]e_1}G^{e_2,c}{}_b{}^{i_1+..+i_4\ l_1}\quad \text{(all three lines } = i_1+..+i_4 \text{ terms)}\\ &-2(\det e)^{1/2}e_c{}^\nu\partial_\nu[\overline{G}_{[a,b]e}\Lambda^{e\tau}]e_\tau{}^{c(2.2.2\ l_1)}\\ &+2\eta_{ab}\Lambda^{e_1}{}_{e_2}(\det e)^{1/2}e^{\nu e_2}\partial_\nu\overline{G}_{[c,e_1]}{}^c\quad (eom\ l_1^{\eta_{ab}})\quad+2\Lambda^{e_1e_2}(\det e)^{1/2}\Omega_{a,b}{}^d\overline{G}_{e_1,e_2d}\ \boxed{l_1\ from\ \delta④}\\ &+\Lambda^{e_1e_2}\overline{G}_{e_1,ae_2}(\det e)^{1/2}\Omega_{d,b}{}^{d②l_1^{(3)}}.\end{aligned}\tag{3.124}$$

We now focus on the 'equation of motion' terms in this result, writing:

$$\begin{aligned}
&\eta_{ab}\Lambda^{e_1e_2}\{4(\det e)^{1/2}[e^{\nu[c}\partial_\nu\overline{G}_{[c,e_1]}{}^{e_2]}] + 2G^{c,}{}_{e_1}{}^{d}\overline{G}_{[c,d]e_2}\ \square_1 + 2G^{c,}{}_{e_2}{}^{d}\overline{G}_{[c,e_1]d}\ \square_2\} \quad (eom)\\
&\quad + \eta_{ab}\Lambda^{e_1}{}_{e_2}\{-2G_{c,}{}^{dc}\overline{G}_{[d,e_1]}{}^{e_2\triangle_1} + G^{d,c}{}_c\overline{G}_{[d,e_1]}{}^{e_2\triangle_2}\} + 2\eta_{ab}\Lambda^{e_1e_2}(\det e)^{1/2}\omega_{e_1,f_1f_2}\overline{G}^{[f_1,f_2]}{}_{e_2}{}^{I_1}\\
&= \eta_{ab}\Lambda^{e_1}{}_{e_2}\{4(\det e)^{1/2}e^{\nu[c}\partial_\nu\overline{G}_{[c,e_1]}{}^{e_2]} + 4G^{[c,e_2]}{}_d\overline{G}_{[c,e_1]}{}^{d}\ \square_2 - 4G^{[c,|d|}{}_c\overline{G}_{[d,e_1]}{}^{e_2]\triangle_1} + 2G^{[d,|c|}{}_c\overline{G}_{[d,e_1]}{}^{e_2]\triangle_2}\}\\
&\quad + 2\eta_{ab}\Lambda^{e_1e_2}(\det e)^{1/2}\omega_{e_1,f_1f_2}\overline{G}^{[f_1,f_2]}{}_{e_2}{}^{I_1} + 2\eta_{ab}\Lambda^{e_1e_2}G^{c,}{}_{e_1}{}^{d}\overline{G}_{[c,d]e_2}\ \square_1 \qquad (3.125)\\
&\quad + 2\eta_{ab}\Lambda^{e_1}{}_{e_2}G^{e_2,c}{}_d\overline{G}_{[c,e_1]}{}^{d}\ \square_2\ l_1 - 2\eta_{ab}\Lambda^{e_1}{}_{e_2}G^{e_2,d}{}_c\overline{G}_{[d,e_1]}{}^{c\triangle_1\ l_1} + \eta_{ab}\Lambda^{e_1}{}_{e_2}G^{e_2,c}{}_c\overline{G}_{[d,e_1]}{}^{d\triangle_2\ l_1}\ .
\end{aligned}$$

The middle line of this last result simplifies as

$$\begin{aligned}
I_1 + \square_1 &= 2\eta_{ab}\Lambda^{e_1e_2}\{(-G_{f_1,(f_2e_1)} + G_{f_2,(f_1e_1)} + G_{e_1,[f_1f_2]})\overline{G}^{[f_1,f_2]}{}_{e_2}{}^{I_1} + G^{c,}{}_{e_1}{}^{d}\overline{G}_{[c,d]e_2}\ \square_1\}\\
&= 2\eta_{ab}\Lambda^{e_1e_2}\{-2G_{c,(de_1)}\overline{G}^{[c,d]}{}_{e_2}{}^{I_1} + G_{c,e_1d}\overline{G}^{[c,d]}{}_{e_2}\ \square_1\} + 2\eta_{ab}\Lambda^{e_1e_2}G_{e_1,[f_1f_2]}\overline{G}^{[f_1,f_2]}{}_{e_2}{}^{I_1\,l_1}\\
&= 2\eta_{ab}\Lambda^{e_1e_2}\{-G_{c,de_1}\overline{G}^{[c,d]}{}_{e_2}{}^{I_1} - G_{c,e_1d}\overline{G}^{[c,d]}{}_{e_2}{}^{I_1} + G_{c,e_1d}\overline{G}^{[c,d]}{}_{e_2}\ \square_1\}\\
&\quad + 2\eta_{ab}\Lambda^{e_1e_2}G_{e_1,[f_1f_2]}\overline{G}^{[f_1,f_2]}{}_{e_2}{}^{I_1\,l_1}\\
&= -2\eta_{ab}\Lambda^{e_1e_2}G_{c,de_1}\overline{G}^{[c,d]}{}_{e_2}{}^{I_1+\square_1} + 2\eta_{ab}\Lambda^{e_1e_2}G_{e_1,[f_1f_2]}\overline{G}^{[f_1,f_2]}{}_{e_2}{}^{I_1\,l_1}\\
&= -4\eta_{ab}\Lambda^{e_1}{}_{e_2}G^{[c,|d|}{}_{e_1}\overline{G}_{[c,d]}{}^{e_2]I_1+\square_1} - 2\eta_{ab}\Lambda^{e_1}{}_{e_2}G^{e_2,d}{}_{e_1}\overline{G}_{[c,d]}{}^{cI_1+\square_1\ l_1}\\
&\quad + 2\eta_{ab}\Lambda^{e_1e_2}G_{e_1,[f_1f_2]}\overline{G}^{[f_1,f_2]}{}_{e_2}{}^{I_1\,l_1}\ . \qquad (3.126)
\end{aligned}$$

We can use this to write equation (3.124) as

$\Lambda^{e_1e_2}$ terms

$$\begin{aligned}
&= \eta_{ab}\Lambda^{e_1}{}_{e_2}\{4(\det e)^{1/2}e^{\nu[c}\partial_\nu\overline{G}_{[c,e_1]}{}^{e_2]} + 4G^{[c,e_2]}{}_d\overline{G}_{[c,e_1]}{}^{d}\ \square_2 - 4G^{[c,|d|}{}_c\overline{G}_{[d,e_1]}{}^{e_2]\triangle_1} + 2G^{[d,|c|}{}_c\overline{G}_{[d,e_1]}{}^{e_2]\triangle_2}\}\\
&\quad + 2\eta_{ab}\Lambda^{e_1e_2}(\det e)^{1/2}\omega_{e_1,f_1f_2}\overline{G}^{[f_1,f_2]}{}_{e_2}{}^{I_1} + 2\eta_{ab}\Lambda^{e_1e_2}G^{c,}{}_{e_1}{}^{d}\overline{G}_{[c,d]e_2}\ \square_1\\
&\quad + 2\eta_{ab}\Lambda^{e_1}{}_{e_2}G^{e_2,c}{}_d\overline{G}_{[c,e_1]}{}^{d}\ \square_2\ l_1 - 2\eta_{ab}\Lambda^{e_1}{}_{e_2}G^{e_2,d}{}_c\overline{G}_{[d,e_1]}{}^{c\triangle_1\ l_1} + \eta_{ab}\Lambda^{e_1}{}_{e_2}G^{e_2,c}{}_c\overline{G}_{[d,e_1]}{}^{d\triangle_2\ l_1}\\
&= \eta_{ab}\Lambda^{e_1}{}_{e_2}\{4(\det e)^{1/2}e^{\nu[c}\partial_\nu\overline{G}_{[c,e_1]}{}^{e_2]} + 4G^{[c,e_2]}{}_d\overline{G}_{[c,e_1]}{}^{d}\ \square_2 - 4G^{[c,|d|}{}_c\overline{G}_{[d,e_1]}{}^{e_2]\triangle_1} + 2G^{[d,|c|}{}_c\overline{G}_{[d,e_1]}{}^{e_2]\triangle_2}\}\\
&\quad - 4\eta_{ab}\Lambda^{e_1}{}_{e_2}G^{[c,|d|}{}_{e_1}\overline{G}_{[c,d]}{}^{e_2]I_1+\square_1} - 2\eta_{ab}\Lambda^{e_1}{}_{e_2}G^{e_2,d}{}_{e_1}\overline{G}_{[c,d]}{}^{cI_1+\square_1\ l_1} + 2\eta_{ab}\Lambda^{e_1e_2}G_{e_1,[f_1f_2]}\overline{G}^{[f_1,f_2]}{}_{e_2}{}^{I_1\,l_1}\\
&\quad + 2\eta_{ab}\Lambda^{e_1}{}_{e_2}G^{e_2,c}{}_d\overline{G}_{[c,e_1]}{}^{d}\ \square_2\ l_1 - 2\eta_{ab}\Lambda^{e_1}{}_{e_2}G^{e_2,d}{}_c\overline{G}_{[d,e_1]}{}^{c\triangle_1\ l_1} + \eta_{ab}\Lambda^{e_1}{}_{e_2}G^{e_2,c}{}_c\overline{G}_{[d,e_1]}{}^{d\triangle_2\ l_1}\\
&= 4\eta_{ab}\Lambda^{e_1}{}_{e_2}\{(\det e)^{\frac{1}{2}}e^{\nu[c}\partial_\nu\overline{G}_{[c,e_1]}{}^{e_2]} + G^{[c,e_2]}{}_d\overline{G}_{[c,e_1]}{}^{d}\ \square_2 - G^{[c,|d|}{}_{e_1}\overline{G}_{[c,d]}{}^{e_2]I_1+\square_1}\\
&\qquad\qquad - G^{[c,|d|}{}_c\overline{G}_{[d,e_1]}{}^{e_2]\triangle_1} + \frac{1}{2}G^{[d,|c|}{}_c\overline{G}_{[d,e_1]}{}^{e_2]\triangle_2}\}\\
&\quad - 2\eta_{ab}\Lambda^{e_1}{}_{e_2}G^{e_2,d}{}_{e_1}\overline{G}_{[c,d]}{}^{cI_1+\square_1\ l_1} + 2\eta_{ab}\Lambda^{e_1e_2}G_{e_1,[f_1f_2]}\overline{G}^{[f_1,f_2]}{}_{e_2}{}^{I_1\,l_1} \qquad (3.127)\\
&\quad + 2\eta_{ab}\Lambda^{e_1}{}_{e_2}G^{e_2,c}{}_d\overline{G}_{[c,e_1]}{}^{d}\ \square_2\ l_1 - 2\eta_{ab}\Lambda^{e_1}{}_{e_2}G^{e_2,d}{}_c\overline{G}_{[d,e_1]}{}^{c\triangle_1\ l_1} + \eta_{ab}\Lambda^{e_1}{}_{e_2}G^{e_2,c}{}_c\overline{G}_{[d,e_1]}{}^{d\triangle_2\ l_1}\ .
\end{aligned}$$

In this form, we can now write $\Lambda^{e_1e_2}$ terms such that an equation of motion, similar to that arising in the $-4\Lambda_{ae}\{...\}$ and $-4\Lambda_{be}\{...\}$ terms previously, arises:

$$
\begin{aligned}
&\Lambda^{e_1e_2} \text{ terms}\\
&= 4\eta_{ab}\Lambda^{e_1}{}_{e_2}\{(\det e)^{\frac{1}{2}} e^{\nu[c}\partial_\nu \overline{G}_{[c,e_1]}{}^{e_2]} + G^{[c,e_2]}{}_d\overline{G}_{[c,e_1]}{}^{d}\ \square_2 - G^{[c,|d|}{}_{e_1}\overline{G}_{[c,d]}{}^{e_2]\,I_1+\square_1}\\
&\qquad - G^{[c,|d|}{}_c\overline{G}_{[d,e_1]}{}^{e_2]\triangle_1} + \frac{1}{2}G^{[d,|c|}{}_c\overline{G}_{[d,e_1]}{}^{e_2]\triangle_2}\}\\
&\quad + \Lambda^{e_1e_2}\varepsilon_{bc}{}^{f_1f_2}(\det e)^{1/2}\omega_{e_1,f_1f_2}E^{e_2,}{}_a{}^e - \Lambda^{e_1e_2}\varepsilon_{bc}{}^{f_1f_2}E_{e_1,f_1f_2}(\det e)^{1/2}\omega^{e_2,}{}_a{}^c\\
&\qquad (\text{from } I_1 + .. + I_4)\\
&\quad + \Lambda^{e_1e_2}\varepsilon_{ac}{}^{f_1f_2}(\det e)^{1/2}\omega_{e_1,f_1f_2}E^{e_2,}{}_b{}^e - \Lambda^{e_1e_2}\varepsilon_{ac}{}^{f_1f_2}E_{e_1,f_1f_2}(\det e)^{1/2}\omega^{e_2,}{}_b{}^c\\
&\qquad (\text{from } i_1 + .. + i_4)\\
&\quad - 2(\det e)^{1/2}e_c{}^\nu\partial_\nu[\overline{G}_{[a,b]e}\Lambda^{e\tau}]e_\tau{}^{c(2.2.2\ l_1)}\\
&\quad + 2\eta_{ab}\Lambda^{e_1}{}_{e_2}(\det e)^{1/2}e^{\nu e_2}\partial_\nu\overline{G}_{[c,e_1]}{}^c \quad (eom\ l_1^{\eta_{ab}})\\
&\quad + 2\Lambda^{e_1e_2}(\det e)^{1/2}\Omega_{a,b}{}^d\overline{G}_{e_1,e_2d}\ \boxed{l_1\ from\ \delta④}\\
&\quad + 2\Lambda^{e_1e_2}\overline{G}_{[b,c]e_1}G^{e_2,c}{}_a{}^{I_1+..+I_4\ l_1}\\
&\quad + 2\Lambda^{e_1e_2}\overline{G}_{[a,c]e_1}G^{e_2,c}{}_b{}^{i_1+..+i_4\ l_1}\\
&\quad + \Lambda^{e_1e_2}\overline{G}_{e_1,ae_2}(\det e)^{1/2}\Omega_{d,b}{}^{d②\,l_1^{(3)}}\\
&\quad - 2\eta_{ab}\Lambda^{e_1}{}_{e_2}G^{e_2,d}{}_{e_1}\overline{G}_{[c,d]}{}^{c\,I_1+\square_1\ l_1} + 2\eta_{ab}\Lambda^{e_1e_2}G_{e_1,[f_1f_2]}\overline{G}^{[f_1,f_2]}{}_{e_2}{}^{I_1\,l_1}\\
&\quad + 2\eta_{ab}\Lambda^{e_1}{}_{e_2}G^{e_2,c}{}_d\overline{G}_{[c,e_1]}{}^d\ \square_2\ l_1 - 2\eta_{ab}\Lambda^{e_1}{}_{e_2}G^{e_2,d}{}_c\overline{G}_{[d,e_1]}{}^{c\triangle_1\ l_1}\\
&\quad + \eta_{ab}\Lambda^{e_1}{}_{e_2}G^{e_2,c}{}_c\overline{G}_{[d,e_1]}{}^{d\triangle_2\ l_1} - \Lambda^{e_1e_2}\overline{G}_{[a,b]e_2}G_{e_1,c}{}^{c\triangle_2\ l_1}\quad .
\end{aligned}
\tag{3.128}
$$

We have now finished processing the $\Lambda^{e_1e_2}$ terms. On the first two lines of equation (3.128) we have a tentative expression for an equation of motion, and this same expression also also occurred in the previous sections as a tentative expression for an equation of motion there. The remaining terms in equation (refe1e2 finished) are all either $l_1$ terms, or involve equations of motion.

### Collecting All Terms

Now that we have the same tenetative expression for an equation of motion, arising as coefficients to the $\Lambda_{ae}$, $\Lambda_{be}$ and $\Lambda^{e_1e_2}$ terms, we collect the results of the previous sections into the following variation:

$$
\begin{aligned}
&\delta\tilde{\mathcal{E}}_{ab}\\
&= -4\Lambda_{ea}\{(\det e)^{1/2}e^{\nu[c}\partial_\nu \overline{G}_{[c,b]}{}^{e]} + G^{[d,e]c}\overline{G}_{[d,b]c} - G^{[c,|d|}{}_b\overline{G}_{[c,d]}{}^{e]}\textcircled{1} - G^{[c,|d|}{}_c\overline{G}_{[d,b]}{}^{e]}\textcircled{2} + \frac{1}{2}G^{[d,|c|}{}_c\overline{G}_{[d,b]}{}^{e]}\textcircled{3}\}\\
&\quad - 2\Lambda_{ae}(\det e)^{1/2}e^{\nu e}\partial_\nu\overline{G}_{[c,b]}{}^c - 2\Lambda_{ea}G^{e,c}{}_d\overline{G}_{[c,b]}{}^d \quad (eom\ l_1^a)\\
&\quad + 2\Lambda_{ea}G^{e,d}{}_b\overline{G}_{[c,d]}{}^{c}\textcircled{1}\ l_1 - 2\Lambda_{ea}G^{e,d}{}_c\overline{G}_{[b,d]}{}^{c}\textcircled{2}\ l_1 + \Lambda_{ea}G^{e,c}{}_c\overline{G}_{[b,d]}{}^{d}\textcircled{3}\ l_1\\
&- 4\Lambda_{eb}\{(\det e)^{1/2}e^{\nu[c}\partial_\nu\overline{G}_{[c,a]}{}^{e]} + G^{[c,e]}{}_d\overline{G}_{[c,a]}{}^d - G^{[d,|c|}{}_a\overline{G}_{[d,c]}{}^{e]\triangle_1} - G^{[c,|d|}{}_c\overline{G}_{[d,a]}{}^{e]\triangle_2} + \frac{1}{2}G^{[d,|c|}{}_c\overline{G}_{[d,a]}{}^{e]\triangle_3}\}\\
&\quad - 2\Lambda_{be}(\det e)^{1/2}e^{\nu e}\partial_\nu\overline{G}_{[c,a]}{}^c - 2\Lambda_{eb}G^{e,c}{}_d\overline{G}_{[c,a]}{}^d \quad (eom\ l_1^b)\\
&\quad + 2\Lambda_{eb}G^{e,d}{}_a\overline{G}_{[c,d]}{}^{c\triangle_1}\ l_1 - 2\Lambda_{eb}G^{e,d}{}_c\overline{G}_{[a,d]}{}^{c\triangle_2}\ l_1 + \Lambda_{eb}G^{e,c}{}_c\overline{G}_{[a,d]}{}^{d\triangle_3}\ l_1\\
&+ 4\eta_{ab}\Lambda^{e_1}_{e_2}\{(\det e)^{\frac{1}{2}}e^{\nu[c}\partial_\nu\overline{G}_{[c,e_1]}{}^{e_2]} + G^{[c,e_2]}{}_d\overline{G}_{[c,e_1]}{}^d\ \square_2 - G^{[c,|d|}{}_{e_1}\overline{G}_{[c,d]}{}^{e_2]I_1+\square_1}\\
&\qquad\qquad - G^{[c,|d|}{}_c\overline{G}_{[d,e_1]}{}^{e_2]\triangle_1} + \frac{1}{2}G^{[d,|c|}{}_c\overline{G}_{[d,e_1]}{}^{e_2]\triangle_2}\}\\
&\quad + \Lambda^{e_1e_2}\varepsilon_{bc}{}^{f_1f_2}(\det e)^{1/2}\omega_{e_1,f_1f_2}E^{e_2,}{}_a{}^e - \Lambda^{e_1e_2}\varepsilon_{bc}{}^{f_1f_2}E_{e_1,f_1f_2}(\det e)^{1/2}\omega^{e_2,}{}_a{}^c \quad (\text{from } I_1 + .. + I_4)\\
&\quad + \Lambda^{e_1e_2}\varepsilon_{ac}{}^{f_1f_2}(\det e)^{1/2}\omega_{e_1,f_1f_2}E^{e_2,}{}_b{}^e - \Lambda^{e_1e_2}\varepsilon_{ac}{}^{f_1f_2}E_{e_1,f_1f_2}(\det e)^{1/2}\omega^{e_2,}{}_b{}^c \quad (\text{from } i_1 + .. + i_4)\\
&\quad - 2(\det e)^{1/2}e_c{}^\nu\partial_\nu[\overline{G}_{[a,b]e}\Lambda^{e\tau}]e_\tau{}^{c(2.2.2\ l_1)}\\
&\quad + 2\eta_{ab}\Lambda^{e_1}{}_{e_2}(\det e)^{1/2}e^{\nu e_2}\partial_\nu\overline{G}_{[c,e_1]}{}^c \quad (eom\ l_1^{\eta ab}) \quad + 2\Lambda^{e_1e_2}(\det e)^{1/2}\Omega_{a,b}{}^d\overline{G}_{e_1,e_2d}\ \boxed{l_1\ from\ \delta\textcircled{4}}\\
&\qquad + 2\Lambda^{e_1e_2}\overline{G}_{[b,c]e_1}G^{e_2,c}{}_a{}^{I_1+..+I_4}\ l_1 + 2\Lambda^{e_1e_2}\overline{G}_{[a,c]e_1}G^{e_2,c}{}_b{}^{i_1+..+i_4}\ l_1\\
&\quad + \Lambda^{e_1e_2}\overline{G}_{e_1,ae_2}(\det e)^{1/2}\Omega_{d,b}{}^{d}\textcircled{2}l_1^{(3)}\\
&\qquad - 2\eta_{ab}\Lambda^{e_1}{}_{e_2}G^{e_2,d}{}_{e_1}\overline{G}_{[c,d]}{}^{cI_1+\square_1}\ l_1 + 2\eta_{ab}\Lambda^{e_1e_2}G_{e_1,[f_1f_2]}\overline{G}^{[f_1,f_2]}{}_{e_2}\ I_1\,l_1\\
&\quad + 2\eta_{ab}\Lambda^{e_1}{}_{e_2}G^{e_2,c}{}_d\overline{G}_{[c,e_1]}{}^d\ \square_2\ l_1 - 2\eta_{ab}\Lambda^{e_1}{}_{e_2}G^{e_2,d}{}_c\overline{G}_{[d,e_1]}{}^{c\triangle_1}\ l_1 + \eta_{ab}\Lambda^{e_1}{}_{e_2}G^{e_2,c}{}_c\overline{G}_{[d,e_1]}{}^{d\triangle_2}\ l_1\\
&\quad - \Lambda^{e_1e_2}\overline{G}_{[a,b]e_2}G_{e_1,c}{}^{c\triangle_2}\ l_1 \quad .
\end{aligned}
\tag{3.129}
$$

Defining the second order equation of motion

$$
\begin{aligned}
\overline{E}'_a{}^b = (\det e)^{1/2}e^{\nu[c}\partial_\nu\overline{G}_{[c,a]}{}^{b]} + G^{[c,b]}{}_d\overline{G}_{[c,a]}{}^d - G^{[c,|d|}{}_a\overline{G}_{[c,d]}{}^{b]}\\
- G^{[c,|d|}{}_c\overline{G}_{[d,a]}{}^{b]} + \frac{1}{2}G^{[d,|c|}{}_c\overline{G}_{[d,a]}{}^{b]}
\end{aligned}
\tag{3.130}
$$

the variation now reads as

$$
\begin{aligned}
&\delta\tilde{\mathcal{E}}_{ab}\\
&= -4\Lambda_{ea}\overline{E}'_b{}^e - 4\Lambda_{eb}\overline{E}'_a{}^e + 4\eta_{ab}\Lambda^{e_1}{}_{e_2}\overline{E}'_{e_1}{}^{e_2}\\
&\qquad + \Lambda^{e_1e_2}\varepsilon_{bc}{}^{f_1f_2}[E^{e_2,}{}_a{}^e(\det e)^{1/2}\omega_{e_1,f_1f_2} - E_{e_1,f_1f_2}(\det e)^{1/2}\omega^{e_2,}{}_a{}^c] \quad (\text{from } I_1+..+I_4)\\
&\qquad + \Lambda^{e_1e_2}\varepsilon_{ac}{}^{f_1f_2}[E^{e_2,}{}_b{}^e(\det e)^{1/2}\omega_{e_1,f_1f_2} - E_{e_1,f_1f_2}(\det e)^{1/2}\omega^{e_2,}{}_b{}^c] \quad (\text{from } i_1+..+i_4)\\
&\quad - 2\Lambda_{ae}(\det e)^{1/2}e^{\nu e}\partial_\nu\overline{G}_{[c,b]}{}^c - 2\Lambda_{ea}G^{e,c}{}_d\overline{G}_{[c,b]}{}^d \quad (eom\ l_1^a)\\
&\quad + 2\Lambda_{ea}G^{e,d}{}_b\overline{G}_{[c,d]}{}^{c\,①\ l_1} - 2\Lambda_{ea}G^{e,d}{}_c\overline{G}_{[b,d]}{}^{c\,②\ l_1} + \Lambda_{ea}G^{e,c}{}_c\overline{G}_{[b,d]}{}^{d\,③\ l_1}\\
&\quad - 2\Lambda_{be}(\det e)^{1/2}e^{\nu e}\partial_\nu\overline{G}_{[c,a]}{}^c - 2\Lambda_{eb}G^{e,c}{}_d\overline{G}_{[c,a]}{}^d \quad (eom\ l_1^b)\\
&\quad + 2\Lambda_{eb}G^{e,d}{}_a\overline{G}_{[c,d]}{}^{c\triangle_1\ l_1} - 2\Lambda_{eb}G^{e,d}{}_c\overline{G}_{[a,d]}{}^{c\triangle_2\ l_1} + \Lambda_{eb}G^{e,c}{}_c\overline{G}_{[a,d]}{}^{d\triangle_3\ l_1}\\
&\quad + 2\eta_{ab}\Lambda^{e_1}{}_{e_2}(\det e)^{1/2}e^{e_2\nu}\partial_\nu\overline{G}_{[c,e_1]}{}^c \quad (eom\ l_1^{\eta_{ab}}) \quad + 2\eta_{ab}\Lambda^{e_1}{}_{e_2}G^{e_2,c}{}_d\overline{G}_{[c,e_1]}{}^d\ \square_2\ l_1\\
&\quad - 2\eta_{ab}\Lambda^{e_1}{}_{e_2}G^{e_2,d}{}_{e_1}\overline{G}_{[c,d]}{}^{c I_1+\square_1\ l_1} + 2\eta_{ab}\Lambda^{e_1}{}_{e_2}G^{e_2,d}{}_c\overline{G}_{[e_1,d]}{}^{c\triangle_1\ l_1} - \eta_{ab}\Lambda^{e_1}{}_{e_2}G^{e_2,c}{}_c\overline{G}_{[e_1,d]}{}^{d\triangle_2\ l_1}\\
&\quad + 2\Lambda^{e_1e_2}\overline{G}_{[b,c]e_1}G^{e_2,c}{}_a{}^{I_1+..+I_4\ l_1} + 2\Lambda^{e_1e_2}\overline{G}_{[a,c]e_1}G^{e_2,c}{}_b{}^{i_1+..+i_4\ l_1} + 2\eta_{ab}\Lambda^{e_1e_2}G_{e_2,[f_1f_2]}\overline{G}^{[f_1,f_2]}{}_{e_1}{}^{I_1\ l_1}\\
&\quad\; - 2(\det e)^{1/2}\delta_\tau{}^\nu\partial_\nu[\overline{G}_{[a,b]e}\Lambda^{e\tau}]^{(2.2.2\ l_1)} \qquad\qquad (3.131)\\
&\quad\; + \Lambda^{e_1e_2}(\det e)^{1/2}\Omega_{d,b}{}^d\overline{G}_{e_1,ae_2}{}^{②l_1^{(3)}} \quad + 2\Lambda^{e_1e_2}(\det e)^{1/2}\Omega_{a,b}{}^d\overline{G}_{e_1,e_2d}\boxed{l_1\ from\ \delta④}\\
&\quad\; - \Lambda^{e_1e_2}G_{e_1,c}{}^c\overline{G}_{[a,b]e_2}{}^{\triangle_2\ l_1} \quad .
\end{aligned}
$$

We now re-write the $l_1$ terms with the variation made explicit as follow:

$$
\begin{aligned}
&= -\delta\{+[(\det e)^{\frac{1}{2}}\hat{e}_a{}^\nu\partial_\nu\overline{G}_{[c,b]}{}^c \ + \hat{G}_{a,}{}^c{}_d\overline{G}_{[c,b]}{}^d - \hat{G}_{a,db}\overline{G}_{[c,d]}{}^{c\,①\ l_1} + \hat{G}_{a,}{}^d{}_c\overline{G}_{[b,d]}{}^{c\,②\ l_1} - \frac{1}{2}\hat{G}_{a,}{}^c{}_c\overline{G}_{[b,d]}{}^{d\,③\ l_1}]\\
&\quad + [(\det e)^{1/2}\hat{e}_b{}^\nu\partial_\nu\overline{G}_{[c,a]}{}^c + \hat{G}_{b,}{}^c{}_d\overline{G}_{[c,a]}{}^d - \hat{G}_{b,}{}^d{}_a\overline{G}_{[c,d]}{}^{c\triangle_1\ l_1} + \hat{G}_{b,}{}^d{}_c\overline{G}_{[a,d]}{}^{c\triangle_2\ l_1} - \frac{1}{2}\hat{G}_{b,}{}^c{}_c\overline{G}_{[a,d]}{}^{d\triangle_3\ l_1}]\\
&\quad - \eta_{ab}[(\det e)^{1/2}\hat{e}^e{}_\nu\partial_\nu\overline{G}_{[c,e]}{}^c + \hat{G}^{e,c}{}_d\overline{G}_{[c,e]}{}^d\ \square_2^{l_1} - \hat{G}^{e,d}{}_e\overline{G}_{[c,d]}{}^{c I_1+\square_1{}^{l_1}} + \hat{G}^{e,d}{}_c\overline{G}_{[e,d]}{}^{c\triangle_1^{l_1}} - \frac{1}{2}\hat{G}^{e,c}{}_c\overline{G}_{[e,d]}{}^{d\triangle_2^{l_1}}]\\
&\quad - [\hat{G}^{e,c}{}_a\overline{G}_{[b,c]e}{}^{I_1+..+I_4\ l_1} + \hat{G}^{e,c}{}_b\overline{G}_{[a,c]e}{}^{i_1+..+i_4\ l_1} + \eta_{ab}\hat{G}^{e,}{}_{[f_1f_2]}\overline{G}^{[f_1,f_2]}{}_e{}^{I_1\ l_1}]\\
&\quad + (\hat{\partial}^e\overline{G}_{[a,b]e} + \hat{G}^{e,}{}_e{}^c\overline{G}_{[a,b]c})^{(2.2.2\ l_1)} + \frac{1}{2}\hat{G}^{e,}{}_c{}^c\overline{G}_{[a,b]e}{}^{\triangle_2\ l_1} \qquad\qquad (3.132)\\
&\quad - \frac{1}{2}(\det e)^{1/2}\Omega_{d,b}{}^d\hat{\overline{G}}^{e,}{}_{ae}{}^{②l_1^{(3)}} - \frac{1}{2}(\det e)^{1/2}\Omega_{a,b}{}^d\hat{\overline{G}}^{e,}{}_{ed}\boxed{l_1\ from\ \delta④}\}.
\end{aligned}
$$

Bringing these terms that are to be varied into the definition of the $l_1$ extended Ricci tensor, we set

$$
\begin{aligned}
&\tilde{\mathcal{E}}'_{ab} \\
&= (\det e)\{e_a{}^{\mu}\partial_{\mu}(\Omega_{\nu,}{}^{bd})e_d{}^{\nu} - \partial_{\nu}(\Omega_{\mu}{}^{bd})e_d{}^{\nu}e_a{}^{\mu} + \Omega_{a,}{}^{b}{}_{c}\Omega_{d,}{}^{cd} - \Omega_{d,}{}^{b}{}_{c}\Omega_{a,}{}^{cd}\} \\
&+ [(\det e)^{1/2}\hat{e}_a{}^{\nu}\partial_{\nu}\overline{G}_{[c,b]}{}^{c} + \hat{G}_{a,}{}^{c}{}_{d}(\det e)^{1/2}\overline{G}_{[c,b]}{}^{d} - \hat{G}_{a,db}\overline{G}_{[c,d]}{}^{c\textcircled{1}\ l_1} + \hat{G}_{a,}{}^{d}{}_{c}\overline{G}_{[b,d]}{}^{c\textcircled{2}\ l_1} - \frac{1}{2}\hat{G}_{a,}{}^{c}{}_{c}\overline{G}_{[b,d]}{}^{d\textcircled{3}\ l_1}] \\
&+ [(\det e)^{1/2}\hat{e}_b{}^{\nu}\partial_{\nu}\overline{G}_{[c,a]}{}^{c} + \hat{G}_{b,}{}^{c}{}_{d}\overline{G}_{[c,a]}{}^{d} - \hat{G}_{b,}{}^{d}{}_{a}\overline{G}_{[c,d]}{}^{c\triangle_1\ l_1} + \hat{G}_{b,}{}^{d}{}_{c}\overline{G}_{[a,d]}{}^{c\triangle_2\ l_1} - \frac{1}{2}\hat{G}_{b,}{}^{c}{}_{c}\overline{G}_{[a,d]}{}^{d\triangle_3\ l_1}] \\
&- \eta_{ab}[(\det e)^{1/2}\hat{e}^{e}{}_{\nu}\partial_{\nu}\overline{G}_{[c,e]}{}^{c} + \hat{G}^{e,c}{}_{d}\overline{G}_{[c,e]}{}^{d}\,{}^{l_1}_{\square_2} - \hat{G}^{e,d}{}_{e}\overline{G}_{[c,d]}{}^{cI_1+\square_1{}^{l_1}} + \hat{G}^{e,d}{}_{c}\overline{G}_{[e,d]}{}^{c\triangle_1^{l_1}} - \frac{1}{2}\hat{G}^{e,c}{}_{c}\overline{G}_{[e,d]}{}^{d\triangle_2^{l_1}}] \\
&- [\hat{G}^{e,c}{}_{a}\overline{G}_{[b,c]e}{}^{I_1+..+I_4\ l_1} + \hat{G}^{e,c}{}_{b}\overline{G}_{[a,c]e}{}^{i_1+..+i_4\ l_1} + \eta_{ab}\hat{G}^{e,}{}_{[f_1f_2]}\overline{G}^{[f_1,f_2]}{}_{e}{}^{I_1\,l_1}] \\
&+ (\hat{\partial}^{e}\overline{G}_{[a,b]e} + \hat{G}^{e,}{}_{e}{}^{c}\overline{G}_{[a,b]c})^{(2.2.2\ l_1)} + \frac{1}{2}\hat{G}^{c,}{}_{e}{}^{e}\overline{G}_{[a,b]c}{}^{\triangle_2\ l_1} \\
&- \frac{1}{2}\hat{\overline{G}}{}^{e,}{}_{ae}(\det e)^{1/2}\Omega_{d,b}{}^{d\textcircled{2}l_1^{(3)}} - \frac{1}{2}(\det e)^{1/2}\Omega_{a,b}{}^{d}\hat{\overline{G}}{}^{e,}{}_{ed}\boxed{l_1\ from\ \delta\textcircled{4}}\ .
\end{aligned}
\tag{3.133}
$$

Its variation is given by

$$
\begin{aligned}
\delta\tilde{\mathcal{E}}'_{ab} &= -4\Lambda_{ea}\overline{E}'_b{}^{e} - 4\Lambda_{eb}\overline{E}'_a{}^{e} + 4\eta_{ab}\Lambda^{e_1}{}_{e_2}\overline{E}'_{e_1}{}^{e_2} \\
&+ \Lambda^{e_1}{}_{e_2}\varepsilon_{bc}{}^{f_1f_2}[E^{e_2,}{}_{a}{}^{e}(\det e)^{1/2}\omega_{e_1,f_1f_2} - E_{e_1,f_1f_2}(\det e)^{1/2}\omega^{e_2,}{}_{a}{}^{c}] \\
&+ \Lambda^{e_1}{}_{e_2}\varepsilon_{ac}{}^{f_1f_2}[E^{e_2,}{}_{b}{}^{e}(\det e)^{1/2}\omega_{e_1,f_1f_2} - E_{e_1,f_1f_2}(\det e)^{1/2}\omega^{e_2,}{}_{b}{}^{c}]
\end{aligned}
\tag{3.134}
$$

This completes the derivation of equations (5.4) and (5.5) proposed in [29]. We discuss the meaning of this result next.

### 3.4.3 Symmetrization of the Dual Graviton Equation of Motion

At this stage we might argue that

$$
E_a{}^{b} = 0 \quad , \quad \overline{E}'_a{}^{b} = 0 \tag{3.135}
$$

are the second order equations of motion of the graviton and dual graviton respectively. However, being the equation of motion for the dual graviton, it should possess the same symmetries as the dual gravity field, and so should be symmetric in its two indices. As defined, $\overline{E}'_a{}^{b}$ in equation (3.130) is not symmetric in $a$ and $b$, which is why it has been denoted with a prime. We thus try to symmetrize the indices of the $\overline{E}'_a{}^{b}$ terms in equation (3.134). To do this we will analyse $\overline{E}'_a{}^{b}$ in equation (3.130).

So far we have overlooked the fact that the $l_1$ extension of the Einstein equation could contain terms $\hat{G}^{b,}{}_{\bullet}X$, where $X$ is any function of the Cartan forms whose derivatives are with respect to the usual level zero space-time coordinates, and $\bullet$ represents any remaining $A_1^{+++}$ indices on $G$. Such terms lead in the variation of the Einstein equation $\mathcal{E}'_{ab}$ to expressions of the form $\Lambda^{be}\overline{G}_{e,\bullet}X$. These terms would result, in the variation of equation (3.134), in an addition to the dual graviton equation $\overline{E}_{ab}$ of a term of the form $G^b{}_{,\bullet}X$. This is a term which contains a space-time derivative whose $b$ index corresponds to the second index on $\overline{E}'_a{}^b$.

Focusing on the first term in equation (3.130) we can re-write it as

$$\begin{aligned}(\det e)^{\frac{1}{2}}e^{\nu[c}\partial_\nu\overline{G}_{[c,a]}{}^{b]} = \frac{1}{4}(\det e)^{\frac{1}{2}}e^{\nu c}\partial_\nu\overline{G}_{c,a}{}^b - \frac{1}{4}(\det e)^{\frac{1}{2}}(e^{\nu c}\partial_\nu\overline{G}_{a,c}{}^b + e^{\nu c}\partial_\nu\overline{G}^{b,}{}_{ac})\\ +\frac{1}{8}(\det e)^{\frac{1}{2}}(e^{\nu b}\partial_\nu\overline{G}_{a,c}{}^c + e_a{}^\nu\partial_\nu\overline{G}^{b,c}{}_c)\\ -\frac{1}{4}(\det e)^{\frac{1}{2}}(e^{\nu b}\partial_\nu\overline{G}_{c,a}{}^c - e^\nu{}_c\partial_\nu\overline{G}^{b,}{}_a{}^c) + \frac{1}{8}(\det e)^{\frac{1}{2}}(e^{\nu b}\partial_\nu\overline{G}_{a,c}{}^c - e_a{}^\nu\partial_\nu\overline{G}^{b,c}{}_c) \quad .\end{aligned} \tag{3.136}$$

The first three terms are symmetric under the interchange of $a$ and $b$. To simplify the last two terms we can rewrite them using the Maurer-Cartan equations. The Cartan form $\mathcal{V} = g^{-1}dg$ obeys the Maurer-Cartan equation

$$d\mathcal{V} = -\mathcal{V}\wedge\mathcal{V} \tag{3.137}$$

so that we find

$$\partial_\mu\mathcal{V}_\nu - \partial_\nu\mathcal{V}_\mu + \mathcal{V}_\mu\mathcal{V}_\nu - \mathcal{V}_\nu\mathcal{V}_\mu = 0. \tag{3.138}$$

Using the form of $\mathcal{V}$ of equation (3.29) we find, amongst other equations, that

$$\begin{aligned}(\det e)^{\frac{1}{2}}e_c{}^\mu\partial_\mu\overline{G}_{d,ab} - (\det e)^{\frac{1}{2}}e_d{}^\mu\partial_\mu\overline{G}_{c,ab} + G_{c,d}{}^e\overline{G}_{e,ab} - G_{d,c}{}^e\overline{G}_{e,ab} - \frac{1}{2}G_{c,e}{}^e\overline{G}_{d,ab}\\ +\frac{1}{2}G_{d,e}{}^e\overline{G}_{c,ab} + G_{c,a}{}^e\overline{G}_{d,eb} + G_{c,b}{}^e\overline{G}_{d,ae} - G_{d,a}{}^e\overline{G}_{c,eb} - G_{d,b}{}^e\overline{G}_{c,ae} = 0 \quad .\end{aligned} \tag{3.139}$$

Using this last equation we can rewrite the last two terms of equation (3.136) as

$$\begin{aligned}-\frac{1}{4}(\det e)^{\frac{1}{2}}[e^{\nu b}\partial_\nu\overline{G}_{c,a}{}^c - e^\nu{}_c\partial_\nu\overline{G}^{b,}{}_a{}^c] = +\frac{1}{4}(G^{b,}{}_c{}^e\overline{G}_{e,a}{}^c - G_{c,}{}^{be}\overline{G}_{e,a}{}^c - \frac{1}{2}G^{b,e}{}_e\overline{G}_{c,a}{}^c\\ +\frac{1}{2}G_{c,e}{}^e\overline{G}^{b,}{}_a{}^c + G^{b,}{}_a{}^e\overline{G}_{c,e}{}^c + G^{b,ce}\overline{G}_{c,ae} - G_{c,a}{}^e\overline{G}^{b,}{}_e{}^c - G_{c,}{}^{ce}\overline{G}^{b,}{}_{ae}) \quad ,\end{aligned} \tag{3.140}$$

and

$$\begin{aligned}\frac{1}{8}(\det e)^{\frac{1}{2}}(e^{\nu b}\partial_\nu\overline{G}_{a,c}{}^c - e_a{}^\nu\partial_\nu\overline{G}^{b,c}{}_c) = -\frac{1}{8}(G^{b,}{}_a{}^e\overline{G}_{e,c}{}^c - G_{a,}{}^{be}\overline{G}_{e,}{}^c{}_c - \frac{1}{2}G^{b,e}{}_e\overline{G}_{a,c}{}^c\\ +\frac{1}{2}G_{a,e}{}^e\overline{G}^{b,c}{}_c + G^{b,}{}_c{}^e\overline{G}_{a,}{}^c{}_e - G_{a,c}{}^e\overline{G}^{b,}{}_e{}^c - G_{a,c}{}^e\overline{G}^{b,c}{}_e) \quad .\end{aligned} \tag{3.141}$$

Using these last two equations, and explicitly writing out the anti-symmetrisations of the $G\bar{G}$ terms, $\overline{E}'_{ab}$ in equation (3.130) can be written as

$$\begin{aligned}\overline{E}'_a{}^b = &\frac{1}{4}(\det e)^{\frac{1}{2}}(e^{\nu c}\partial_\nu\overline{G}_{c,a}{}^b - e^{\nu c}\partial_\nu\overline{G}_{a,c}{}^b - e^{\nu c}\partial_\nu\overline{G}^{b,}{}_{ac} + \frac{1}{2}e^{\nu b}\partial_\nu\overline{G}_{a,c}{}^c + \frac{1}{2}e_a{}^\nu\partial_\nu\overline{G}^{b,c}{}_c) \\ &+\frac{1}{8}\Big[G^{c,b}{}_e(2\overline{G}_{c,a}{}^e - 2\overline{G}^{e,}{}_{ac} - 2\overline{G}_{a,c}{}^e) + G^{b,}{}_c{}^e(-2\overline{G}_{a,}{}^c{}_e + 2\overline{G}_{e,a}{}^c + 2\overline{G}^{c,}{}_{ae}) \\ &\quad + G^{e,c}{}_e(-2\overline{G}_{c,a}{}^b + 2\overline{G}_{a,c}{}^b) + G^{d,c}{}_a(-2\overline{G}_{d,c}{}^b + 2\overline{G}_{c,d}{}^b) + G^{b,c}{}_a(2\overline{G}_{d,c}{}^d - 2\overline{G}_{c,d}{}^d) \\ &\quad + G^{d,c}{}_c(\overline{G}_{d,a}{}^b - \overline{G}_{a,d}{}^b + \overline{G}^{b,}{}_{ad}) + G^{b,c}{}_c(-\overline{G}_{d,a}{}^d + \overline{G}_{a,d}{}^d + \frac{1}{2}\overline{G}_{a,d}{}^d - \overline{G}_{d,a}{}^d) \\ &\quad + (G_{a,}{}^{be}\overline{G}_{e,c}{}^c) + G^{b,}{}_a{}^e(-\overline{G}_{e,c}{}^c + 2\overline{G}_{c,e}{}^c) + G_{a,c}{}^e(\overline{G}^{b,}{}_e{}^c + \overline{G}^{b,c}{}_e) \\ &\quad + G_{c,a}{}^e(-2\overline{G}^{b,}{}_e{}^c) + G_{c,}{}^{ce}(-2\overline{G}^{b,}{}_{ae}) + G_{a,e}{}^e(-\frac{1}{2}\overline{G}^{b,c}{}_c)\Big] \quad . \end{aligned} \tag{3.142}$$

It is useful to number the fourteen expressions in $(\ldots)$ brackets as $1, 2, \ldots, 14$.

We now add terms to both sides of equation (3.134), the variation of the $l_1$ extension of the Einstein expression $\tilde{\mathcal{E}}'_{ab}$, and to the dual graviton expressions that live on the right-hand side of equation (3.134), now written in the form of equation (3.142). The terms of equation (3.142) can be divided in to three types

- (a) terms containing a $G_{b,\bullet}$,
- (b) terms containing a $G_{a,\bullet}$,
- (c) remaining terms.

Type (a) terms are all of the form $G_{b,\bullet}X$ and so can immediately be removed by adding terms to $\mathcal{E}'_a{}^b$ as explained previously. These terms occur in equation (3.142) as terms 3, 6, 8, 10, 11, 12, 13, 14 as well as the last expression in term 7. We remove these terms by adding their negatives to $\overline{E}'_{ab}$. That is, we simply add to $\overline{E}'_a{}^b$ in equation (3.142) the expression

$$\begin{aligned} -Y^{b,}{}_a := &-\frac{1}{4}G^{b,}{}_c{}^e(-\overline{G}_{a,}{}^c{}_e + \overline{G}_{e,a}{}^c + \overline{G}^{c,}{}_{ae}) + \frac{1}{4}G^{b,c}{}_a(\overline{G}_{c,d}{}^d - \overline{G}_{d,c}{}^d) \\ &- \frac{1}{8}G^{b,c}{}_c(-2\overline{G}_{d,a}{}^d + \frac{3}{2}\overline{G}_{a,d}{}^d) - \frac{1}{8}G^{b,}{}_a{}^e(-\overline{G}_{e,c}{}^c + 2\overline{G}_{c,e}{}^c) - \frac{1}{4}G_{a,c}{}^e\overline{G}^{b,}{}_e{}^c \\ &+ \frac{1}{4}G_{c,a}{}^e\overline{G}^{b,}{}_e{}^c + \frac{1}{4}G_{c,}{}^{ce}\overline{G}^{b,}{}_{ae} + \frac{1}{16}G_{a,e}{}^e\overline{G}^{b,c}{}_c - \frac{1}{8}G^{d,c}{}_c\overline{G}^{b,}{}_{ad} \quad . \end{aligned} \tag{3.143}$$

In practice this is of course done by analysing the three $-4\Lambda_{bp}\overline{E}'_q{}^b$-type terms on the right-hand side of equation (3.134). These type (a) terms in equation (3.142) arise on the right-hand side of equation (3.134) in the form of three $-4\Lambda_{bp}(+Y^{b,}{}_q)$-type terms. These can be written as $-2\delta(+\hat{Y}_{p,a})$, involving level one derivatives. The effect is that we then modify the Einstein equation on the left-hand side of equation (3.134) by adding the three $-2(-\hat{Y}_{p,a})$-type terms to $\tilde{\mathcal{E}}'_{ab}$.

Type (b) terms by definition contain a $G_{a,\bullet}X$ factor and occur in equation (3.142) as the terms 2 (only the last expression), 4 (only the last expression), 7 (only the middle expression) and 9. These terms are given by

$$+\frac{1}{8}(-2G^{c,b}{}_e\overline{G}_{a,c}{}^e + 2G^{e,c}{}_e\overline{G}_{a,c}{}^b - G^{d,c}{}_c\overline{G}_{a,d}{}^b + G_{a,}{}^{be}\overline{G}_{e,c}{}^c) \quad . \tag{3.144}$$

For each of these terms we can add to the dual graviton equation a corresponding term with the $a$ and $b$ indices swapped, since it contains a $G_{b,\bullet}X$ factor, so that in effect we can simply symmetrise type (b) terms by hand. The effect is that we add the terms

$$+\frac{1}{8}(-2G^{c,}{}_{ae}\overline{G}^{b,}{}_c{}^e + 2G^{e,c}{}_e\overline{G}^{b,}{}_{ca} - G^{d,c}{}_c\overline{G}^{b,}{}_{da} + G^{b,}{}_a{}^e\overline{G}_{e,c}{}^c) \tag{3.145}$$

to the dual graviton equation. The net effect of this is that we find in $\bar{E}_a{}^b$ the $a, b$ symmetric terms

$$\begin{aligned} &+\frac{1}{4}(G^{c,b}{}_e\overline{G}_{a,c}{}^e + G^{c,}{}_{ae}\overline{G}^{b,}{}_c{}^e) + \frac{1}{4}(G^{e,c}{}_e\overline{G}_{a,c}{}^b + G^{e,c}{}_e\overline{G}^{b,}{}_{ca}) \\ &-\frac{1}{8}(G^{d,c}{}_c\overline{G}_{a,d}{}^b + G^{d,c}{}_c\overline{G}^{b,}{}_{da}) + \frac{1}{8}(G_{a,}{}^{be}\overline{G}_{e,c}{}^c + G^{b,}{}_a{}^e\overline{G}_{e,c}{}^c) \quad . \end{aligned} \tag{3.146}$$

In practice, this is carried out by again analysing equation (3.134) as was explained for the Type (a) terms, resulting in more modifications to the $l_1$-extended Einstein equation on the left-hand side.

The terms of the type (c) are given by

$$+\frac{1}{2}(G^{c,be}\overline{G}_{[c,e]a} - G^{c,d}{}_a\overline{G}_{[c,d]}{}^b) - \frac{1}{4}G^{e,c}{}_e\overline{G}_{c,a}{}^b + \frac{1}{8}G^{d,c}{}_c\overline{G}_{d,a}{}^b \tag{3.147}$$

The last two terms are symmetric in $a \leftrightarrow b$ while the first two terms can be written as

$$\begin{aligned} &+\frac{1}{2}(G^{c,bd}\overline{G}_{[c,d]a} + G^{c,}{}_a{}^d\overline{G}_{[c,d]}{}^b) - \frac{1}{4}\varepsilon^{cde_1e_2}\overline{G}_{[e_1,e_2]a}\overline{G}_{[c,d]}{}^b \\ &\qquad + \frac{1}{2}(E_{a,}{}^{cd} - \frac{1}{2}G_{a,}{}^{[c,d]})\overline{G}_{[c,d]}{}^b \quad . \end{aligned} \tag{3.148}$$

The first and second terms in this expression are symmetric, while the third term is the gravity-dual gravity duality relation and the fourth term can be viewed as a modulo transformation to which this duality relation holds. While the first two terms contribute to the dual gravity equation of motion, the last two terms can be reinterpreted as terms that explicitly occur in the variation of the $l_1$ extended gravity equation of motion $\tilde{\mathcal{E}}'_{ab}$.

Carrying out these steps explicitly in equation (3.134) results in the following expression for the $l_1$ extended Ricci tensor

$$
\begin{aligned}
\mathcal{E}_{ab} := {} & (\det e)\mathcal{R}_{ab} + \frac{1}{2}[G^{c,}{}_{be}\hat{\overline{G}}_{a,c}{}^{e} - G^{e,c}{}_{e}\hat{\overline{G}}_{a,cb} + \frac{1}{2}G^{d,c}{}_{c}\hat{\overline{G}}_{a,db} - \frac{1}{2}\hat{G}_{a,b}{}^{e}\overline{G}_{e,c}{}^{c} \\
& + \hat{G}_{a,c}{}^{e}(-\overline{G}_{b,}{}^{c}{}_{e} + \overline{G}_{e,b}{}^{c} + \overline{G}^{c,}{}_{be}) + \frac{1}{2}G^{d,c}{}_{c}\hat{\overline{G}}_{a,bd} - 2\hat{G}_{a,}{}^{c}{}_{b}\overline{G}_{[c,d]}{}^{d} \\
& - G_{c,}{}^{ce}\hat{\overline{G}}_{a,be} + \frac{1}{2}\hat{G}_{a,}{}^{c}{}_{c}(-2\overline{G}_{d,b}{}^{d} + \frac{3}{2}\overline{G}_{b,d}{}^{d}) + \frac{1}{2}\hat{G}_{a,b}{}^{e}(-\overline{G}_{e,c}{}^{c} + 2\overline{G}_{c,e}{}^{c}) \\
& - G_{c,b}{}^{e}\hat{\overline{G}}_{a,e}{}^{c} + G_{b,c}{}^{e}\hat{\overline{G}}_{a,e}{}^{c} - \frac{1}{4}G_{b,e}{}^{e}\hat{\overline{G}}_{a,}{}^{c}{}_{c} - 4c_2(\hat{G}_{a,}{}^{c}{}_{c}\overline{G}_{b,d}{}^{d} + G_{b,}{}^{c}{}_{c}\hat{\overline{G}}_{a,d}{}^{d})] \\
& + \frac{1}{2}[G^{c,}{}_{ae}\hat{\overline{G}}_{b,c}{}^{e} - G^{e,c}{}_{e}\hat{\overline{G}}_{b,ca} + \frac{1}{2}G^{d,c}{}_{c}\hat{\overline{G}}_{b,da} - \frac{1}{2}\hat{G}_{b,a}{}^{e}\overline{G}_{e,c}{}^{c} \\
& + \hat{G}_{b,c}{}^{e}(-\overline{G}_{a,}{}^{c}{}_{e} + \overline{G}_{e,a}{}^{c} + \overline{G}^{c,}{}_{ae}) + \frac{1}{2}G^{d,c}{}_{c}\hat{\overline{G}}_{b,ad} - 2\hat{G}_{b,}{}^{c}{}_{a}\overline{G}_{[c,d]}{}^{d} \\
& - G_{c,}{}^{ce}\hat{\overline{G}}_{b,ae} + \frac{1}{2}\hat{G}_{b,}{}^{c}{}_{c}(-2\overline{G}_{d,a}{}^{d} + \frac{3}{2}\overline{G}_{a,d}{}^{d}) + \frac{1}{2}\hat{G}_{b,a}{}^{e}(-\overline{G}_{e,c}{}^{c} + 2\overline{G}_{c,e}{}^{c}) \\
& - G_{c,a}{}^{e}\hat{\overline{G}}_{b,e}{}^{c} + G_{a,c}{}^{e}\hat{\overline{G}}_{b,e}{}^{c} - \frac{1}{4}G_{a,e}{}^{e}\hat{\overline{G}}_{b,}{}^{c}{}_{c} - 4c_2(\hat{G}_{b,}{}^{c}{}_{c}\overline{G}_{a,d}{}^{d} + G_{a,}{}^{c}{}_{c}\hat{\overline{G}}_{b,d}{}^{d})] \\
& - \frac{1}{2}\eta_{ab}[G^{c,}{}_{e_1e_2}\hat{\overline{G}}^{e_1,}{}_{c}{}^{e_2} - G^{e_1,c}{}_{e_1}\hat{\overline{G}}^{e_2,}{}_{ce_2} + \frac{1}{2}G^{d,c}{}_{c}\hat{\overline{G}}^{e,}{}_{de} - \frac{1}{2}\hat{G}^{e_1,}{}_{e_1}{}^{e_2}\overline{G}_{e_2,c}{}^{c} \\
& + \hat{G}^{e_1,}{}_{c}{}^{e_2}(-\overline{G}_{e_1,}{}^{c}{}_{e_2} + \overline{G}_{e_2,e_1}{}^{c} + \overline{G}^{c,}{}_{e_1e_2}) + \frac{1}{2}G^{d,c}{}_{c}\hat{\overline{G}}^{e,}{}_{ed} - 2\hat{G}^{e,c}{}_{e}\overline{G}_{[c,d]}{}^{d} \\
& - G_{c,}{}^{ce_1}\hat{\overline{G}}^{e_2,}{}_{e_2e_1} + \frac{1}{2}\hat{G}^{e,c}{}_{c}(-2\overline{G}_{d,e}{}^{d} + \frac{3}{2}\overline{G}_{e,d}{}^{d}) + \frac{1}{2}\hat{G}^{e_1,}{}_{e_1}{}^{e_2}(-\overline{G}_{e_2,c}{}^{c} + 2\overline{G}_{c,e_2}{}^{c}) \\
& - G_{c,e_1}{}^{e_2}\hat{\overline{G}}^{e_1,}{}_{e_2}{}^{c} + G_{e_1,c}{}^{e_2}\hat{\overline{G}}^{e_1,}{}_{e_2}{}^{c} - \frac{1}{4}G_{e_1,e_2}{}^{e_2}\hat{\overline{G}}^{e_1,c}{}_{c} - 4c_2(\hat{G}^{e,c}{}_{c}\overline{G}_{e,d}{}^{d} + G_{e,}{}^{c}{}_{c}\hat{\overline{G}}^{e,}{}_{d}{}^{d})]
\end{aligned}
\tag{3.149}
$$

Its variation is given by

$$
\begin{aligned}
\delta\mathcal{E}_{ab} = {} & -4\Lambda_{ea}\overline{E}_{b}{}^{e} - 4\Lambda_{eb}\overline{E}_{a}{}^{e} + 4\eta_{ab}\Lambda^{e_1}{}_{e_2}\overline{E}_{e_1}{}^{e_2} \\
& + \Lambda^{e_1e_2}\varepsilon_{bc}{}^{f_1f_2}(E_{e_2,a}{}^{e}(\det e)^{\frac{1}{2}}\omega_{e_1,f_1f_2} - E_{e_1,f_1f_2}(\det e)^{\frac{1}{2}}\omega_{e_2,a}{}^{c}) \\
& + \Lambda^{e_1e_2}\varepsilon_{ac}{}^{f_1f_2}(E_{e_2,b}{}^{e}(\det e)^{\frac{1}{2}}\omega_{e_1,f_1f_2} - E_{e_1,f_1f_2}(\det e)^{\frac{1}{2}}\omega_{e_2,b}{}^{c}) \\
& - 2\Lambda_{ea}(E_{b,}{}^{cd} - \frac{1}{2}G_{b,}{}^{[c,d]})\overline{G}_{[c,d]}{}^{e} - 2\Lambda_{eb}(E_{a,}{}^{cd} - \frac{1}{2}G_{a,}{}^{[c,d]})\overline{G}_{[c,d]}{}^{e} \\
& + 2\eta_{ab}\Lambda^{e_1}{}_{e_2}(E_{e_1,}{}^{cd} - \frac{1}{2}G_{e_1,}{}^{[c,d]})\overline{G}_{[c,d]}{}^{e_2} \quad ,
\end{aligned}
\tag{3.150}
$$

where the non-symmetric $\overline{E}'_{a}{}^{b}$ has replaced by the symmetric second order dual graviton

equation of motion:

$$\begin{aligned}
\overline{E}_a{}^b := &\frac{1}{4}(\det e)^{\frac{1}{2}}(e^{\nu c}\partial_\nu \overline{G}_{c,a}{}^b - e^{\nu c}\partial_\nu \overline{G}_{a,}{}^b{}_c - e^{\nu c}\partial_\nu \overline{G}^{b,}{}_{ac} + \frac{1}{2}e^{\nu b}\partial_\nu \overline{G}_{a,c}{}^c + \frac{1}{2}e_a{}^\nu \partial_\nu \overline{G}^{b,c}{}_c) \\
&- \frac{1}{4}G^{e,c}{}_e \overline{G}_{c,a}{}^b + \frac{1}{8}G^{d,c}{}_c \overline{G}_{d,a}{}^b \\
&- \frac{1}{4}\varepsilon^{cde_1e_2}\overline{G}_{[e_1,e_2]a}\overline{G}_{[c,d]}{}^b + \frac{1}{2}(G^{c,bd}\overline{G}_{[c,d]a} + G^{c,}{}_a{}^d \overline{G}_{[c,d]}{}^b) \\
&- \frac{1}{4}(G^{c,}{}_{ae}\overline{G}^{b,}{}_c{}^e + G^{c,b}{}_e \overline{G}_{a,c}{}^e) + \frac{1}{4}G^{e,c}{}_e(\overline{G}^{b,}{}_{ca} + \overline{G}_{a,c}{}^b) \\
&- \frac{1}{8}G^{d,c}{}_c(\overline{G}^{b,}{}_{da} + \overline{G}_{a,d}{}^b) + \frac{1}{8}(G^{b,}{}_a{}^e + G_{a,}{}^{be})\overline{G}_{e,c}{}^c \\
&= 0.
\end{aligned} \tag{3.151}$$

This is the symmetric dual graviton equation of motion in four dimensions, in $A_1^{+++}$ tangent space indices, proposed in equation (5.17) of [29].

There is one potential ambiguity in these results. This ambiguity arises because some of the terms in $\overline{E}'_a{}^b$ are of the form $G_{a,\bullet}\overline{G}_{b,\bullet}$ which is both of type (a) and of type (b). Listed as they appear in $\overline{E}'_a{}^b$, these terms arise from terms 1, 7, 10, 13:

$$-\frac{1}{4}G^{b,}{}_c{}^e \overline{G}_{a,}{}^c{}_e + \frac{1}{4}G_{a,c}{}^e \overline{G}^{b,c}{}_e + \frac{3}{16}G^{b,c}{}_c \overline{G}_{a,d}{}^d - \frac{1}{16}G_{a,c}{}^c \overline{G}^{b,d}{}_d \quad . \tag{3.152}$$

Such terms can be treated as either type (a) terms, terms which can immediately be removed, or as type (b) terms, terms that must be symmetrised by adding extra terms that must then be removed.

The net effect is that we can add the terms to the dual graviton equation that are of the form

$$\begin{aligned}
&+ c_1(G_{b,c}{}^e \overline{G}_{a,ce} + G_{a,c}{}^e \overline{G}_{b,ce}) \\
&+ c_2(G_{b,c}{}^c \overline{G}_{a,d}{}^d + G_{a,c}{}^c \overline{G}_{b,d}{}^d) \quad ,
\end{aligned} \tag{3.153}$$

for constants $c_1$ and $c_2$. One important check of the above dual graviton equation (3.152) is that it is Lorentz covariant under the transformations

$$\begin{aligned}
\delta G_{a,bc} &= \Lambda_a{}^e G_{e,bc} + \Lambda_b{}^e G_{a,ec} + \Lambda_c{}^e G_{a,be} + e_a{}^\mu \partial_\mu \Lambda^{cb} \quad , \\
\delta \overline{G}_{a,bc} &= \Lambda_a{}^e \overline{G}_{e,bc} + \Lambda_b{}^e \overline{G}_{a,ec} + \Lambda_c{}^e \overline{G}_{a,be} \quad .
\end{aligned} \tag{3.154}$$

Such ambiguous terms are apparently not needed, and so have been left out of equation (3.152). By considering the $I_c(A_1^{+++})$ variation of the nonlinear dual graviton equation (3.152), one could resolve whether the ambiguous terms of equation (3.153) should be included or not. In the next section we consider this at linearised level.

### 3.4.4 $A_1^{+++}$ Dual Graviton in World Indices

We now illustrate how to re-write the $A_1^{+++}$ Nonlinear Dual Graviton equation of motion in world indices, starting from equation (3.152). The first term in (3.152) simplifies as

$$\begin{aligned}
&\frac{1}{4}(\det e)^{\frac{1}{2}} e^{\nu c}\partial_\nu[\overline{G}_{c,a}{}^b] \\
&\quad= \frac{1}{4}(\det e)^{\frac{1}{2}} e^{\nu c}\partial_\nu[(\det e)^{\frac{1}{2}} e_c{}^\mu \overline{G}_{\mu,a}{}^b] \\
&\quad= \frac{1}{8}(\det e)^{\frac{1}{2}} G^{c,}{}_e{}^e e_c{}^\mu \overline{G}_{\mu,a}{}^b - \frac{1}{4}(\det e)^{\frac{1}{2}} e^{\nu c}[G_{\nu,c}{}^\mu]\overline{G}_{\mu,a}{}^b - \frac{1}{4}(\det e)^{\frac{1}{2}} e_c{}^\mu G^{c,ad}\overline{G}_{\mu,d}{}^b \\
&\quad- \frac{1}{4}(\det e)^{\frac{1}{2}} e_c{}^\mu G^{c,bd}\overline{G}_{\mu,ad} + \frac{1}{4}(\det e) e^{\nu c} e_c{}^\mu e_a{}^\rho e^{b\kappa}\partial_\nu \overline{G}_{\mu,\rho\kappa} \quad ,
\end{aligned} \tag{3.155}$$

where we use

$$e_a{}^\mu \partial_\mu (\det e)^{\frac{1}{2}} = \frac{1}{2}(\det e)^{\frac{1}{2}} e_a{}^\mu (e^c{}_\rho \partial_\mu e_c{}^\rho) = \frac{1}{2} G_{a,c}{}^c \quad . \tag{3.156}$$

The second term simplifies as

$$\begin{aligned}
-\frac{1}{4}(\det e)^{\frac{1}{2}} e^{\nu c}\partial_\nu \overline{G}_{a,}{}^b{}_c &= -\frac{1}{4}(\det e)^{\frac{1}{2}} e^{\mu c}\partial_\nu[(\det e)^{\frac{1}{2}} e_a{}^\mu \overline{G}_{\mu,}{}^b{}_c] \\
&= -\frac{1}{8}(\det e)^{\frac{1}{2}} e^{\nu c} G^{c,}{}_e{}^e e_a{}^\mu \overline{G}_{\mu,}{}^b{}_c + \frac{1}{4}(\det e)^{\frac{1}{2}} e^{\nu c} G_{\nu,a}{}^\mu \overline{G}_{\mu,}{}^b{}_c \\
&+ \frac{1}{4}(\det e)^{\frac{1}{2}} e_a{}^\mu G^{c,bd} e_c{}^\lambda \overline{G}_{\mu,d\lambda} + \frac{1}{4}(\det e)^{\frac{1}{2}} e_a{}^\mu e^{b\kappa} G^{c,}{}_c{}^d \overline{G}_{\mu,\kappa d} \\
&- \frac{1}{4}(\det e) e^{\nu c} e_a{}^\mu e^{b\kappa} e_c{}^\lambda \partial_\nu \overline{G}_{\mu,\kappa\lambda} \quad .
\end{aligned} \tag{3.157}$$

The third term is this with $a \leftrightarrow b$

$$\begin{aligned}
-\frac{1}{4}(\det e)^{\frac{1}{2}} e^{\nu c}\partial_\nu \overline{G}^{b,}{}_{ac} &= -\frac{1}{8}(\det e)^{\frac{1}{2}} e^{\nu c} G^{c,}{}_e{}^e e^{b\mu} \overline{G}_{\mu,ac} + \frac{1}{4}(\det e)^{\frac{1}{2}} e^{\nu c} G_{\nu,}{}^{b\mu} \overline{G}_{\mu,ac} \\
&+ \frac{1}{4}(\det e)^{\frac{1}{2}} e^{b\mu} G^{c,}{}_a{}^d e_c{}^\lambda \overline{G}_{\mu,d\lambda} + \frac{1}{4}(\det e)^{\frac{1}{2}} e^{b\mu} e_a{}^\kappa G^{c,}{}_c{}^d \overline{G}_{\mu,\kappa d} \\
&- \frac{1}{4}(\det e) e^{\nu c} e^{b\mu} e_a{}^\kappa e_c{}^\lambda \partial_\nu \overline{G}_{\mu,\kappa\lambda} \quad .
\end{aligned} \tag{3.158}$$

The fourth term simplifies as

$$\begin{aligned}
\frac{1}{8}(\det e)^{\frac{1}{2}} e^{\nu b}\partial_\nu[\overline{G}_{a,c}{}^c] &= \frac{1}{8}(\det e)^{\frac{1}{2}} e^{\nu b}\partial_\nu[(\det e)^{\frac{1}{2}} e_a{}^\mu \overline{G}_{\mu,c}{}^c] \\
&= \frac{1}{16}(\det e)^{\frac{1}{2}} G^{b,}{}_e{}^e e_a{}^\mu \overline{G}_{\mu,c}{}^c - \frac{1}{8}(\det e)^{\frac{1}{2}} G^{b,}{}_a{}^\mu \overline{G}_{\mu,c}{}^c \\
&- \frac{1}{8}(\det e)^{\frac{1}{2}} e_a{}^\mu G^{b,}{}_c{}^\lambda e^{c\kappa} \overline{G}_{\mu,\lambda\kappa} - \frac{1}{8}(\det e)^{\frac{1}{2}} e_a{}^\mu e_c{}^\lambda G^{b,c\kappa} \overline{G}_{\mu,\lambda\kappa} \\
&+ \frac{1}{8}(\det e) e^{\nu b} e_a{}^\mu e_c{}^\lambda e^{c\kappa} \partial_\nu \overline{G}_{\mu,\lambda\kappa} \quad .
\end{aligned} \tag{3.159}$$

Finally, the fifth term is this with $a \leftrightarrow b$

$$\begin{aligned}
\frac{1}{8}(\det e)^{\frac{1}{2}} e_a{}^\nu \partial_\nu[\overline{G}^{b,c}{}_c] &= \frac{1}{16}(\det e)^{\frac{1}{2}} G_{a,e}{}^e e^{b\mu} \overline{G}_{\mu,c}{}^c - \frac{1}{8}(\det e)^{\frac{1}{2}} G_{a,}{}^{b\mu} \overline{G}_{\mu,c}{}^c \\
&- \frac{1}{8}(\det e)^{\frac{1}{2}} e^{b\mu} G_{a,c}{}^\lambda e^{c\kappa} \overline{G}_{\mu,\lambda\kappa} - \frac{1}{8}(\det e)^{\frac{1}{2}} e^{b\mu} e_c{}^\lambda G_{a,}{}^{c\kappa} \overline{G}_{\mu,\lambda\kappa} \\
&+ \frac{1}{8}(\det e) e^\nu{}_a e^{b\mu} e_c{}^\lambda e^{c\kappa} \partial_\nu \overline{G}_{\mu,\lambda\kappa}
\end{aligned} \tag{3.160}$$

Collecting first all the $\partial\overline{G}$ type terms we find

$$\begin{aligned}
\partial\overline{G}_a{}^b := &\frac{1}{8}(\det e)[2e^{\nu c}e_c{}^\mu e_a{}^\rho e^{b\kappa}\partial_\nu\overline{G}_{\mu,\rho\kappa} - 2e^{\nu c}e_a{}^\mu e^{b\kappa}e_c{}^\lambda\partial_\nu\overline{G}_{\mu,\kappa\lambda} - 2e^{\nu c}e^{b\mu}e_a{}^\kappa e_c{}^\lambda\partial_\nu\overline{G}_{\mu,\kappa\lambda} \\
&\qquad + e^{\nu b}e_a{}^\mu e_c{}^\lambda e^{c\kappa}\partial_\nu\overline{G}_{\mu,\lambda\kappa} + e^\nu{}_a e^{b\mu}e_c{}^\lambda e^{c\kappa}\partial_\nu\overline{G}_{\mu,\lambda\kappa}] \\
&= \frac{1}{8}(\det e)g^{\lambda\nu}e_a{}^\rho e^{b\kappa}[2\partial_\nu(\overline{G}_{\lambda,\kappa\rho} - \overline{G}_{\kappa,\lambda\rho}) - 2\partial_\nu\overline{G}_{\rho,\kappa\lambda} + \partial_\kappa\overline{G}_{\rho,\lambda\nu} + \partial_\rho\overline{G}_{\kappa,\lambda\nu}] \\
&= \frac{1}{8}(\det e)g^{\lambda\nu}e_a{}^\rho e^{b\kappa}[4\partial_\nu\overline{G}_{[\lambda,\kappa]\rho} - 2(\partial_\rho\overline{G}_{\lambda,\kappa\nu} - \partial_\rho\overline{G}_{\kappa,\lambda\nu})] \\
&= \frac{1}{2}(\det e)g^{\lambda\nu}e_a{}^\rho e^{b\kappa}[\partial_\nu\overline{G}_{[\lambda,\kappa]\rho} - \partial_\rho\overline{G}_{[\lambda,\kappa]\nu}] \\
&= (\det e)g^{\lambda\nu}e_a{}^\rho e^{b\kappa}\partial_{[\nu|}\overline{G}_{[\lambda,\kappa]|\rho]} \quad .
\end{aligned} \tag{3.161}$$

Continuing in this manner one can derive

$$\begin{aligned}
\overline{E}_{\mu\nu} = &g^{\rho\sigma}\partial_{[\sigma|}\overline{F}_{[\rho,\nu]|\mu]} + \frac{1}{4}g^{\rho\sigma}G_{\tau,\rho}{}^\tau(\overline{G}_{\nu,\mu\sigma} + \overline{G}_{\mu,\sigma\nu} - \overline{G}_{\sigma,\mu\nu}) \\
&+ \frac{1}{4}g^{\rho\sigma}G_{\rho,\tau}{}^\tau(-\overline{G}_{\nu,\mu\sigma} - \overline{G}_{\mu,\nu\sigma} + \overline{G}_{\sigma,\mu\nu}) + \frac{1}{4}g^{\rho\sigma}G_{\rho,\sigma}{}^\tau(-\overline{G}_{\tau,\mu\nu} + \overline{G}_{\mu,\nu\tau} + \overline{G}_{\nu,\mu\tau}) \\
&- \frac{1}{4}g^{\rho\sigma}G_{\nu,\rho}{}^\tau\overline{G}_{\mu,\tau\sigma} - g^{\rho\sigma}\frac{1}{4}G_{\mu,\rho}{}^\tau\overline{G}_{\nu,\tau\sigma} + \frac{1}{16}g^{\rho\sigma}G_{\nu,\tau}{}^\tau\overline{G}_{\mu,\rho\sigma} + \frac{1}{16}g^{\rho\sigma}G_{\mu,\tau}{}^\tau\overline{G}_{\nu,\rho\sigma} \\
&- \frac{1}{4}(\det e)^{-1}\varepsilon^{\tau_1\tau_2\tau_3\tau_4}\overline{G}_{[\tau_1,\tau_2]\mu}\overline{G}_{[\tau_3,\tau_4]\nu} \quad ,
\end{aligned} \tag{3.162}$$

where $\overline{F}_{\mu,\nu_1\nu_2} := \partial_\mu A_{\nu_1\nu_2}$. This is the symmetric dual graviton equation of motion in four dimensions, in space-time (world) indices, proposed in equation (5.22) of [29].

### 3.4.5 Linearised Dual Graviton Equation Variation

In this section we carry out the $I_c(A_1^{+++})$ variation of the nonlinear dual graviton equation at linearised level. The dual graviton transforms under the $I_c(A_1^{+++})$ variations into terms involving the graviton and dual dual-graviton, thus we can expect that the resulting variation may involve second order derivatives of the gravity and dual gravity equations, along with derivatives of the first order duality relations. At the linearised level the dual graviton equation is given by

$$\overline{E}_a{}^b{}_{(lin)} = \partial^{[d}\overline{G}_{[d,a]}{}^{b]} \quad . \tag{3.163}$$

By directly varying (3.163), the $l_1$ extension $\mathcal{E}_a{}^b{}_{(lin)}$ of this equation can be determined as

$$\begin{aligned}
\overline{\mathcal{E}}_a{}^b{}_{(lin)} = &\overline{E}_a{}^b{}_{(lin)} + \frac{1}{2}(\hat{\partial}_a(G^s)^{[d,}{}_d{}^{b]} + \hat{\partial}^b(G^s)_{[d,a]}{}^d) + \frac{1}{4}\hat{\partial}^c(G^{b,}{}_{(ac)} - (G^s)_{a,}{}^b{}_c + 2G^{b,}{}_{[ac]}) \\
&+ \frac{1}{4}\hat{\partial}^c(\varepsilon_{abc}{}^e\overline{G}_{[e,d]}{}^d - 2G^{[d,}{}_{da,}{}^{b]}{}_c - 2G^{d,}{}_{[d}{}^{b,}{}_{a]c} - G^{b,}{}_{ad,}{}^d{}_c) \quad .
\end{aligned} \tag{3.164}$$

Its variation under $I_c(A_1^{+++})$ gives

$$\begin{aligned}\delta\overline{\mathcal{E}}_a{}^b =& \frac{1}{2}\Lambda_a{}^c R_c{}^b + \frac{1}{2}R_a{}^c\Lambda^b{}_c + 3\partial^b E_{dac_1,}{}^d{}_{c_2}\Lambda^{c_1c_2} + \partial^b E_{c_2,ac_1}\Lambda^{c_1c_2} \\ &- \frac{3}{2}\partial^d(E_{dac_1,bc_2} + E_{dbc_1,ac_2})\Lambda^{c_1c_2} + \frac{3}{2}\partial^d E_{abc_1,dc_2}\Lambda^{c_1c_2} \\ &- \frac{1}{4}\varepsilon_{abc_1}{}^e\partial^d\overline{E}_{d,e}{}^{c_2}\Lambda^{c_1}{}_{c_2} + \varepsilon_{abc_1}{}^e\overline{E}_e{}^{c_2}\Lambda^{c_1}{}_{c_2} \quad .\end{aligned} \tag{3.165}$$

In these equation we have set $G^s_{a,bc} = G_{a,(bc)}$ and

$$E_{a_1a_2a_3,bc} = \frac{1}{3!}\varepsilon_{a_1a_2a_3}{}^e\overline{E}_{e,bc} \quad , \tag{3.166}$$

where $\overline{E}_{e,bc}$ is the dual gravity - dual dual gravity duality relation (3.71). Here all of the Cartan forms and duality relations are in linearised form. Setting $\overline{\mathcal{E}}_a{}^b = 0$ we find, at the linearised level, the gravity equation of motion $E_a{}^b = R_a{}^b = 0$ and the dual gravity - dual dual gravity duality relation $E_{a,b_1b_2} \doteq 0$ which can also be written in the form of equation (3.166). This shows that the variation of the dual gravity equation of motion transforms into quantities that we already know to vanish, and so the full nonlinear variation can be expected to fix the above ambiguous terms.

# Chapter 4

# $E_{11}$, Gravity and Dual Gravity

## 4.1 Motivation for $E_{11}$

As discussed in the introduction, the empirical rules of group oxidation suggest that the dimensional reduction of eleven-dimensional supergravity down to $D$ dimensions possesses an $E_{11-D}$ symmetry, and that the scalars can be described by a realization based on $E_{11-D}$ with respect to the maximal compact subgroup of $E_{11-D}$. A generalization of this to eleven dimensions, incorporating all the bosonic fields of supergravity, was proposed in [78], based closely on the results of [77][1]. In this chapter we will discuss this generalization.

Motivated, in part, by the nonlinear realization of gravity taken in [6] discussed in section 2.3), along with the desire to incorporate contributions from the three-form $A_{a_1a_2a_3}$ and six-form $A_{a_1\ldots a_6}$ gauge fields of the first-order formulation of eleven-dimensional supergravity [77], an extension of $\mathrm{GL}(11,\mathbb{R})$ to produce a new Lie algebra (in the same manner as that discussed in Section A.8) referred to as

$$G_{11} = \{K^a{}_b\ ,\ R^{a_1a_2a_3}\ ,\ R^{a_1\ldots a_6},\ P_a\} \tag{4.1}$$

was postulated, with $R^{a_1a_2a_3}$ and $R^{a_1\ldots a_6}$ naturally transforming as the rank three and rank six fundamental tensor representations of $\mathrm{SL}(11,\mathbb{R})$. As discussed in Appendix (A.8), the question now arises as to how one should define the commutators between $R^{a_1a_2a_3}, R^{a_1\ldots a_6}$ and $P_a$. The general index structure allows for no non-trivial $[R^{a_1a_2a_3}, P_b]$, $[R^{a_1\ldots a_6}, P_b]$ or $[R^{a_1a_2a_3}, R^{b_1\ldots b_6}]$ commutators based on $G_{11}$. The index structure of $[R^{a_1a_2a_3}, R^{a_4a_5a_6}]$ allows for only one non-trivial possibility in $G_{11}$, and introduces a free normalization factor chosen to be 2 (this appears natural in constructing the explicit nonlinear realization). We thus consider

$$\begin{aligned} &[K^a{}_b, K^c{}_d] = \delta^c{}_b K^a{}_d - \delta^a{}_d K^c{}_b \ , \ [K^a{}_b, P_c] = -\delta_c{}^a P_b \ , \\ &[K^a{}_b, R^{c_1c_2c_3}] = 3\delta^{[c_1}{}_b R^{|a|c_2c_3]} \ , \ [K^a{}_b, R^{c_1..c_6}] = 6\delta^{[c_1}{}_b R^{|a|c_2...c_6]} \ , \\ &[R^{a_1a_2a_3}, R^{a_4a_5a_6}] = 2R^{a_1...a_6} \ , \ [R^{a_1a_2a_3}, R^{b_1...b_6}] = 0 \ , \ [R^{a_1a_2a_3}, P_b] = [R^{a_1...a_6}, P_b] = 0 \ . \end{aligned} \tag{4.2}$$

[1] Very readable accounts of these results are given in [2] and [9].

The group element was then taken with $\mathrm{SO}(1,10)$ generated by $J_{ab} = 2\eta_{[a|e}K^e{}_{|b]}$ as the local $H$ subgroup, and the group element written (without use of $H$) as

$$g(x) = e^{x^a P_a} e^{h_a{}^b K^a{}_b} e^{\frac{1}{3!}A_{a_1a_2a_3}R^{a_1a_2a_3}} e^{\frac{1}{6!}A_{a_1...a_6}R^{a_1...a_6}} \ . \tag{4.3}$$

Following the discussion given earlier in (2.3), covariant derivatives constructed from the Cartan form coefficients arising in the Cartan form $g^{-1}dg$ derived using (4.3) can then be constructed, together with conditions on the associated coefficients so that these derivatives are the same covariant derivatives as those arising from a nonlinear realization of the conformal group.

The simultaneous nonlinear realization was shown to be powerful enough to identify the spin connection of general relativity, and field strengths $F_{[a_1,a_2a_3a_4]}$ and $F_{[a_1,a_2...a_7]}$, as the quantities transforming simultaneously under conformal and $G_{11}$ groups, and linearly local Lorentz transformations. This resulted in the unique equations

$$F_{a_1...a_4} = \frac{1}{7!}\varepsilon_{a_1...a_4}{}^{c_1...c_7}F_{c_1...c_7} \ . \tag{4.4}$$

However, invariant second order equations of motion involving the Riemann tensor could only be considered up to an ambiguous constant $C$

$$R_{\mu\nu b}{}^c e_c{}^\nu e_a{}^\mu - \frac{1}{2}\eta_{ab}R_{\mu\nu d}{}^c e_c{}^\nu e^{d\mu} = \frac{C}{4}(F_{ac_1c_2c_3}F_b{}^{c_1c_2c_3} - \frac{1}{6}\eta_{ab}F_{c_1...c_4}F^{c_1...c_4}) \ , \tag{4.5}$$

which implied a supersymmetric extension would be necessary to fix the constant $C$ to be $C = 1$.

However, considerations of dimensional reduction and the associated exceptional symmetries indicate that the above discussion is only part of a larger story. At this stage we take one of two approaches. The first approach is to explicitly construct an extension of $G_{11}$ which contains the Lie algebras of $E_7$ and $E_8$ as subalgebras. This construction turns out to give $E_{11}$ and was first derived in [78], with a readable account of this procedure given in [2], which we would simply just repeat here. Alternatively, a second approach is to argue that, since $G_{11}$ does not contain $E_7$ or $E_8$ as subalgebras, and the empirical rules of group oxidation suggest, since $E_{11}$ contains all the $E_{11-D}$ symmetry groups found via dimensional reduction, that $E_{11}$ may be a symmetry of eleven-dimensional supergravity. If an analysis of $E_{11}$ turns out to contain $G_{11}$ as a subalgebra, it subsumes the above results.

In fact, the generator $P_a$, being associated to the coordinates $x^a$ rather than fields, behaves slightly differently, so it should be excluded and instead treated as arising from the extended $G_{11}/\{P_a\}$ algebra. This leads to the idea of constructing the vector representation of the enlarged algebra, which when it is shown to be $E_{11}$, results in the vector representation of $E_{11}$, containing $P_a$ along with the central charges $Z^{a_1a_2}, Z^{a_1\dots a_5}$ of eleven-dimensional supergravity, now treated as coordinates in a larger 'generalized space-time', and more coordinates listed in the next section.

## 4.2 The Kac-Moody Algebra $E_{11}$

We now consider the Kac-Moody algebra $E_{11}$ [78], decomposed with respect to its $A_{10}$ subalgebra, and will then consider its $l_1$ representation. The $E_{11}$ Kac-Moody algebra is defined by the Dynkin diagram

$$\begin{array}{ccccccccccccccccccc} & & & & & & & & & & & & & & \otimes\ 11 & & & & \\ & & & & & & & & & & & & & & | & & & & \\ \bullet & - & \bullet & - & \bullet & - & \bullet & - & \bullet & - & \bullet & - & \bullet & - & \bullet & - & \bullet & - & \bullet \\ 1 & & 2 & & 3 & & 4 & & 5 & & 6 & & 7 & & 8 & & 9 & & 10 \end{array}$$

Here node 11 is labelled using a $\otimes$ to signify that we will be considering the decomposition of $E_{11}$ into representations of the remaining simple Lie algebra when this node is deleted, namely $A_{10} = \mathrm{SL}(11)$.

Using the program [49] we can find the representations of $E_{11}$ in terms of $A_{10}$ to level three as:

Table 4.1: $A_{10}$ representations in $E_{11}$

| $l$ | $A_{10}$ Irrep | $E_{11}$ Root | $\alpha^2$ | $d_r$ | $\mu$ | $R^\alpha$ |
|---|---|---|---|---|---|---|
| 0 | 0 0 0 0 0 0 0 0 0 0 | 0 0 0 0 0 0 0 0 0 0 0 | 0 | 1 | 1 | $D$ |
| 0 | 1 0 0 0 0 0 0 0 0 1 | -1 -1 -1 -1 -1 -1 -1 -1 -1 -1 0 | 2 | 120 | 1 | $\hat{K}^a{}_b$ |
| 1 | 0 0 0 0 0 0 0 1 0 0 | 0 0 0 0 0 0 0 0 0 0 1 | 2 | 165 | 1 | $R^{a_1a_2a_3}$ |
| 2 | 0 0 0 0 1 0 0 0 0 0 | 0 0 0 0 0 1 2 3 2 1 2 | 2 | 462 | 1 | $R^{a_1...a_6}$ |
| 3 | 0 0 1 0 0 0 0 0 0 1 | 0 0 0 1 2 3 4 5 3 1 3 | 2 | 1760 | 1 | $R^{a_1...a_8,b}$ |

The table implies that at levels zero, one, two, three we have the GL(11) generators

$$K^a{}_b \ , \quad R^{a_1a_2a_3} \quad , \quad R^{a_1..a_6} \quad , \quad R^{a_1..a_8,b} \ , \tag{4.6}$$

respectively, where as in the case of $A_1^{+++}$, we have defined $D = \sum_{c=1}^{11} K^c{}_c$ for $K^a{}_b$ a generator of GL(11), and used $D$ to define the combinations $K^a{}_b = \hat{K}^a{}_b + \frac{1}{11}\delta^a{}_b D$. The generators $R^{a_1a_2a_3}$ and $R^{a_1..a_6}$ are anti-symmetric in their indices, while $R^{a_1..a_8,b}$ is anti-symmetric in the $a_1,..,a_8$ indices. The tensor $R^{a_1..a_8,b}$ must satisfy irreducibility conditions, otherwise we could construct a new tensor from it with different index symmetries by contracting against $\varepsilon_{c_1c_2a_1...a_8b}$, and so it would not provide an irreducible representation. From the table, the dimension of the representation associated to the tensor $R^{a_1..a_8,b}$, is $d_r = 1760$. Analysed directly, the tensor has has $\binom{11}{8} \cdot 11 = 165 \cdot 11 = 1815$ terms from its index structure, so there are $1815 - 1760 = 55$ constraints i.e. irreducibility conditions. These irreducibility conditions are given by

$$R^{[a_1..a_8,b]} = 0 \ , \tag{4.7}$$

expressing $\binom{11}{9} = 55$ constraints. The negative level generators

$$R_{a_1a_2a_3} \ , \ R_{a_1...a_6} \ , \ R_{a_1..a_8,b} \ , \ \ldots \tag{4.8}$$

possess similar properties, for example $R_{[a_1..a_8,b]} = 0$.

The algebra of these generators can be determined by postulating the most general expressions consistent with the index symmetries, and then establishing consistency via the Jacobi identities [85]. Since the higher level generators form tensor representations of

GL(11), the following commutators are immediate

$$\begin{aligned}
[K^a{}_b, K^c{}_d] &= \delta^c{}_b K^a{}_d - \delta^a{}_d K^c{}_b \quad , \\
[K^a{}_b, R^{c_1\ldots c_3}] &= 3\delta^{[c_1}{}_b R^{|a|c_2c_3]} \quad , \quad [K^a{}_b, R_{c_1\ldots c_3}] = 3\delta^a{}_{[c_1} R_{|b|c_2c_3]} \quad , \\
[K^a{}_b, R^{c_1\ldots c_6}] &= 6\delta^{[c_1}{}_b R^{|a|c_2\ldots c_6]} \quad , \quad [K^a{}_b, R_{c_1\ldots c_6}] = -6\delta^a{}_{[c_1} R_{|b|c_2\ldots c_6]} \quad , \\
[K^a{}_b, R^{c_1\ldots c_8,d}] &= 8\delta^{[c_1}{}_b R^{|a|c_2\ldots c_8],d} + \delta^a{}_b R^{c_1\ldots c_8,a} \quad , \\
[K^a{}_b, R_{c_1\ldots c_8,d}] &= -8\delta^a{}_{[c_1} R_{|b|c_2\ldots c_8],d} - \delta^a{}_b R_{c_1\ldots c_8,a} \quad ,
\end{aligned} \tag{4.9}$$

and the higher level generators can be determined to act as

$$\begin{aligned}
[R^{a_1a_2a_3}, R^{a_4a_5a_6}] &= 2R^{a_1..a_6} \quad , \quad [R_{a_1a_2a_3}, R_{a_4a_5a_6}] = 2R_{a_1..a_6} \quad , \\
[R^{a_1..a_6}, R^{b_1b_2b_3}] &= 3R^{a_1..a_6[b_1b_2,b_3]} \quad , \quad [R_{a_1..a_6}, R_{b_1b_2b_3}] = 3R_{a_1..a_6[b_1b_2,b_3]} \quad , \\
[R^{a_1\ldots a_3}, R_{b_1\ldots b_3}] &= 18\delta^{[a_1a_2}_{[b_1b_2} K^{a_3]}{}_{b_3]} - 2\delta^{a_1a_2a_3}_{b_1b_2b_3} D \ , \ [R_{b_1\ldots b_3}, R^{a_1\ldots a_6}] = \frac{5!}{2}\delta^{[a_1a_2a_3}_{b_1b_2b_3} R^{a_4a_5a_6]} \quad , \\
[R^{a_1\ldots a_6}, R_{b_1\ldots b_6}] &= -5! \cdot 3 \cdot 3\delta^{[a_1\ldots a_5}_{[b_1\ldots b_5} K^{a_6]}{}_{b_6]} + 5!\delta^{a_1\ldots a_6}_{b_1\ldots b_6} D \quad , \\
[R_{a_1\ldots a_3}, R^{b_1\ldots b_8,c}] &= 8 \cdot 7 \cdot 2(\delta^{[b_1b_2b_3}_{[a_1a_2a_3} R^{b_4\ldots b_8]c} - \delta^{[b_1b_2|c|}_{[a_1a_2a_3} R^{b_3\ldots b_8]}) \quad , \\
[R_{a_1\ldots a_6}, R^{b_1\ldots b_8,c}] &= \frac{7! \cdot 2}{3}(\delta^{[b_1\ldots b_6}_{[a_1\ldots a_6} R^{b_7b_8]c} - \delta^{c[b_1\ldots b_5}_{[a_1\ldots a_6} R^{b_6b_7b_8]}) \quad ,
\end{aligned} \tag{4.10}$$

where $\delta^{a_1..a_p}_{b_1..b_p}$ are generalized Kronecker delta functions, normalised with a $1/p$ factor, for example $\delta^{a_1a_2}_{b_1b_2} = \frac{1}{2}(\delta^{a_1}_{b_1}\delta^{a_2}_{b_2} - \delta^{a_2}_{b_1}\delta^{a_1}_{b_2})$.

The first fundamental representation of $E_{11}$, the 'vector representation' or $l_1$ representation, can be found by the same method as that for $A_1^{+++}$ in the previous chapter. Generalizing to $E_{12}$, considering the level decomposition with respect to the zero'th and 11'th nodes, the program [49] results in the following generators:

Table 4.2: $A_{10}$ representations in $E_{12}$

| $l$ | $A_{10}$ Irrep | $E_{12}$ Root | $a^2$ | $d_r$ | $\mu$ | $Z^\alpha$ |
|---|---|---|---|---|---|---|
| 0 0 | 0 0 0 0 0 0 0 0 0 0 | 0 0 0 0 0 0 0 0 0 0 0 0 | 0 | 1 | 2 | |
| 0 0 | 1 0 0 0 0 0 0 0 0 1 | 0 -1 -1 -1 -1 -1 -1 -1 -1 -1 -1 0 | 2 | 120 | 1 | |
| 1 0 | 1 0 0 0 0 0 0 0 0 0 | 1 0 0 0 0 0 0 0 0 0 0 0 | 2 | 11 | 1 | $P_a$ |
| 0 1 | 0 0 0 0 0 0 0 1 0 0 | 0 0 0 0 0 0 0 0 0 0 0 1 | 2 | 165 | 1 | |
| 1 1 | 0 0 0 0 0 0 0 0 1 0 | 1 1 1 1 1 1 1 1 1 0 0 1 | 2 | 55 | 1 | $Z^{a_1a_2}$ |
| 0 2 | 0 0 0 0 1 0 0 0 0 0 | 0 0 0 0 0 0 1 2 3 2 1 2 | 2 | 462 | 1 | |
| 1 2 | 0 0 0 0 0 1 0 0 0 0 | 1 1 1 1 1 1 1 2 3 2 1 2 | 2 | 462 | 1 | $Z^{a_1\ldots a_5}$ |
| 0 3 | 0 0 1 0 0 0 0 0 0 1 | 0 0 0 0 1 2 3 4 5 3 1 3 | 2 | 1760 | 1 | |
| 1 3 | 0 0 1 0 0 0 0 0 0 0 | 1 1 1 1 2 3 4 5 6 4 2 3 | 0 | 165 | 1 | $Z^{a_1\ldots a_8}$ |
| 1 3 | 0 0 0 1 0 0 0 0 0 1 | 1 1 1 1 1 2 3 4 5 3 1 3 | 2 | 3465 | 1 | $Z^{a_1\ldots a_7,b}$ |
| 2 3 | 0 0 0 1 0 0 0 0 0 0 | 2 2 2 2 2 3 4 5 6 4 2 3 | 2 | 330 | 1 | |

In this table, $l=(l_1,l_2)$, where $l_1$ is the level of node 0, and $l_2$ is the level of node 11. The vector representation of $E_{11}$ is given by the level $l_1=1$ generators in this table, given to low levels in the $l_2$ node by the generators

$$P_a \ , \ Z^{a_1a_2} \ , \ Z^{a_1..a_5} \ , \ Z^{a_1\ldots a_8} \ , \ Z^{a_1\ldots a_7,b} \ , \ \ldots \ . \tag{4.11}$$

Here $P_a$ introduces Minkowski space-time. However, now one finds the central charges of eleven-dimensional supergravity $Z^{a_1a_2}$ and $Z^{a_1\cdots a_5}$, which are interpreted as being associated to 'higher coordinates' in a 'generalized space-time', and $Z^{a_1\cdots a_7,b}$ is irreducible:

$$Z^{[a_1\ldots a_7,b]}=0. \tag{4.12}$$

These generators are taken to commute with one another, and possess the following commutation relations with the $E_{11}$ generators [85]

$$\begin{aligned}
&[K^a{}_b,P_c]=-\delta^a{}_cP_b+\frac{1}{2}\delta^a{}_bP_c \ , \ [K^a{}_b,Z^{c_1c_2}]=2\delta^{[c_1}{}_bZ^{|a|c_2]}+\frac{1}{2}\delta^a{}_bZ^{c_1c_2} \ , \\
&[K^a{}_b,Z^{c_1\ldots c_5}]=5\delta^{[c_1}{}_bZ^{|a|c_2\ldots c_5]}+\frac{1}{2}\delta^a{}_bZ^{c_1\ldots c_5} \ , \ [R^{a_1a_2a_3},P_b]=3\delta^{[a_1}{}_bZ^{a_2a_3]} \ , \\
&[R^{a_1a_2a_3},Z^{b_1b_2}]=Z^{a_1a_2a_3b_1b_2} \ , \ [R^{a_1a_2a_3},Z^{b_1\ldots b_5}]=Z^{b_1\ldots b_5[a_1a_2,a_3]}+Z^{b_1\ldots b_5a_1a_2a_3} \ , \\
&[R^{a_1\ldots a_6},P_b]=-3\delta^{[a_1}{}_bZ^{\ldots a_6]} \ , \ [R^{a_1\ldots a_6},Z^{b_1b_2}]=-Z^{b_1b_2[a_1\ldots a_5,a_6]}-Z^{b_1b_2a_1\ldots a_6} \ , \\
&[R_{a_1a_2a_3},P_b]=0 \ , \ [R_{a_1a_2a_3},Z^{b_1b_2}]=6\delta^{b_1b_2}_{[a_1a_2}P_{a_3]} \ , \ [R_{a_1a_2a_3},Z^{b_1\ldots b_5}]=\frac{5!}{2}\delta^{[b_1b_2b_3}_{a_1a_2a_3}Z^{b_4b_5]} \ .
\end{aligned} \tag{4.13}$$

We now construct the Cartan involution invariant subgroup of $E_{11}$. The map $I_c$ defined by

$$\begin{aligned}
I_c(K^a{}_b)&=-\eta^{ac}K^d{}_c\eta_{db} \ , \\
I_c(R^{a_1a_2a_3})&=-\eta^{a_1b_1}..\eta^{a_3b_3}R_{b_1b_2b_3} \ , \\
I_c(R^{a_1..a_6})&=\eta^{a_1b_1}..\eta^{a_6b_6}R_{b_1..b_6} \ , \\
I_c(R^{a_1..a_8,b})&=-\eta^{a_1c_1}..\eta^{a_8c_8}\eta^{bd}R_{c_1..c_8,d} \ , \ \ldots \ ,
\end{aligned} \tag{4.14}$$

satisfies $I_c^2 = I$ and so defines an involution on $E_{11}$ referred to as the Cartan involution.

The following combinations

$$\begin{aligned} J_{ab} &= \eta_{ac} K^c{}_b - \eta_{bc} K^c{}_a \quad , & S_{a_1a_2a_3} &= R^{b_1b_2b_3}\eta_{b_1a_1}\eta_{b_2a_2}\eta_{b_3a_3} - R_{a_1a_2a_3} \quad , \\ S_{a_1..a_6} &= R^{b_1..b_6}\eta_{b_1a_1}..\eta_{b_6a_6} - R_{a_1..a_6} \quad , & S_{a_1..a_8,b} &= R^{c_1..c_8,d}\eta_{c_1a_1}..\eta_{c_8a_8}\eta_{db} - R_{a_1..a_8,b} \quad , \end{aligned} \tag{4.15}$$

are invariant under $I_c$ and so generate what is referred to as the Cartan involution invariant subalgebra $I_c(E_{11})$ of $E_{11}$. Using the above commutation relations, these generators can be shown to satisfy the algebra

$$\begin{aligned} [S^{a_1a_2a_3}, S_{b_1b_2b_3}] &= -18\delta^{[a_1a_2}_{[b_1b_2} J^{a_3]}{}_{b_3]} + 2S^{a_1a_2a_3}{}_{b_1b_2b_3} \quad , \\ [S_{a_1a_2a_3}, S^{b_1\ldots b_6}] &= -\frac{5!}{2}\delta^{[b_1b_2b_3}_{a_1a_2a_3} S^{b_4b_5b_6]} - 3S^{b_1\ldots b_6}{}_{[a_1a_2,a_3]} \quad , \\ [S^{a_1a_2a_3}, P_b] &= 3\delta^{[a_1}_b Z^{a_1a_3]} \quad , \\ [S_{a_1a_2a_3}, Z^{b_1b_2}] &= Z_{a_1a_2a_3}{}^{b_1b_2} - 6\delta^{b_1b_2}_{[a_1a_2} P_{a_3]} \quad , \\ [S_{a_1a_2a_3}, Z^{b_1\ldots b_5}] &= Z^{b_1\ldots b_5}{}_{[a_1a_2,a_3]} + Z^{b_1\ldots b_5}{}_{a_1a_2a_3} - \frac{5!}{2}\delta^{[b_1\ldots b_3}_{a_1a_2a_3} Z^{b_4b_5]} \quad . \end{aligned} \tag{4.16}$$

## 4.3 Nonlinear Realization of $E_{11} \otimes_s l_1$

We now construct the nonlinear realization of $E_{11} \otimes_s l_1$ with respect to the $I_c(E_{11})$ subgroup. This is defined to be a set of equations of motion (constructed in terms of group elements $g \in E_{11} \otimes_s l_1$), which are invariant under the transformations

$$g \;\; \rightarrow \;\; g_0 g, \;\; g_0 \in E_{11} \otimes_s l_1 \;, \;\; \text{and} \;\; g \;\; \rightarrow \;\; gh \;, \;\; h \in I_c(E_{11}) \quad , \tag{4.17}$$

and can be constructed from the Cartan form. A given group element $g \in E_{11} \otimes_s l_1$ can be taken to be of the form

$$g = g_l g_E \quad , \tag{4.18}$$

where we set

$$\begin{aligned} g_E &= \ldots e^{h_{a_1..a_8,b}R^{a_1..a_8,b}} e^{A_{a_1..a_6}R^{a_1..a_6}} e^{A_{a_1a_2a_3}R^{a_1a_2a_3}} e^{h_a{}^b K^a{}_b} = \Pi_{\underline{\alpha}} e^{A_{\underline{\alpha}} R^{\underline{\alpha}}} \quad , \\ g_l &= e^{x^a P_a} e^{x_{ab}Z^{ab}} e^{x_{a_1..a_5}Z^{a_1..a_5}} \ldots = \Pi_A e^{z^A l_A} \quad , \end{aligned} \tag{4.19}$$

noting that the terms in $g_E$ corresponding to the negative level generators were removed by fixing a gauge, up to level zero local Lorentz transformations in $I_c(E_{11})$, as was done for $A_1^{+++}$.

The dynamics of $E_{11}$ can be constructed from the Cartan forms

$$\mathcal{V} \equiv g^{-1}dg = \mathcal{V}_E + \mathcal{V}_l \quad , \tag{4.20}$$

where we set

$$\begin{aligned} \mathcal{V}_E &= g_E^{-1}dg_E = dz^\Pi G_{\Pi,\underline{\alpha}} R^{\underline{\alpha}} \quad , \\ \mathcal{V}_l &= g_E^{-1}(g_l^{-1}dg_l)g_E = g_E^{-1}dz\cdot l g_E = dz^\Pi E_\Pi{}^A l_A \quad . \end{aligned} \tag{4.21}$$

Here $\mathcal{V}_E$ belongs to the $E_{11}$ algebra and $\mathcal{V}_l$ to the space of generators of the $l_1$ representation.

The term $\mathcal{V}_E$ can be calculated and written in the form

$$\mathcal{V}_E = G_a{}^b K^a{}_b + G_{a_1\ldots a_3} R^{a_1\ldots a_3} + G_{a_1\ldots a_6} R^{a_1\ldots a_6} + G_{a_1\ldots a_8,b} R^{a_1\ldots a_8,b} + \ldots \ , \tag{4.22}$$

where

$$\begin{aligned} G_a{}^b &= (e^{-1}de)_a{}^b \ , \\ G_{a_1\ldots a_3} &= e_{a_1}{}^{\mu_1}\ldots e_{a_3}{}^{\mu_3} dA_{\mu_1\ldots\mu_3} \ , \\ G_{a_1\ldots a_6} &= e_{a_1}{}^{\mu_1}\ldots e_{a_6}{}^{\mu_6}(dA_{\mu_1\ldots\mu_6} - A_{[\mu_1\ldots\mu_3}dA_{\mu_4\ldots\mu_6]}) \ , \\ G_{a_1\ldots a_8,b} &= e_{a_1}{}^{\mu_1}\ldots e_{a_8}{}^{\mu_8} e_b{}^\nu (dh_{\mu_1\ldots\mu_8,\nu} - A_{[\mu_1\ldots\mu_3}dA_{\mu_4\mu_5\mu_6}A_{\mu_7\mu_8]\nu} + 3A_{[\mu_1\ldots\mu_6}dA_{\mu_7\mu_8]\nu} \\ &\qquad + A_{[\mu_1\ldots\mu_3}dA_{\mu_4\mu_5\mu_6}A_{\mu_7\mu_8\nu]} - 3A_{[\mu_1\ldots\mu_6}dA_{\mu_7\mu_8\nu]}) \ , \end{aligned} \tag{4.23}$$

where we use $e_\mu{}^a = (e^h)_\mu{}^a$ and the last two terms in $G_{a_1..a_8,b}$ ensure the irreducibility requirement $G_{[a_1..a_8,b]} = 0$.

The Cartan forms $\mathcal{V}_E$ and $\mathcal{V}_l$ are invariant under rigid transformations of (4.17), but under the local $I_c(E_{11})$ transformations of ((4.17)) they change as

$$\begin{aligned} \mathcal{V}_E &\to h^{-1}\mathcal{V}_E h + h^{-1}dh \quad , \\ \mathcal{V}_l &\to h^{-1}\mathcal{V}_l h \quad . \end{aligned} \tag{4.24}$$

At level zero they transform as local Lorentz transformations, while at level one they transform under a group element $h = 1 - \Lambda^{a_1a_2a_3}S_{a_1a_2a_3}$, with $\mathcal{V}_E$ transforming as

$$\delta\mathcal{V}_E = [\Lambda^{a_1a_2a_3}S_{a_1a_2a_3}, \mathcal{V}_E] - S^{a_1a_2a_3}d\Lambda_{a_1a_2a_3} \ , \tag{4.25}$$

which expands to give

$$\begin{aligned} \delta G_a{}^b &= 18\Lambda^{c_1c_2b}G_{c_1c_2a} - 2\delta_a{}^b\Lambda^{c_1c_2c_3}G_{c_1c_2c_3} \ , \\ \delta G_{a_1a_2a_3} &= -\frac{5!}{2}G_{b_1b_2b_3a_1a_2a_3}\Lambda^{b_1b_2b_3} - 3G^c{}_{[a_1}\Lambda_{|c|a_2a_3]} - d\Lambda_{a_1a_2a_3} \ , \\ \delta G_{a_1\ldots a_6} &= 2\Lambda_{[a_1a_2a_3}G_{a_4a_5a_6]} - 8\cdot 7\cdot 2G_{b_1b_2b_3[a_1\ldots a_5,a_6]}\Lambda^{b_1b_2b_3} + 8\cdot 7\cdot 2G_{b_1b_2[a_1\ldots a_5a_6,b_3]}\Lambda^{b_1b_2b_3} \ , \\ \delta G_{a_1\ldots a_8,b} &= -3G_{[a_1\ldots a_6}\Lambda_{a_7a_8]b} + 3G_{[a_1\ldots a_6}\Lambda_{a_7a_8b]} \ . \end{aligned} \tag{4.26}$$

As in the $A_1^{+++}$ case, the $d\Lambda_{a_1a_2a_3}$ is chosen so as to preserve the gauge which eliminates the negative level generators in (4.19), implying the constraint

$$d\Lambda^{a_1a_2a_3} - 3G_e{}^{[a_1|}\Lambda^{e|a_2a_3]} = 0 \ , \tag{4.27}$$

which is solved by $\Lambda^{a_1a_2a_3} = \Lambda^{\tau_1\tau_2\tau_3} e_{\tau_1}{}^{a_1} e_{\tau_2}{}^{a_2} e_{\tau_3}{}^{a_3}$ where $\Lambda^{\tau_1\tau_2\tau_3}$ is treated as constant. This lets us re-write the three-form variation as

$$\delta G_{a_1a_2a_3} = -\frac{5!}{2} G_{b_1b_2b_3a_1a_2a_3}\Lambda^{b_1b_2b_3} - 6G_{(c[a_1)}\Lambda^{c}{}_{a_2a_3]}. \tag{4.28}$$

We now determine how the above transformations affect the components $G_{A,\underline{\alpha}} = E_A{}^{\Pi} G_{\Pi,\underline{\alpha}}$. To do this we note that the generalised vielbein $E_\Pi{}^A$, can be determined from its definition in (4.21) to level one by the same procedure as in the $A_1^{+++}$ case, and turns out to be [67] [66]

$$E_\Pi{}^A = (\det e)^{-\frac{1}{2}} \begin{bmatrix} e_\mu{}^a & -3e_\mu{}^c A_{cb_1b_2} \\ 0 & (e^{-1})_{[b_1}{}^{\mu_1}(e^{-1})_{b_2]}{}^{\mu_2} \end{bmatrix} . \tag{4.29}$$

The variation of (4.29) can then be determined by the same method as in the $A_1^{+++}$ case and one finds [84] [70]

$$\begin{aligned} \delta G_{a,\bullet} &= -3G^{b_1b_2}{}_{,\bullet}\Lambda_{b_1b_2a} \ , \\ \delta G^{a_1a_2}{}_{,\bullet} &= 6\Lambda^{a_1a_2b} G_{b,\bullet} \ , \end{aligned} \tag{4.30}$$

where $\bullet$ is any collection of $E_{11}$ indices. The total variation of $G_{\alpha,\bullet}$ is then computed by varying with respect to both $l_1$ and tangent indices, for example

$$\begin{aligned} \delta G_{a_1,a_2a_3a_4} &= \delta[(E^{-1})_{a_1}{}^{\Pi} G_{\Pi,a_2a_3a_4}] \\ &= -3\Lambda_{b_1b_2a_1} G^{b_1b_2}{}_{,b_1b_2a_2a_3a_4} - \frac{5!}{2} G_{a_1,b_1b_2b_3a_2a_3a_4}\Lambda^{b_1b_2b_3} - 6G_{a_1,(d[a_2)}\Lambda^{d}{}_{a_3a_4]}. \end{aligned} \tag{4.31}$$

## 4.4 Derivation of First Order Duality Relations

Given its fundamental importance, we now review the construction of the dynamics associated to the nonlinear realization of $E_{11} \otimes_s l_1$ over $I_c(E_{11})$ [68] [69]. The process is similar to that for $A_1^{+++}$ in the previous chapter. However, as a contrast, we now begin not from the graviton Cartan forms at level zero, but rather the three-form Cartan form at level one, and postulate a first order duality relation between the 3-form and 6-form at level two. At level zero it must transform covariantly, thus it is natural to consider

$$D_{a_1..a_4} = G_{[a_1,a_2a_3a_4]} + c\varepsilon_{a_1..a_4}{}^{b_1..b_7} G_{b_1,b_2..b_7} \quad , \tag{4.32}$$

for some constant $c$. This is the most general equation of motion containing first derivatives that connects only the 3-form and 6-form. We now construct its $l_1$-extension. We can consider an abstract $l_1$ extension of the above equation into

$$\begin{aligned}\mathcal{D}_{a_1..a_4} &= \mathcal{G}_{a_1a_2a_3a_4} + c\varepsilon_{a_1..a_4}{}^{b_1..b_7}\mathcal{G}_{b_1,b_2..b_7} + c_3 G_{[a_1a_2,a_3a_4]} \\ &= (G_{[a_1,a_2a_3a_4]} + c_1 G^{b_1b_2}{}_{,b_1b_2a_1..a_4}) + c\varepsilon_{a_1..a_4}{}^{b_1..b_7}(G_{[b_1,b_2..b_7]} + c_2 G^{c_1c_2}{}_{,c_1c_2[b_1..b_6,b_7]}) \\ &\quad + c_3 G_{[a_1a_2,a_3a_4]}.\end{aligned} \tag{4.33}$$

We can fix the coefficient $c_1$ by noting (ignoring level one derivatives)

$$\delta G_{[a_1,a_2a_3a_4]} = -\frac{5!}{2} G_{[a_1,|b_1b_2b_3|a_2a_3a_4]}\Lambda^{b_1b_2b_3} - 6G_{[a_1,(a_2|d|)}\Lambda^d{}_{a_3a_4]} \tag{4.34}$$

is not fully anti-symmetric in the indices of the six form derivative. We thus choose $c_1$ so that the result is anti-symmetrized

$$\begin{aligned}\delta\mathcal{G}_{[a_1,a_2a_3a_4]} &= \delta(G_{[a_1,a_2a_3a_4]} + c_1 G^{b_1b_2}{}_{,b_1b_2a_1..a_4}) \\ &= -\frac{5!}{2} G_{[a_1,|b_1b_2b_3|a_2a_3a_4]}\Lambda^{b_1b_2b_3} - 6G_{[a_1,(a_2|d|)}\Lambda^d{}_{a_3a_4]} + c_1 6\Lambda^{b_1b_2b_3} G_{b_3,b_1b_2a_1..a_4} \\ &= -6G_{[a_1,(a_2|d|)}\Lambda^d{}_{a_3a_4]} + 4\cdot 15 G_{[a_1,a_2a_3a_4]b_1b_2b_3}\Lambda^{b_1b_2b_3} + c_1 6\Lambda^{b_1b_2b_3} G_{[b_1,b_2b_3]a_1..a_4} \\ &= -6G_{[a_1,(a_2|d|)}\Lambda^d{}_{a_3a_4]} + 4\cdot 15 G_{[a_1,a_2a_3a_4]b_1b_2b_3}\Lambda^{b_1b_2b_3} + 3\cdot 15\Lambda^{b_1b_2b_3} G_{[b_1,b_2b_3]a_1..a_4} \\ &= -6G_{[a_1,(a_2|d|)}\Lambda^d{}_{a_3a_4]} + 7\cdot 15 G_{[a_1,a_2a_3a_4b_1b_2b_3]}\Lambda^{b_1b_2b_3} \\ &= -6G_{[a_1,(a_2|d|)}\Lambda^d{}_{a_3a_4]} + 105 G_{[a_1,a_2a_3a_4b_1b_2b_3]}\Lambda^{b_1b_2b_3}.\end{aligned} \tag{4.35}$$

This result can be further simplified

$$\begin{aligned}
\delta\mathcal{G}_{a_1a_2a_3a_4} &= -6G_{[a_1,(a_2|d|)}\Lambda^d{}_{a_3a_4]} + 105G_{[a_1,a_2a_3a_4b_1b_2b_3]}\Lambda^{b_1b_2b_3} \\
&= -6G_{[a_1,(a_2|d|)}\Lambda^d{}_{a_3a_4]} - \frac{105}{4!7!}\varepsilon_{a_1,a_2a_3a_4b_1b_2b_3c_1..c_4}\varepsilon^{c_1...c_4e_1..e_7}G_{e_1..e_7}\Lambda^{b_1b_2b_3} \\
&= -6G_{[a_1,(a_2|d|)}\Lambda^d{}_{a_3a_4]} - \frac{7!/(2\cdot 4!)}{4!7!}\varepsilon_{a_1,a_2a_3a_4b_1b_2b_3c_1..c_4}\varepsilon^{c_1...c_4e_1..e_7}G_{e_1..e_7}\Lambda^{b_1b_2b_3} \\
&= -6G_{[a_1,(a_2|d|)}\Lambda^d{}_{a_3a_4]} - \frac{1}{2\cdot 4!\cdot 4!}\varepsilon_{a_1,a_2a_3a_4b_1b_2b_3c_1..c_4}\varepsilon^{c_1...c_4e_1..e_7}G_{e_1..e_7}\Lambda^{b_1b_2b_3}.
\end{aligned} \tag{4.36}$$

Similarly, to fix $c_2$ in

$$\mathcal{G}_{b_1,b_2..b_7} = G_{[b_1,b_2..b_7]} + c_2G^{c_1c_2}{}_{,c_1c_2[b_1..b_6,b_7]} \tag{4.37}$$

we note from

$$\begin{aligned}
\delta G_{[b_1,b_2..b_7]} &= 2\Lambda_{[b_2b_3b_4}G_{b_1,b_5b_6b_7]} - 336G_{[b_1,|c_1c_2c_3|b_2..b_6,b_7]}\Lambda^{c_1c_2c_3} \quad , \\
2\Lambda_{[b_2b_3b_4}G_{b_1,b_5b_6b_7]} &= 2\delta^{c_1c_2c_3c_4c_5c_6c_7}_{b_2b_3b_4b_1b_5b_6b_7}\Lambda_{c_1c_2c_3}G_{c_4,c_5c_6c_7} = 2\delta^{c_1c_2c_3c_4c_5c_6c_7}_{b_2b_3b_4b_1b_5b_6b_7}G_{c_4,c_5c_6c_7}\Lambda_{c_1c_2c_3} \\
&= 2\delta^{c_4c_5c_6c_7c_1c_2c_3}_{b_2b_3b_4b_1b_5b_6b_7}G_{c_4,c_5c_6c_7}\Lambda_{c_1c_2c_3} = -2\delta^{c_4c_5c_6c_7c_1c_2c_3}_{b_1b_2b_3b_4b_5b_6b_7}G_{c_4,c_5c_6c_7}\Lambda_{c_1c_2c_3} \\
&= -2G_{[b_1,b_2b_3b_4}\Lambda_{b_5b_6b_7]} \quad , \\
336G_{[b_1,|c_1c_2c_3|b_2..b_6,b_7]}\Lambda^{c_1c_2c_3} &= -\frac{1}{2}336G_{[b_1,b_2..b_6b_7][c_1c_2,c_3]}\Lambda^{c_1c_2c_3} \\
&= -168G_{[b_1,b_2..b_6b_7]c_1c_2,c_3}\Lambda^{c_1c_2c_3} \quad , \\
\delta G_{[b_1,b_2..b_7]} &= 2\Lambda_{[b_2b_3b_4}G_{b_1,b_5b_6b_7]} - 336G_{[b_1,|c_1c_2c_3|b_2..b_6,b_7]}\Lambda^{c_1c_2c_3} \\
&= -2G_{[b_1,b_2b_3b_4}\Lambda_{b_5b_6b_7]} + 168G_{[b_1,b_2..b_6b_7]c_1c_2,c_3}\Lambda^{c_1c_2c_3} \quad ,
\end{aligned} \tag{4.38}$$

that $\delta G_{[b_1,b_2..b_7]}$ is not anti-symmetric in all indices and so we $l_1$ extend to get

$$\begin{aligned}
\delta\mathcal{G}_{b_1,b_2..b_7} &= \delta G_{[b_1,b_2..b_7]} + c_2\delta G^{c_1c_2}{}_{,c_1c_2[b_1..b_6,b_7]} \\
&= -2G_{[b_1,b_2b_3b_4}\Lambda_{b_5b_6b_7]} + 168G_{[b_1,b_2..b_6b_7]c_1c_2,c_3}\Lambda^{c_1c_2c_3} + c_26\Lambda^{c_1c_2c_3}G_{c_3,c_1c_2[b_1..b_6,b_7]} \\
&= -2G_{[b_1,b_2b_3b_4}\Lambda_{b_5b_6b_7]} + 168(G_{[b_1,b_2..b_6b_7]c_1c_2,c_3} + G_{c_1,c_2c_3[b_1..b_6,b_7]})\Lambda^{c_1c_2c_3} \quad .
\end{aligned} \tag{4.39}$$

We have thus found the coefficients $c_1, c_2$ in $\mathcal{G}_{a_1a_2a_3a_4}$ and $\mathcal{G}_{b_1,b_2..b_7}$:

$$\begin{aligned}
\mathcal{G}_{a_1a_2a_3a_4} &= G_{[a_1,a_2a_3a_4]} + \frac{15}{2}G^{b_1b_2}{}_{,b_1b_2a_1..a_4} \quad , \\
\mathcal{G}_{b_1,b_2..b_7} &= G_{[b_1,b_2..b_7]} + 28G^{c_1c_2}{}_{,c_1c_2[b_1..b_6,b_7]} \quad ,
\end{aligned} \tag{4.40}$$

and we found their variations to be

$$\begin{aligned}
\delta\mathcal{G}_{a_1a_2a_3a_4} &= -6G_{[a_1,(a_2|d|)}\Lambda^d{}_{a_3a_4]} + 105G_{[a_1,a_2a_3a_4b_1b_2b_3]}\Lambda^{b_1b_2b_3} \quad , \\
\delta\mathcal{G}_{b_1,b_2..b_7} &= -2G_{[b_1,b_2b_3b_4}\Lambda_{b_5b_6b_7]} + 168(G_{[b_1,b_2..b_6b_7]c_1c_2,c_3} + G_{c_1,c_2c_3[b_1..b_6,b_7]})\Lambda^{c_1c_2c_3} \quad .
\end{aligned} \tag{4.41}$$

This lets us find the variation of

$$\mathcal{D}_{a_1..a_4} = \mathcal{G}_{a_1a_2a_3a_4} + c\varepsilon_{a_1..a_4}{}^{b_1..b_7}\mathcal{G}_{b_1,b_2..b_7} + c_3 G_{[a_1a_2,a_3a_4]} \tag{4.42}$$

to be

$$\begin{aligned}
\delta\mathcal{D}_{a_1..a_4} &= \delta\mathcal{G}_{a_1a_2a_3a_4} + c\delta\varepsilon_{a_1..a_4}{}^{b_1..b_7}\mathcal{G}_{b_1,b_2..b_7} + c_3\delta G_{[a_1a_2,a_3a_4]} \\
&= -6G_{[a_1,(a_2|d|)}\Lambda^d{}_{a_3a_4]} + 105 G_{[a_1,a_2a_3a_4b_1b_2b_3]}\Lambda^{b_1b_2b_3} \\
&+ c\varepsilon_{a_1..a_4}{}^{b_1..b_7}\{-2G_{[b_1,b_2b_3b_4}\Lambda_{b_5b_6b_7]} + 168(G_{[b_1,b_2..b_6b_7]c_1c_2,c_3} + G_{c_1,c_2c_3[b_1..b_6,b_7]})\Lambda^{c_1c_2c_3}\} \\
&+ c_3\delta G_{[a_1a_2,a_3a_4]} \\
&= 105G_{[a_1,a_2a_3a_4b_1b_2b_3]}\Lambda^{b_1b_2b_3} - 2c\varepsilon_{a_1..a_4}{}^{b_1..b_7}G_{[b_1,b_2b_3b_4}\Lambda_{b_5b_6b_7]} \\
&+ 168c\varepsilon_{a_1..a_4}{}^{b_1..b_7}(G_{[b_1,b_2..b_6b_7]c_1c_2,c_3} + G_{c_1,c_2c_3[b_1..b_6,b_7]})\Lambda^{c_1c_2c_3} \\
&+ c_3\delta G_{[a_1a_2,a_3a_4]} - 6G_{[a_1,(a_2|d|)}\Lambda^d{}_{a_3a_4]} \quad .
\end{aligned} \tag{4.43}$$

The first two terms combine as

$$\begin{aligned}
&105G_{[a_1,a_2a_3a_4b_1b_2b_3]}\Lambda^{b_1b_2b_3} - 2c\varepsilon_{a_1..a_4}{}^{b_1..b_7}G_{[b_1,b_2b_3b_4}\Lambda_{b_5b_6b_7]} \\
&= -\frac{1}{2\cdot 4!\cdot 4!}\varepsilon_{a_1,a_2a_3a_4b_1b_2b_3c_1..c_4}\varepsilon^{c_1...c_4e_1..e_7}G_{e_1,e_2..e_7}\Lambda^{b_1b_2b_3} - 2c\varepsilon_{a_1..a_4}{}^{b_1..b_7}G_{[b_1,b_2b_3b_4}\Lambda_{b_5b_6b_7]} \\
&= -2c(\frac{1}{c(2\cdot 4!)^2}\varepsilon_{a_1,a_2a_3a_4b_1..b_4}{}^{b_5b_6b_7}\varepsilon^{b_1...b_4e_1..e_7}G_{e_1,e_2..e_7}\Lambda_{b_5b_6b_7} + \varepsilon_{a_1..a_4}{}^{b_1..b_7}G_{b_1,b_2b_3b_4}\Lambda_{b_5b_6b_7}) \\
&= -2c(\frac{1}{c(2\cdot 4!)^2}\varepsilon_{a_1a_2a_3a_4}{}^{b_1..b_7}\varepsilon_{b_1...b_4}{}^{e_1..e_7}G_{e_1,e_2..e_7} + \varepsilon_{a_1..a_4}{}^{b_1..b_7}G_{b_1,b_2b_3b_4})\Lambda_{b_5b_6b_7} \\
&= -2c\varepsilon_{a_1a_2a_3a_4}{}^{b_1..b_7}(G_{b_1,b_2b_3b_4} + \frac{1}{c(2\cdot 4!)^2}\varepsilon_{b_1...b_4}{}^{e_1..e_7}G_{e_1,e_2..e_7})\Lambda_{b_5b_6b_7} \quad .
\end{aligned} \tag{4.44}$$

Since we expect equations of motion to transform into themselves, this allows us to fix $c$ by requiring

$$\begin{aligned}
D_{b_1..b_4} &= G_{[b_1,b_2b_3b_4]} + c\varepsilon_{b_1...b_4}{}^{e_1..e_7}G_{e_1,e_2..e_7} \\
&= G_{[b_1,b_2b_3b_4]} + \frac{1}{c(2\cdot 4!)^2}\varepsilon_{b_1...b_4}{}^{e_1..e_7}G_{e_1,e_2..e_7} \quad ,
\end{aligned} \tag{4.45}$$

which tells us that $c^2 = \frac{1}{(2\cdot 4!)^2}$ i.e. $c = \pm\frac{1}{2\cdot 4!}$ and we choose the minus sign:

$$c = -\frac{1}{2\cdot 4!}. \tag{4.46}$$

This tells us that the three-form duality relation $D_{a_1..a_4} = 0$ equation should be

$$D_{a_1..a_4} = G_{[a_1,a_2a_3a_4]} - \frac{1}{2\cdot 4!}\varepsilon_{a_1..a_4}{}^{b_1..b_7}G_{b_1,b_2..b_7} = 0. \tag{4.47}$$

We now find that $\delta\mathcal{D}_{a_1..a_4}$ becomes

$$
\begin{aligned}
\delta\mathcal{D}_{a_1..a_4} &= 105G_{[a_1,a_2a_3a_4b_1b_2b_3]}\Lambda^{b_1b_2b_3} - 2c\varepsilon_{a_1..a_4}{}^{b_1..b_7}G_{[b_1,b_2b_3b_4}\Lambda_{b_5b_6b_7]}\\
&+ 168c\varepsilon_{a_1..a_4}{}^{b_1..b_7}(G_{[b_1,b_2..b_6b_7]c_1c_2,c_3} + G_{c_1,c_2c_3[b_1..b_6,b_7]})\Lambda^{c_1c_2c_3}\\
&+ c_3\delta G_{[a_1a_2,a_3a_4]} - 6G_{[a_1,(a_2|d|)}\Lambda^{d}{}_{a_3a_4]}\\
&= -2\frac{-1}{2\cdot 4!}\varepsilon_{a_1a_2a_3a_4}{}^{b_1..b_7}(G_{b_1,b_2b_3b_4} - \frac{1}{\frac{1}{2\cdot 4!}(2\cdot 4!)^2}\varepsilon_{b_1...b_4}{}^{e_1..e_7}G_{e_1,e_2..e_7})\Lambda_{b_5b_6b_7}\\
&- \frac{168}{2\cdot 4!}\varepsilon_{a_1..a_4}{}^{b_1..b_7}(G_{[b_1,b_2..b_6b_7]c_1c_2,c_3} + G_{c_1,c_2c_3[b_1..b_6,b_7]})\Lambda^{c_1c_2c_3}\\
&+ c_3\delta G_{[a_1a_2,a_3a_4]} - 6G_{[a_1,(a_2|d|)}\Lambda^{d}{}_{a_3a_4]}\\
&= \frac{1}{4!}\varepsilon_{a_1a_2a_3a_4}{}^{b_1..b_7}(G_{b_1,b_2b_3b_4} - \frac{1}{2\cdot 4!}\varepsilon_{b_1...b_4}{}^{e_1..e_7}G_{e_1,e_2..e_7})\Lambda_{b_5b_6b_7}\\
&- \frac{7}{2}\varepsilon_{a_1..a_4}{}^{b_1..b_7}(G_{[b_1,b_2..b_6b_7]c_1c_2,c_3} + G_{c_1,c_2c_3[b_1..b_6,b_7]})\Lambda^{c_1c_2c_3}\\
&+ c_3\delta G_{[a_1a_2,a_3a_4]} - 6G_{[a_1,(a_2|d|)}\Lambda^{d}{}_{a_3a_4]}.
\end{aligned}
\tag{4.48}
$$

We next deal with the last line, calling it $C$:

$$
\begin{aligned}
C &= c_3 6\delta G_{[a_1a_2,a_3a_4]} - 6G_{[a_1,(a_2|d|)}\Lambda^{d}{}_{a_3a_4]}\\
&= c_3 6\Lambda_{[a_1a_2}{}^{c}G_{|c,|a_3a_4]} - 6G_{[a_1,(a_2|c|)}\Lambda^{c}{}_{a_3a_4]}\\
&= c_3 6\Lambda_{[a_3a_4}{}^{c}G_{|c,|a_1a_2]} - 6G_{[a_1,(a_2|c|)}\Lambda^{c}{}_{a_3a_4]}\\
&= c_3 6G_{c,[a_1a_2}\Lambda^{c}{}_{a_3a_4]} - 3G_{[a_1,(a_2|c|)}\Lambda^{c}{}_{a_3a_4]} + 3G_{[a_2,(a_1|c|)}\Lambda^{c}{}_{a_3a_4]}\\
&= 3(-G_{[a_1,(a_2|c|)} + G_{[a_2,(a_1|c|)} + G_{c,[a_1a_2})\Lambda^{c}{}_{a_3a_4]}\\
&= 3(\det e)^{1/2}(\det e)^{-1/2}(-G_{[a_1,(a_2|c|)} + G_{[a_2,(a_1|c|)} + G_{c,[a_1a_2})\Lambda^{c}{}_{a_3a_4]}\\
&= 3(\det e)^{1/2}\omega_{c,[a_1a_2}\Lambda^{c}{}_{a_3a_4]} \quad , \quad \text{where}\\
\omega_{c,[a_1a_2}\Lambda^{c}{}_{a_3a_4]} &= (\det e)^{-1/2}(-G_{[a_1,(a_2|c|)} + G_{[a_2,(a_1|c|)} + G_{c,[a_1a_2})\Lambda^{c}{}_{a_3a_4]}\\
\omega_{c,a_1a_2} &= (\det e)^{-1/2}(-G_{a_1,(a_2|c|)} + G_{a_2,(a_1|c|)} + G_{c,[a_1a_2]}).
\end{aligned}
\tag{4.49}
$$

We now have that the variation becomes

$$
\begin{aligned}
\delta\mathcal{D}_{a_1..a_4} &= \frac{1}{4!}\varepsilon_{a_1a_2a_3a_4}{}^{b_1..b_7}(G_{b_1,b_2b_3b_4} - \frac{1}{2\cdot 4!}\varepsilon_{b_1...b_4}{}^{e_1..e_7}G_{e_1,e_2..e_7})\Lambda_{b_5b_6b_7}\\
&- \frac{7}{2}\varepsilon_{a_1..a_4}{}^{b_1..b_7}(G_{[b_1,b_2..b_6b_7]c_1c_2,c_3} + G_{c_1,c_2c_3[b_1..b_6,b_7]})\Lambda^{c_1c_2c_3}\\
&+ c_3\delta G_{[a_1a_2,a_3a_4]} - 6G_{[a_1,(a_2|d|)}\Lambda^{d}{}_{a_3a_4]}\\
&= \frac{1}{4!}\varepsilon_{a_1a_2a_3a_4}{}^{b_1..b_7}D_{b_1b_2b_3b_4}\Lambda_{b_5b_6b_7}\\
&- \frac{7}{2}\varepsilon_{a_1..a_4}{}^{b_1..b_7}(G_{[b_1,b_2..b_6b_7]c_1c_2,c_3} + G_{c_1,c_2c_3[b_1..b_6,b_7]})\Lambda^{c_1c_2c_3}\\
&+ 3(\det e)^{1/2}\omega_{c,[a_1a_2}\Lambda^{c}{}_{a_3a_4]}\ .
\end{aligned}
\tag{4.50}
$$

We finally deal with the middle line, calling it $B$:

$$\begin{aligned} B &= -\frac{7}{2}\varepsilon_{a_1..a_4}{}^{b_1..b_7}(G_{[b_1,b_2..b_6b_7]c_1c_2,c_3} + G_{c_1,c_2c_3[b_1..b_6,b_7]})\Lambda^{c_1c_2c_3} \\ &= -\frac{3}{4}\varepsilon_{[a_1a_2|}{}^{b_1..b_9}G_{b_1,b_2..b_9,c}\Lambda^c{}_{|a_3a_4]}\ , \end{aligned} \tag{4.51}$$

This allows us to combine $B$ an $C$ into

$$\begin{aligned} B+C &= 3(\det e)^{1/2}\omega_{c,[a_1a_2}\Lambda^c{}_{a_3a_4]} - \frac{3}{4}\varepsilon_{[a_1a_2|}{}^{b_1..b_9}G_{b_1,b_2..b_9,c}\Lambda^c{}_{|a_3a_4]} \\ &= 3\{(\det e)^{1/2}\omega_{c,[a_1a_2} - \frac{1}{4}\varepsilon_{[a_1a_2|}{}^{b_1..b_9}G_{b_1,b_2..b_9,c}\}\Lambda^c{}_{|a_3a_4]} \\ &= 3D_{c,[a_1a_2}\Lambda^c{}_{|a_3a_4]} \end{aligned} \tag{4.52}$$

where we have defined

$$D_{a,b_1b_2} := (\det e)^{1/2}\omega_{a,b_1b_2} - \frac{1}{4}\varepsilon_{b_1b_2}{}^{c_1..c_9}G_{c_1,c_2..c_9,a}. \tag{4.53}$$

This is the duality relation between the graviton and dual graviton in an $E_{11}$ context. Thus the final variation is

$$\delta\mathcal{D}_{a_1..a_4} = \frac{1}{4!}\varepsilon_{a_1a_2a_3a_4}{}^{b_1..b_7}D_{b_1b_2b_3b_4}\Lambda_{b_5b_6b_7} + 3D_{c,[a_1a_2}\Lambda^c{}_{|a_3a_4]}. \tag{4.54}$$

A similar computation can be done starting from $D_{a,b_1b_2}$ [66].

# 4.5 Second Order Gravity and Dual Gravity Equations of Motion

## 4.5.1 Einstein Gravity Equation from $E_{11}$

In the previous section we have consider the $I_c(E_{11})$ variation of the $l_1$ extension of the three form duality relation (4.47). Another variation involving this duality relation can be taken, a modification of the duality relation so as to eliminate the six form. This is achieved as follows, where in this section we simply state many results and give references. By converting (4.47) to world indices, and taking derivatives as shown, the six form can be eliminated to produce a second order equation for the three form as [70]

$$\begin{aligned} E^{\mu_1\mu_2\mu_3} &= \partial_\nu[(\det e)^{1/2}D^{\nu\mu_1\mu_2\mu_3}] \qquad (4.55) \\ &= \partial_\nu[(\det e)^{1/2}G^{[\nu,\mu_1\mu_2\mu_3]}) + \frac{1}{2\cdot 4!}(\det e)^{-1}\varepsilon^{\mu_1\mu_2\mu_3\tau_1\ldots\tau_8}G_{[\tau_1,\tau_2\tau_3\tau_4]}G_{[\tau_5,\tau_6\tau_7\tau_8]} = 0\ . \end{aligned}$$

This equation can be put back into tangent indices $E^{a_1a_2a_3}$, and the variation of a suitable $l_1$ extension can be shown to give [69]

$$\delta\mathcal{E}^{a_1a_2a_3} = \frac{3}{2}E_b{}^{[a_1}\Lambda^{|b|a_2a_3]} + \dots \quad , \tag{4.56}$$

where ... indicates more contributions depending on $D_{a_1..a_4}$, and

$$E_a{}^b = (\det e)R_a{}^b - 48G_{[a,c_1c_2c_3]}G^{[b,c_1c_2c_3]} + 4\delta_a{}^bG_{[c_1,c_2c_3c_4]}G^{[c_1,c_2c_3c_4]} \quad , \tag{4.57}$$

is the Einstein equation of motion for the graviton in supergravity. The Einstein equation can also be $l_1$ extended

$$\begin{aligned}\mathcal{E}_a{}^b = &(\det e)\mathcal{R}_a{}^b - 48G_{[a,c_1c_2c_3]}G^{[b,c_1c_2c_3]} + 4\delta_a{}^bG_{[c_1,c_2c_3c_4]}G^{[c_1,c_2c_3c_4]}\\ &- 360G^{d_1d_2,}{}_{d_!d_2ac_1c_2c_3}G^{[b,c_1c_2c_3]} - 360G^{d_1d_2,}{}_{d_1d_2}{}^{bc_1c_2c_3}G_{[a,c_1c_2c_3]}\\ &+ 60\delta_a{}^bG^{d_1d_2,}{}_{d_1d_2c_1c_2c_3c_4}G^{[c_1,c_2c_3c_4]} - 12G_{c_1c_2,ac_3}G^{[b,c_1c_2c_3]} + 3G_{c_1c_2,d}{}^dG_{[a,}{}^{bc_1c_2]}\\ &- 6(\det e)e_a{}^\lambda e^{b\mu}\partial_{[\mu}\{(\det e)^{-1/2}G^{\tau_1\tau_2,}{}_{\tau_1\tau_2\lambda]}\}\\ &- (\det e)^{1/2}\omega_{c,b}{}^cG^{d_1d_2,}{}_{d_1d_2a} - 3(\det e)^{1/2}\omega_{a,b}{}^cG^{d_1d_2,}{}_{d_1d_2c} \quad ,\end{aligned} \tag{4.58}$$

where $(\det e)\mathcal{R}_a{}^b$ is the Ricci curvature in terms of the $E_{11}$ $l_1$-extended spin connection, and its variation can be shown to be [69]

$$\begin{aligned}\delta\mathcal{E}_a{}^b = &- 36E_{ac_1c_2}\Lambda^{bc_1c_2} - 36E^{bc_1c_2}\Lambda_{ac_1c_2} + 8\delta_a{}^b\Lambda^{c_1c_2c_3}E_{c_1c_2c_3}\\ &- 2\varepsilon_{ac_1\dots c_7d_1d_2d_3}G^{[b,c_1c_2c_3]}D^{c_4\dots c_7}\Lambda^{d_1d_2d_3} - 2\varepsilon^{bc_1\dots c_7d_1d_2d_3}G_{[a,c_1c_2c_3]}D_{c_4\dots c_7}\Lambda_{d_1d_2d_3}\\ &+ \frac{1}{3}\delta_a{}^b\varepsilon^{c_1\dots c_8d_1d_2d_3}G_{c_1,c_2c_3c_4}D_{c_5\dots c_8}\Lambda_{d_1d_2d_3} \quad .\end{aligned} \tag{4.59}$$

Thus $E_{11}$ exactly reproduces the equations of motion of the bosonic sector of eleven-dimensional supergravity when we neglect the effects of the higher level derivatives in these equations.

### 4.5.2 Dual Gravity Equation from $E_{11}$

Instead of projecting out the six form field, we can project out the three form by first re-writing the three form duality relation (4.47) in terms of the six-form as

$$D_{a_1\dots a_7} := G_{[a_1,a_2\dots a_7]} + \frac{2}{7!}\varepsilon_{a_1\dots a_7}{}^{b_1\dots b_4}G_{b_1,b_2b_3b_4} = 0 \quad . \tag{4.60}$$

In world indices the six form equation of motion reads as

$$E_{\mu_1\dots\mu_7} \equiv G_{[\mu_1,\dots\mu_7]} + \frac{2}{7!}(\det e)^{-1}\epsilon_{\mu_1\dots\mu_7}{}^{\nu_1\dots\nu_4}G_{\nu_1,\nu_2\nu_3\nu_4} = 0 \quad . \tag{4.61}$$

Taking the derivative of this equation one finds

$$E^{\mu_1\dots\mu_6} \equiv \partial_\nu\{(\det e)^{\frac{1}{2}} G^{[\nu,\mu_1\dots\mu_6]}\} = 0 \quad . \tag{4.62}$$

In terms of tangent indices this is given by

$$\begin{aligned} E_{a_1\dots a_6} &:= (\det e)^{\frac{1}{2}} e^{d\mu}\partial_\mu G_{[d,a_1\dots a_6]} + \frac{1}{2} G^{d}{}_{,b}{}^{b} G_{[d,a_1\dots a_6]} \\ &\quad - G^{d}{}_{,}{}^{c}{}_{d} G_{[c,a_1\dots a_6]} - 6\, G^{d}{}_{,}{}^{c}{}_{[a_1|} G_{[d,c|a_2\dots a_6]]} = 0 \quad . \end{aligned} \tag{4.63}$$

In [72] the $l_1$ extension of this term was found to be

$$\begin{aligned} \hat{\mathcal{E}}^{a_1\dots a_6} = e^{[a_1}_{\mu_1}..e^{a_6]}_{\mu_6} \Big( & \partial_\nu\Big((\det e)^{\frac{1}{2}} G^{[\nu,\mu_1\dots\mu_6]}\Big) - 8\,\partial_\nu((\det e)^{\frac{1}{2}} G^{\tau_1\tau_2,\nu\mu_1\dots\mu_6}{}_{\tau_1,\tau_2}) \\ & + \frac{1}{7}(\det e)^{-\frac{1}{2}} \partial^{\mu_1\mu_2}((\det e)^{\frac{1}{2}} G^{\mu_3,\mu_4\mu_5\mu_6}) - 72\,(\det e)^{\frac{1}{2}} G^{[\nu,\mu_1\dots\mu_6\sigma_1\lambda]}{}_{,\tau} Q^{\tau}{}_{\sigma_1,\nu\lambda} \\ & - 36\, e_{\tau_1}{}^{b_1} e_{\tau_2}{}^{b_2} e_{\rho_1 b_1} e_{\rho_2 b_2} \partial^{\rho_1\rho_2}((\det e)^{\frac{1}{2}} G^{[\nu,\mu_1\dots\mu_6\tau_1\tau_2]}{}_{,\nu})\Big) \\ & - 3 G^{c_1c_2}{}_{,c_1c_2e} G^{[e,a_1\dots a_6]} - 18\, G^{c[a_1|}{}_{,cd_1d_2} G^{[d_1,d_2|a_2\dots a_6]]} = 0. \end{aligned} \tag{4.64}$$

We have put a hat on the symbol for the six form equation as this will not be our final result. Its $I_c(E_{11})$ variation was found to be [72]

$$\begin{aligned} \delta\hat{\mathcal{E}}^{a_1\dots a_6} &= 432\,\Lambda_{c_1c_2c_3} \hat{E}^{a_1\dots a_6c_1c_2,\,c_3} + \frac{8}{7}\Lambda^{[a_1a_2a_3} E^{a_4a_5a_6]} \\ &+ \frac{2}{105} G_{[e_5,c_1c_2c_3]} \epsilon^{a_1\dots a_6e_1\dots e_5} E_{e_1\dots e_4} \Lambda^{c_1c_2c_3} \\ &- \frac{3}{35} G_{[e_5,e_6c_1c_2]} \Lambda^{c_1c_2[a_1} \epsilon^{a_2\dots a_6]e_1\dots e_6} E_{e_1\dots e_4} + \frac{1}{420}\,\epsilon_{c_1}{}^{a_1\dots a_6b_1\dots b_4}\,\omega_{c_2,\,b_1b_2}\, E_{c_3,\,b_3b_4}\, \Lambda^{c_1c_2c_3} \quad . \end{aligned} \tag{4.65}$$

Here the expression $\hat{E}^{a_1\dots a_6c_1c_2,\,c_3}$ is the $E_{11}$ analogue of the $A_1^{+++}$ dual graviton equation $\overline{E}'_a{}^b$ of equation (3.130). In world indices it reads as

$$\overline{E}'^{\mu_1\dots\mu_8}{}_{,\nu} := \partial_{[\tau|}\left[(\det e)^{\frac{1}{2}} \overline{G}^{[\tau,\mu_1\dots\mu_8]}{}_{,|\nu]}\right] \quad . \tag{4.66}$$

As in the $A_1^{+++}$ case, this expression is not irreducible. In [30] a derivation of the nonlinear dual graviton equation of motion which satisfies the $E_{11}$ irreducibility conditions was provided. We summarize the derivation as follows.

The dual graviton field $h_{a_1\dots a_8,b}$ satisfies $h_{[a_1\dots a_8,b]} = 0$, thus we demand that the true dual graviton equation of motion expression $E_{a_1\dots a_8,b}$ satisfying

$$E_{[a_1\dots a_8,b]} = 0 \quad . \tag{4.67}$$

We can further modify the term $\hat{E}^{a_1\ldots a_6 c_1 c_2,c_3}$ in the six form variation (4.65) by adding additional $l_1$-extension terms of the form $\Lambda^{\bullet\tau} f_\bullet \partial_\tau g_\bullet$ type terms, where $\bullet$ signifies indices have been suppressed.

To begin we re-write equation (4.66) in terms of tangent space indices:

$$\begin{aligned}\hat{E}^{a_1..a_8}{}_b &= (\det e)^{\frac{1}{2}} e_{[c|}{}^\nu \partial_\nu G^{[c,a_1\ldots a_8]}{}_{,|b]} - 8G_{[c|,e}{}^{a_1} G^{[c,ea_2\ldots a_8]}{}_{,|b]} + G_{[c,b]}{}^e G^{[c,a_1\ldots a_8]}{}_{,e} \\ &\quad - G_{[c,|e}{}^c G^{[e,a_1\ldots a_8]}{}_{,|b]} + \frac{1}{2} G_{[c|,e}{}^e G^{[c,a_1\ldots a_8]}{}_{,|b]} \\ &:= \hat{E}^{(1)a_1\ldots a_8}{}_{,b} + \hat{E}^{(2)a_1\ldots a_8}{}_{,b} \quad .\end{aligned} \tag{4.68}$$

In the last line we have re-written the dual graviton equation as a sum of two parts, the first of which, $\hat{E}^{(1)a_1\ldots a_8}{}_{,b}$, contains terms of the generic form $\partial G_{1,8,1}$, while the second part, $\hat{E}^{(2)a_1\ldots a_8}{}_{,b}$, contains terms of the generic form $G_{1,1,1}G_{1,8,1}$. Here, $G_{1,8,1}$ and $G_{1,1,1}$ denote the dual gravity and gravity Cartan forms, respectively. We assume that the $a_1 \ldots a_8$ indices in any equation are completely antisymmetrised.

We begin by processing the $G_{1,1,1}G_{1,8,1}$ terms, which can be written as

$$\begin{aligned}&\hat{E}^{(2)a_1\ldots a_8}{}_{,b} \\ &= \Big(-4G_{c,e}{}^{a_1} G^{[c,ea_2\ldots a_8]}{}_{,b} + \frac{1}{2} G_{c,b}{}^e G^{[c,a_1\ldots a_8]}{}_{,e} - \frac{1}{2} G_{e,c}{}^e G^{[c,a_1\ldots a_8]}{}_{,b} + \frac{1}{4} G_{c,e}{}^e G^{[c,a_1\ldots a_8]}{}_{,b}\Big) \\ &\quad + \Big(4G_{b,e}{}^{a_1} G^{[c,ea_2\ldots a_8]}{}_{,c} - \frac{1}{2} G_{b,c}{}^e G^{[c,a_1\ldots a_8]}{}_{,e} + \frac{1}{2} G_{b,c}{}^e G^{[c,a_1\ldots a_8]}{}_{,e} - \frac{1}{4} G_{b,e}{}^e G^{[c,a_1\ldots a_8]}{}_{,c}\Big) \\ &:= A_1 + A_2 + A_3 + A_4 \\ &\quad + \Big(4G_{b,e}{}^{a_1} G^{[c,ea_2\ldots a_8]}{}_{,c} - \frac{1}{2} G_{b,c}{}^e G^{[c,a_1\ldots a_8]}{}_{,e} + \frac{1}{2} G_{b,c}{}^e G^{[c,a_1\ldots a_8]}{}_{,e} - \frac{1}{4} G_{b,e}{}^e G^{[c,a_1\ldots a_8]}{}_{,c}\Big) \quad .\end{aligned} \tag{4.69}$$

Here $A_1$, $A_2$, $A_3$ and $A_4$ are the terms in the first bracket in the order in which they occur. The terms on the last line of this equation have $b$ as their first index, and since these are contracted with the parameter $\Lambda^{c_1 c_2 b}$ they are $l_1$ terms. Thus they can be removed from the dual graviton equation by adding terms to the six form equation of motion. We now consider $A_4$. This can be written as

$$\begin{aligned}A_4 = \frac{1}{4\cdot 9}\{G_{c,e}{}^e G^{c,a_1\ldots a_8}{}_{,b}\} &+ \frac{1}{4\cdot 9} G_{c,e}{}^e \{8G^{a_1,a_2\ldots a_8 c}{}_{,b} - G_{b,}{}^{a_1\ldots a_8,c}\} \\ &+ \frac{1}{4\cdot 9}\Big[G_{c,e}{}^e G_{b,}{}^{a_1\ldots a_8,c}\Big].\end{aligned} \tag{4.70}$$

The first term in equation (4.70) is symmetric since it satisfies the irreducibility condition $G_{c,[a_1\ldots a_8,b]} = 0$. Until specified otherwise, in this section we will use curly {} brackets to denote quantities that are symmetric for bookkeeping purposes.

The second term in equation (4.70) is also symmetric. We can see this by bringing $c$ out of the $a_1 \dots a_8, c$ brackets in the irreducibility condition $G_{b,[a_1\dots a_8,c]} = 0$ to re-write the expression in the bracket as

$$8G^{a_1,a_2\dots a_8c,b} - G^{b,a_1\dots a_8,c} = 8G^{a_1,a_2\dots a_8c,b} - 8G^{b,a_1\dots a_7c,a_8} \quad , \tag{4.71}$$

so that if we now anti-symmetrize in the indices $a_1, \dots a_8$ and $b$ this expression vanishes. To be clear the last term in this expression is $G^{b,[a_1\dots a_7|c|,a_8]}$ since we are assuming the $a_1 \dots a_8$ indices are anti-symmetrized.

The final term in equation (4.70) can be removed from the dual graviton equation of motion as it is an $l_1$ term. We will place all $l_1$ terms in square brackets $[\,\dots\,]$ for bookkeeping purposes.

We now consider the term $A_3$. It can similarly be written as

$$\begin{aligned} A_3 = -\frac{1}{2\cdot 9}\{G_{e,c}{}^{e}G^{c,a_1\dots a_8}{}_{,b}\} - \frac{1}{2\cdot 9}\{8G_{e,c}{}^{e}G^{a_1,a_2\dots a_8c}{}_{,b} - G_{b,}{}^{a_1\dots a_8,c}\} \\ - \frac{1}{2\cdot 9}\Big[G_{c,e}{}^{e}G_{b,}{}^{a_1\dots a_8,c}\Big] \quad . \end{aligned} \tag{4.72}$$

The first two terms are symmetric and the last term is an $l_1$ term. The terms $A_1$ and $A_2$ are not symmetric, we will return to them below.

We will now consider the $\partial G_{1,8,1}$ terms in $\hat{E}^{(1)a_1\dots a_8}{}_{,b}$ from equation (4.68). We can write these terms as

$$\begin{aligned} \hat{E}^{(1)a_1\dots a_8}{}_{,b} =& \frac{1}{2\cdot 9}\{(\det e)^{\frac{1}{2}}e_c{}^{\mu}\partial_\mu G^{c,a_1\dots a_8}{}_{,b}\} + \frac{1}{2\cdot 9}\{(\det e)^{\frac{1}{2}}e^{c\mu}\partial_\mu(8G_{a_1,a_2\dots a_8c,b} - G_{b,a_1\dots a_8,c})\} \\ &- \frac{4}{9}\{(\det e)^{\frac{1}{2}}(e_b{}^{\mu}\partial_\mu G^{a_1,a_2\dots a_8c}{}_{,c} - e_{[b}{}^{\mu}\partial_\mu G^{a_1,a_2\dots a_8]c}{}_{,c})\} \\ &- \frac{1}{2\cdot 9}(\det e)^{\frac{1}{2}}(e_b{}^{\mu}\partial_\mu G_{c,}{}^{a_1\dots a_8,c} - e_c{}^{\mu}\partial_\mu G_{b,}{}^{a_1\dots a_8,c}) \\ &- \frac{4}{9\cdot 9}(\det e)^{\frac{1}{2}}(e_b{}^{\mu}\partial_\mu G^{a_1,a_2\dots a_8c}{}_{,c} - e^{a_1\mu}\partial_\mu G^{b,a_2\dots a_8c}{}_{,c}) \\ &- \frac{4\cdot 7}{9\cdot 9}(\det e)^{\frac{1}{2}}e^{a_1\mu}\partial_\mu G^{a_2,a_3\dots a_8bc}{}_{,c}. \end{aligned} \tag{4.73}$$

The first and third terms in equation (4.73) are symmetric as can be seen by observation. The second term is symmetric on using irreducibility as was done for $A_4$ above. The remaining three terms are not symmetric.

Let us consider the fourth term in (4.73). The first term in this expression is an $l_1$ term. The second term is not of this form as there is a derivative in between the parameter $\Lambda^{c_1c_2b}$ and the Cartan form $G_{b,\bullet}$, and so we re-write it as

$$-\frac{1}{2\cdot 9}(\det e)^{\frac{1}{2}}\left[e_b{}^\mu\partial_\mu G_{c,}{}^{a_1\ldots a_8,c}\right] - e_c{}^\mu\partial_\mu G_{b,}{}^{a_1\ldots a_8,c}) \tag{4.74}$$
$$= -\frac{1}{2\cdot 9}(\det e)^{\frac{1}{2}}\left[e_b{}^\mu\partial_\mu G_{c,}{}^{a_1\ldots a_8,c}\right] - (e_c{}^\mu\partial_\mu e_b{}^\nu)G_{\nu,}{}^{a_1\ldots a_8,c} - \left[e_b{}^\nu(e_c{}^\mu)\partial_\mu G_{\nu,}{}^{a_1\ldots a_8,c}\right] \ .$$

Recall that, for a term to be an $l_1$ term it must, when multiplied by the parameter $\Lambda^{c_1c_2b}$, contain a factor of the form $\Lambda^{c_1c_2b}G_{b,\bullet}$. Here, the first term in equation (4.74) is an $l_1$ term. To see why the third term is an $l_1$ term we note that on multiplying this term by the parameter $\Lambda^{c_1c_2b}$ we find the factor $e_{\rho_1}{}^{c_1}e_{\rho_1}{}^{c_2}\Lambda^{\rho_1\rho_2\nu}$. Using the fact that the parameter with upper world indices is a constant we can simply take it past the derivative to find that this term is thus an $l_1$ term. The second term of equation (4.74) can be written as

$$M_{10} := -\frac{1}{2\cdot 9}G_{c,b}{}^eG_{e,}{}^{a_1\ldots a_8} \quad , \tag{4.75}$$

and will be needed later on.

The fifth term in equation (4.73) can be treated in a similar way and we find that

$$-\frac{4}{9\cdot 9}(\det e)^{\frac{1}{2}}\left(\left[e_b{}^\mu\partial_\mu G^{a_1,a_2\ldots a_8c,}{}_c\right] - \left[e_b{}^\mu e^{a_1\nu}\partial_\nu G_{\mu,}{}^{a_2\ldots a_8c,}{}_c)\right]\right) - \frac{4}{9\cdot 9}G^{a_1}{}_{,be}G^{e,a_2\ldots a_8c,}{}_c$$
$$:= -\frac{4}{9\cdot 9}(\det e)^{\frac{1}{2}}\left(\left[e_b{}^\mu\partial_\mu G^{a_1,a_2\ldots a_8c,}{}_c\right] - \left[e_b{}^\mu e^{a_1\nu}\partial_\nu G_{\mu,}{}^{a_2\ldots a_8c,}{}_c)\right]\right) + M_{11} \tag{4.76}$$

In this equation we find two $l_1$ terms and one term, denoted as $M_{11}$, which will be needed later.

The last term in equation (4.73) cannot be analysed in this way. Similar to the approach we took in the $A_1^{+++}$ case, we now use the Maurer-Cartan equations of $E_{11}$ for the dual graviton Cartan form which we now derive.

The Cartan forms of $E_{11}$ are contained in the expression $\mathcal{V} = g_E^{-1}dg_E$ where $g_E$ is the group element of $E_{11}$. It obviously obeys the usual Maurer-Cartan equation $d\mathcal{V}+\mathcal{V}\wedge\mathcal{V} = 0$. The precise expression for $\mathcal{V}$ in terms of the Cartan forms is given by equation (4.22). Using the $E_{11}$ algebra one can show that the last term in equation (4.73) is given by

$$-\frac{4\cdot 7}{9\cdot 9}(\det e)^{\frac{1}{2}}e_{[a_1}{}^\mu\partial_\mu G_{a_2\ldots a_8]bc,}{}^c$$
$$= -\frac{4\cdot 7}{9.9}\Big(\frac{1}{2}G_{a_1,e}{}^eG_{a_2,a_3\ldots a_8bc,}{}^c - G_{a_1,a_2}{}^eG_{e,a_3\ldots a_8bc,}{}^c - \frac{8\cdot 8}{9}G_{a_1,[a_3|}{}^eG_{a_2,e|a_4\ldots a_8bc],}{}^c$$
$$+\frac{8}{9}G_{a_1,}{}^{ce}G_{a_2,e[a_3\ldots a_8b,c]} - \frac{7\cdot 8}{9}G_{a_1,[a_3|}{}^eG_{a_2,e|a_4\ldots a_8b}{}^c{}_{,c]} - \frac{8}{9}G_{a_1,}{}^{ce}G_{a_2,[a_3\ldots a_8bc],e} \tag{4.77}$$
$$+\frac{8}{9}G_{a_1,[a_3|}{}^eG_{a_2|,a_4\ldots a_8bc]}{}^c{}_{,e} - 2G_{a_1,[a_3a_4a_5}G_{|a_2|,a_6a_7a_8bc]}{}^c + 2G_{a_1,[a_3a_4|}{}^cG_{a_2|,a_5\ldots a_8bc]}\Big)$$
$$:= M_1 + M_2 + \ldots + M_8 + M_9 \ .$$

Here $M_1, \dots, M_9$ denote the expressions in the order in which they occur.

Our next task is to evaluated the terms $M_1, \dots, M_9$. Let us first consider $M_1$. We can rewrite this as

$$\begin{aligned} M_1 &= -\frac{2\cdot 7}{9\cdot 9} G_{a_1,e}{}^{e} G_{a_2,a_3\dots a_8 bc,}{}^{c} \\ &= -\frac{2\cdot 7}{9\cdot 9}\{G_{a_1,e}{}^{e} G_{a_2,a_3\dots a_8 bc,}{}^{c} - \frac{1}{2} G_{b,e}{}^{e} G_{a_1,a_2\dots a_8 c,}{}^{c} + \frac{1}{2} G_{a_1,e}{}^{e} G_{b,a_2\dots a_8 c,}{}^{c}\} \\ &+ \frac{2\cdot 7}{9\cdot 9}\Big[\frac{1}{2} G_{b,e}{}^{e} G_{a_1,a_2\dots a_8 c,}{}^{c} - \frac{1}{2} G_{a_1,e}{}^{e} G_{b,a_2\dots a_8 c,}{}^{c}\Big] \quad . \end{aligned} \tag{4.78}$$

The first term is symmetric while the terms in the last line are $l_1$ terms and so can be removed. The first term is symmetric because we can write it as

$$\begin{aligned} &\{G_{a_1,e}{}^{e} G_{a_2,a_3\dots a_8 bc,}{}^{c} - \frac{1}{2} G_{b,e}{}^{e} G_{a_1,a_2\dots a_8 c,}{}^{c} + \frac{1}{2} G_{a_1,e}{}^{e} G_{b,a_2\dots a_8 c,}{}^{c}\} \\ &= \frac{9}{2}\{G_{a_1,e}{}^{e} G_{a_2,a_3\dots a_8 bc,}{}^{c} - G_{[a_1|,e}{}^{e} G_{|a_2,a_3\dots a_8 b]c,}{}^{c}\} \quad . \end{aligned} \tag{4.79}$$

The term $M_2$ can be combined with the term $M_{11}$ in equation (4.76) to give

$$\begin{aligned} M_2 + M_{11} =& \frac{4}{9\cdot 9}\{7 G_{[a_1,a_2}{}^{e} G_{|e|,a_3\dots a_8]bc,}{}^{c} - G_{[a_1,|b}{}^{e} G_{e|,a_2\dots a_8]c,}{}^{c} - 8 G_{b,[a_1}{}^{e} G_{|e|,a_2\dots a_8]c,}{}^{c}\} \\ &+ \frac{4\cdot 8}{9\cdot 9}\Big[G_{b,[a_1}{}^{e} G_{|e|,a_2\dots a_8]c,}{}^{c}\Big] \quad . \end{aligned} \tag{4.80}$$

The first term is symmetric by the same reasoning used in equation (4.79), while the last term is an $l_1$ term and so can be removed.

We now consider $M_6 + M_7$. This can be written as

$$\begin{aligned} M_6 + M_7 &= -\frac{4\cdot 7}{9\cdot 9}(-\frac{8}{9} G_{a_1,}{}^{ce} G_{a_2,[a_3\dots a_8 bc],e} + \frac{8}{9} G_{a_1,[a_3|}{}^{e} G_{a_2|,a_4\dots a_8 bc]}{}^{c}{}_{,e}) \\ &= +\frac{4\cdot 7}{9\cdot 9} G_{a_1,c}{}^{e} G_{a_2,a_3\dots a_8 bc}{}^{c}{}_{,e} \\ &= +\frac{4\cdot 7}{9\cdot 9}\{G_{a_1,c}{}^{e} G_{a_2,a_3\dots a_8 bc,e} - \frac{1}{2} G_{b,c}{}^{e} G_{a_1,a_2\dots a_8 c,e} + \frac{1}{2} G_{a_1,c}{}^{e} G_{b,a_2\dots a_8 c,e}\} \\ &+ \frac{2\cdot 7}{9\cdot 9}\Big[G_{b,c}{}^{e} G_{a_1,a_2\dots a_8 c,e} - G_{a_1,c}{}^{e} G_{b,a_2\dots a_8 c,e}\Big] \quad . \end{aligned} \tag{4.81}$$

The terms in the first bracket are symmetric and the terms in the second line can be removed as they are $l_1$ terms.

The terms $M_3, M_4$ and $M_5$ in equation (4.77) can be evaluated as

$$
\begin{aligned}
&M_3 + M_4 + M_5 \\
&= -\frac{4\cdot 7}{9\cdot 9}\Big(6G_{a_1,a_2}{}^{e}G_{a_3,a_4\ldots a_8bec,}{}^{c} - G_{a_1,b}{}^{e}G_{a_2,a_3\ldots a_8ec,}{}^{c} + G_{a_1,c}{}^{e}G_{a_2,a_3\ldots a_8eb,}{}^{c}\Big) \\
&= -\frac{4\cdot 7}{9\cdot 9}\{6G_{a_1,a_2}{}^{e}G_{a_3,a_4\ldots a_8bec,}{}^{c} - G_{a_1,b}{}^{e}G_{a_2,a_3\ldots a_8ec,}{}^{c} - 8G_{b,a_1}{}^{e}G_{a_2,a_3\ldots a_8ec,}{}^{c} \\
&\qquad + G_{a_1,a_2}{}^{e}G_{b,ea_3\ldots a_8c,}{}^{c}\} \qquad (4.82)\\
&+ \frac{4\cdot 7}{9\cdot 9}\{G_{a_1,c}{}^{e}G_{a_2,a_3\ldots a_8be,c} - \frac{1}{2}G_{b,c}{}^{e}G_{a_1,a_2\ldots a_8e,c} + \frac{1}{2}G_{a_1,c}{}^{e}G_{b,a_2\ldots a_8e,c}\} \\
&+ \frac{4\cdot 7}{9\cdot 9}\Big[-8G_{b,a_1}{}^{e}G_{a_2,a_3\ldots a_8ec,}{}^{c} + G_{a_1,a_2}{}^{e}G_{b,ea_3\ldots a_8c,}{}^{c} + \frac{1}{2}G_{b,c}{}^{e}G_{a_1,a_2\ldots a_8e,c} \\
&\qquad - \frac{1}{2}G_{a_1,c}{}^{e}G_{b,a_2\ldots a_8e,c}\Big]
\end{aligned}
$$

The curly bracket terms are symmetric while the remaining terms are $l_1$ terms that will be removed.

Finally, the terms $M_8$ and $M_9$ can be evaluated as

$$
\begin{aligned}
M_8 + M_9 = &-\frac{1}{9}(-2G_{a_1,ba_2}{}^{c}G_{a_3,a_4\ldots a_8}{}^{c} + 5G_{a_1,a_2a_3}{}^{c}G_{a_4,a_5\ldots a_8bc} \\
&\qquad - 8G_{b,a_1a_2}{}^{c}G_{a_3,a_4\ldots a_8c} - G_{a_1,a_2a_3}{}^{c}G_{b,a_4\ldots a_8c}) \qquad (4.83)\\
&-\frac{1}{9}\Big[8G_{b,a_1a_2}{}^{c}G_{a_3,a_4\ldots a_8c} + G_{a_1,a_2a_3}{}^{c}G_{b,a_4\ldots a_8c}\Big]
\end{aligned}
$$

The first term is symmetric and the last term is an $l_1$ term that can be removed.

The only terms we have not processed so far are the terms $A_1$ and $A_2$ of equation (4.69) and the term $M_{10}$ of equation (4.77). The combination $A_2 + M_{10}$ can be written as

$$
\begin{aligned}
A_2 + M_{10} = &G^{[c,}{}_{b}{}^{e]}G_{[c,a_1,ea_2\ldots a_8],e} - \frac{4}{9}\{G^{c,}{}_{b}{}^{e}G_{a_1,ea_2\ldots a_8,c} + G^{c,}{}_{a_1}{}^{e}G_{b,ea_2\ldots a_8,c}\} \\
&+ \frac{4}{9}\Big[G^{c,}{}_{a_1}{}^{e}G_{b,ea_2\ldots a_8,c}\Big] \quad .
\end{aligned} \qquad (4.84)
$$

The first term of this expression we will be needed later on, the middle term is a symmetric term, while the last term is an $l_1$ term that will be removed.

We now consider $A_1$. This can be written as

$$
\begin{aligned}
A_1 &= -4G_{[c,e]a_1}G^{[c,ea_2\ldots a_8]}{}_{,b} \\
&= -4(G_{[c,e]a_1} + G_{[c|,a_1|e]})G^{[c,ea_2\ldots a_8]}{}_{,b} + 4G_{[c|,a_1|e]}G^{[c,ea_2\ldots a_8]}{}_{,b} \qquad (4.85)\\
&:= A_{1.1} + A_{1.2} \quad .
\end{aligned}
$$

We can reformulate $A_{1.1}$ as

$$
A_{1.1} = 4(E^{a_1,}{}_{ce} - G^{a_1,}{}_{[ce]})G^{[c,ea_2\ldots a_8]}{}_{,b} + \{\varepsilon_{ce}{}^{f_1\ldots f_9}G_{[f_1,f_2\ldots f_9]}G^{[c,ea_2\ldots a_8]}{}_{,b}\} \quad , \qquad (4.86)
$$

where we used the gravity-dual gravity relation from equation (4.53). The last term is symmetric.

The term $A_{1.2}$ from equation (4.85) can be rewritten as

$$\begin{aligned}
A_{1.2} =& - G^{[c,}{}_{b}{}^{e]} G_{[c,a_1 \ldots a_8],e} \\
&+ \frac{4}{9}\{ \Big( G_{c,a_1}{}^{e} G^{c,}{}_{ea_2 \ldots a_8,b} - G_{e,a_1}{}^{c} G_{c,}{}^{e}{}_{a_2 \ldots a_8,b} \Big) + \frac{1}{8}( G_{c,b}{}^{e} G^{c,}{}_{a_1 \ldots a_8,e} - G_{e,b}{}^{c} G_{c,a_1 \ldots a_8,}{}^{e} ) \} \\
&+ \{ \frac{7 \cdot 4}{9} G^{[c,}{}_{a_1}{}^{e]} G_{a_2,cea_3 \ldots a_8,b} - \frac{8}{9} G_{[c|,b}{}^{|e]} G_{a_1,ca_2 \ldots a_8,e} \} \\
&+ \frac{7 \cdot 4}{9 \cdot 9} \Big[ G^{[c,}{}_{a_1}{}^{e]} G_{b,cea_2 \ldots a_7,a_8} \Big] - \frac{8}{9 \cdot 9} \Big[ G_{[c|,a_1|e]} G_{b,ca_2 \ldots a_8,e} \Big] \ .
\end{aligned} \tag{4.87}$$

Examining equation (4.87), the first term cancels the first term in $A_2 + M_{10}$ of equation (4.84). The curly brackets terms are symmetric, while the terms in the final two brackets are $l_1$ terms that will be removed. In checking that the second term is symmetric it is useful to use the irreducibility of the dual gravity Cartan form and the identity

$$G_{a_1,ca_2 \ldots a_8,e} - G_{a_1,ea_2 \ldots a_8,c} = 7 G_{a_1,cea_2 \ldots a_7,a_8} . \tag{4.88}$$

In addition to the above terms there are ambiguous terms which can be added to the dual graviton equation of motion that are symmetric but at the same time are $l_1$ terms. Such terms contain a $b$ index as the first index on one of the two Cartan forms and an $a_1$ as the first index of the other Cartan form. To ensure the remaining eleven indices are consistent we need two summed over indices. The possible ambiguous terms are given by

$$\begin{aligned}
& c_1 ( G_{b,e}{}^{e} G_{a_1,a_2 \ldots a_8 c,}{}^{c} + G_{a_1,e}{}^{e} G_{b,a_2 \ldots a_8 c,}{}^{c} ) \\
& c_2 ( G_{b,}{}^{ec} G_{a_1,a_2 \ldots a_8 (e,c)} + G_{a_1,}{}^{ec} G_{b,a_2 \ldots a_8 (e,c)} ) \\
& c_3 ( G_{b,}{}^{ec} G_{a_1,a_2 \ldots a_8 [e,c]} + G_{a_1,}{}^{ec} G_{b,a_2 \ldots a_8 [e,c]} ) \\
& c_4 ( G_{b,a_2}{}^{c} G_{a_1,a_3 \ldots a_8 ce,}{}^{e} + G_{a_1,a_2}{}^{c} G_{b,a_3 \ldots a_8 ce,}{}^{e} ) \quad ,
\end{aligned} \tag{4.89}$$

where $c_1, \ldots, c_4$ are constants.

To find the dual graviton equation of motion we now collect up all the symmetric parts given in the curly brackets in the previous section. Dropping our curly bracket convention, and making the anti-symmetry of the $a_1 \ldots a_8$ indices explicit, the dual graviton equation is

given by

$$
\begin{aligned}
E_{a_1\ldots a_8,b} =& \frac{1}{9\cdot 2}(\det e)^{\frac{1}{2}}e^{c\mu}\partial_\mu G_{c,a_1\ldots a_8,b} + \frac{1}{2\cdot 9}(\det e)^{\frac{1}{2}}e^{c\mu}\partial_\mu(8G_{[a_1,a_2\ldots a_8]c,b} - G_{b,a_1\ldots a_8,c}) \\
& - \frac{4}{9\cdot 9}(\det e)^{\frac{1}{2}}(8e_b{}^\mu\partial_\mu G_{[a_1,a_2\ldots a_8]c,}{}^c + e_{[a_1|}{}^\mu\partial_\mu G_{b|,a_2\ldots a_8]c,}{}^{,c} - 7e_{[a_1}{}^\mu\partial_{|\mu|}G_{a_2,a_3\ldots a_8]bc,}{}^c) \\
& - \frac{2\cdot 7}{9\cdot 9}(G_{[a_1,|e|}{}^e G_{a_2,a_3\ldots a_8]bc,}{}^c - \frac{1}{2}G_{b,e}{}^e G_{[a_1,a_2\ldots a_8]c,}{}^c + \frac{1}{2}G_{[a_1,|e}{}^e G_{b|,a_2\ldots a_8]c,}{}^c) \\
& + \frac{1}{9\cdot 4}G^{c,e}{}_e G_{c,a_1\ldots a_8,b} + \frac{1}{9\cdot 4}G^{c,e}{}_e(8G_{[a_1,a_2\ldots a_8]c,b} - G_{b,a_1\ldots a_8,c}) \\
& - \frac{1}{2\cdot 9}G^{e,c}{}_e(G_{c,a_1\ldots a_8,b} + 8G_{[a_1,a_2\ldots a_8]c,b} - G_{b,a_1\ldots a_8,c}) \\
& + \frac{4\cdot 7}{9\cdot 9}\{(G_{a_1,}{}^{ce} + G_{a_1,}{}^{ec})G_{a_2,a_3\ldots a_8 be,c} - \frac{1}{2}(G_{b,}{}^{ce} + G_{b,}{}^{ec})G_{a_1,a_2\ldots a_8 e,c} \\
& + \frac{1}{2}(G_{a_1,}{}^{ce} + G_{a_1,}{}^{ec})G_{b,a_2\ldots a_8 e,c}\} \\
& - \frac{1}{9}(-2G_{[a_1,|b|a_2}{}^c G_{a_3,a_4\ldots a_8]}{}^c + 5G_{[a_1,a_2a_3}{}^c G_{a_4,a_5\ldots a_8]bc} - 8G_{b,[a_1a_2}{}^c G_{a_3,a_4\ldots a_8]c} \\
& - G_{[a_1,a_2a_3}{}^c G_{|b|,a_4\ldots a_8]c}) \\
& + \frac{4}{9\cdot 9}(7G_{[a_1,a_2}{}^e G_{|e|,a_3\ldots a_8]bc,}{}^c - G_{[a_1,|b}{}^e G_{e|,a_2\ldots a_8]c,}{}^c - 8G_{b,[a_1}{}^e G_{|e|,a_2\ldots a_8]c,}{}^c) \\
& + \varepsilon^{c_1c_2e_1\ldots e_9}G_{e_1,e_2\ldots e_9,[a_1|}G_{[c_1,c_2|a_2\ldots a_8]],b} \qquad (4.90) \\
& - \frac{4\cdot 7}{9\cdot 9}(6G_{[a_1,a_2}{}^e G_{a_3,a_4\ldots a_8]bec,}{}^c - G_{[a_1,|b|}{}^e G_{a_2,a_3\ldots a_8]ec,}{}^c - 8G_{b,[a_1}{}^e G_{a_2,a_3\ldots a_8]ec,}{}^c \\
& + G_{[a_1,a_2}{}^e G_{|b,e|a_3\ldots a_8]c,}{}^c) \\
& - \frac{4}{9}G_{c,[a_1}{}^e G_{|b,e|a_2\ldots a_8],}{}^c + \frac{4}{9}(G_{c,[a_1}{}^e G^{c,}{}_{|e|a_2\ldots a_8],b} - G_{e,[a_1}{}^c G_{|c|,}{}^e{}_{a_2\ldots a_8],b}) \\
& + \frac{1}{9\cdot 2}(G_{c,b}{}^e G^{c,}{}_{[a_1\ldots a_8],e} - G_{e,b}{}^c G_{c,a_1\ldots a_8,}{}^e) + \frac{7\cdot 4}{9}G^{[c,}{}_{[a_1}{}^{e]}G_{a_2,|ce|a_3\ldots a_8],b} \\
& + \frac{4}{9}G_{c,b}{}^e G_{[a_1,a_2\ldots a_8]}{}^c{}_{,e} \quad .
\end{aligned}
$$

Under a Lorentz transformation, the Cartan forms transform as

$$
\begin{aligned}
\delta G_{a,bc} &= \Lambda_a{}^e G_{e,bc} + \Lambda_b{}^e G_{a,ec} + \Lambda_c{}^e G_{a,be} + e_a{}^\mu\partial_\mu\Lambda^{cb} \\
\delta\overline{G}_{a,b_1\ldots b_8,c} &= \Lambda_a{}^e\bar{G}_{e,b_1\ldots b_8,c} + \Lambda_{b_1}{}^e\bar{G}_{a,eb_2\ldots b_8,c} + \ldots + \Lambda_{b_8}{}^e\bar{G}_{a,b_1\ldots b_7e,c} + \Lambda_c{}^e\bar{G}_{a,b_1\ldots b_8,e}.
\end{aligned} \qquad (4.91)
$$

Using these equations one can verify that dual graviton equation (4.90) is Lorentz invariant, an important consistency check, noting that this calculation works without needing the terms (4.89), as in the case of $A_1^{+++}$.

To finish the calculation we must, as in the $A_1^{+++}$ case, list the changes to the $l_1$-extended six form of equation (4.64), and the changes to the variation of this $l_1$-extended six form given in equation (4.65). The $l_1$-extended six form equation of motion (4.64) is given by

$$\begin{aligned}
\mathcal{E}^{a_1\dots a_6} &= \hat{\mathcal{E}}^{a_1\dots a_6} + \frac{432}{6}\Big( -4G_{c_1c_2,e}{}^{[a_1}G^{|[d,e|a_2\dots a_6c_1c_2]]}{}_{,d} + \frac{1}{2}G_{c_1c_2,d}{}^{e}G^{[d,a_1\dots a_6c_1c_2]}{}_{,e} \\
&- \frac{1}{2}G_{c_1c_2,d}{}^{e}G^{[d,a_1\dots a_6c_1c_2]}{}_{,e} + \frac{1}{4}G_{c_1c_2,e}{}^{e}G^{[d,a_1\dots a_6c_1c_2]}{}_{,d} \\
&+ \frac{1}{2\cdot 9}(\det e)^{\frac{1}{2}}e_{c_1c_2}{}^{\Pi}(\partial_\Pi G_{d,}{}^{a_1\dots a_6c_1c_2,d} - e_d{}^{\nu}\partial_\nu G_{\Pi,}{}^{a_1\dots a_6c_1c_2,d}) \\
&+ \frac{4}{9\cdot 9}(\det e)^{\frac{1}{2}}e_{c_1c_2}{}^{\Pi}(\partial_\Pi G^{[a_1,a_2\dots a_6c_1c_2]d}{}_{,d} - e^{[a_1|\nu|}\partial_\nu G_{\Pi,}{}^{a_2\dots a_6c_1c_2]d}{}_{,d}) \\
&+ \frac{7}{9\cdot 9}G_{c_1c_2,e}{}^{e}G^{[a_1,a_2\dots a_6c_1c_2]d,}{}_{d} - \frac{7}{9\cdot 9}G^{[a_1|}{}_{,e}{}^{e}G_{c_1c_2,}{}^{|a_2\dots a_6c_1c_2]d,}{}_{d} \\
&- \frac{1}{4\cdot 9}G_{d,e}{}^{e}G_{c_1c_2,}{}^{a_1\dots a_6c_1c_2,d} + \frac{1}{2\cdot 9}G_{e,d}{}^{e}G_{c_1c_2,}{}^{a_1\dots a_6c_1c_2,d} \\
&+ \frac{4\cdot 7}{9\cdot 9}(8G_{c_1c_2,}{}^{[a_1|e|}G^{a_2,}{}_{e}{}^{a_3\dots a_6c_1c_2]d}{}_{,d} - G^{[a_1,a_2|e|}G_{c_1c_2,e}{}^{a_3\dots a_6c_1c_2]d}{}_{,d}) \\
&- \frac{8}{9}G_{c_1c_2,d}{}^{[a_1a_3}G^{a_2,a_4a_5a_6c_1c_2]d} + \frac{1}{9}G^{[a_1,}{}_{d}{}^{a_3a_4}G_{c_1c_2,}{}^{a_2a_5a_6c_1c_2]d} \\
&- \frac{4\cdot 8}{9\cdot 9}G_{c_1c_2,}{}^{[a_1|e|}G_{e,}{}^{a_2\dots a_6c_1c_2]d}{}_{,d} - \frac{8}{9\cdot 9}G_{[d,}{}^{[a_1}{}_{e]}G_{c_1c_2,}{}^{|d|a_2\dots a_6c_1c_2],e} \\
&+ \frac{4\cdot 7}{9\cdot 9}G_{[d,}{}^{[a_1}{}_{e]}G_{c_1c_2,}{}^{|de|a_2\dots a_6c_1c_2],a_8} - \frac{4}{9}G_{d,}{}^{[a_1}{}_{e}G_{c_1c_2,e}{}^{a_2\dots a_6c_1c_2],d} \\
&+ \frac{2\cdot 7}{9\cdot 9}\Big(G_{b,}{}^{ce} + G_{b,}{}^{ec})G_{[a_1,a_2\dots a_8]c,e} - (G_{[a_1|,}{}^{ce} + G_{[a_1,|}{}^{ec})G_{b,|a_2\dots a_8]c,e}\Big)\Big).
\end{aligned} \tag{4.92}$$

Similarly, the $l_1$-extended variation (4.65) now reads as

$$\begin{aligned}
\delta\mathcal{E}^{a_1\dots a_6} &= 432\,\Lambda_{c_1c_2c_3}\,E^{a_1\dots a_6c_1c_2,\,c_3} + 4\cdot 432\Lambda_{c_1c_2c_3}(E^{[a_1|,}{}_{de} - G^{[a_1|,}{}_{[de]})G^{d,e|a_2\dots a_6c_1c_2],c_3} \\
&+ \frac{8}{7}\Lambda^{[a_1a_2a_3}E^{a_4a_5a_6]} + \frac{2}{105}G_{[e_5,c_1c_2c_3]}\epsilon^{a_1\dots a_6e_1\dots e_5}E_{e_1\dots e_4}\Lambda^{c_1c_2c_3} \\
&- \frac{3}{35}G_{[e_5,e_6c_1c_2]}\Lambda^{c_1c_2[a_1}\epsilon^{a_2\dots a_6]e_1\dots e_6}E_{e_1\dots e_4} + \frac{1}{420}\,\epsilon_{c_1}{}^{a_1\dots a_6b_1\dots b_4}\,\omega_{c_2,\,b_1b_2}\,E_{c_3,\,b_3b_4}\,\Lambda^{c_1c_2c_3}\quad .
\end{aligned} \tag{4.93}$$

This result has the same form as that of (4.65), but now with the hats removed and an additional term involving the gravity-dual gravity relation (4.53), noting that a similar term arose in the case of $A_1^{+++}$ in equation (4.90).

The nonlinear dual graviton equation (4.90) simplifies when written in terms of world

indices. The result is

$$\begin{aligned}
E_{\mu_1\ldots\mu_8,\tau} &= g^{\nu\kappa}\partial_{[\nu|}F_{[\kappa,\mu_1\ldots\mu_8],|\tau]} - \frac{1}{9}g^{\nu\kappa}\hat{G}_{\tau,\rho}{}^{\rho}\hat{G}_{[\mu_1,\mu_2\ldots\mu_8]\nu,\kappa} - \frac{1}{9}g^{\nu\kappa}\hat{G}_{[\mu_1|,\rho}{}^{\rho}\hat{G}_{\tau,|\mu_2\ldots\mu_8]\nu,\kappa} \\
&+ \frac{1}{2}g^{\nu\kappa}\hat{G}_{\nu,\rho}{}^{\rho}\hat{G}_{[\kappa,\mu_1\ldots\mu_8],\tau} - \frac{1}{2\cdot 9}g^{\nu\kappa}\hat{G}_{\nu,\rho}{}^{\rho}\hat{G}_{\tau,\mu_1\ldots\mu_8,\kappa} - \hat{G}_{\nu,}{}^{(\kappa\nu)}\hat{G}_{[\kappa,\mu_1\ldots\mu_8],\tau} \\
&+ \frac{1}{9}\hat{G}_{\nu,}{}^{(\kappa\nu)}\hat{G}_{\tau,\mu_1\ldots\mu_8,\kappa} + \frac{4}{9}\hat{G}_{\tau,}{}^{(\nu\kappa)}\hat{G}_{[\mu_1,\mu_2\ldots\mu_8]\nu,\kappa} + \frac{4}{9}\hat{G}_{[\mu_1|,}{}^{(\nu\kappa)}\hat{G}_{\tau,|\mu_2\ldots\mu_8]\nu,\kappa} \\
&+ (\det e)^{-1}\varepsilon^{\kappa_1\kappa_2\nu_1\ldots\nu_9}\hat{G}_{\nu_1,\nu_2\ldots\nu_9,[\mu_1|}\hat{G}_{[\kappa_1,\kappa_2|\mu_2\ldots\mu_8]],\tau} + g^{\nu\kappa}\hat{G}_{\tau,[\mu_1\mu_2|\nu}\hat{G}_{|\mu_3,\mu_4\ldots\mu_8]\kappa} \\
&+ \frac{1}{9}g^{\nu\kappa}(\hat{G}_{\nu,[\mu_1\mu_2\mu_3|}\hat{G}_{\tau,|\mu_4\ldots\mu_8]\kappa} - \hat{G}_{\nu,[\mu_1\mu_2|\kappa}G_{\tau,|\mu_3\ldots\mu_8]} - \hat{G}_{\tau,[\mu_1\mu_2\mu_3|}\hat{G}_{\nu,|\mu_4\ldots\mu_8]\kappa} \\
&+ \hat{G}_{\tau,[\mu_1\mu_2|\kappa}\hat{G}_{\nu,|\mu_3\ldots\mu_8]}) \quad ,
\end{aligned} \tag{4.94}$$

where we defined

$$\begin{aligned}
\hat{G}_{\tau,\mu}{}^{\nu} &= (\partial_\tau e_\rho{}^b)e_b{}^\nu, \ \hat{G}_{\tau,\mu_1\mu_2\mu_3} = \partial_\tau A_{\mu_1\mu_2\mu_3}, \\
\hat{G}_{\tau,\mu_1\ldots\mu_6} &= (\partial_\tau A_{\mu_1\ldots\mu_6} - A_{[\mu_1\mu_2\mu_3|}\partial_\tau A_{|\mu_4\mu_5\mu_6]}) \\
F_{\tau,\mu_1\ldots\mu_8,\nu} &= (\partial_\tau h_{\mu_1\ldots\mu_8,\nu} - A_{[\mu_1\mu_2\mu_3|}\partial_\tau A_{|\mu_4\mu_5\mu_6}A_{\mu_7\mu_8]\nu} + 2\partial_\tau A_{[\mu_1\ldots\mu_6}A_{\mu_7\mu_8]\nu} \\
&\quad + 2\partial_\tau A_{[\mu_1\ldots\mu_5\nu}A_{\mu_6\mu_7\mu_8]}) \quad .
\end{aligned} \tag{4.95}$$

Since $E_{11}$ contains $A_8^{+++}$, the Kac-Moody algebra describing pure gravity in eight dimensions, we can obtain the dual graviton equation in eight dimensions by simply setting to zero the three form and six form gauge fields in equation (4.90). Future work could involve considering the $I_c(E_{11})$ variation of the dual gravity equation (4.90), as a means to produce the level four equations of motion associated to the nonlinear realization. This would resolve whether the ambiguous terms (4.89) should be included or not.

# Chapter 5

# Irreducible Representations of $I_c(E_{11}) \otimes_s l_1$

## 5.1 Incorporating Branes

In the previous chapter we have seen how the nonlinear realization of $E_{11} \otimes_s l_1$ with respect to the Cartan involution invariant subgroup $I_c(E_{11})$ led uniquely to the equations of motion of the bosonic sector of eleven-dimensional supergravity when one excludes the effects of higher level derivatives (after $I_c(E_{11})$ variations). Thus far, our discussion has focused on the dynamics of the fields associated to the $g_E$ factor in the $E_{11}$ group element $g = g_E g_l$. We now consider the problem of establishing dynamics for the $g_l$ factor, leading to the dynamics of branes.

Recall that in eleven-dimensional supergravity, the central charges $Z^{a_1 a_2}$ and $Z^{a_1 \cdots a_5}$ are associated to the M2 and M5 branes [19] and so can be referred to as brane charges. Furthermore, the presence, for example, of an M2 brane in space-time is known to break the initial $\mathrm{ISO}(1,10)$ Poincaré invariance to $P_3 \otimes \mathrm{SO}(8)$ where $P_3$ is the $(1,2)$ Poincaré group, while an M5 brane results in $P_6 \times \mathrm{SO}(5)$.

From an $E_{11}$ perspective, the above brane charges fit naturally into the $l_1$ representation (4.11) containing an infinite number of brane charges. The generator $P_a$ is interpreted as producing the charge of a point particle, $Z^{a_1 a_2}$ and $Z^{a_1 \cdots a_5}$ the M2 and M5 branes, and the higher $l_1$ brane charges associated to higher branes predicted by E theory.

It has been proposed in [88] that dynamics for the branes of E theory can be described by the nonlinear realization of $E_{11} \otimes_s l_1$ with respect to not $I_c(E_{11})$ but a subgroup $\mathcal{H}$ of $I_c(E_{11})$. The nonlinear realization must be invariant under the transformation (4.17)

$$g \rightarrow g_0 g \ , \ g_0 \in E_{11} \otimes_s l_1 \ , \ \text{and} \ g \rightarrow gh \ , \ h \in \mathcal{H}. \tag{5.1}$$

The group element involves all brane charges simultaneously, and so from this perspective we see that different choices for the subgroups $\mathcal{H}$ must lead to dynamics for different branes. The necessity for considering subgroups $\mathcal{H}$ of $I_c(E_{11})$ is a natural generalization of the fact that the M2 and M5 brane examples above break the initial $\mathrm{SO}(1,10)$ symmetry (which is at level zero of $I_c(E_{11})$). The group element in this case can now be parametrized in the form

$$\begin{gathered} g = g_l g_h g_E \ , \\ g_l = e^{x^A l_A} \ , \ g_h = e^{\varphi_{\underline{\alpha}} S^{\underline{\alpha}}} \ , \ g_E = e^{A_{\underline{\alpha}} R^{\underline{\alpha}}} \ , \end{gathered} \tag{5.2}$$

where $l_A$ is a generator of the $l_1$ representation of $E_{11}$, $R^{\underline{\alpha}}$ is a generator of $E_{11}$, and $S^{\underline{\alpha}}$ is a generator of $I_c(E_{11})$ that does not live in $\mathcal{H}$.

The fields $A_{\underline{\alpha}}$ and $\varphi$ depend on the generalized coordinates $x^A$. We can introduce brane parameters $\xi^{\underline{\alpha}}$ which describe the world-volume of a brane sitting in the generalised space-time $x^A$. Thus $\xi^{\underline{\alpha}}$ includes parameters such as $\xi^a, \xi_{a_1 a_2}, \ldots$.

In the absence of background fields $A_{\underline{\alpha}}$, the above nonlinear realization reduces to the nonlinear realization of $I_c(E_{11}) \otimes_s l_1$ with respect to a subgroup $\mathcal{H} \subset I_c(E_{11})$. In this case, the group elements transform under a rigid transformation as

$$g_l \rightarrow g_0 g_l g_0^{-1} \ \text{and} \ g_h \rightarrow g_0 g_h \ , \tag{5.3}$$

and under a local $h \in \mathcal{H}$ transformation as

$$g_l \rightarrow g_l \ , \ \text{and} \ g_h \rightarrow g_h h \ . \tag{5.4}$$

The Cartan form $\mathcal{V} = g^{-1} dg$ associated to $g = g_l g_h$ reads as

$$\mathcal{V} = \mathcal{V}_l + \mathcal{V}_h \ , \tag{5.5}$$

where $\mathcal{V}_h = g_h^{-1} dg_h$ lives in $I_c(E_{11})$, and we can express $\mathcal{V}_l$ as

$$\mathcal{V}_l = g_h^{-1}(g_l^{-1} dg_l) g_h = g_h^{-1}(dx^A l_A) g_h = d\xi^{\underline{\alpha}} \nabla_{\underline{\alpha}} z^A l_A \ . \tag{5.6}$$

## 5.2 Wigner Method of Irreducible Representations

A systematic method for constructing the subgroups $\mathcal{H}$ of $I_c(E_{11}) \otimes_S l_1$ is required. A method modelled on the Wigner method of induced representations [90], first applied to the Poincaré Group $\mathrm{SO}(1,3) \otimes_s T_4$, for $T_4$ the translations, has been suggested in [89]. This is natural because both groups are semi-direct products, and indeed $I_c(E_{11}) \otimes_S l_1$ contains the eleven-dimensional Poincaré group when restricted to the lowest level generators $J_{ab}$ and $P_a$. We thus develop this method as a generalisation of that applied to the Poincaré group in [89], [76]. In this approach we assume that there exists a subgroup $\mathcal{H}$ of $I_c(E_{11})$ called the little group for which its irreducible representations are known. We will then 'induce' a representation of $I_c(E_{11}) \otimes_s l_1$ in terms of the known representations of the subgroup $\mathcal{H} \otimes_s l_1$, hence this is also known as the method of induced representations [60, Ch. 3]. Since the $l_1$ representation is commutative, we can consider simultaneous eigenfunctions of a representation $U$ of the $l_1$ generators, denoting the $l_1$ generators by $L_A$ and their eigenvalues on eigenfunctions $\Psi_I$, treated as functions of the fields $\varphi$ outside of $\mathcal{H}$, thus taken at the origin 0, by $l_a$:

$$U(L_A)\Psi_I(0) = l_A \Psi_I(0). \tag{5.7}$$

To describe the action of a representation of $\mathcal{H}$ on $\Psi_I(0)$, we note that, in general, given a representation $U$ of some group acting via a left action on a general field $\psi_I(x)$ we must set $U(g)\psi_I(x) = D(g^{-1})_I{}^J \psi_J(g^{-1}x)$ so that the group multiplication property $U(g)U(h)\psi_I(x) = U(g)D(h^{-1})\psi_I(h^{-1}x) = U(gh)\psi_I(x)$ holds. Thus at the origin we can set

$$U(h)\Psi_I(0) = D(h^{-1})_I{}^J \Psi_J(0) \quad , \tag{5.8}$$

where $h \in \mathcal{H}$. Now, since elements of $I_c(E_{11})$ can be written as $g = e^{\chi \cdot S} h$, with $h \in \mathcal{H}$ and $\chi \cdot S = \chi_{\underline{\alpha}} S^{\underline{\alpha}}$, it is natural to associate fields of the full irreducible representation $I_c(E_{11}) \otimes_s l_1$ to group elements outside of $\mathcal{H} \otimes_s l_1$ through the 'boost'

$$\Psi_I(\varphi) = U(e^{\varphi \cdot S})\Psi_I(0) \quad . \tag{5.9}$$

Thus, $\Psi_I(\varphi)$ describes a field on $I_c(E_{11})/\mathcal{H}$, and this may be compared to equation (2.6) from Section 2.1.

We now consider the action of $U(L_A)$ on a general $\Psi_I(\varphi)$:

$$\begin{aligned} U(L_A)\Psi_I(\varphi) &= U(L_A)U(e^{\varphi\cdot S})\Psi_I(0) = U(e^{\varphi\cdot S})[U(e^{-\varphi\cdot S})U(L_A)U(e^{\varphi\cdot S})]\Psi_I(0) \\ &= U(e^{\varphi\cdot S})D[e^{\varphi\cdot S}]_A{}^B U(L_B)\Psi_I(0) = D[e^{\varphi\cdot S}]_A{}^B l_B U(e^{\varphi\cdot S})\Psi_I(0) \\ &= D[e^{\varphi\cdot S}]_A{}^B l_B \Psi_I(\varphi) \ . \end{aligned} \tag{5.10}$$

This shows that the eigenvalue of a boosted field is a boosted eigenvalue.

We now show that, as expected, the action of a group element $g_0 \in I_c(E_{11})$ on a field in the representation transforms in accordance with the transformation law $U(g_0)\psi(x) = D(g_0^{-1})_I{}^J \psi_J(g^{-1}x)$ discussed in general above. Using the fact that in $I_c(E_{11})/\mathcal{H}$ we must have $g_0 e^{\varphi\cdot S} = e^{\varphi'\cdot S}h$ for $h \in \mathcal{H}$, we find

$$\begin{aligned} U(g_0)\Psi_I(\varphi) &= U(g_0)U(e^{\varphi\cdot S})\Psi_I(0) = U(g_0 e^{\varphi\cdot S})\Psi(0)_I \\ &= U(e^{\varphi'\cdot S})U(h)\Psi_I(0) = U(e^{\varphi'\cdot S})D(h^{-1})_I{}^J\Psi(0)_J \\ &= D(h^{-1})_I{}^J\Psi_J(\varphi') \ . \end{aligned} \tag{5.11}$$

This is the $E_{11}$ 'momentum space' analogue of the Wigner method applied to the Poincaré group. A position space analogue can also be established using a generalized Fourier inversion

$$\tilde{\Psi}_I(x^A) = \int d\varphi e^{l_A(\varphi)x^A}\Psi_I(\varphi) = \int d\varphi U(e^{L_A x^A})\Psi_I(\varphi) = \int d\varphi e^{x^A D_A{}^B(e^{\varphi\cdot S})l_B(0)}\Psi_I(\varphi), \tag{5.12}$$

where the relation $l_A(\varphi)x^A = x^A D_A{}^B(e^{\varphi\cdot S})l_B(0)$ is completely determined by the group theory, i.e. the element $e^{\varphi\cdot S}$ and representation eigenvalues $l_B(0)$ associated to the irreducible representation of $\mathcal{H}$. Similarly, in position space $U(g_0)\tilde{\Psi}_I(x^A)$ results in

$$U(g_0)\tilde{\Psi}_I(x^A) = \int d\varphi e^{x'^A l_A(\varphi')}D(h^{-1})_I{}^J\Psi_J(\varphi) \tag{5.13}$$

on using $g_0 e^{x^A L_A} e^{\varphi\cdot S} = e^{x'^A L_A} e^{\varphi'\cdot S}h$. The operator $L_A$ on $\tilde{\Psi}(x^A)$ now acts as a differential operator $\partial_A$ in position space. Since the field $\Psi_I(\varphi)$ is only associated to $I_c(E_{11})/\mathcal{H}$, we must embed it in a covariant representation of all of $I_c(E_{11})$.

Finally we note that Casimir operators from the universal enveloping algebra can also be used to label the irreducible representations. An $E_{11}$ Casimir generalizing the Poincaré algebra mass-shell relation $p^2 = -m^2$ is known [89]

$$L^2 = P_a P^a + 2Z^{a_1a_2}Z_{a_1a_2} + Z^{a_1\ldots a_5}Z_{a_1\ldots a_5} + \ldots \tag{5.14}$$

As an example of these steps, consider the case of a massive particle in the Poincaré group. The little group is defined to be the subgroup which preserves the choice of 'brane charges' $p_a(0) = (m, 0, 0, 0)$, which is seen to be $\mathcal{H} = \mathrm{SO}(3) = \{J_{ij}\}$ with $a = 0, i$, and $i = 1, 2, 3$. The representations of the little group SO(3) are known, and we can consider the spin one representation $\psi_i(0)$. This can be boosted to the field $\psi_i(\varphi) = e^{\varphi^{0i}J_{0i}}\psi_i(0)$ defined on SO(1,3)/SO(3). The position space representation is then given by $\tilde{\Psi}(x^a) = \int d\varphi^{0i} e^{x^a P_a}\psi_i(\varphi)$ where we have $x^a P_a = x^a D_a{}^b(e^{\varphi^{0i}J_{0i}})p_b(0) = x^a D_a{}^0(e^{\varphi^{0i}J_{0i}})m = x^a \Lambda_a{}^0(\varphi)m$. To embed this into a representation of all of SO(1,3) we define a field $A_a(p)$ satisfying $A_a(0) = 0$, $A_i(0) = \varphi_i(0)$. We can unify these conditions into $p^a(0)A_a(0) = 0$ , which then boosts to $p^a A_a(p) = 0$ or in position space $\partial^a A_a = 0$. Similarly the Casimir reads as $(p^2(0) + m^2)A_a(0) = 0$, and in position space as $(\partial^2 + m^2)A_a(x) = 0$. This is similar to the conventional approach [4, Sec. 14] in which an SO(3) vector $\psi_i(0)$ can be interpreted either as the space components of a four-vector $A_a(0)$, or the space components of an anti-symmetric tensor $F_{ab}(0)$, via $\psi_i(0) = \frac{1}{2}\varepsilon_{ijk}F_{jk}(0)$, in the rest frame. Thus, in an arbitrary frame, $A_a(p)$ and $F_{ab}(p)$ are constrained to be related to one another, which is taken as $iF_{ab} = 2p_{[a}A_{b]}$ and $im^2 A_a = p^b F_{ba}$. Applying $p^a$ to both sides results in $p^a A_a = 0$. Similarly inserting the first into the second reproduces $(p^2 + m^2)A_a = 0$.

In this and the next chapter we will mainly be concerned with the construction of the little group $\mathcal{H}$ regarding the above steps. As we have seen in the case of the massive point particle of the Poincaré group, the little group is determined by having to preserve some condition placed on the brane charges in the $l_1$ representation. This idea extends to the case of $I_c(E_{11})$ as follows.

In order to try to find the little group, we thus seek the subalgebra of $I_c(E_{11})$ which leaves specific choices of the charges invariant. We thus consider group elements $g \in I_c(E_{11})$ which expand infinitesimally as

$$g = I - (\Lambda^{\underline{a}\,\underline{b}}J_{\underline{a}\,\underline{b}} + \Lambda^{\underline{a_1}\,\underline{a_2}\,\underline{a_3}}S_{\underline{a_1}\,\underline{a_2}\,\underline{a_3}} + \Lambda^{\underline{a_1}\cdots\underline{a_6}}S_{\underline{a_1}\cdots\underline{a_6}} + \Lambda^{\underline{a_1}\cdots\underline{a_8},\underline{a_9}}S_{\underline{a_1}\cdots\underline{a_8},\underline{a_9}} + \ldots) \quad , \tag{5.15}$$

where $\underline{a}, \underline{b}, ... = 0, 1, ..., 10$. From $g^{-1}l_A g = D(g)_A{}^B l_B = l'_A$ expanded infinitesimally, the requirement that these group elements preserve a choice of brane charges $l_A^{(0)}$ is equivalent to solving

$$\delta l_A = [\Lambda^{\underline{\alpha}}S_{\underline{\alpha}}, l_A]|_{l_A = l_A^{(0)}} = 0 \quad , \tag{5.16}$$

where we first compute the above commutators for each generator $l_A$ in the $l_1$ representation, set the result to zero, insert the values $l_A^{(0)}$ we wish to preserve, and then determine the conditions on the $\Lambda^{\underline{\alpha}}$. Using the $E_{11}$ algebra of Section 4.2, we can explicitly compute the above variations for the charges $P_{\underline{b}}, Z^{\underline{b}_1\underline{b}_2}, \ldots$, given up to level three, as [89]

$$\begin{aligned}\delta P_{\underline{b}} &= 2\Lambda^{\underline{a}_1}{}_{\underline{b}} P_{\underline{a}_1} + 3\Lambda_{\underline{b}\,\underline{c}_2\underline{c}_3} Z^{\underline{c}_2\underline{c}_3} - 3\Lambda_{\underline{b}\,\underline{c}_2\cdots\underline{c}_6} Z^{\underline{c}_2\cdots\underline{c}_6} - \tfrac{4}{3}\Lambda_{\underline{c}_1\cdots\underline{c}_8,\underline{b}} Z^{\underline{c}_1\cdots\underline{c}_8} \\ &\quad + \tfrac{4}{3}\Lambda_{\underline{b}\,\underline{c}_2\cdots\underline{c}_8,\underline{d}} Z^{\underline{c}_2\cdots\underline{c}_8\underline{d}} + \tfrac{4}{3}\Lambda_{\underline{b}\,\underline{c}_2\cdots\underline{c}_8,\underline{d}} Z^{\underline{c}_2\cdots\underline{c}_8,\underline{d}} + \ldots \quad ,\end{aligned} \tag{5.17}$$

and

$$\begin{aligned}\delta Z^{\underline{b}_1\underline{b}_2} &= -4\Lambda^{[\underline{b}_1}{}_{\underline{c}} Z^{|\underline{c}|\underline{b}_2]} + \Lambda_{\underline{c}_1\underline{c}_2\underline{c}_3} Z^{\underline{c}_1\underline{c}_2\underline{c}_3\underline{b}_1\underline{b}_2} - 6\Lambda^{\underline{b}_1\underline{b}_2\underline{a}_3} P_{\underline{a}_3} - \Lambda_{\underline{c}_1\cdots\underline{c}_6} Z^{\underline{c}_1\cdots\underline{c}_6\underline{b}_1\underline{b}_2} \\ &\quad + \tfrac{1}{3}\Lambda_{\underline{c}_1\cdots\underline{c}_6} Z^{\underline{c}_1\cdots\underline{c}_6[\underline{b}_1,\underline{b}_2]} - \tfrac{16}{135}\Lambda_{\underline{c}_1\cdots\underline{c}_8,\underline{d}} Z^{\underline{b}_1\underline{b}_2\underline{d}\,\underline{c}_1\cdots\underline{c}_5,\underline{c}_6\underline{c}_7\underline{c}_8} + \tfrac{4}{63}\Lambda_{\underline{c}_1\cdots\underline{c}_8,\underline{d}} \hat{Z}^{\underline{b}_1\underline{b}_2\underline{c}_1\cdots\underline{c}_7,\underline{c}_8\,\underline{d}} \\ &\quad + \tfrac{16}{189}\Lambda_{\underline{c}_1\cdots\underline{c}_8,\underline{d}} Z^{\underline{b}_1\underline{b}_2\underline{d}\,\underline{c}_1\cdots\underline{c}_6,\underline{c}_7\underline{c}_8} - \tfrac{16}{189}\Lambda_{\underline{c}_1\cdots\underline{c}_8,\underline{d}} Z^{\underline{b}_1\underline{b}_2\underline{c}_1\cdots\underline{c}_7,\underline{c}_8\underline{d}} + \tfrac{1}{42}\Lambda_{\underline{c}_1\cdots\underline{c}_8,\underline{d}} Z_{(1)}^{\underline{b}_1\underline{b}_2\underline{d}\,\underline{c}_1\cdots\underline{c}_7,\underline{c}_8} \\ &\quad - \tfrac{1}{42}\Lambda_{\underline{c}_1\cdots\underline{c}_8,\underline{d}} Z_{(1)}^{\underline{b}_1\underline{b}_2\underline{c}_1\cdots\underline{c}_8,\underline{d}} - \tfrac{1}{6}\Lambda_{\underline{c}_1\cdots\underline{c}_8,\underline{d}} Z_{(2)}^{\underline{b}_1\underline{b}_2\underline{d}\,\underline{c}_1\cdots\underline{c}_7,\underline{c}_8} + \tfrac{1}{6}\Lambda_{\underline{c}_1\cdots\underline{c}_8,\underline{d}} Z_{(2)}^{\underline{b}_1\underline{b}_2\underline{c}_1\cdots\underline{c}_8,\underline{d}} + \ldots \quad ,\end{aligned} \tag{5.18}$$

and

$$\begin{aligned}\delta Z^{\underline{b}_1\cdots\underline{b}_5} &= -10\Lambda^{[\underline{b}_1}{}_{\underline{c}} Z^{|\underline{c}|\underline{b}_2\cdots\underline{b}_5]} + \Lambda_{\underline{c}_1\underline{c}_2\underline{c}_3} Z^{\underline{b}_1\cdots\underline{b}_5\underline{c}_1\underline{c}_2\underline{c}_3} + \Lambda_{\underline{c}_1\underline{c}_2\underline{c}_3} Z^{\underline{b}_1\cdots\underline{b}_5\underline{c}_1\underline{c}_2,\underline{c}_3} - 60 Z^{[\underline{b}_1\underline{b}_2}\Lambda^{\underline{b}_3\underline{b}_4\underline{b}_5]} \\ &\quad - \Lambda_{\underline{c}_1\cdots\underline{c}_6} Z^{\underline{c}_1\cdots\underline{c}_6\underline{b}_1\cdots\underline{b}_5} + \tfrac{4}{189}\Lambda_{\underline{c}_1\cdots\underline{c}_6} Z^{\underline{c}_1\cdots\underline{c}_6[\underline{b}_1\underline{b}_2\underline{b}_3,\underline{b}_4\underline{b}_5]} - \tfrac{40}{441}\Lambda_{\underline{c}_1\cdots\underline{c}_6} Z^{\underline{c}_1\cdots\underline{c}_6[\underline{b}_1\underline{b}_2\underline{b}_3,\underline{b}_4\underline{b}_5]} \\ &\quad - \tfrac{55}{336}\Lambda_{\underline{c}_1\cdots\underline{c}_6} Z_{(1)}^{\underline{c}_1\cdots\underline{c}_6[\underline{b}_1\cdots\underline{b}_4,\underline{b}_5]} + \tfrac{5}{16}\Lambda_{\underline{c}_1\cdots\underline{c}_6} Z_{(2)}^{\underline{c}_1\cdots\underline{c}_6[\underline{b}_1\cdots\underline{b}_4,\underline{b}_5]} - 360\Lambda^{\underline{b}_1\cdots\underline{b}_5\underline{a}_6} P_{\underline{a}_6} + \ldots \quad ,\end{aligned} \tag{5.19}$$

and

$$\begin{aligned}\delta Z^{\underline{b}_1\cdots\underline{b}_8} &= -16\Lambda^{[\underline{b}_1}{}_{\underline{c}} Z^{|\underline{c}|\underline{b}_2\cdots\underline{b}_8]} + \Lambda_{\underline{c}_1\underline{c}_2\underline{c}_3} Z^{\underline{c}_1\underline{c}_2\underline{c}_3\underline{b}_1\cdots\underline{b}_8} + \tfrac{4}{135}\Lambda_{\underline{c}_1\underline{c}_2\underline{c}_3} Z^{\underline{c}_1\underline{c}_2\underline{c}_3[\underline{b}_1\cdots\underline{b}_5,\underline{b}_6\underline{b}_7\underline{b}_8]} \\ &\quad - \tfrac{20}{63}\Lambda_{\underline{c}_1\underline{c}_2\underline{c}_3} Z^{\underline{c}_1\underline{c}_2\underline{c}_3[\underline{b}_1\cdots\underline{b}_6,\underline{b}_7\underline{b}_8]} - \Lambda_{\underline{c}_1\underline{c}_2\underline{c}_3} Z_{(2)}^{\underline{c}_1\underline{c}_2\underline{c}_3[\underline{b}_1\cdots\underline{b}_7,\underline{b}_8]} + 42\Lambda^{[\underline{b}_1\underline{b}_2\underline{b}_3} Z^{\underline{b}_4\cdots\underline{b}_8]} \\ &\quad + 2520\Lambda^{[\underline{a}_1\cdots\underline{a}_6} Z^{\underline{b}_7\underline{b}_8]} + \tfrac{9\cdot 6720}{8}\Lambda^{\underline{b}_1\cdots\underline{b}_8,\underline{e}} P_{\underline{e}} + \ldots \quad ,\end{aligned} \tag{5.20}$$

and

$$\begin{aligned}\delta Z^{\underline{b}_1\cdots\underline{b}_7,\underline{d}} &= -14\Lambda^{[\underline{b}_1}{}_{\underline{c}} Z^{|\underline{c}|\underline{b}_2\cdots\underline{b}_7],\underline{d}} - 2\Lambda^{\underline{d}}{}_{\underline{c}} Z^{\underline{b}_1\cdots\underline{b}_7,\underline{c}} + \Lambda_{\underline{c}_1\underline{c}_2\underline{c}_3} Z^{\underline{c}_1\underline{c}_2\underline{c}_3\underline{d}[\underline{b}_1\cdots\underline{b}_4,\underline{b}_5\underline{b}_6\underline{b}_7]} \\ &\quad + \tfrac{1}{2}\Lambda_{\underline{c}_1\underline{c}_2\underline{c}_3} Z^{\underline{d}[\underline{b}_1\cdots\underline{b}_5|\underline{c}_1\underline{c}_2,\underline{c}_3|\underline{b}_6\underline{b}_7]} + \Lambda_{\underline{c}_1\underline{c}_2\underline{c}_3} Z^{\underline{c}_1\underline{c}_2\underline{c}_3\underline{d}[\underline{b}_1\cdots\underline{b}_5,\underline{b}_6\underline{b}_7]} - \tfrac{3}{7}\Lambda_{\underline{c}_1\underline{c}_2\underline{c}_3} Z^{\underline{d}[\underline{b}_1\cdots\underline{b}_6|\underline{c}_1\underline{c}_2,\underline{c}_3|\underline{b}_7]} \\ &\quad - \Lambda_{\underline{c}_1\underline{c}_2\underline{c}_3} \hat{Z}^{\underline{d}[\underline{b}_1\cdots\underline{b}_6|\underline{c}_1\underline{c}_2,\underline{c}_3|\underline{b}_7]} - \Lambda_{\underline{c}_1\underline{c}_2\underline{c}_3} Z_{(1)}^{\underline{c}_1\underline{c}_2\underline{c}_3\underline{d}[\underline{b}_1\cdots\underline{b}_6,\underline{b}_7]} - \tfrac{3}{8}\Lambda_{\underline{c}_1\underline{c}_2\underline{c}_3} Z_{(1)}^{\underline{d}\underline{b}_1\cdots\underline{b}_7\underline{c}_1\underline{c}_2,\underline{c}_3} \\ &\quad - \tfrac{945}{4} Z^{[\underline{b}_1\cdots\underline{b}_5}\Lambda^{\underline{b}_6\underline{b}_7]\underline{d}} - \tfrac{945}{4} Z^{[\underline{b}_1\cdots\underline{b}_4|\underline{d}|}\Lambda^{\underline{b}_5\underline{b}_6\underline{b}_7]} + 5670\Lambda^{[\underline{b}_1\cdots\underline{b}_5|\underline{d}|} Z^{\underline{b}_6\underline{b}_7]} \\ &\quad - 5670\Lambda^{[\underline{b}_1\cdots\underline{b}_6} Z^{\underline{b}_7]\underline{d}} - 7560\Lambda^{\underline{b}_1\cdots\underline{b}_7\underline{d},\underline{a}_1} P_{\underline{a}_1} - 60480\Lambda^{\underline{a}_1\underline{b}_1\cdots\underline{b}_7,\underline{d}} P_{\underline{a}_1} + \ldots \quad .\end{aligned} \tag{5.21}$$

## 5.3 Massive Particle

We now seek to construct the little group of a massive particle in E theory. We thus set all higher brane charges $l_A^{(0)}$ to zero apart from those involving $P_{\underline{\alpha}}$. In the massive particle case we choose

$$P_{\underline{b}} = (m, 0, \ldots, 0) \ , \ l_A = 0 \ \text{ otherwise } \ , \ (\underline{b} = 0, 1, ..., 10) \ , \tag{5.22}$$

for which the Casimir $L^2$ possesses the value $-m^2$. Considering this choice in (5.17) — (5.22) results in the variations

$$\begin{aligned} \delta P_{\underline{b}} = 2m\Lambda^0{}_{\underline{b}} \ , \ \delta Z^{\underline{b}_1\underline{b}_2} = -6m\Lambda^{0\underline{b}_1\underline{b}_2} \ , \ \delta Z^{\underline{b}_1\cdots\underline{b}_5} = 360m\Lambda^{0\underline{b}_1\cdots\underline{b}_5}, \\ \delta Z^{\underline{b}_1\cdots\underline{b}_8} = \frac{9\cdot 6720}{8} m\Lambda^{\underline{b}_1\cdots\underline{b}_8,0} \ , \ \delta Z^{\underline{b}_1\cdots\underline{b}_7,\underline{d}} = -7560m\Lambda^{\underline{b}_1\cdots\underline{b}_7 d,0} - 60480m\Lambda^{0\underline{b}_1\cdots\underline{b}_7,\underline{d}} \ . \end{aligned} \tag{5.23}$$

On setting these terms to zero we find that the $\Lambda^{\underline{\alpha}}$'s are constrained to satisfy

$$\Lambda^{0\underline{b}} = 0 \ , \ \Lambda^{0\underline{b}_1\underline{b}_2} = 0 \ , \ \Lambda^{0\underline{b}_1\cdots\underline{b}_5} = 0 \ , \ \Lambda^{0\underline{b}_1\cdots\underline{b}_7,\underline{d}} = 0 \ , \ \Lambda^{\underline{b}_1\cdots\underline{b}_8,0} = 0 \ , \ \ldots \tag{5.24}$$

so that any $\Lambda^{\underline{\alpha}}$ with one index equal to zero is taken to be zero, and so more generally the subalgebra $\mathcal{H}$ is $I_c(E_{11})$ in which every generator with an index equal to zero is set to zero, resulting in

$$\mathcal{H} = I_c(E_{10}). \tag{5.25}$$

A representation of $I_c(E_{10})$ is provided by the Cartan involution odd generators of $E_{10}$

$$T_{ij} = \eta_{ik}K^k{}_j + \eta_{jk}K^k{}_i \ , \ T_{i_1i_2i_3} = R^{k_1k_2k_3}\eta_{k_1i_1}\eta_{k_2i_2}\eta_{k_3i_3} + R_{i_1i_2i_3}, \ \ldots \ , \tag{5.26}$$

with $i, j = 1, \ldots, 10$. These generators form a representation of $E_{10}$ since the commutator of a Cartan involution even generator with an odd generator of produces an odd generator, due to the involution property $I_c(AB) = I_c(A)I_c(B)$ for generators $A, B$. The fields associated to these generators

$$h_{ij}(0) \ , \ A_{i_1i_2i_3}(0) \ , \ \ldots \ , \tag{5.27}$$

where $h_{ij}(0)$ is symmetric in its indices, $h_{(ij)}(0) = h_{ij}(0)$, while $A_{i_1i_2i_3}(0)$ and higher level fields have anti-symmetric indices in each block of indices as usual.

## 5.4 M2 and M5 Branes

For the M2 and M5 branes the subgroups $\mathcal{H}$ that lead to brane dynamics have been found, independently of the above Wigner method, in [88] to take the forms

$$\begin{aligned}\mathcal{H}_{M2} = \{ & J_{a_1a_2} \ , \ J_{a'_1a'_2} \ , \ \hat{S} = \epsilon^{a_1a_2a_3}S_{a_1a_2a_3} \ , \ L_{ab'} = 2J_{ab'} + \epsilon_{ae_1e_2}S^{e_1e_2}{}_{b'} \ , \ S_{ab'_1b'_2} \ , \\ & \hat{S}_{a'_1a'_2a'_3} = S_{a'_1a'_2a'_3} + \frac{1}{3}\epsilon^{e_1e_2e_3}S_{e_1e_2e_3a'_1a'_2a'_3} \ , \ S_{b'_1\ldots b'_6} \ , \ S_{a_1b'_1\ldots b'_5} \ , \ S_{a_1a_2b'_1\ldots b'_4} \ , \ \ldots\}\end{aligned} \tag{5.28}$$

where $a_1, a_2, \ldots = 0, 1, 2$, $a'_1, a'_2, \ldots = 3, 4 \ldots, 10$, and

$$\begin{aligned}\mathcal{H}_{M5} = \{ & J_{a_1a_2} \ , \ J_{a'_1a'_2} \ , \ L_{ab'} = J_{ab'} + \frac{2}{5!}\epsilon_{ae_1\ldots e_5}S^{e_1\ldots e_5}{}_{b'} \ , \\ & S_{-a_1a_2a_3} := \tfrac{1}{2}(S_{a_1a_2a_3} - \tfrac{1}{3!}\epsilon_{a_1a_2a_3b_1b_2b_3}S^{b_1b_2b_3}) \ , \ S = \frac{1}{6!}\epsilon^{a_1\ldots a_6}S_{a_1\ldots a_6} \ , \ \ldots\}\end{aligned} \tag{5.29}$$

where $a_1, a_2, \ldots = 0, \ldots 5$, $a'_1, a'_2 = 6, \ldots 10$. We now consider some specific choice of brane charges and examine how closely the result agrees with the above algebras. In the case of the M2 brane we consider the choices

$$P_{\underline{a}} = (m, 0, \ldots, 0) \quad , \quad Z^{12} = em \quad , \quad Z^{\alpha} = 0 \ \text{ otherwise} \ . \tag{5.30}$$

We note that the Casimir $L^2$ vanishes in the case $e^2 = 1$. The variations (5.17) — (5.22) reduce to

$$\begin{aligned}\delta P_{\underline{b}} &= 2m\Lambda^0{}_{\underline{b}} + 6em\Lambda_{\underline{b}12} \ , \ \delta Z^{\underline{b}_1\underline{b}_2} = -6m\Lambda^{0\underline{b}_1\underline{b}_2} - 4em\Lambda^{[\underline{b}_1|}{}_i\epsilon^{ij}\delta_j^{|\underline{b}_2]} \ , \\ \delta Z^{\underline{b}_1\cdots\underline{b}_5} &= 360m\Lambda^{0\underline{b}_1\cdots\underline{b}_5} - 60em\epsilon^{ij}\delta_{ij}^{[\underline{b}_1\underline{b}_2}\Lambda^{\underline{b}_3\underline{b}_4\underline{b}_5]} \ , \\ \delta Z^{\underline{b}_1\cdots\underline{b}_8} &= \frac{9\cdot 6720}{8}m\Lambda^{\underline{b}_1\cdots\underline{b}_8,0} + 2520em\epsilon^{ij}\delta_{ij}^{[\underline{b}_1\underline{b}_2}\Lambda^{\underline{b}_3\cdots\underline{b}_8]} \ , \\ \delta Z^{\underline{b}_1\cdots\underline{b}_7,\underline{d}} &= 5670m(e\Lambda^{[\underline{b}_1\cdots\underline{b}_5|\underline{d}|}\epsilon^{ij}\delta_{ij}^{\underline{b}_6\underline{b}_7]} - e\Lambda^{[\underline{b}_1\cdots\underline{b}_6}\epsilon^{ij}\delta_{ij}^{\underline{b}_7]\underline{d}} - m\Lambda^{\underline{b}_1\cdots\underline{b}_7\underline{d},0}) - 8.7560m\Lambda^{0\underline{b}_1\cdots\underline{b}_7,\underline{d}} \ , \ \ldots\end{aligned} \tag{5.31}$$

Here and below we set $i, j = 1, 2$ and $a'_1, \ldots = 3, \ldots, 10$. The transformations that preserve the chosen values of the charges are found by setting these variations to zero. This results in the conditions on the $\Lambda^{\underline{\alpha}}$'s

$$\begin{aligned}&\Lambda^{0i} = 0 \ , \ \Lambda^{ab'} = -\frac{3}{2}\epsilon^{ac_1c_2}e\Lambda_{c_1c_2}{}^{b'} \ , \ \Lambda^{ij} \neq 0 \ , \ \Lambda^{a'_1a'_2} \neq 0 \ , \\ &\Lambda^{a_1a_2a_3} = 0 \ , \ \Lambda^{a'_1a'_20} = 0 \ , \ \Lambda^{a'_1a'_2i} \neq 0 \ , \ \Lambda^{b'_1b'_2b'_3} = -\frac{10}{e}\epsilon_{a_1a_2a_3}\Lambda^{a_1a_2a_3b'_1b'_2b'_3} \ , \\ &\Lambda^{a'_1\ldots a'_6} = -\frac{28}{e}\varepsilon_{b_1b_2b_3}\Lambda^{b_1b_2b_3a'_1\ldots a'_5,a'_6} = \frac{14}{e}\varepsilon_{b_1b_2b_3}\Lambda^{b_1b_2a'_1\ldots a'_6,b_3} \ , \\ &\Lambda^{0a'_1\ldots a'_5} = 0 \ , \ \Lambda^{ia'_1\ldots a'_5} = \frac{56}{6e}\varepsilon_{b_1b_2b_3}\Lambda^{b_1b_2b_3ia'_1\ldots'_5} \ , \ \Lambda^{ia'_1\ldots a'_5} \neq 0 \ , \ \Lambda^{0ia'_1\ldots a'_4} = 0 \ , \ \Lambda^{ija'_1\ldots a'_4} \neq 0 \ .\end{aligned} \tag{5.32}$$

These relations are consistent only when $e^2 = 1$ and so we choose $e = 1$. This is because the case $\delta P_a = 0$ leads to $\Lambda^{ab'} = -\frac{3}{2}\varepsilon^{ae_1e_2}e\Lambda_{e_1e_2}{}^{b'}$, while the case $\delta Z^{b_1b_2} = 0$ leads to $\Lambda^{e_1e_2b'} = \frac{1}{3}\varepsilon^{e_1e_2a}e\Lambda_a{}^{b'}$, and inserting one into the other fixes $e^2 = 1$.

Inserting the relations (5.32) into the group element (5.15), we find the subalgebra $\mathcal{H}$ that preserves the above choice of charges is given by

$$\mathcal{H} = \{J_{ij}, J_{a'_1a'_2},\ \hat{L}_{a_1a'_2},\ S_{ia'_1a'_2},\ \hat{S}_{b'_1b'_2b'_3},\ S_{ia'_1a'_2..a'_5},\ S_{ija'_1a'_2..a'_4},\ \hat{S}_{ia'_1...a'_5a'_6},\ \hat{S}_{a'_1..a'_6},\ \ldots\} \quad , \tag{5.33}$$

where we set

$$\begin{aligned} \hat{L}_{a_1a'_2} &= 2J_{a_1a'_2} + \epsilon_{a_1}{}^{e_1e_2}S_{e_1e_2a'_2} \quad , \quad \hat{S}_{b'_1b'_2b'_3} = S_{b'_1b'_2b'_3} + \frac{1}{3}\varepsilon^{e_1e_2e_3}S_{e_1e_2e_3b'_1b'_2b'_3}, \\ \hat{S}_{ia'_1...a'_5} &= S_{ia'_1...a'_5} - \frac{1}{2}\varepsilon^{b_1b_2b_3}S_{ia'_1...a'_5b_1b_2,b_3} \quad , \quad \hat{S}_{a'_1...a'_6} = S_{a'_1...a'_6} - \frac{1}{2}\varepsilon^{b_1b_2b_3}S_{a'_1...a'_6b_1b_2,b_3}, \ldots \end{aligned} \tag{5.34}$$

The generators of $I_c(E_{11})$ not in $\mathcal{H}$ are those associated to the $\Lambda^{\underline{\alpha}} = 0$ values

$$J_{0i} \ , \quad S_{0i_1i_2} \ , \ S_{a_1a_2a_3} \ , \ S_{0a'_1...a'_5} \ , \ S_{i_1i_2a'_1...a'_4} \ , \ \ldots \tag{5.35}$$

This shows that the subalgebra $\mathcal{H}$ in equation (5.34) is almost, but not exactly, equal to the subalgebra $\mathcal{H}_{M2}$ in equation (5.28), because these last generators were excluded by the method. This means that a slight generalisation of the conditions (5.30) is needed in order to fully reproduce the subalgebra (5.28).

We next consider the analogous results for the M5 algebra, with the choice of charges

$$P_{\underline{\alpha}} = (m, 0, \ldots, 0) \quad , \quad Z^{12345} = em \tag{5.36}$$

for which the Casimir $L^2$ also vanishes when $e = \pm 1$. The variations (5.17) — (5.22) reduce in this case to

$$\begin{aligned} \delta P_{\underline{b}} &= 2m(\Lambda^0{}_{\underline{b}} - 180e\Lambda_{\underline{b}1...5}) \quad , \quad \delta Z^{\underline{b}_1\underline{b}_2} = \delta^{\underline{e}_1\underline{e}_2\underline{e}_3\underline{b}_1\underline{b}_2}_{i_1...i_5}\varepsilon^{i_1...i_5}em\Lambda_{\underline{e}_1\underline{e}_2\underline{e}_3} - 6m\Lambda^{0\underline{b}_1\underline{b}_2}, \\ \delta Z^{\underline{b}_1\cdots\underline{b}_5} &= 10m(36\Lambda^{0\underline{b}_1\cdots\underline{b}_5} - e\varepsilon^{i_1...i_5}\Lambda^{[\underline{b}_1}{}_{\underline{c}}\delta^{|\underline{c}|\underline{b}_2\cdots\underline{b}_5]}_{i_1...i_5}) \\ \delta Z^{\underline{b}_1\cdots\underline{b}_8} &= 42m(180\Lambda^{\underline{b}_1\cdots\underline{b}_8,0} + e\varepsilon^{i_1...i_5}\Lambda^{[\underline{b}_1\underline{b}_2\underline{b}_3}\delta^{\underline{b}_4\cdots\underline{b}_8]}_{i_1...i_5}) \\ \delta Z^{\underline{b}_1\cdots\underline{b}_7,\underline{d}} &= -\frac{945}{4}[em\varepsilon^{i_1...i_5}(\delta^{[\underline{b}_1\cdots\underline{b}_5}_{i_1...i_5}\Lambda^{\underline{b}_6\underline{b}_7]\underline{d}} + \delta^{[\underline{b}_1\cdots\underline{b}_4|\underline{d}|}_{i_1...i_5}\Lambda^{\underline{b}_5\underline{b}_6\underline{b}_7]}) + 32(\Lambda^{\underline{b}_1\cdots\underline{b}_7\underline{d},0} + 8m\Lambda^{0\underline{b}_1\cdots\underline{b}_7,\underline{d}})] \ . \end{aligned} \tag{5.37}$$

Here and below we set $i, j = 1, \ldots, 5$ and $a'_1, \ldots = 6, \ldots, 10$. On setting these conditions to zero we find the following constraints on the $\Lambda^{\underline{\alpha}}$ terms:

$$\begin{aligned}
\Lambda^{0i} &= 0,\ \Lambda^{i_1 i_2} \neq 0,\ \Lambda^{ab'} = -\frac{3}{2e}\varepsilon^{a}{}_{c_1\ldots c_5}\Lambda^{c_1\ldots c_5 b'},\ \Lambda^{a'b'} \neq 0,\\
\Lambda^{a_1a_2a_3} &= \frac{e}{3!}\varepsilon^{a_1a_2a_3b_1b_2b_3}\Lambda_{b_1b_2b_3},\ \Lambda^{0ia'} = 0,\ \Lambda^{i_1i_2a'} \neq 0,\ \Lambda^{0a'_1a'_2} = 0,\\
\Lambda^{ia'_1a'_2} &= -\frac{14}{3e}\varepsilon_{b_1\ldots b_6}\Lambda^{b_1\ldots b_6 a'_1 a'_2, i},\ \Lambda^{a'_1a'_2a'_3} = -\frac{7}{e}\varepsilon_{b_1\ldots b_6}\Lambda^{b_1\ldots b_6 a'_1a'_2,a'_3}.
\end{aligned} \tag{5.38}$$

As in the M2 case, these relations are consistent only when $e^2 = 1$ and we choose $e = 1$. Placing these conditions into the group element (5.15) produces the algebra

$$\begin{aligned}
\mathcal{H} = \{J_{i_1i_2}\ ,\ J_{a'_1a'_2},\ L_{a_1a'_2},\ S_{-a_1a_2a_3},\ S_{i_1i_2a'_3},\ \hat{S}_{ia'_2a'_3},\ \hat{S}_{a'_1a'_2a'_3},\ S_{i_1\ldots i_4a'_5a'_6},\\
S_{i_1i_2i_3a'_1a'_2a'_3},\ S_{i_1i_2a'_1\ldots a'_4},\ S_{ia'_1\ldots a'_5},\ S_{a'_1\ldots a'_6}, \ldots\}\ ,
\end{aligned} \tag{5.39}$$

where we defined

$$\begin{aligned}
L_{a_1a'_2} &= 2(J_{a_1a'_2} + \frac{2}{5!}e\varepsilon_{a_1}{}^{c_1\ldots c_5}S_{c_1\ldots c_5a'_2})\quad ,\quad S_{-a_1a_2a_3} = \frac{1}{2}(S_{a_1a_2a_3} - \frac{1}{3!}e\varepsilon_{a_1a_2a_3b_1b_2b_3}S^{b_1b_2b_3}),\\
\hat{S}_{ia'_2a'_3} &= S_{ia'_2a'_3} - \frac{1}{5!}\varepsilon^{b_1\ldots b_6}S_{b_1\ldots b_6[i,a'_2a'_3]}\quad ,\quad \hat{S}_{a'_1a'_2a'_3} = S_{a'_1a'_2a'_3} - \frac{1}{5!}\varepsilon^{b_1\ldots b_6}S_{b_1\ldots b_6[a'_1a'_2,a'_3]}.
\end{aligned} \tag{5.40}$$

The generators of $I_c(E_{11})$ not in $\mathcal{H}$ in the case of the M5 brane are given by

$$J_{0i},\ S_{a_1a_2a_3},\ S_{0ib'},\ S_{0a'_1a'_2},\ S_{a_1\ldots a_6},\ S_{0i_1i_2i_3b'_1b'_2}, S_{0i_1i_2b'_1b'_2b'_3},\ S_{0ib'_1\ldots b'_4},\ S_{0a'_1\ldots a'_5},\ \ldots \tag{5.41}$$

Thus, as in the M2 case, the subalgebra (5.39) is again almost, but not exactly, equal to the subalgebra (5.29), and so a slight generalisation of the conditions (5.36) is needed to reproduce the subalgebra (5.29).

## 5.5 Massless Particle

In the case of a massless particle we consider the choice

$$P_0 = -m\quad ,\quad P_{10} = m\quad ,\quad Z^{\alpha} = 0\ \ \text{otherwise}\ . \tag{5.42}$$

This choice satisfies $L^2 = 0$. In light-cone coordinates $V^{\pm} = \frac{1}{\sqrt{2}}(V^{10} \pm V^0), V^i,\ i = 1, \ldots, 9$, for which the metric $\eta_{\underline{a}\underline{b}}$ takes the form $\eta_{+-} = 1$, $\eta_{ij} = \delta_{ij}$, becomes

$$P_{-} = \sqrt{2}m. \tag{5.43}$$

The parameters found by inserting the above brane charges into (5.17) — (5.22) and setting the results to zero read, in light-cone coordinates, as

$$\Lambda_{+a} = 0\ ,\ \Lambda_{+-} = 0\ ,\ \Lambda_{+a_1a_2} = 0\ ,\ \Lambda_{+a_1\ldots a_5} = 0\ ,\ \Lambda_{+a_1\ldots a_7} = 0\ ,\ \Lambda_{+a_1\ldots a_6,b} = 0\ ,\ \ldots \tag{5.44}$$

so that any $\Lambda^{\underline{\alpha}}$ with a lowered index + vanishes. Inserting these parameters into the group element (5.15) we find the subalgebra preserving the choice of brane charge to be those generators with no lowered − index

$$\mathcal{H} = \{J_{+i}\ \ ,\ \ J_{ij}\ \ ,\ \ S_{+ij}\ \ ,\ \ S_{i_1i_2i_3}\ \ ,\ \ S_{+i_1\ldots i_5}\ \ ,\ \ \ldots\}\ \ ,\ \ i = 1,\ldots,9. \tag{5.45}$$

All commutators between generators with + indices also vanish:

$$[J_{+i}, J_{+j}] = [J_{+i}, S_{+jk}] = [S_{+ij}, S_{+kl}] = \ldots = 0\ \ , \tag{5.46}$$

where we assume this holds to all orders. Thus the generators with a + index can be neglected, and indeed in any irreducible representation these operators must be trivially realized, so the algebra of generators with no + index close and generate $I_c(E_9)$.

Thus one is interested in representations of $I_c(E_9)$, formed from $E_{11}$ generators with the indices restricted to $i = 1,\ldots,9$:

$$K^i{}_j\ \ ,\ R^{i_1i_2i_3}\ ,\ R_{i_1i_2i_3}\ \ ,\ \ R^{i_1\ldots i_6}\ \ ,\ \ R^{i_1\ldots i_8,j}\ \ ,\ \ R^{i_1\ldots i_9,j_1j_2j_3}\ \ ,\ R^{i_1\ldots i_9,j_1\ldots j_6}\ \ ,\ \ \ldots \tag{5.47}$$

where we note for example at level four there is no $E_9$ analogue of the $E_{11}$ generators $R^{a_1\ldots a_{10},b_1b_2}$ or $R^{a_1\ldots a_{11},b}$. The algebra $\mathcal{H} = I_c(E_9)$ is generated by

$$\begin{aligned}
J_{ij} &= \eta_{ik}K^k{}_j - \eta_{jk}K^k{}_i \quad , \\
S_{i_1i_2i_3} &= R^{j_1j_2j_3}\eta_{j_1i_1}\eta_{j_2i_2}\eta_{j_3i_3} - R_{i_1i_2i_3} \quad , \\
S_{i_1..i_6} &= R^{j_1..j_6}\eta_{j_1i_1}..\eta_{j_6i_6} + R_{i_1..i_6} \quad , \\
S_{i_1..i_8,k} &= R^{j_1..j_8,l}\eta_{j_1i_1}..\eta_{j_8i_8}\eta_{lk} - R_{i_1..i_8,k} \quad , \\
S_{i_1..i_9,j_1j_2j_3} &= R^{m_1..m_9,l_1l_2l_3}\eta_{m_1i_1}..\eta_{m_9i_9}\eta_{l_1j_1}\eta_{l_2j_2}\eta_{l_3j_3} + R_{i_1..i_9,j_1j_2j_3} \quad .
\end{aligned} \tag{5.48}$$

We can thus construct a representation of $\mathcal{H}$ using the Cartan involution odd generators

$$\begin{aligned}
T_{ij} &= \eta_{ik}K^k{}_j + \eta_{jk}K^k{}_i \quad , \\
T_{i_1i_2i_3} &= R^{j_1j_2j_3}\eta_{j_1i_1}\eta_{j_2i_2}\eta_{j_3i_3} + R_{i_1i_2i_3} \quad , \\
T_{i_1..i_6} &= R^{j_1..j_6}\eta_{j_1i_1}..\eta_{j_6i_6} - R_{i_1..i_6} \quad , \\
T_{i_1..i_8,k} &= R^{j_1..j_8,l}\eta_{j_1i_1}..\eta_{j_8i_8}\eta_{lk} + R_{i_1..i_8,k} \quad , \\
T_{i_1..i_9,j_1j_2j_3} &= R^{m_1..m_9,l_1l_2l_3}\eta_{m_1i_1}..\eta_{m_9i_9}\eta_{l_1j_1}\eta_{l_2j_2}\eta_{l_3j_3} - R_{i_1..i_9,j_1j_2j_3} \quad .
\end{aligned} \tag{5.49}$$

Associated to these generators are fields which we view as the $\varphi = 0$ case of $\Psi$

$$\Psi(0) = \{h_{ij} \ , \ A_{i_1 i_2 i_3} \ , \ A_{i_1 \ldots i_6} \ , \ A_{i_1 \ldots i_8, j} \ , A_{i_1 \ldots i_9, j_1 j_2 j_3} \ , \ \ldots\}. \tag{5.50}$$

## 5.6 $I_c(E_9)$ Duality Relations, Degrees of Freedom of a Massless Particle

Given that the graviton is a massless particle, we will consider the linearised dynamics associated to a massless particle in the representation of the previous section.

In eleven-dimensional supergravity in the rest frame [85], the graviton is a rank two symmetric traceless tensor representation of SO(9). Thus, from (5.50) we impose, since ${h^i}_i$ is a scalar under $I_c(E_9)$, the condition that $h_{ij}$ be traceless, ${h^i}_i = 0$. On doing this we see that $h_{ij}$ and $A_{i_1 i_2 i_3}$ possess the $(45-1)+84 = 128$ degrees of freedom of the graviton and three-form of eleven-dimensional supergravity. This indicates that all degrees of freedom in the higher fields may be determined in terms of the fields $h_{ij}$ and $A_{i_1 i_2 i_3}$ due to the existence of an infinite number of duality relations, some of which we will establish in this section.

We thus treat the fields in the linear representation (5.50) as though they were Cartan forms

$$\mathcal{V} = h_{ij} T^{ij} + A_{i_1 i_2 i_3} T^{i_1 i_2 i_3} + A_{i_1..i_6} T^{i_1..i_6} + h_{i_1..i_8,j} T^{i_1..i_8,j} + A_{i_1..i_9,j_1 j_2 j_3} T^{i_1..i_9,j_1 j_2 j_3} + .. \tag{5.51}$$

and consider the variation of the Cartan forms under the local $I_c(E_9)$ transformations

$$S_{k_1 k_2 k_3} = R^{p_1 p_2 p_3} \eta_{p_1 k_1} \eta_{p_2 k_2} \eta_{p_3 k_3} - R_{k_1 k_2 k_3} \ , \tag{5.52}$$

which can be computed via $\delta \mathcal{V} = \Lambda^{ijk}[S_{ijk}, \mathcal{V}]$ using the $E_{11}$ algebra to level four [66] to give

$$\begin{aligned}
\delta h_{ij} &= 18 \Lambda_{(i|k_1 k_2|} {A_{j)}}^{k_1 k_2} - 2 \eta_{ij} \Lambda^{k_1 k_2 k_3} A_{k_1 k_2 k_3} \ , \\
\delta A_{i_1 i_2 i_3} &= -3 \cdot 2 {h_{[i_1}}^c \Lambda_{i_2 i_3]c} + 60 {A_{i_1 i_2 i_3}}^{k_1 k_2 k_3} \Lambda_{k_1 k_2 k_3} \ , \\
\delta A_{i_1..i_6} &= 2 \Lambda_{[i_1 i_2 i_3} A_{i_4 i_5 i_6]} + 112 \Lambda^{k_1 k_2 k_3} (h_{i_1..i_6 k_1 k_2, k_3} + h_{[i_1..i_5|k_1 k_2 k_3|, i_6]}) \ , \\
\delta h_{i_1..i_8,j} &= 2(\Lambda_{[i_1 i_2 i_3} A_{i_4..i_8]j} - \Lambda_{j[i_1 i_2} A_{i_3..i_8]}) \\
&\quad - 12 \cdot 8 \cdot 3 \Lambda^{k_1 k_2 k_3} (A_{i_1 \ldots i_8 k_1, k_2 k_3 j} + A_{[i_1 \ldots i_7|j k_1, k_2 k_3|i_8]}) \ , \ldots \ .
\end{aligned} \tag{5.53}$$

We now postulate the following duality relation between the three-form and six form up to an arbitrary coefficient $c$

$$E_{i_1i_2i_3} = A_{i_1i_2i_3} + c\varepsilon_{i_1i_2i_3}{}^{j_1..j_6}A_{j_1..j_6} = 0. \tag{5.54}$$

The variation of (5.54) under (5.53) fixes the coefficient $c$ to satisfy $c = \pm\frac{1}{12}$. We arbitrarily choose the + sign and find the duality relation

$$D_{i_1i_2i_3} = A_{i_1i_2i_3} + \frac{1}{12}\varepsilon_{i_1i_2i_3}{}^{j_1..j_6}A_{j_1..j_6} = 0. \tag{5.55}$$

Similarly, under (5.53) the duality relation (5.55) varies into

$$\delta E_{i_1i_2i_3} = \frac{1}{6}\Lambda_{j_1j_2j_3}\varepsilon_{i_1i_2i_3}{}^{j_1..j_6}E_{j_4j_5j_6} - 6E_{k_3[i_1}\Lambda_{i_2i_3]}{}^{k_3}, \tag{5.56}$$

where we defined the linearised gravity - dual gravity duality equation of motion

$$E_{ij} = h_{ij} - \frac{1}{4}\varepsilon_j{}^{r_1..r_8}h_{r_1..r_8,i} = 0. \tag{5.57}$$

In this relation, the tracelessness of $h_{ij}$ is compatible with the irreducibility of $h_{i_1...i_8,j}$, and the positioning of the indices ensures a correspondence with $E_{11}$ (5.72) below.

We now vary the duality relation of equation (5.57) to find that

$$\begin{aligned}\delta E_{ij} = &-\frac{1}{2}\frac{1}{3\cdot 5!}\,\varepsilon_i{}^{r_1..r_8}(\Lambda_{r_1r_2r_3}\varepsilon_{r_4..r_8j}{}^{k_1k_2k_3}E_{k_1k_2k_3} - \Lambda_{jr_1r_2}\varepsilon_{r_3..r_8}{}^{k_1k_2k_3}E_{k_1k_2k_3}) \\ &+ 12\cdot 2\cdot 3\,\Lambda^{k_1k_2k_3}\varepsilon_i{}^{r_1..r_8}(E_{r_1..r_8k_1,k_2k_3j} - E_{jr_1..r_7k_1,k_2k_3r_8}) = 0\end{aligned} \tag{5.58}$$

where we defined

$$E_{i_1..i_9,j_1j_2j_3} = A_{i_1..i_9,j_1j_2j_3} + \frac{1}{9!}\,\varepsilon_{i_1..i_9}A_{j_1j_2j_3} = 0. \tag{5.59}$$

The variation of $E_{ij}$ thus contains the previously-derived duality relation of equation (5.55), and a new duality relation (5.59) relating the three-form $A_{i_1i_2i_3}$ and nine-three form $A_{i_1...i_9,j_1j_2j_3}$.

These duality relations

$$E_{ij} = 0\ ,\ E_{i_1i_2i_3} = 0\ ,\ E_{i_1...i_9,j_1j_2j_3} = 0\ , \tag{5.60}$$

are just the first of an infinite tower of duality relations, and it shows that the fields at higher levels, two, three, four, etc... can be expressed in terms of the first two fields

$$h_{ij}\ ,\ A_{i_1i_2i_3}\ . \tag{5.61}$$

For example the six form is related to the three form by equation (5.55) and the dual graviton $h_{i_1...i_9,j}$ is related to the graviton by equation (5.57), etc...

Since all the higher level fields carry the same indices of these fields up to multiple sets of blocks of nine indices, we can expect that in all of these higher level duality relations the blocks of indices are carried by $\varepsilon_{i_1..i_9}$'s in a way that is similar to how they appear in equation (5.59).

This shows that the massless irreducible representation contains the fields of equation (5.61), thus as a result it contains the $44 + 84 = 128$ bosonic degrees of freedom of eleven-dimensional supergravity.

Another way to express the fields of equation (5.61) is in terms of an $E_8$ multiplet, or since we are working in the Cartan involution invariant subalgebra, an $I_C(E_8) = \mathrm{SO}(16)$ multiplet. To do this we set $i = (1, i')$ with $i' = 2, \dots, 9$, and use the duality relations to express the fields with a 1 index in terms of fields, none of whose indices involve a 1. For example we can express $A_{1i'_2i'_3}$ in terms of $A_{i'_1...i'_6}$ by using the duality relation of equation (5.55),

$$A_{1i'_2i'_3} = -\frac{1}{12}\varepsilon_{1i'_2i'_3}{}^{j'_1...j'_6}A_{j'_1...j'_6} \ , \tag{5.62}$$

and similarly $h_{1j'}$ can be expressed in terms of $h_{i'_1...i'_8,j'}$ by using the duality relation (5.57). Thus, we write the fields of equation (5.63) as SO(16) multiplets in the form

$$h_{i'j'} \quad , \quad A_{i'_1i'_2i'_3} \quad , \quad A_{i'_1...i'_6} \quad , \quad h_{j'} \equiv \frac{1}{8!}\epsilon^{i'_1...i'_8}h_{i'_1...i'_8,j'} \quad , \quad i', j', \ldots = 3, \ldots, 10. \tag{5.63}$$

The fields of equation (5.63) still give the correct $36 + 56 + 28 + 8 = 128$ degrees of freedom. Since $h_{i'}{}^{i'} + h_9{}^9 = 0$ still holds, and because we have not included $h_9{}^9$, we do not take $h_{i'}{}^{i'}$ equal to zero.

We close this section by showing that the above linearised duality relations can be obtained by linearising the first order $E_{11}$ duality relations (4.47) and (4.53). In light-cone coordinates, at the linearised level, we assume $\partial_-$ is invertible, so that the independent components of the linearised Cartan forms

$$\begin{aligned} G_{a_1..a_4} &= \partial_{[a_1}A_{a_2a_3a_4]} \quad , \\ G_{a_1..a_7} &= \partial_{[a_1}A_{a_2..a_7]} \quad , \end{aligned} \tag{5.64}$$

are given by

$$\begin{aligned} G_{-i_1..i_3} &= \tfrac{1}{4}\partial_-\hat{A}_{i_1i_2i_3} \quad , \\ G_{-i_1..i_6} &= \tfrac{1}{7}\partial_-\hat{A}_{i_1..i_6} \quad , \end{aligned} \tag{5.65}$$

where we have defined

$$\begin{aligned} \hat{A}_{i_1i_2i_3} &= A_{i_1i_2i_3} - 3\frac{1}{\partial_-}\partial_{[i_1}A_{|-|i_2i_3]} \quad , \\ \hat{A}_{i_1..i_6} &= A_{i_1..i_6} - 6\frac{1}{\partial_-}\partial_{[i_1}A_{|-|i_2..i_7]} \quad . \end{aligned} \tag{5.66}$$

On setting $\varepsilon^{+-i_1...i_9} = \varepsilon^{i_1...i_9}$, the duality relation $D_{a_1...a_4} = 0$ (4.47) in linearised form becomes

$$D_{-i_1i_2i_3} = \tfrac{1}{4}\partial_- D_{i_1i_2i_3} \tag{5.67}$$

where we have set

$$D_{i_1i_2i_3} = \hat{A}_{i_1i_2i_3} + \frac{1}{12}\varepsilon_{i_1i_2i_3}{}^{j_1...j_6}\hat{A}_{j_1...j_6} = 0. \tag{5.68}$$

This is equation (5.55). Similarly, the $E_{11}$ duality relation $D_{a,b_1b_2}$ of equation (4.53) in linearised form in light-cone coordinates reduces to

$$D_{i,-j} = -\partial_- E_{i,j} \doteq 0 \quad , \tag{5.69}$$

where we have set

$$E_{i,j} = \hat{h}^s_{ij} - \frac{1}{4}\varepsilon_j{}^{k_1...k_8}\hat{h}_{k_1...k_8,i} \quad , \tag{5.70}$$

using the definition

$$\begin{aligned} \hat{h}^s_{ij} &= h^s_{ij} - \frac{\partial_j}{\partial_-}h^s_{-i} - \frac{\partial_i}{\partial_-}h^s_{-j} + \frac{\partial_i\partial_j}{\partial_-^2}h^s_{--} \quad , \\ \hat{h}_{k_1...k_8,i} &= h_{k_1...k_8,i} - 7\tfrac{1}{\partial_-}\partial_{[k_1}h_{|-|k_2...k_8],j} \quad . \end{aligned} \tag{5.71}$$

Equation (5.70) is just equation (5.57) in terms of the fields (5.71).

As a consistency check, one can directly linearise both sides of the variation of the duality relation $D_{a_1...a_4}$ from equation (4.54), in light-cone coordinates, to find

$$\begin{aligned} &\frac{1}{4}\partial_-\delta[\hat{A}_{i_1i_2i_3} + \frac{1}{12}\varepsilon_{i_1i_2i_3}{}^{j_1..j_6}\hat{A}_{j_1..j_6}] \\ &\qquad = -\frac{1}{4!}\varepsilon_{i_1i_2i_3}{}^{j_1..j_6}\partial_- E_{j_1j_2j_3}\Lambda_{j_4j_5j_6} - 3\frac{1}{2}\partial_- E_{k[i_1}\Lambda^k{}_{i_2i_3]} \ . \end{aligned} \tag{5.72}$$

Pulling off $\partial_-$, this exactly reproduces the variation (5.56) of (5.55) under (5.53).

## 5.7 Annihilating Ideal Associated With The Massless Particle

In the previous section we have shown that the higher level fields can be expressed in terms of the fields $h_{ij}$ and $A_{i_1 i_2 i_3}$ at levels zero and one through the existence of duality relations such as (5.54), (5.57) and (5.59). In this section we now show that the massless irreducible representation is annihilated by an infinite set of generators and this corresponds to the existence of the infinite set of duality relations. We begin by considering the generator of the form

$$N_{k_1 k_2 k_3} = S_{k_1 k_2 k_3} + c_1 \; \varepsilon_{k_1 k_2 k_3}{}^{r_1 \ldots r_6} S_{r_1 \ldots r_6} \tag{5.73}$$

for a constant $c_1$. Under the action of this generator the fields transform under $\Lambda^{k_1 k_2 k_3} N_{k_1 k_2 k_3}$ according to

$$\begin{aligned} \delta\mathcal{V} &= \delta h_{ij} T^{ij} + \delta A_{i_1 i_2 i_3} T^{i_1 i_2 i_3} + \delta A_{i_1 .. i_6} T^{i_1 .. i_6} + \delta h_{i_1 .. i_8, j} T^{i_1 .. i_8, j} \\ &\quad + \delta A_{i_1 .. i_9, j_1 j_2 j_3} T^{i_1 .. i_9, j_1 j_2 j_3} + \delta A_{i_1 .. i_{10}, j_1 j_2} T^{i_1 .. i_{10}, j_1 j_2} + \delta A_{i_1 .. i_{11}, j} T^{i_1 .. i_{11}, j} + .. \\ &= [\Lambda^{k_1 k_2 k_3} N_{k_1 k_2 k_3}, \mathcal{V}] \quad . \end{aligned} \tag{5.74}$$

Continuing this computation results in the variations

$$\begin{aligned} \delta h_{ij} &= 18 \Lambda_{(i|k_1 k_2|} A_{j)}{}^{k_1 k_2} - 2\eta_{ij} \Lambda^{k_1 k_2 k_3} A_{k_1 k_2 k_3} \\ &\quad - 5! c_1 (\cdot 3 \cdot 3 \; \Lambda^{k_1 k_2 k_3} \varepsilon_{k_1 k_2 k_3}{}^{r_1 \ldots r_5}{}_{(i} A_{j) r_1 .. r_5} + \eta_{ij} \Lambda^{k_1 k_2 k_3} \varepsilon_{k_1 k_2 k_3}{}^{r_1 \ldots r_6} A_{r_1 .. r_6}) \\ \delta A_{i_1 i_2 i_3} &= -3 \cdot 2 h_{[i_1}{}^{j} \Lambda_{i_2 i_3] j} + 60 A_{i_1 i_2 i_3}{}^{k_1 k_2 k_3} \Lambda_{k_1 k_2 k_3} - \frac{5!}{2} c_1 \; \Lambda^{k_1 k_2 k_3} A^{r_1 r_2 r_3} \varepsilon_{k_1 k_2 k_3 r_1 r_2 r_3}{}^{i_1 i_2 i_3} \\ &\quad - \frac{7! \cdot 2}{3} c_1 \; \Lambda^{k_1 k_2 k_3} (\varepsilon_{k_1 k_2 k_3 r_1 \ldots r_6} A^{r_1 .. r_6}{}_{[i_1 i_2, i_3]} - \varepsilon_{k_1 k_2 k_3 r_1 \ldots r_5 j} A_{i_1 i_2 i_3}{}^{r_1 .. r_5, j}) \\ \delta A^{i_1 \ldots i_6} &= 2 \Lambda^{[i_1 i_2 i_3} A^{i_4 i_5 i_6]} + 112 h^{i_1 .. i_6 k_1 k_2, k_3} \Lambda_{k_1 k_2 k_3} + 112 h^{[i_1 .. i_5 | k_1 k_2 k_3 |, i_6]} \tilde{\Lambda}_{k_1 k_2 k_3} \\ &\quad + 12 \; c_1 \; \Lambda^{k_1 k_2 k_3} \varepsilon_{k_1 k_2 k_3 j}{}^{[i_1 \ldots i_5} h^{i_6] j} \\ &\quad - 3 \cdot 7! c_1 \; \Lambda^{k_1 k_2 k_3} \varepsilon_{k_1 k_2 k_3 r_1 \ldots r_6} A^{[i_1 .. i_4 | [r_1 .. r_5, r_6] | i_5 i_6]} \\ &\quad - 15 \cdot 7! c_1 \; \Lambda^{k_1 k_2 k_3} \varepsilon_{k_1 k_2 k_3 r_1 \ldots r_4 r_5 r_6} A^{[r_1 .. r_4 | [i_1 .. i_5, i_6] | r_5 r_6]} \\ &\quad - 4 \cdot 7! c_1 \Lambda^{k_1 k_2 k_3} \varepsilon_{k_1 k_2 k_3 r_1 \ldots r_6} A^{r_1 .. r_6 [i_1 i_2 i_3, i_4 i_5 i_6]} \\ \delta h_{i_1 \ldots i_8, j} &= 2 (\Lambda_{[i_1 i_2 i_3} A_{i_4 .. i_8] j} - \tilde{\Lambda}_{j [i_1 i_2} A_{i_3 .. i_8]}) - 12 \cdot 8 \cdot 3 \Lambda^{k_1 k_2 k_3} (A_{i_1 .. i_8 k_1, k_2 k_3 j} + A_{[i_1 .. i_7 | j k_1, k_2 k_3 | i_8]}) \\ &\quad + 3 c_1 \; \Lambda^{k_1 k_2 k_3} \varepsilon_{k_1 k_2 k_3}{}^{[i_1 \ldots i_6} A^{i_7 i_8] j} - 3 c_1 \; \Lambda^{k_1 k_2 k_3} \varepsilon_{k_1 k_2 k_3}{}^{[i_1 \ldots i_6} A^{i_7 i_8 j]} \quad . \end{aligned} \tag{5.75}$$

The level four commutators listed in the appendices of [66] and [56] are needed to derive this calculation. On setting $c_1 = -\frac{1}{3\cdot 5!}$, the generator of equation (5.73) becomes

$$N_{k_1k_2k_3} = S_{k_1k_2k_3} - \frac{1}{3\cdot 5!}\, \varepsilon_{k_1k_2k_3}{}^{r_1\ldots r_6}S_{r_1\ldots r_6} \quad , \tag{5.76}$$

With this choice the variations of equation (5.75) can be expressed in terms of the duality relations defined in equations (5.54), (5.57) and (5.59) and become equal to zero

$$\begin{aligned} \delta h_{ij} &= 18\Lambda_{(i}{}^{k_1k_2}E_{j)k_1k_2} - 2\eta_{ij}\Lambda^{k_1k_2k_3}E_{k_1k_2k_3} \\ &= 0 \ , \\ \delta A_{i_1i_2i_3} &= \frac{1}{3!}\varepsilon_{i_1i_2i_3k_1k_2k_3r_1r_2r_3}\Lambda^{k_1k_2k_3}E^{r_1r_2r_3} - 6\Lambda_{[i_1i_2}{}^{k}E_{i_3]k} \\ &= 0 \ , \\ \delta A_{i_1\ldots i_6} &= \frac{1}{30}\Lambda^{k_1k_2k_3}\varepsilon_{k_1k_2k_3[i_1\ldots i_5|}{}^{j}E_{|i_6]j} + 28\Lambda_{k_1k_2k_3}\varepsilon^{k_1k_2k_3r_1\ldots r_6}E_{r_1..r_6[i_1i_2i_3,i_4i_5i_6]} \\ &\quad - 140\Lambda_{k_1k_2k_3}\varepsilon^{k_1k_2k_3r_1\ldots r_4r_5r_6}E_{i_1..i_6[r_1r_2r_3,r_4r_5r_6]} \\ &= 0 \ . \end{aligned} \tag{5.77}$$

In a similar fashion one can consider a generator of the form

$$N_{ij} = J_{ij} + c_2\varepsilon_{[i}{}^{r_1\ldots r_8}S_{|r_1\ldots r_8|,j]} \quad , \tag{5.78}$$

and consider $\delta\mathcal{V}$ under $\Lambda^{ij}N_{ij}$. This computation fixes $c_2 = \frac{2}{8!}$ so that

$$N_{ij} = J_{ij} + \frac{2}{8!}\varepsilon_{[i}{}^{r_1\ldots r_8}S_{|r_1\ldots r_8|,j]} \quad . \tag{5.79}$$

The variation of the graviton field under this transformation gives

$$\delta h_{ij} = 2\Lambda_i{}^{k}E_{kj} + 2\Lambda_j{}^{k}E_{ki} = 0. \tag{5.80}$$

This suggests that one can find an infinite number of generators which annihilate the entire representation. At the next level we would expect to find the generator

$$\tilde{N}_{i_1i_2i_3} = S_{i_1i_2i_3} + \frac{1}{9!}\epsilon^{j_1\ldots j_9}S_{j_1\ldots j_9,}{}^{i_1i_2i_3} \tag{5.81}$$

The coefficient in this generator can be determined directly as before, or from ensuring closure in the algebra given below. The pattern of the higher level generators that annihilate the representation that, to a given generator one simply adds another generator, up to a coefficient, which possess an additional block of nine anti-symmetric indices which are contracted against the $\epsilon$ symbol. The coefficient can then be fixed by the methods just mentioned.

The set of all generators that annihilate the representation must form a subalgebra, which we denote by $I$, of $I_c(E_9)$ and we will now show that this is true for the lowest level generators. The algebra of these lowest level generators is given by

$$[N^{i_1i_2i_3}, N_{j_1j_2j_3}] = 2N^{i_1i_2i_3}{}_{j_1j_2j_3} - 9\cdot 4\delta^{[i_1i_2}_{[j_1j_2}N^{i_3]}{}_{j_3]} - \frac{1}{2}\varepsilon^{i_1i_2i_3}{}_{j_1j_2j_3k_1k_2k_3}N^{k_1k_2k_3} \quad , \tag{5.82}$$

where we have written

$$N_{i_1\ldots i_6} = S_{i_1..i_6} + \frac{1}{12}\varepsilon_{i_1..i_6}{}^{k_1k_2k_3}S_{k_1k_2k_3} = \frac{1}{12}\varepsilon_{i_1..i_6}{}^{k_1k_2k_3}N_{k_1k_2k_3} \quad . \tag{5.83}$$

The generators that annihilate the irreducible representation also form an ideal in $I_c(E_9)$. Recall that an ideal $I$ of a Lie algebra $G$ is just a subalgebra consisting of elements $X \in I$ such that

$$[X, Y] \in I \tag{5.84}$$

holds for all $Y \in G$. One finds that

$$\begin{aligned} [S^{i_1i_2i_3}, N_{j_1j_2j_3}] &= -\frac{1}{6}\varepsilon^{i_1i_2i_3}{}_{j_1j_2j_3k_1k_2k_3}N^{k_1k_2k_3} - 18\delta^{[i_1i_2}_{[j_1j_2}N^{i_3]}{}_{j_3]} \quad , \\ [S_{i_1i_2i_3}, N_{jk}] &= 3\eta_{[i_1|j}N_{k|i_2i_3]} + 3\eta_{[i_1|j}N_{k|i_2i_3]}. \end{aligned} \tag{5.85}$$

Since the generator $S^{i_1i_2i_3}$ leads, under repeated commutators, to the whole of $I_c(E_9)$, it follows that the equations (5.85), together with their higher level analogues, imply that the generators in the subalgebra $I$ are an ideal of $I_c(E_9)$.

Starting with the level zero graviton field $h_{ij}(0)$ in the massless irreducible representation, we can then find all fields in the representation via the action of the generator $S^{i_1i_2i_3}$. Thus, if all the elements of $I$ annihilate this lowest field, then they will annihilate all fields in the representation because these generators form an ideal in $I_c(E_9)$. Thus we have shown [31] that the reduced number of degrees of freedom in the massless irreducible representation can ultimately be expressed in terms of the existence of an ideal $I$ in $I_c(E_9)$. This ideal only contains generators associated with the graviton ($N_{ij}$) and the three form ($N_{i_1i_2i_3}$).

# Chapter 6

# Irreducible Representations of the IIA String

## 6.1 Introduction

In [77] the bosonic sector of the IIA supergravity theory was expressed as a nonlinear realization based on an algebra similar to the algebra $G_{11}$ discussed in Section 4.1. The generators are given by $G_{IIA} = \{K^a{}_b, P_a, R, R_{a_1 \ldots a_q} \mid a, b = 0, \ldots, 9 \ , \ q = 1, 2, 3, 5, 6, 7, 8\}$ satisfying the algebra of IGL(10) and commutators given in that reference. In [78] it was shown that, just as the generators of $G_{11}$ were generalized to $E_{11}$, the generators of $G_{IIA}$ could be identified with certain generators of $E_{11}$, and in [81] the common origin of the IIA and IIB theories from $E_{11}$, involving all the generators of $E_{11}$, was found. In [82] this algebra was then re-expressed in terms of representations of $\mathrm{SO}(10, 10)$ by deleting node 10, referred to as the IIA node, and is reviewed in [85, Sec. 17.5]. When the IIA algebra and its $l_1$ representation are written in terms of $\mathrm{SO}(10, 10)$ representations, one *independently* finds, at lowest level in the vector representation, the doubled coordinates $x^a, y_a$ of doubled field theory [35] first introduced in [21], [23], [23], [64], [65] and interpreted as arising from treating left/right-movers along the string as independent degrees of freedom [65].

In this chapter we will begin from the IIA algebra in this latter form, introduced directly in the next section, and establish the algebra needed in the study of the irreducible representations of the IIA theory to low levels. In the following section we will then construct a little algebra for a choice of charges described there.

## 6.2 The IIA Algebra

The IIA theory arises from the nonlinear realization of $E_{11} \otimes_s l_1$ by deleting node 10 in the $E_{11}$ Dynkin diagram and then decomposing the $E_{11}$ algebra as representations of the remaining SO(10, 10) algebra. We will refer to node 10 as the IIA node, and node 11 as the $E_{11}$ node. By further deleting node 11 and decomposing the algebra with respect to the remaining SL(10) algebra we can identify the resulting generators with those of $E_{11}$ decomposed with respect to the $E_{11}$ node from earlier chapters. The Dynkin diagram associated to this decomposition, with deleted nodes indicated by a $\oplus$, is given by

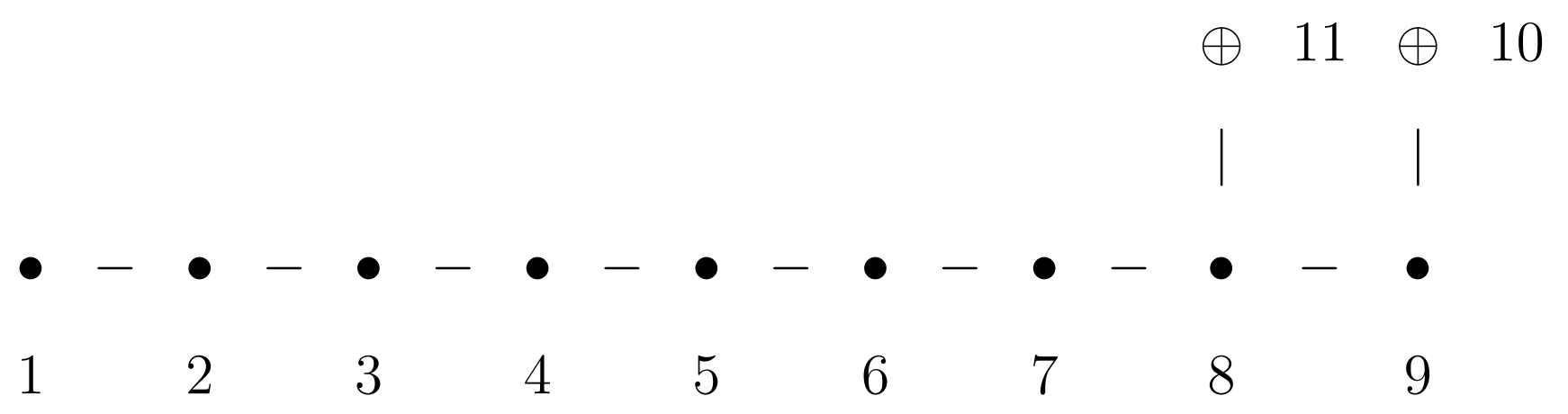

The list of generators resulting from this decomposition can be found using the program [49]. Given the size of this listing we will only give the results. The generators are now organized in terms of two levels $(l_{(1)}, l_{(2)})$, which are associated to the deleted nodes ten and eleven respectively, and we can refer to the first level as the IIA level, or just the level, and $l_{(2)}$ as the $E_{11}$ level. The number of up minus down indices on a given generator is equal to $l_{(1)} + 2l_{(2)}$, and this can be used to determine the $E_{11}$ level $l_{(2)}$ in a generator in the list below once the IIA level is known[1].

The choice of decomposition $E_{11}$ is immaterial, however, and so one can also find the generators in the above IIA decomposition directly by starting from the $E_{11}$ algebra in its eleven dimension, deleting just the node eleven, and interpreting the generators as IIA generators. We will denote the latter generators with a hat ˆ below, and give the association between the generators of the two decompositions of $E_{11}$ explicitly. At level zero ($l_{(1)} = 0$) we find [82]

$$K^{\underline{a}}{}_{\underline{b}}\ ,\quad \tilde{R}\ ,\quad R^{\underline{ab}}\ ,\quad R_{\underline{ab}}\ , \tag{6.1}$$

where $\underline{a}, \underline{b}, \ldots = 0, 1, \ldots, 9$. The relation between these generators and those in the $E_{11}$ node

[1]For example, below we will find $R^{\underline{a}_1\cdots\underline{a}_9,\underline{b}}$ at level $l_{(1)} = 2$, thus the associated underlying $E_{11}$ generator satisfies $10 = 2 + 2 \cdot l_{(2)}$ and so must be at $E_{11}$ level four, hence we will find $R^{\underline{a}_1\cdots\underline{a}_9,\underline{b}} = \hat{R}^{\underline{a}_1\cdots\underline{a}_9 11,\underline{b}11}$.

decomposition are given by

$$K^{\underline{a}}{}_{\underline{b}} = \hat{K}^{\underline{a}}{}_{\underline{b}} + \frac{1}{6}\delta^{\underline{a}}{}_{\underline{b}}\tilde{R}\ ,\quad \tilde{R} = -\sum_{\underline{a}=0}^{9}\hat{K}^{\underline{a}}{}_{\underline{a}} + 2\hat{K}^{11}{}_{11}\ , R^{\underline{ab}} = \hat{R}^{\underline{ab}11}\ ,\quad R_{\underline{ab}} = \hat{R}_{\underline{ab}11}. \tag{6.2}$$

We will often use the number 11 instead of 10 to denote the eleventh dimension.

We now list the generators to (IIA) level two. At level one

$$\begin{aligned} R^{\underline{a}} = \hat{K}^{\underline{a}}{}_{11}\ ,\ R^{\underline{a}_1\underline{a}_2\underline{a}_3} = \hat{R}^{\underline{a}_1\underline{a}_2\underline{a}_3}\ ,\ R^{\underline{a}_1\cdots\underline{a}_5} = \hat{R}^{\underline{a}_1\cdots\underline{a}_5 11}\ ,\\ R^{\underline{a}_1\cdots\underline{a}_7} = \hat{R}^{\underline{a}_1\cdots\underline{a}_7 11,11}\ ,\ R^{\underline{a}_1\cdots\underline{a}_9} = \hat{R}^{\underline{a}_1\cdots\underline{a}_9 11,11\,11}\ , \end{aligned} \tag{6.3}$$

and at level minus one

$$\begin{aligned} R_{\underline{a}} = \hat{K}^{11}{}_{\underline{a}}\ ,\ R_{\underline{a}_1\underline{a}_2\underline{a}_3} = \hat{R}_{\underline{a}_1\underline{a}_2\underline{a}_3}\ ,\ R_{\underline{a}_1\cdots\underline{a}_5} = \hat{R}_{\underline{a}_1\cdots\underline{a}_5 11}\ ,\\ R_{\underline{a}_1\cdots\underline{a}_7} = \hat{R}_{\underline{a}_1\cdots\underline{a}_7 11,11}\ ,\ R_{\underline{a}_1\cdots\underline{a}_9} = \hat{R}_{\underline{a}_1\cdots\underline{a}_9 11,11\,11}\ . \end{aligned} \tag{6.4}$$

At level two we find

$$\begin{aligned} R^{\underline{a}_1\cdots\underline{a}_6} = \hat{R}^{\underline{a}_1\cdots\underline{a}_6}\ ,\ R^{\underline{a}_1\cdots\underline{a}_8} = \hat{R}^{\underline{a}_1\cdots\underline{a}_8,11}\ ,\ R^{\underline{a}_1\cdots\underline{a}_7,\underline{b}} = \hat{R}^{\underline{a}_1\cdots\underline{a}_7 11,\underline{b}}\ ,\\ R_{(1)}^{\underline{a}_1\cdots\underline{a}_{10}} = \hat{R}^{\underline{a}_1\cdots\underline{a}_{10},(11\,11)}\ ,\ R_{(2)}^{\underline{a}_1\cdots\underline{a}_{10}} = \hat{R}^{\underline{a}_1\cdots\underline{a}_{10}11,11}\ ,\\ R^{\underline{a}_1\cdots\underline{a}_9,\underline{b}} = \hat{R}^{\underline{a}_1\cdots\underline{a}_9 11,\underline{b}11}\ ,\ R^{\underline{a}_1\cdots\underline{a}_8,\underline{b}_1\underline{b}_2} = \hat{R}^{\underline{a}_1\cdots\underline{a}_8 11,\underline{b}_1\underline{b}_2 11}\ ,\\ R^{\underline{a}_1\cdots\underline{a}_{10},\underline{b}_1\underline{b}_2} = \hat{R}^{\underline{a}_1\cdots\underline{a}_{10}11,\underline{b}_1\underline{b}_2 11,11}\ ,\ R^{\underline{a}_1\cdots\underline{a}_9,\underline{b}_1\underline{b}_2\underline{b}_3} = \hat{R}^{\underline{a}_1\cdots\underline{a}_9 11,\underline{b}_1\underline{b}_2\underline{b}_3 11,11}\ ,\\ R^{\underline{a}_1\cdots\underline{a}_{10},\underline{b}_1\cdots\underline{b}_4} = \hat{R}^{\underline{a}_1\cdots\underline{a}_{10}11,\underline{b}_1\cdots\underline{b}_4 11,(11,11)}\ , \end{aligned} \tag{6.5}$$

and at level minus two we have

$$\begin{aligned} R_{\underline{a}_1\cdots\underline{a}_6} = \hat{R}_{\underline{a}_1\cdots\underline{a}_6}\ ,\ R_{\underline{a}_1\cdots\underline{a}_8} = \hat{R}_{\underline{a}_1\cdots\underline{a}_8,11}\ ,\ R_{\underline{a}_1\cdots\underline{a}_7,\underline{b}} = \hat{R}_{\underline{a}_1\cdots\underline{a}_7 11,\underline{b}}\ ,\\ R_{(1)\underline{a}_1\cdot\cdot\underline{a}_{10}} = \hat{R}_{\underline{a}_1\cdot\cdot\underline{a}_{10},(11\,11)}\ ,\quad R_{(2)\underline{a}_1\cdot\cdot\underline{a}_{10}} = \hat{R}_{\underline{a}_1\cdot\cdot\underline{a}_{10}11,11}\ ,\\ R_{\underline{a}_1\cdot\cdot\underline{a}_9,\underline{b}} = \hat{R}_{\underline{a}_1\cdot\cdot\underline{a}_9 11,\underline{b}11}\ ,\quad R_{\underline{a}_1\cdot\cdot\underline{a}_8,\underline{b}_1\underline{b}_2} = \hat{R}_{\underline{a}_1\cdot\cdot\underline{a}_8 11,\underline{b}_1\underline{b}_2 11}\ ,\\ R_{\underline{a}_1\cdots\underline{a}_{10},\underline{b}_1\underline{b}_2} = \hat{R}_{\underline{a}_1\cdots\underline{a}_{10}11,\underline{b}_1\underline{b}_2 11,11}\ ,\quad R_{\underline{a}_1\cdots\underline{a}_9,\underline{b}_1\underline{b}_2\underline{b}_3} = \hat{R}_{\underline{a}_1\cdots\underline{a}_9,\underline{b}_1\underline{b}_2\underline{b}_3 11,11}\ ,\\ R_{\underline{a}_1\cdots\underline{a}_{10},\underline{b}_1\cdots\underline{b}_4} = \hat{R}_{\underline{a}_1\cdots\underline{a}_{10}11,\underline{b}_1\cdots\underline{b}_4\ 11,(11,11)}\ . \end{aligned} \tag{6.6}$$

Again, the indices of in any given block are anti-symmetric. A generator with ten anti-symmetric indices is found with multiplicity two, and these are distinguished above by the subscripts (1) and (2). The algebra of these generators [82] can be computed directly using the $E_{11}$ algebra from Section 4.2 to level three, and using the algebra in the appendices of [66] for commutators involving level four $E_{11}$ generators, and is listed in appendix (C).

We can also examine the level decomposition of $E_{11}$ with respect to the IIA node alone. Given that the table is small, we can give it explicitly

Table 6.1: $D_{10}$ representations in $E_{11}$

| $l$ | $D_{10}$ Irrep | $E_{11}$ Root | $a^2$ | $d_r$ | $\mu$ | $R^{\underline{\alpha}}$ |
|---|---|---|---|---|---|---|
| 0 | 0 0 0 0 0 0 0 0 0 0 | 0 0 0 0 0 0 0 0 0 0 0 | 0 | 1 | 1 | $\chi$ |
| 0 | 0 1 0 0 0 0 0 0 0 0 | 1 2 2 2 2 2 2 2 1 0 1 | 2 | 190 | 1 | $J_{\underline{ab}}$ |
| 1 | 0 0 0 0 0 0 0 0 1 0 | 1 2 3 4 5 6 7 8 5 1 4 | 2 | 512 | 1 | $\Psi$ |
| 2 | 0 0 0 0 0 0 0 0 0 0 | 1 2 3 4 5 6 7 8 5 2 4 | -2 | 1 | 1 | $\phi$ |
| 2 | 0 0 0 1 0 0 0 0 0 0 | 2 4 6 8 9 10 11 12 7 2 6 | 2 | 4845 | 1 | $A^{a_1\cdots a_4}$ |

The generators at level zero belong to the adjoint representation of $\mathrm{SO}(10,10)$, and when the $E_{11}$ node is deleted the above listing of generators at level zero is analogous to the decomposition of $\mathrm{SO}(2D)$ in terms of its $\mathrm{SU}(D)\otimes\mathrm{U}(1)$ subalgebra [60, Ch. 8], [18, Ch. 11]. At level one we find 512 generators and the irreducible representation is associated to one of the spinor nodes in the Dynkin diagram of $\mathrm{SO}(10,10)$, thus this representation describes the Weyl spinor representation of $\mathrm{SO}(10,10)$, and similarly at level minus one which we have not listed in the table. Similarly at level two the generators belong to the rank-four tensor, and singlet, representations of $\mathrm{SO}(10,10)$ having dimensions 4845, and 1, respectively.

We now consider the Cartan involution invariant subalgebra in the IIA theory. The action of the $E_{11}$ Cartan involution $I_c$ on the IIA generators can be determined from its action on the $E_{11}$ generators $R^\alpha$, where in eleven dimensions we have $I_c(\hat{R}^\alpha) = -(-1)^{l+1}\hat{R}_\alpha$, where $l$ is the level of $R^\alpha$, except at level zero where $I_c(\hat{K}^{\underline{a}}{}_{\underline{b}}) = -\eta^{\underline{ac}}\eta_{\underline{bd}}\hat{K}^{\underline{d}}{}_{\underline{c}}$. Acting on the (IIA) level zero algebra generators we thus find that

$$I_c(K^{\underline{a}}{}_{\underline{b}}) = -K^{\underline{b}}{}_{\underline{a}} \quad , \quad I_c(\tilde{R}) = -\tilde{R} \quad , \quad I_c(R^{\underline{a}_1\underline{a}_2}) = -\eta^{\underline{a}_1\underline{b}_1}\eta^{\underline{a}_2\underline{b}_2}R_{\underline{b}_1\underline{b}_2} \ . \tag{6.7}$$

On the level one generators we find

$$\begin{aligned} I_c(R^{\underline{a}}) &= -\eta^{\underline{ab}}R_{\underline{b}} \quad , \quad I_c(R^{\underline{a}_1\underline{a}_2\underline{a}_3}) = -\eta^{\underline{a}_1\underline{a}_2\underline{a}_3|\underline{b}_1\underline{b}_2\underline{b}_3}R_{\underline{b}_1\underline{b}_2\underline{b}_3} \ , \\ I_c(R^{\underline{a}_1\cdots\underline{a}_5}) &= +\eta^{\underline{a}_1\cdots\underline{a}_5|\underline{b}_1\cdots\underline{b}_5}R_{\underline{b}_1\cdots\underline{b}_5} \ , \ I_c(R^{\underline{a}_1\cdots\underline{a}_7}) = -\eta^{\underline{a}_1\cdots\underline{a}_7|\underline{b}_1\cdots\underline{b}_7}R_{\underline{b}_1\cdots\underline{b}_7} \ , \\ I_c(R^{\underline{a}_1\cdots\underline{a}_9}) &= +\eta^{\underline{a}_1\cdots\underline{a}_9|\underline{b}_1\cdots\underline{b}_9}R_{\underline{b}_1\cdots\underline{b}_9} \ , \end{aligned} \tag{6.8}$$

and on the level two generators we find

$$
\begin{aligned}
I_c(R^{\underline{a}_1\cdots\underline{a}_6}) &= +\eta^{\underline{a}_1\cdots\underline{a}_6|\underline{b}_1\cdots\underline{b}_6}R_{\underline{b}_1\cdots\underline{b}_6}\ ,\ I_c(R^{\underline{a}_1\cdots\underline{a}_8}) = -\eta^{\underline{a}_1\cdots\underline{a}_8|\underline{b}_1\cdots\underline{b}_8}R_{\underline{b}_1\cdots\underline{b}_8}\ ,\\
I_c(R^{\underline{a}_1\cdot\cdot\underline{a}_7,\underline{b}}) &= -\eta^{\underline{a}_1\cdot\cdot\underline{a}_7,\underline{b}|\underline{c}_1\cdot\cdot\underline{c}_7\underline{d}}R_{\underline{c}_1\cdot\cdot\underline{c}_7,\underline{d}}\ ,\ I_c(R_{(1)}^{\underline{a}_1\cdot\cdot\underline{a}_{10}}) = +\eta^{\underline{a}_1\cdot\cdot\underline{a}_{10}|\underline{b}_1\cdot\cdot\underline{b}_{10}}R_{(1)\underline{b}_1\cdot\cdot\underline{b}_{10}}\ ,\\
I_c(R_{(2)}^{\underline{a}_1\cdot\cdot\underline{a}_{10}}) &= +\eta^{\underline{a}_1\cdot\cdot\underline{a}_{10}|\underline{b}_1\cdot\cdot\underline{b}_{10}}R_{(2)\underline{b}_1\cdot\cdot\underline{b}_{10}}\ ,\ I_c(R^{\underline{a}_1\cdots\underline{a}_9,\underline{b}}) = +\eta^{\underline{a}_1\cdot\cdot\underline{a}_9\underline{b}|\underline{c}_1\cdot\cdot\underline{b}_9\underline{d}}R_{\underline{c}_1\cdots\underline{c}_9,\underline{d}}\ ,\\
I_c(R^{\underline{a}_1\cdots\underline{a}_8,\underline{b}_1\underline{b}_2}) &= \eta^{\underline{a}_1\cdot\cdot\underline{b}_2|\underline{c}_1\cdot\cdot\underline{d}_2}R_{\underline{c}_1\cdots\underline{c}_8,\underline{d}_1\underline{d}_2},I_c(R^{\underline{a}_1\cdots\underline{a}_{10},\underline{b}_1\underline{b}_2}) = -\eta^{\underline{a}_1\cdot\cdot\underline{b}_2|\underline{c}_1\cdot\cdot\underline{d}_2}R_{\underline{c}_1\cdots\underline{c}_{10},\underline{d}_1\underline{d}_2},\\
I_c(R^{\underline{a}_1\cdots\underline{a}_9,\underline{b}_1\underline{b}_2\underline{b}_3}) &= -\eta^{\underline{a}_1\cdots\underline{b}_3|\underline{c}_1\cdots\underline{d}_3}R_{\underline{c}_1\cdots\underline{c}_9,\underline{d}_1\underline{d}_2\underline{d}_3},\\
I_c(R^{\underline{a}_1\cdots\underline{a}_{10},\underline{b}_1\cdots\underline{b}_4}) &= -\eta^{\underline{a}_1\cdots\underline{b}_4|\underline{c}_1\cdots\underline{d}_4}R_{\underline{c}_1\cdots\underline{c}_9,\underline{d}_1\cdots\underline{d}_4}
\end{aligned}
\tag{6.9}
$$

where we have set $\eta^{\underline{a}_1\underline{a}_2\cdots|\underline{b}_1\underline{b}_2\cdots} := \eta^{\underline{a}_1\underline{b}_1}\eta^{\underline{a}_2\underline{b}_2}\ldots$.

We can now construct the Cartan involution invariant subalgebra in the IIA theory using involution invariant combinations of the above generators. At level zero the combinations

$$
J_{\underline{a}_1\underline{a}_2} = \eta_{\underline{a}_1\underline{c}}K^{\underline{c}}{}_{\underline{a}_2} - \eta_{\underline{a}_2\underline{c}}K^{\underline{c}}{}_{\underline{a}_1}\ ,S_{\underline{a}_1\underline{a}_2} = R^{\underline{c}_1\underline{c}_2}\eta_{\underline{c}_1\underline{a}_1}\eta_{\underline{c}_2\underline{a}_2} - R_{\underline{a}_1\underline{a}_2}\ , \tag{6.10}
$$

are invariant under $I_c$. Similarly at level one we find

$$
\begin{aligned}
S_{\underline{a}} &= R^{\underline{c}}\eta_{\underline{c}\underline{a}} - R_{\underline{a}}\ ,\ S_{\underline{a}_1\underline{a}_2\underline{a}_3} = R^{\underline{c}_1\underline{c}_2\underline{c}_3}\eta_{\underline{c}_1\underline{c}_2\underline{c}_3|\underline{a}_1\underline{a}_2\underline{a}_3} - R_{\underline{a}_1\underline{a}_2\underline{a}_3}\ ,\\
S_{\underline{a}_1\cdots\underline{a}_5} &= R^{\underline{c}_1\cdots\underline{c}_5}\eta_{\underline{c}_1\cdots\underline{c}_5|\underline{a}_1\cdots\underline{a}_5} + R_{\underline{a}_1\cdots\underline{a}_5}\ ,\\
S_{\underline{a}_1\cdots\underline{a}_7} &= R^{\underline{c}_1\cdots\underline{c}_7}\eta_{\underline{c}_1\cdots\underline{c}_7|\underline{a}_1\cdots\underline{a}_7} - R_{\underline{a}_1\cdots\underline{a}_7}\ ,\\
S_{\underline{a}_1\cdots\underline{a}_9} &= R^{\underline{c}_1\cdots\underline{c}_9}\eta_{\underline{c}_1\cdots\underline{c}_9|\underline{a}_1\cdots\underline{a}_9} + R_{\underline{a}_1\cdots\underline{a}_9}\ ,
\end{aligned}
\tag{6.11}
$$

and at level two we find

$$
\begin{aligned}
S_{\underline{a}_1\cdots\underline{a}_6} &= R^{\underline{c}_1\cdots\underline{c}_6}\eta_{\underline{c}_1\cdots\underline{c}_6|\underline{a}_1\cdots\underline{a}_6} + R_{\underline{a}_1\cdots\underline{a}_6}\ ,\\
S_{\underline{a}_1\cdots\underline{a}_8} &= R^{\underline{c}_1\cdots\underline{c}_8}\eta_{\underline{c}_1\cdots\underline{c}_8|\underline{a}_1\cdots\underline{a}_8} - R_{\underline{a}_1\cdots\underline{a}_8}\ ,\\
S_{\underline{a}_1\cdots\underline{a}_7,\underline{b}} &= R^{\underline{c}_1\cdots\underline{c}_7,\underline{d}}\eta_{\underline{c}_1\cdots\underline{c}_7,\underline{d}|\underline{a}_1\cdots\underline{a}_7,\underline{b}} - R_{\underline{a}_1\cdots\underline{a}_7,\underline{b}}\ ,\\
S_{(1)\underline{a}_1\cdots\underline{a}_{10}} &= R_{(1)}^{\underline{c}_1\cdots\underline{c}_{10}}\eta_{\underline{c}_1\cdots\underline{c}_{10}|\underline{a}_1\cdots\underline{a}_{10}} + R_{(1)\underline{a}_1\cdots\underline{a}_{10}}\ ,\\
S_{(2)\underline{a}_1\cdots\underline{a}_{10}} &= R_{(2)}^{\underline{c}_1\cdots\underline{c}_{10}}\eta_{\underline{c}_1\cdots\underline{c}_{10}|\underline{a}_1\cdots\underline{a}_{10}} + R_{(2)\underline{a}_1\cdots\underline{a}_{10}}\ ,\\
S_{\underline{a}_1\cdots\underline{a}_9,\underline{b}} &= R^{\underline{c}_1\cdots\underline{c}_9,\underline{d}}\eta_{\underline{c}_1\cdots\underline{c}_9,\underline{d}|\underline{a}_1\cdots\underline{a}_9,\underline{b}} + R_{\underline{a}_1\cdots\underline{a}_9,\underline{b}}\ ,\\
S_{\underline{a}_1\cdots\underline{a}_8,\underline{b}_1\underline{b}_2} &= R^{\underline{c}_1\cdots\underline{c}_8,\underline{d}_1\underline{d}_2}\eta_{\underline{c}_1\cdots\underline{d}_2|\underline{a}_1\cdots\underline{b}_2} + R_{\underline{a}_1\cdots\underline{a}_8,\underline{b}_1\underline{b}_2}\ ,\\
S_{\underline{a}_1\cdots\underline{a}_{10},\underline{b}_1\underline{b}_2} &= R^{\underline{c}_1\cdots\underline{c}_{10},\underline{d}_1\underline{d}_2}\eta_{\underline{c}_1\cdots\underline{d}_2|\underline{a}_1\cdots\underline{b}_2} - R_{\underline{a}_1\cdots\underline{a}_{10},\underline{b}_1\underline{b}_2}\ ,\\
S_{\underline{a}_1\cdots\underline{a}_9,\underline{b}_1\underline{b}_2\underline{b}_3} &= R^{\underline{c}_1\cdots\underline{c}_9,\underline{d}_1\underline{d}_2\underline{d}_3}\eta_{\underline{c}_1\cdots\underline{d}_3|\underline{a}_1\cdots\underline{b}_3} - R_{\underline{a}_1\cdots\underline{a}_9,\underline{b}_1\underline{b}_2\underline{b}_3}\ ,\\
S_{\underline{a}_1\cdots\underline{a}_{10},\underline{b}_1\cdots\underline{b}_4} &= R^{\underline{c}_1\cdots\underline{c}_{10},\underline{d}_1\cdots\underline{d}_4}\eta_{\underline{c}_1\cdots\underline{d}_4|\underline{a}_1\cdots\underline{b}_4} - R_{\underline{a}_1\cdots\underline{a}_{10},\underline{b}_1\cdots\underline{b}_4}\ .
\end{aligned}
\tag{6.12}
$$

are all invariant under $I_c$.

The commutators between the above $I_c$ invariant generators can be found by re-expressing them in terms of their eleven dimensional formulation and using the $I_c(E_{11})$ algebra from Section 4.2 to level three, and using the algebra in the appendices of [66] for commutators involving level four $E_{11}$ generators. At level zero we find that

$$\begin{aligned}
&[J^{\underline{a}_1\underline{a}_2}, J_{\underline{b}_1\underline{b}_2}] = -4\delta^{[\underline{a}_1}{}_{[\underline{b}_1} J^{\underline{a}_2]}{}_{\underline{b}_2]} \ , \ [J^{\underline{a}_1\underline{a}_2}, S_{\underline{b}_1\underline{b}_2}] = -4\delta^{[\underline{a}_1}{}_{[\underline{b}_1} S^{\underline{a}_2]}{}_{\underline{b}_2]} \ , \\
&[S^{\underline{a}_1\underline{a}_2}, S_{\underline{b}_1\underline{b}_2}] = -4\delta^{[\underline{a}_1}{}_{[\underline{b}_1} J^{\underline{a}_2]}{}_{\underline{b}_2]} \ .
\end{aligned} \tag{6.13}$$

The commutators are given by

$$\begin{aligned}
&[J_{\underline{a}_1\underline{a}_2}, S_{\underline{b}}] = -2\eta_{\underline{b}[\underline{a}_1} S_{\underline{a}_2]}, \ [J^{\underline{a}_1\underline{a}_2}, S_{\underline{b}_1\underline{b}_2\underline{b}_3}] = -2\cdot 3\delta^{[\underline{a}_1}{}_{[\underline{b}_1} S^{\underline{a}_2]}{}_{\underline{b}_2\underline{b}_3]} \ , \\
&[J^{\underline{a}_1\underline{a}_2}, S_{\underline{b}_1\ldots\underline{b}_5}] = -2\cdot 5\delta^{[\underline{a}_1}_{[\underline{b}_1} S^{\underline{a}_2]}{}_{\underline{b}_2\ldots\underline{b}_5]} \ , \ [J^{\underline{a}_1\underline{a}_2}, S_{\underline{b}_1..\underline{b}_7}] = 16 S_{[\underline{b}_1..\underline{b}_6}{}^{[\underline{a}_1}\delta^{\underline{a}_2]}{}_{\underline{b}_7]} \ , \\
&[S_{\underline{a}_1\underline{a}_2}, S_{\underline{b}}] = S_{\underline{a}_1\underline{a}_2\underline{b}} \ , \ [S^{\underline{a}_1\underline{a}_2}, S_{\underline{b}_1\underline{b}_2\underline{b}_3}] = +6\delta^{\underline{a}_1\underline{a}_2}_{[\underline{b}_1\underline{b}_2} S_{\underline{b}_3]} - 2S^{\underline{a}_1\underline{a}_2}{}_{\underline{b}_1\underline{b}_2\underline{b}_3} \ , \\
&[S^{\underline{a}_1\underline{a}_2}, S_{\underline{b}_1\ldots\underline{b}_5}] = +10\delta^{\underline{a}_1\underline{a}_2}_{[\underline{b}_1\underline{b}_2} S_{\underline{b}_3\underline{b}_4\underline{b}_5]} - S^{\underline{a}_1\underline{a}_2}{}_{\underline{b}_1\ldots\underline{b}_5} \ , \\
&[S^{\underline{a}_1\underline{a}_2}, S_{\underline{b}_1\ldots\underline{b}_7}] = 42 S_{[\underline{b}_1\ldots\underline{b}_5}\delta^{\underline{a}_1\underline{a}_2}_{\underline{b}_6\underline{b}_7]} \ , \ldots
\end{aligned} \tag{6.14}$$

as well as

$$\begin{aligned}
&[S^{\underline{a}}, S_{\underline{b}}] = -J^{\underline{a}}{}_{\underline{b}} \ , \ [S^{\underline{a}}, S_{\underline{b}_1\underline{b}_2\underline{b}_3}] = -\frac{3}{2}\delta^{\underline{a}}_{[\underline{b}_1} S_{\underline{b}_2\underline{b}_3]} \ , \\
&[S^{\underline{a}}, S_{\underline{b}_1\ldots\underline{b}_5}] = -5\delta^{\underline{a}}_{[\underline{b}_1} S_{\underline{b}_2\ldots\underline{b}_5]} \ , \ [S^{\underline{a}}, S_{\underline{b}_1\ldots\underline{b}_7}] = -2S^{\underline{a}}{}_{\underline{b}_1\ldots\underline{b}_7} + 7S^{\underline{a}}{}_{[\underline{b}_1\ldots\underline{b}_6,\underline{b}_7]} \ , \\
&[S^{\underline{a}_1\underline{a}_2\underline{a}_3}, S_{\underline{b}_1\underline{b}_2\underline{b}_3}] = -18 J^{[\underline{a}_1}{}_{[\underline{b}_1}\delta^{\underline{a}_2\underline{a}_3]}_{\underline{b}_2\underline{b}_3]} + 2S^{\underline{a}_1\underline{a}_2\underline{a}_3}{}_{\underline{b}_1\underline{b}_2\underline{b}_3} \ , \\
&[S^{\underline{a}_1\underline{a}_2\underline{a}_3}, S_{\underline{b}_1\ldots\underline{b}_5}] = -30 S_{[\underline{b}_1\underline{b}_2}\delta^{\underline{a}_1\underline{a}_2\underline{a}_3}_{\underline{b}_3\underline{b}_4\underline{b}_5]} + S^{\underline{a}_1\underline{a}_2\underline{a}_3}{}_{\underline{b}_1\ldots\underline{b}_5} - 5S^{\underline{a}_1\underline{a}_2\underline{a}_3}{}_{[\underline{b}_1\ldots\underline{b}_4,\underline{b}_5]} \ , \\
&[S^{\underline{a}_1\underline{a}_2\underline{a}_3}, S_{\underline{b}_1\ldots\underline{b}_7}] = 0 \ , \ [S^{\underline{a}_1\cdots\underline{a}_5}, S_{\underline{b}_1\ldots\underline{b}_5}] = -5\cdot 30 J^{[\underline{a}_1}{}_{[\underline{b}_1}\delta^{\underline{a}_2\cdots\underline{a}_5]}_{\underline{b}_2\ldots\underline{b}_5]} \ , \\
&[S^{\underline{a}_1\cdots\underline{a}_5}, S_{\underline{b}_1\ldots\underline{b}_7}] = 9\cdot 70 S_{[\underline{b}_1\underline{b}_2}\delta^{\underline{a}_1..\underline{a}_5}_{\underline{b}_3..\underline{b}_7]} \ , \ [S^{\underline{a}_1\cdots\underline{a}_7}, S_{\underline{b}_1\ldots\underline{b}_7}] = -7\cdot 7\cdot 180 J^{[\underline{a}_1}{}_{[\underline{b}_1}\delta^{\underline{a}_2\cdots\underline{a}_7]}_{\underline{b}_2\ldots\underline{b}_7]}, \ldots
\end{aligned} \tag{6.15}$$

We now consider how the vector representation of the IIA theory. These generators can again be related to the $l_1$ generators in eleven dimensions, which we will again denoted by a hat. The generators in the $l_1$ representation of the IIA theory, at (IIA) level zero, are [82]

$$P_{\underline{a}} = \hat{P}_{\underline{a}} \ , \ Q^{\underline{a}} = -\hat{Z}^{\underline{a}11} \ . \tag{6.16}$$

where we have taken a different normalization in $Q^{\underline{a}}$ to that in used in [82].

The generators in the $l_1$ representation at (IIA) level one are given by

$$\begin{aligned}
&Z = \hat{P}_{11} \ , \ Z^{\underline{a}_1\underline{a}_2} = \hat{Z}^{\underline{a}_1\underline{a}_2} \ , \ Z^{\underline{a}_1\cdots\underline{a}_4} = \hat{Z}^{\underline{a}_1\cdots\underline{a}_4 11} \ , \ Z^{\underline{a}_1\cdots\underline{a}_6} = \hat{Z}^{\underline{a}_1\cdots\underline{a}_6 11,11} \ , \\
&Z^{\underline{a}_1\cdots\underline{a}_8} = \hat{Z}^{\underline{a}_1\cdots\underline{a}_8} \ , \ Z^{\underline{a}_1\cdots\underline{a}_{10}} = \hat{Z}^{\underline{a}_1\cdots\underline{a}_{10}11} \ ,
\end{aligned} \tag{6.17}$$

and at level two as

$$
\begin{aligned}
&Z^{\underline{a}_1\cdots\underline{a}_5} = \hat{Z}^{\underline{a}_1\cdots\underline{a}_5} \ , \ Z_{(1)}^{\underline{a}_1\cdots\underline{a}_7} = \hat{Z}^{\underline{a}_1\cdots\underline{a}_7,11} \ , \ Z_{(2)}^{\underline{a}_1\cdots\underline{a}_7} = \hat{Z}^{\underline{a}_1\cdots\underline{a}_7 11} \ , \\
&Z^{\underline{a}_1\cdots\underline{a}_6,\underline{b}} = \hat{Z}^{\underline{a}_1\cdots\underline{a}_6 11,\underline{b}} \ , \\
&Z^{\underline{a}_1\cdots\underline{a}_7,\underline{b}_1\underline{b}_2} = \hat{Z}^{\underline{a}_1\cdots\underline{a}_8 11,\underline{b}_1\underline{b}_2 11} \ , \ Z_{(1)}^{\underline{a}_1\cdots\underline{a}_9} = \hat{Z}_{(1)}^{\underline{a}_1\cdots\underline{a}_9 11,11} \ , \ Z_{(2)}^{\underline{a}_1\cdots\underline{a}_9} = \hat{Z}_{(2)}^{\underline{a}_1\cdots\underline{a}_9 11,11} \ , \\
&Z_{(3)}^{\underline{a}_1\cdots\underline{a}_9} = \hat{Z}^{\underline{a}_1\cdots\underline{a}_9,(11\,11)} \ , \ Z_{(1)}^{\underline{a}_1\cdots\underline{a}_8,\underline{b}} = \hat{Z}^{\underline{a}_1\cdots\underline{a}_8 11,\underline{b}11} \ , \ Z_{(2)}^{\underline{a}_1\cdots\underline{a}_8,\underline{b}} = \hat{Z}^{\underline{a}_1\cdots\underline{a}_8 11,(\underline{b}11)} \ , \\
&Z^{\underline{a}_1\cdots\underline{a}_7,\underline{b}_1\underline{b}_2} = \hat{Z}^{\underline{a}_1\cdots\underline{a}_7 11,\underline{b}_1\underline{b}_2 11} \ , \ Z_{(1)}^{\underline{a}_1\cdots\underline{a}_{10},\underline{b}} = \hat{Z}_{(1)}^{\underline{a}_1\cdots\underline{a}_{10}11,\underline{b}\,11,11} \ , \\
&Z_{(2)}^{\underline{a}_1\cdots\underline{a}_{10},\underline{b}} = \hat{Z}_{(2)}^{\underline{a}_1\cdots\underline{a}_{10}11,\underline{b}\,11,11} \ , \ Z_{(3)}^{\underline{a}_1\cdots\underline{a}_{10},\underline{b}} = \hat{Z}^{\underline{a}_1\cdots\underline{a}_{10}11,(\underline{b}\,11,11)} \ , \\
&Z_{(1)}^{\underline{a}_1\cdots\underline{a}_9,\underline{b}_1\underline{b}_2} = \hat{Z}_{(1)}^{\underline{a}_1\cdots\underline{a}_9 11,\underline{b}_1\underline{b}_2\,11,11} \ , \ Z_{(2)}^{\underline{a}_1\cdots\underline{a}_9,\underline{b}_1\underline{b}_2} = \hat{Z}_{(2)}^{\underline{a}_1\cdots\underline{a}_9 11,\underline{b}_1\underline{b}_2\,11,11} \ , \\
&Z^{\underline{a}_1\cdots\underline{a}_8,\underline{b}_1\underline{b}_2\underline{b}_3} = \hat{Z}^{\underline{a}_1\cdots\underline{a}_8 11,\underline{b}_1\underline{b}_2\underline{b}_3\,11,11} \ , \ Z^{\underline{a}_1\cdots\underline{a}_9,\underline{b}_1\cdots\underline{b}_4} = \hat{Z}^{\underline{a}_1\cdots\underline{a}_9 11,\underline{b}_1\cdots\underline{b}_4\,11,(11,11)} \ , \\
&Z_{(1)}^{\underline{a}_1\cdots\underline{a}_{10},\underline{b}_1\underline{b}_2\underline{b}_3} = \hat{Z}_{(1)}^{\underline{a}_1\cdots\underline{a}_{10}11,\underline{b}_1\underline{b}_2\underline{b}_3\,11,(11,11)} \ , \ Z_{(2)}^{\underline{a}_1\cdots\underline{a}_{10},\underline{b}_1\underline{b}_2\underline{b}_3} = \hat{Z}_{(2)}^{\underline{a}_1\cdots\underline{a}_{10}11,\underline{b}_1\underline{b}_2\underline{b}_3\,11,(11,11)} \ .
\end{aligned}
\tag{6.18}
$$

Again the indices in a given block are assumed anti-symmetric, and generators with lowered subscripts (1), (2), (3) arise from underlying generators that occur with multiplicity greater than one.

The commutators of the $l_1$ generators with those of the Cartan involution invariant subalgebra are again easily computed using the relationship to the $E_{11}$ generators and the $E_{11}$ algebra in Section 4.2 and the appendices of [66]. These commutators are given as

$$
\begin{aligned}
&[J_{\underline{a}_1\underline{a}_2}, P_{\underline{c}}] = -2\eta_{\underline{c}[\underline{a}_1} P_{\underline{a}_2]} \ , \ [J^{\underline{a}_1\underline{a}_2}, Q^{\underline{c}}] = -2\eta^{\underline{c}[\underline{a}_1} Q^{\underline{a}_2]} \ , \\
&[S^{\underline{a}_1\underline{a}_2}, P_{\underline{c}}] = -2\delta_{\underline{c}}^{[\underline{a}_1} Q^{\underline{a}_2]} \ , \ [S_{\underline{a}_1\underline{a}_2}, Q^{\underline{c}}] = -2\delta^{\underline{c}}_{[\underline{a}_1} P_{\underline{a}_2]} \ ,
\end{aligned}
\tag{6.19}
$$

and

$$
\begin{aligned}
&[J_{\underline{a}_1\underline{a}_2}, Z] = 0 \ , \ [J_{\underline{a}_1\underline{a}_2}, Z^{\underline{b}_1\underline{b}_2}] = -2\cdot 2\delta^{[\underline{b}_1}_{[\underline{a}_1} Z_{\underline{a}_2]}{}^{\underline{b}_2]} \ , \\
&[J_{\underline{a}_1\underline{a}_2}, Z^{\underline{b}_1\cdots\underline{b}_4}] = -4\cdot 2\delta^{[\underline{b}_1}_{[\underline{a}_1} Z_{\underline{a}_2]}{}^{\underline{b}_2\underline{b}_3\underline{b}_4]} \ , [J^{\underline{a}_1\underline{a}_2}, Z^{\underline{b}_1\cdots\underline{b}_6}] = -6\cdot 2\delta^{[\underline{b}_1}_{[\underline{a}_1} Z_{\underline{a}_2]}{}^{\underline{b}_2\cdots\underline{b}_6]}, \\
&[S^{\underline{a}_1\underline{a}_2}, Z] = Z^{\underline{a}_1\underline{a}_2} \ , \ [S^{\underline{a}_1\underline{a}_2}, Z_{\underline{b}_1\underline{b}_2}] = Z^{\underline{a}_1\underline{a}_2}{}_{\underline{b}_1\underline{b}_2} - 2\delta^{a_1a_2}_{\underline{b}_1\underline{b}_2} Z \ , \\
&[S_{\underline{a}_1\underline{a}_2}, Z^{\underline{b}_1\cdots\underline{b}_4}] = -12 Z^{[\underline{b}_1\underline{b}_2}\delta^{\underline{b}_3\underline{b}_4]}_{\underline{a}_1\underline{a}_2} + \frac{1}{3} Z^{\underline{b}_1\cdots\underline{b}_4\underline{a}_1\underline{a}_2} \ , \\
&[S_{\underline{a}_1\underline{a}_2}, Z^{\underline{b}_1\cdots\underline{b}_6}] = -2\cdot 45 Z^{[\underline{b}_1\cdots\underline{b}_4}\delta^{\underline{b}_5\underline{b}_6]}_{\underline{a}_1\underline{a}_2} \ , \ldots
\end{aligned}
\tag{6.20}
$$

and

$$
\begin{aligned}
&[S_{\underline{a}}, P_{\underline{c}}] = -\eta_{\underline{ac}} Z, \ [S^{\underline{a}_1\underline{a}_2\underline{a}_3}, P_{\underline{b}}] = 3\delta_{\underline{b}}^{[\underline{a}_1} Z^{\underline{a}_2\underline{a}_3]}, \ [S^{\underline{a}_1\cdots\underline{a}_5}, P_{\underline{b}}] = -\frac{5}{2}\delta_{\underline{b}}^{[\underline{a}_1} Z^{\underline{a}_2\cdot\cdot\underline{a}_5]} \ , \\
&[S^{\underline{a}_1\cdots\underline{a}_7}, P_{\underline{b}}] = \frac{7}{6}\delta_{\underline{b}}^{[\underline{a}_1} Z^{\underline{a}_2\cdots\underline{a}_7]} \ , \ [S^{\underline{a}}, Q^{\underline{b}}] = Z^{\underline{ab}} \ , \ [S^{\underline{a}_1\underline{a}_2\underline{a}_3}, Q^{\underline{b}}] = -Z^{\underline{a}_1\underline{a}_2\underline{a}_3\underline{b}} \ , \\
&[S^{\underline{a}_1\cdots\underline{a}_5}, Q^{\underline{b}}] = \frac{1}{6} Z^{\underline{a}_1\cdots\underline{a}_5\underline{b}} \ , \ [S^{\underline{a}_1\cdots\underline{a}_7}, Q^{\underline{b}}] = 0 \ , \ldots
\end{aligned}
\tag{6.21}
$$

and

$$
\begin{aligned}
[S_{\underline{a}}, Z] &= P_{\underline{a}}\ ,\ [S_{\underline{a}}, Z^{\underline{b}_1\underline{b}_2}] = -2\delta^{[\underline{b}_1}_{\underline{c}} Q^{\underline{b}_2]}\ ,\ [S^{\underline{a}}, Z^{\underline{b}_1\cdots\underline{b}_4}] = Z^{\underline{a}\underline{b}_1\cdots\underline{b}_4}\ ,\\
[S^{\underline{a}}, Z^{\underline{b}_1\cdots\underline{b}_6}] &= Z^{\underline{a}\underline{b}_1\cdots\underline{b}_6}_{(1)} + Z^{\underline{b}_1\cdots\underline{b}_6,\underline{a}}\ ,\ [S_{\underline{a}_1\underline{a}_2\underline{a}_3}, Z] = 0\ ,\\
[S_{\underline{a}_1\underline{a}_2\underline{a}_3}, Z^{\underline{b}_1\underline{b}_2}] &= -6\delta^{\underline{b}_1\underline{b}_2}_{[\underline{a}_1\underline{a}_2} P_{\underline{a}_3]} + Z_{\underline{a}_1\underline{a}_2\underline{a}_3}{}^{\underline{b}_1\underline{b}_2}\ ,\\
[S_{\underline{a}_1\underline{a}_2\underline{a}_3}, Z^{\underline{b}_1\cdots\underline{b}_4}] &= 4!\delta^{[\underline{b}_1\underline{b}_2\underline{b}_3}_{\underline{a}_1\underline{a}_2\underline{a}_3} Q^{\underline{b}_4]} - Z_{(2)\underline{a}_1\underline{a}_2\underline{a}_3}{}^{\underline{b}_1\cdots\underline{b}_4} + Z^{\underline{b}_1\cdots\underline{b}_4}{}_{[\underline{a}_1\underline{a}_2,\underline{a}_3]}\ ,\\
[S_{\underline{a}_1\underline{a}_2\underline{a}_3}, Z^{\underline{b}_1\cdots\underline{b}_6}] &= 0\ ,\ [S^{\underline{a}_1\cdots\underline{a}_5}, Z] = \frac{1}{2} Z^{\underline{a}_1\cdots\underline{a}_5}\ ,\\
[S^{\underline{a}_1\cdots\underline{a}_5}, Z^{\underline{b}_1\underline{b}_2}] &= -Z^{\underline{a}_1\cdots\underline{a}_5\underline{b}_1\underline{b}_2}_{(2)} - \frac{1}{3} Z^{\underline{a}_1\cdots\underline{a}_5[\underline{b}_1,\underline{b}_2]}\ ,\\
[S_{\underline{a}_1\cdots\underline{a}_5}, Z^{\underline{b}_1\cdots\underline{b}_4}] &= 60 P_{[\underline{a}_1}\delta^{\underline{b}_1\cdots\underline{b}_4}_{\underline{a}_2\cdots\underline{a}_5]}\ ,\ [S_{\underline{a}_1\cdots\underline{a}_5}, Z^{\underline{b}_1\cdots\underline{b}_6}] = \frac{6!}{2} Q^{[\underline{b}_1}\delta^{\underline{b}_2\cdots\underline{b}_6]}_{\underline{a}_1\cdots\underline{a}_5}\ ,\\
[S^{\underline{a}_1\cdots\underline{a}_7}, Z] &= -\frac{1}{6} Z^{\underline{a}_1\cdots\underline{a}_7}_{(1)} - \frac{3}{2} Z^{\underline{a}_1\cdots\underline{a}_7}_{(2)}\ ,\\
[S^{\underline{a}_1\cdots\underline{a}_7}, Z^{\underline{b}_1\underline{b}_2}] &= 0\ ,\\
[S^{\underline{a}_1\cdots\underline{a}_7}, Z^{\underline{b}_1\cdots\underline{b}_4}] &= 0\ ,\\
[S^{\underline{a}_1\cdots\underline{a}_7}, Z^{\underline{b}_1\cdots\underline{b}_6}] &= \frac{3\cdot 7!}{2} P_{[\underline{a}_1}\delta^{\underline{b}_1\cdots\underline{b}_6}_{\underline{a}_2\cdots\underline{a}_7]}\ .
\end{aligned}
\tag{6.22}
$$

A general element of $I_c(E_{11})$ expands as

$$
h = I - \Lambda^{\underline{\alpha}} S_{\underline{\alpha}} \tag{6.23}
$$

where in terms of its IIA decomposition we have

$$
\begin{aligned}
\Lambda^{\underline{\alpha}} S_{\underline{\alpha}} =& \Lambda^{\underline{a}_1\underline{a}_2} J_{\underline{a}_1\underline{a}_2} + \tilde{\Lambda}^{\underline{a}_1\underline{a}_2} S_{\underline{a}_1\underline{a}_2} + \Lambda^{\underline{a}} S_{\underline{a}} + \Lambda^{\underline{a}_1\underline{a}_2\underline{a}_3} S_{\underline{a}_1\underline{a}_2\underline{a}_3} + \Lambda^{\underline{a}_1\cdots\underline{a}_5} S_{\underline{a}_1\cdots\underline{a}_5}\\
&+ \Lambda^{\underline{a}_1\cdots\underline{a}_6} S_{\underline{a}_1\cdots\underline{a}_6} + \Lambda^{\underline{a}_1\cdots\underline{a}_7} S_{\underline{a}_1\cdots\underline{a}_7} + \Lambda^{\underline{a}_1\cdots\underline{a}_7,\underline{b}} S_{\underline{a}_1\cdots\underline{a}_7,\underline{b}},\ + \Lambda^{\underline{a}_1\cdots\underline{a}_8} S_{\underline{a}_1\cdots\underline{a}_8} + \ldots
\end{aligned}
\tag{6.24}
$$

The transformation of the $l_1$ representation under $I_c(E_{11})$ transformations is given by

$$
h^{-1} l_\Pi h = D(h)_\Pi{}^\Lambda l_\Lambda = l'_\Pi\ , \tag{6.25}
$$

and infinitesimally this reads as

$$
\delta l_A = [\Lambda^{\underline{\alpha}} S_{\underline{\alpha}}, l_A]. \tag{6.26}
$$

Evaluating this equation on the $l_1$ generators we find, at (IIA) level zero:

$$
\begin{aligned}
\delta P_{\underline{b}} &= -2\Lambda_{\underline{b}}{}^{\underline{e}} P_{\underline{e}} - 2\tilde{\Lambda}_{\underline{b}\underline{e}} Q^{\underline{e}} - \Lambda_{\underline{b}} Z + 3\Lambda_{\underline{b}\underline{e}_1\underline{e}_2} Z^{\underline{e}_1\underline{e}_2} - \frac{5}{2}\Lambda_{\underline{b}\underline{e}_1\cdots\underline{e}_4} Z^{\underline{e}_1\cdots\underline{e}_4} + \ldots\\
\delta Q^{\underline{b}} &= -2\Lambda^{\underline{b}}{}_{\underline{e}} Q^{\underline{e}} - 2\tilde{\Lambda}^{\underline{b}\underline{e}} P_{\underline{e}} + \Lambda_{\underline{e}} Z^{\underline{e}\underline{b}} - \Lambda_{\underline{e}_1\underline{e}_2\underline{e}_3} Z^{\underline{e}_1\underline{e}_2\underline{e}_3\underline{b}} + \frac{1}{6}\Lambda_{\underline{e}_1\cdots\underline{e}_5} Z^{\underline{e}_1\cdots\underline{e}_5\underline{b}} + \ldots\ ;
\end{aligned}
\tag{6.27}
$$

and at level one

$$\begin{aligned}
\delta Z &= \Lambda^{\underline{e}} P_{\underline{e}} + \tilde{\Lambda}_{\underline{e}_1\underline{e}_2} Z^{\underline{e}_1\underline{e}_2} + \frac{1}{2}\Lambda_{\underline{e}_1\cdots\underline{e}_5} Z^{\underline{e}_1\cdots\underline{e}_5} + \dots \\
\delta Z^{\underline{b}_1\underline{b}_2} &= -6\Lambda^{\underline{e}\underline{b}_1\cdots\underline{b}_4} P_{\underline{e}} - 2\Lambda^{[\underline{b}_1} Q^{\underline{b}_2]} - 2\Lambda^{\underline{b}_1\underline{b}_2} Z + 4\Lambda^{[\underline{b}_1}{}_{\underline{e}} Z^{\underline{b}_2]\underline{e}} + \tilde{\Lambda}_{\underline{e}_1\underline{e}_2} Z^{\underline{e}_1\underline{e}_2\underline{b}_1\underline{b}_2} \\
&\quad + \Lambda_{\underline{e}_1\underline{e}_2\underline{e}_3} Z^{\underline{e}_1\underline{e}_2\underline{e}_3\underline{b}_1\underline{b}_2} - \Lambda_{\underline{e}_1\cdots\underline{e}_5} Z_{(2)}^{\underline{e}_1\cdots\underline{e}_5\underline{b}_1\underline{b}_2} - \frac{1}{3}\Lambda_{\underline{e}_1\cdots\underline{e}_5} Z^{\underline{e}_1\cdots\underline{e}_5[\underline{b}_1,\underline{b}_2]} + \dots \\
\delta Z^{\underline{b}_1\cdots\underline{b}_4} &= 60\Lambda^{\underline{e}\underline{b}_1\cdots\underline{b}_4} P_{\underline{e}} + 24\Lambda^{[\underline{b}_1\underline{b}_2\underline{b}_3} Q^{\underline{b}_4]} - 8\Lambda_{\underline{e}}{}^{[\underline{b}_1} Z^{\underline{b}_2\underline{b}_3\underline{b}_4]\underline{e}} - 12\tilde{\Lambda}^{[\underline{b}_1\underline{b}_2} Z^{\underline{b}_3\underline{b}_4]} \\
&\quad + \frac{1}{3}\Lambda_{\underline{e}_1\underline{e}_2} Z^{\underline{e}_1\underline{e}_2\underline{b}_1\cdots\underline{b}_4} + \Lambda_{\underline{e}} Z^{\underline{e}\underline{b}_1\cdots\underline{b}_4} - \Lambda_{\underline{e}_1\underline{e}_2\underline{e}_3} Z_{(2)}^{\underline{e}_1\underline{e}_2\underline{e}_3\underline{b}_1\cdots\underline{b}_4} + \Lambda_{\underline{e}_1\underline{e}_2\underline{e}_3} Z^{\underline{b}_1\cdots\underline{b}_4\underline{e}_1\underline{e}_2,\underline{e}_3} + \dots \\
\delta Z^{\underline{b}_1\cdots\underline{b}_6} &= 8\cdot 135 Q^{[\underline{b}_1}\Lambda^{\underline{b}_2\cdots\underline{b}_6]} - 8\cdot 7\cdot 135\Lambda^{\underline{e}\underline{b}_1\cdots\underline{b}_6} P_{\underline{e}} - 6\cdot 15 Z^{[\underline{b}_1\cdots\underline{b}_4}\tilde{\Lambda}^{\underline{b}_5\underline{b}_6]} \\
&\quad + \Lambda_{\underline{e}} Z^{\underline{e}\underline{b}_1\cdots\underline{b}_6} + \Lambda_{\underline{e}} Z^{\underline{b}_1\cdots\underline{b}_6,\underline{e}} - 6\cdot 2\Lambda_{\underline{e}}{}^{[\underline{b}_1} Z^{\underline{b}_2\cdots\underline{b}_6]\underline{e}} + \dots \quad ,
\end{aligned} \tag{6.28}$$

where we only list results up the generators $Z^{\underline{b}_1\cdots\underline{b}_6}$. At level two we similarly find

$$\begin{aligned}
\delta Z^{\underline{b}_1\cdots\underline{b}_5} &= \frac{6!}{2}\Lambda^{\underline{e}\underline{b}_1\cdots\underline{b}_5} P_{\underline{e}} - 60\Lambda^{\underline{b}_1\cdots\underline{b}_5} Z - 60 Z^{[\underline{b}_1\underline{b}_2}\Lambda^{\underline{b}_3\underline{b}_4\underline{b}_5]} - 5 Z^{[\underline{b}_1\cdots\underline{b}_4}\Lambda^{\underline{b}_5]} + 10\Lambda_{\underline{e}}{}^{[\underline{b}_1} Z^{\underline{b}_2\cdots\underline{b}_5]\underline{e}} \\
&\quad + \tilde{\Lambda}_{\underline{e}_1\underline{e}_2} Z_{(2)}^{\underline{e}_1\underline{e}_2\underline{b}_1\cdots\underline{b}_5} + \frac{1}{3}\tilde{\Lambda}_{\underline{e}_1\underline{e}_2} Z_{(1)}^{\underline{b}_1\cdots\underline{b}_5\underline{e}_1\underline{e}_2} - \frac{2}{3}\tilde{\Lambda}_{\underline{e}_1\underline{e}_2} Z^{\underline{b}_1\cdots\underline{b}_5[\underline{e}_1,\underline{e}_2]} + \dots \\
\delta Z_{(1)}^{\underline{b}_1\cdots\underline{b}_7} &= 6\cdot 7\cdot 135 Q^{[\underline{b}_1}\Lambda^{\underline{b}_2\cdots\underline{b}_7]} - 8\cdot 7\cdot 7\cdot 135\Lambda^{\underline{e}\underline{b}_1\cdots\underline{b}_7} P_{\underline{e}} - 7\cdot 7\cdot 135\Lambda^{\underline{e}[\underline{b}_1\cdots\underline{b}_6,\underline{b}_7]} P_{\underline{e}} \\
&\quad + 7\cdot 135 Z^{[\underline{b}_1\underline{b}_2}\Lambda^{\underline{b}_3\cdots\underline{b}_7]} + 7\cdot 7\cdot 135 Z^{\underline{b}_1\cdots\underline{b}_7} Z + \dots \\
\delta Z_{(2)}^{\underline{b}_1\cdots\underline{b}_7} &= -2\cdot 3\cdot 7\cdot 15 Q^{[\underline{b}_1}\Lambda^{\underline{b}_2\cdots\underline{b}_7]} - 8\cdot 7\cdot 15\Lambda^{\underline{e}\underline{b}_1\cdots\underline{b}_7} P_{\underline{e}} + 7\cdot 7\cdot 15\Lambda^{\underline{e}[\underline{b}_1\cdots\underline{b}_6,\underline{b}_7]} P_{\underline{e}} \\
&\quad + 8\cdot 7\cdot 15\Lambda^{\underline{b}_1\cdots\underline{b}_7,\underline{e}} P_{\underline{e}} + 3\cdot 7\cdot 15 Z^{[\underline{b}_1\underline{b}_2}\Lambda^{\underline{b}_3\cdots\underline{b}_7]} + 9\cdot 7\cdot 15\Lambda^{\underline{b}_1\cdots\underline{b}_7} Z + \dots \\
\delta Z^{\underline{b}_1\cdots\underline{b}_6,\underline{c}} &= 135\cdot 12 Q^{[\underline{b}_1}\Lambda^{\underline{b}_2\cdots\underline{b}_6]\underline{c}} - 135\cdot 6 Q^{\underline{c}}\Lambda^{\underline{b}_1\cdots\underline{b}_6} - 135\cdot 8\cdot 7\Lambda^{\underline{e}\underline{b}_1\cdots\underline{b}_6\underline{c}} P_{\underline{e}} \\
&\quad + 135\cdot 6\cdot 7\Lambda^{\underline{e}\underline{c}[\underline{b}_1\cdots\underline{b}_5,\underline{b}_6]} P_{\underline{e}} - 135\cdot 7\cdot 7\Lambda^{\underline{e}\underline{b}_1\cdots\underline{b}_6,\underline{c}} P_{\underline{e}} \\
&\quad + 135\cdot 7\Lambda^{\underline{b}_1\cdots\underline{b}_6\underline{c}} Z - 135\cdot 5 Z^{[\underline{b}_1\underline{b}_2}\Lambda^{\underline{b}_3\cdots\underline{b}_6]\underline{c}} - 135\cdot 6 Z^{[\underline{b}_1|\underline{c}|}\Lambda^{\underline{b}_2\cdots\underline{b}_6]\underline{c}} + \dots
\end{aligned} \tag{6.29}$$

where we only list up to and including transformations of $Z^{\underline{b}_1\cdots\underline{b}_6,\underline{c}}$.

The subalgebra $I_c(E_{11})$ can be interpreted as a local tangent space [57], and in eleven dimensions a Casimir $I_c(E_{11})$ is [89]

$$L^2_{E_{11}} = L_A L_B K^{AB} = P_{\tilde{a}} P^{\tilde{a}} + \frac{1}{2} Z_{\tilde{a}_1\tilde{a}_2} Z^{\tilde{a}_1\tilde{a}_2} + \frac{1}{5!} Z_{\tilde{a}_1\ldots\tilde{a}_5} Z^{\tilde{a}_1\ldots\tilde{a}_5} + \dots \quad , \tag{6.30}$$

where $\tilde{a}, \tilde{b}, \dots = 0, \dots, 10$ and $K^{AB}$ is the tangent group metric. The corresponding IIA invariant can be found by re-expressing this in terms of IIA generators and one finds

$$L^2_{IIA} = P_{\underline{a}} P^{\underline{a}} + Q_{\underline{a}} Q^{\underline{a}} + ZZ + \frac{1}{2} Z_{\underline{a}_1\underline{a}_2} Z^{\underline{a}_1\underline{a}_2} + \frac{1}{4!} Z_{\underline{a}_1\cdots\underline{a}_4} Z^{\underline{a}_1\cdots\underline{a}_4} + \frac{1}{5!} Z_{\underline{a}_1\cdots\underline{a}_5} Z^{\underline{a}_1\cdots\underline{a}_5} + \dots \tag{6.31}$$

## 6.3 The String Little Algebra

We now construct the string little algebra. To do this we must begin by selecting a choice of brane charges to be preserved by the little algebra. At IIA level zero we find the translations $P_{\underline{a}}$ and the string charge $Q^{\underline{a}}$, and although it may seem plausible, given the discussion of the previous chapter, to consider values of the form

$$P_{\underline{a}} = (m, 0, \dots, 0) \quad , \quad Q^{\underline{a}} = (0, m, 0, \dots, 0) \tag{6.32}$$

where $\underline{a}, \underline{b}, \dots = 0, 1, \dots, 9$, the difference is that, unlike the point particle, a string has a world volume which has a symmetry $\mathrm{SO}(1,1)$ that is broken by the above choices. We thus require that this symmetry be preserved by our brane charge choice and so we take

$$P_a = -\varepsilon_{ab} Q^b \;\text{ or equivalently }\; Q^a = -\varepsilon^{ab} P_b \;\; , \;\; l_A = 0 \;\text{ otherwise} \;\; , \tag{6.33}$$

where we set $a, b, \dots = 0, 1$, and $i_1, i_2, ... = 2, ..., 10$.

The adoption of a covariant relation between the brane charges, rather than a choice of some particular values, is a new and important difference in constructing irreducible representations of $I_c(E_{11}) \otimes_s l_1$, corresponding to branes, as opposed to the particular choices chosen in the case of the Poincaré algebra, corresponding to massive and massless point particles. The choice of brane charges in equation (6.33) obeys $L^2_{IIA} = 0$, using equation (6.31), and also obeys the condition $P_a Q^a = 0$ which means that it obeys the $I_c(E_{11})$ half BPS conditions discussed in [83].

We now construct the little algebra $\mathcal{H}$ which preserves the brane charge choices of equation (6.33). Inserting these into the $\delta l_A$ variations of the previous section we find at level zero

$$\begin{aligned} \delta P_{\underline{b}} &= -2(\Lambda_{\underline{b}}{}^e - \tilde{\Lambda}_{\underline{b}c}\varepsilon^{ce}) P_e \;\; , \\ \delta Q^{\underline{b}} &= 2(\Lambda^{\underline{b}}{}_c \varepsilon^{ce} - \tilde{\Lambda}^{\underline{b}e}) P_e \;\; , \end{aligned} \tag{6.34}$$

at level one:

$$\begin{aligned} \delta Z &= \Lambda^e P_e \;\; , \\ \delta Z^{\underline{b}_1 \underline{b}_2} &= -6(\Lambda^{\underline{b}_1 \underline{b}_2 e} - \frac{1}{3}\Lambda^{[\underline{b}_1}\delta^{\underline{b}_2]}{}_c \varepsilon^{ce}) P_e \;, \\ \delta Z^{\underline{b}_1 \cdots \underline{b}_4} &= 12(5\Lambda^{e\underline{b}_1 \cdots \underline{b}_4} - 2\Lambda^{[\underline{b}_1 \underline{b}_2 \underline{b}_3}\delta^{\underline{b}_4]}{}_c \varepsilon^{ce}) P_e \;, \\ \delta Z^{\underline{b}_1 \cdots \underline{b}_6} &= -8 \cdot 135(\delta^{[\underline{b}_1}{}_c \Lambda^{\underline{b}_2 \cdots \underline{b}_6]}\varepsilon^{ce} + 7\Lambda^{e\underline{b}_1 \cdots \underline{b}_6}) P_e \;\; , \end{aligned} \tag{6.35}$$

and at level two:

$$
\begin{aligned}
\delta Z^{\underline{b}_1\cdots\underline{b}_5} &= \frac{6!}{2}\Lambda^{e\underline{b}_1\cdots\underline{b}_5}P_e, \delta \underline{Z}^{\underline{b}_1\cdots\underline{b}_7} = 7\cdot 15(6\delta^{[\underline{b}_1}{}_p\Lambda^{\underline{b}_2\cdots\underline{b}_7]}\varepsilon^{pe} - 8\Lambda^{e\underline{b}_1\cdots\underline{b}_7} + 7\Lambda^{e[\underline{b}_1\cdots\underline{b}_6,\underline{b}_7]})P_e, \\
\delta Z^{\underline{b}_1\cdots\underline{b}_6,\underline{c}} &= 135(-12\delta^{[\underline{b}_1}{}_p\Lambda^{\underline{b}_2\cdots\underline{b}_6]\underline{c}}\varepsilon^{pe} + 6\delta^{\underline{c}}{}_p\Lambda^{\underline{b}_1\cdots\underline{b}_6}\varepsilon^{pe} - 8\cdot 7\Lambda^{e\underline{b}_1\cdots\underline{b}_6\underline{c}} + 6\cdot 7\Lambda^{e\underline{c}[\underline{b}_1\cdots\underline{b}_5,\underline{b}_6]} \\
&\qquad - 7\cdot 7\Lambda^{e\underline{b}_1\cdots\underline{b}_6,\underline{d}})P_e \\
\delta Z^{\underline{b}_1\cdots\underline{b}_7} &= 7\cdot 135(-6\delta^{[\underline{b}_1}{}_p\Lambda^{\underline{b}_2\cdots\underline{b}_7]}\varepsilon^{pe} - 8\cdot 7\Lambda^{e\underline{b}_1\cdots\underline{b}_7} - 7\Lambda^{e[\underline{b}_1\cdots\underline{b}_6,\underline{b}_7]})P_e \quad ,
\end{aligned}
\tag{6.36}
$$

where we have used equation (6.33) to express $Q^a$ in terms of $P_a$, and have not listed some of the more complicated charge variations.

Requiring that the variations of the charges vanish, while preserving (6.33), results in restrictions on the $\Lambda^{\underline{\alpha}}$'s, and so the overall $\Lambda^{\underline{\alpha}}S_{\underline{\alpha}}$ transformation. We first consider the variations of the level IIA zero charges $P_{\underline{b}}$ and $Q^{\underline{b}}$. These charges do not separately have to vanish, we only require that

$$\delta(P_a + \varepsilon_{ab}Q^b) = 0. \tag{6.37}$$

Inserting (6.33) for the case of $\underline{b} = b$ into this equation we see $\Lambda^{ab}$ and $\tilde{\Lambda}^{ab}$ must satisfy

$$2(-\Lambda_a{}^b + \varepsilon_{ac}\Lambda^c{}_d\varepsilon^{db})P_b + 2(\tilde{\Lambda}_{ac}\varepsilon^{cb} - \varepsilon_{ac}\tilde{\Lambda}^{cb})P_b = 0. \tag{6.38}$$

This equation vanishes identically since in two dimensions we may re-write $\Lambda^{ab}$ as $\Lambda^{ab} = \varepsilon^{ab}\Lambda$ and $\tilde{\Lambda}^{ab} = \varepsilon^{ab}\tilde{\Lambda}$. Thus there are no restrictions on either $\Lambda^{ab}$ or $\tilde{\Lambda}^{ab}$. This symmetry results in the expected SO$(1,1)$ Lorentz world sheet symmetry that we insisted the charge conditions preserve. However, these conditions also preserve the transformations corresponding to $S_{ab}$.

We can now set the variations of all remaining brane charges to zero, $\delta l_A = 0$ for $l_A \neq P_a, Q^a$, which gives further restrictions on the allowed $I_c(E_{11})$ transformation. At level zero we find from $\delta P_i = 0, \delta Q^i = 0$ the conditions

$$\Lambda^{ab} \neq 0 \ , \ \Lambda^{ij} \neq 0 \ , \ \tilde{\Lambda}^{ab} \neq 0 \ , \ \tilde{\Lambda}^{ai} = \varepsilon^{ac}\Lambda_c{}^i \ , \ \tilde{\Lambda}^{ij} \neq 0 \ . \tag{6.39}$$

While varying just the charges $Z, Z^{\underline{b}_1\underline{b}_2}, Z^{\underline{b}_1\cdots\underline{b}_4}$ and $Z^{\underline{b}_1\cdots\underline{b}_6}$, at level one leads to the conditions

$$
\begin{aligned}
&\Lambda^a = 0, \quad \Lambda^{ai_1i_2} = 0, \quad \Lambda^{abi} = \frac{1}{6}\varepsilon^{ab}\Lambda^i, \quad \Lambda^{ai_1\ldots i_4} = 0, \\
&\Lambda^{abi_1i_2i_3} = \frac{2}{5\cdot 4}\varepsilon^{ab}\Lambda^{i_1i_2i_3}, \quad \Lambda^{i_1\ldots i_5} \neq 0, \ \Lambda^{ei_1\ldots i_6} = 0, \\
&\Lambda^{abi_1\ldots i_5} = \frac{1}{7\cdot 6}\varepsilon^{ab}\Lambda^{i_1\ldots i_5}, \quad \Lambda^{i_1\ldots i_7} \neq 0 \quad .
\end{aligned}
\tag{6.40}
$$

At level two the computations become more difficult but the conditions include

$$
\begin{aligned}
&\Lambda^{ai_1\ldots i_5} = 0, \quad \Lambda^{abi_1\ldots i_4} = 0, \quad \Lambda^{i_1\ldots i_6} \neq 0, \\
&\Lambda^{ai_1\ldots i_7} = 0, \quad \Lambda^{ai_1\ldots i_6,j} = 0, \quad \Lambda^{i_1\ldots i_7,j} \neq 0, \quad \Lambda^{i_1\ldots i_8} \neq 0 \ .
\end{aligned}
\tag{6.41}
$$

The little algebra generators are now found by inserting the above conditions on the parameters $\Lambda^{\underline{\alpha}}$ into the generator of equation (6.24), and we find

$$
S_a \ , \ S_{ai_1i_2} \ , \ S_{ai_1\ldots i_4} \ , \ S_{ai_1\ldots i_6} \ , \ S_{a_1a_2i_1\ldots i_4} \ , \ S_{ai_1\ldots i_7} \notin \mathcal{H} \ , \tag{6.42}
$$

and also that $\mathcal{H}$ contains

$$
J_{ij} \ , S_{ij} \ , J_{ab} \ , S_{ab} \ , \ L^{(0)}_{ai} = J_{ai} + \varepsilon_a{}^e S_{ei} \ , \tag{6.43}
$$

at level zero and

$$
\begin{aligned}
&L^{(1)}_{i} = S_i + \frac{1}{2}\varepsilon^{e_1e_2}S_{e_1e_2i} \ , \quad L^{(1)}_{i_1i_2i_3} = S_{i_1i_2i_3} + \varepsilon^{e_1e_2}S_{e_1e_2i_1i_2i_3} \ , \\
&L^{(1)}_{i_1\ldots i_5} = S_{i_1\ldots i_5} + \frac{1}{2}\varepsilon^{e_1e_2}S_{e_1e_2i_1\ldots i_5} \ ,
\end{aligned}
\tag{6.44}
$$

at level one, where the superscript refers to the IIA level to which a given generator belongs.

Finding further relations between the $\Lambda^{\underline{\alpha}}$'s resulting from variations of the higher level charges will becomes more and more difficult, however one can now use the fact that the little group algebra $\mathcal{H}$ must close to find in this way new elements simply by calculating their commutators, and the following generators will be found below to arise in this way. At IIA level one we find the generator

$$
L^{(1)}_{i_1\ldots i_7} = S_{i_1\ldots i_7} - \frac{1}{2}\varepsilon^{a_1a_2}S_{a_1a_2i_1\ldots i_7} \ , \tag{6.45}
$$

and similarly at level two we find the generators

$$
\begin{aligned}
L^{(2)}_{i_1\ldots i_6} &= S_{i_1\ldots i_6} + \frac{1}{2}\varepsilon^{a_1a_2}S_{i_1\ldots i_6a_1a_2} - \varepsilon^{a_1a_2}S_{i_1\ldots i_6a_1,a_2} - \frac{1}{8}\varepsilon^{a_1a_2}\varepsilon^{b_1b_2}S_{i_1\ldots i_6a_1a_2,b_1b_2} \ , \\
L^{(2)}_{i_1..i_8} &= S_{i_1..i_8} + \varepsilon^{a_1a_2}S_{i_1..i_8a_1,a_2} - \frac{1}{6}\varepsilon^{a_1a_2}S_{i_1..i_8,a_1a_2} - \frac{1}{2}\varepsilon^{a_1a_2}S_{(1)a_1a_2i_1..i_8} \\
&\quad - \frac{4}{9}\varepsilon^{a_1a_2}S_{(2)a_1a_2i_1..i_8} \ , \\
L^{(2)}_{i_1..i_7,j} &= S_{i_1..i_7,j} - \frac{1}{2}\varepsilon^{a_1a_2}S_{a_1a_2i_1..i_7,j} - \frac{1}{2}\varepsilon^{a_1a_2}S_{i_1..i_7a_1,a_2j} - \frac{1}{12}\varepsilon^{a_1a_2}S_{i_1..i_7j,a_1a_2} \\
&\quad - \frac{1}{18}\varepsilon^{a_1a_2}S_{(2)a_1a_2i_1..i_7j} \ .
\end{aligned}
\tag{6.46}
$$

The above results imply that the string little algebra is given by

$$
\begin{aligned}
\mathcal{H} = \{ &J_{ab} \ , \ L^{(0)}_{ai} \ , \ J_{ij} \ , \ S_{ab} \ , \ S_{ij} \ ; \ L^{(1)}_{i} \ , \ L^{(1)}_{i_1i_2i_3} \ , \ L^{(1)}_{i_1\ldots i_5} \ , \ L^{(1)}_{i_1\ldots i_7} \ ; \\
&L^{(2)}_{i_1\ldots i_6} \ , \ L^{(2)}_{i_1\ldots i_8} \ , \ L^{(2)}_{i_1\ldots i_7,j} \ , \ \ldots \}
\end{aligned}
\tag{6.47}
$$

where the IIA levels have been separated by a ; semi-colon, and ... denotes higher level generators.

We will now compute the algebra of the generators of $\mathcal{H}$. The commutators between $\mathcal{H}$ generators arising from IIA level zero generators are given by

$$\begin{aligned}
&[J^{ab}, J_{cd}] = -4\delta^{[a}_{[c} J^{b]}{}_{d]}, \quad [J^{ab}, L^{(0)}_{ci}] = -2\delta^{[a}_{c} L^{b](0)}{}_{i}, \quad [J^{ab}, J_{ij}] = 0, \\
&[J^{ab}, S_{cd}] = -4\delta^{[a}_{[c} S^{b]}{}_{d]}, \quad [J^{ab}, S_{ij}] = 0, \quad [L^{ai(0)}, L^{(0)}_{bj}] = 0, \quad [L^{ai(0)}, J_{jk}] = 2\delta^{i}_{[j} L^{a(0)}{}_{k]}, \\
&[L^{ai(0)}, S_{bc}] = 2\varepsilon^{a}{}_{[b} L^{(0)}_{c]i} \ , \ [L^{ai(0)}, S_{jk}] = 2\varepsilon^{a}{}_{e}\delta^{i}_{[j} L^{e(0)}{}_{k]} \quad , \quad [J^{ij}, J_{kl}] = -4\delta^{[i}_{[k} J^{j]}{}_{l]}, \qquad (6.48) \\
&[J^{ij}, S_{cd}] = 0 \quad , \quad [J^{ij}, S_{kl}] = 0 \quad , \quad [S^{ab}, S_{cd}] = -4\delta^{[a}{}_{[c} S^{b]}{}_{d]} \quad , \\
&[S^{ab}, S_{ij}] = 0 \quad , \quad [S^{ij}, S_{kl}] = -4\delta^{[i}_{[k} J^{j]}{}_{l]} \quad .
\end{aligned}$$

The commutators between $\mathcal{H}$ generators constructed from IIA level zero generators with those constructed from IIA level one generators is given by

$$\begin{aligned}
&[J_{ab}, L^{(1)}_{i}] = 0 \quad , \quad [J_{ab}, L^{(1)}_{i_1 i_2 i_3}] = 0 \quad , \quad [J_{ab}, L^{(1)}_{i_1 \ldots i_5}] = 0 \quad , \quad [J_{ab}, L^{(1)}_{i_1 \ldots i_7}] = 0 \quad , \\
&[L^{(0)}_{ai}, L^{(1)}_{j}] = 0 \ , \ [L^{(0)}_{ai}, L^{(1)}_{j_1 j_2 j_3}] = 0 \ , \ [L^{(0)}_{ai}, L^{(1)}_{j_1 \ldots j_5}] = 0 \ , \ [L^{(0)}_{ai}, L^{(1)}_{j_1 \ldots j_7}] = 0 \quad , \\
&[J^{ij}, L^{(1)}_{k}] = -2\delta^{[i}_{k} L^{j](1)} \ , \ [J^{ij}, L^{(1)}_{k_1 k_2 k_3}] = -6\delta^{[i}_{[k_1} L^{j](1)}{}_{k_2 k_3]} \quad , \\
&[J^{ij}, L^{(1)}_{k_1 \ldots k_5}] = -10\delta^{[i}_{[k_1} L^{j](1)}{}_{k_2 \ldots k_5]} \quad , \quad [J^{ij}, L^{(1)}_{k_1 \ldots k_7}] = -14\delta^{[i}_{[k_1} L^{j](1)}{}_{k_2 \ldots k_7]} \quad , \\
&[S^{ab}, L^{(1)}_{i}] = \varepsilon^{ab} L^{(1)}_{i} \quad , \quad [S^{ab}, L^{(1)}_{i_1 i_2 i_3}] = \varepsilon^{ab} L^{(1)}_{i_1 i_2 i_3} \quad , \quad [S^{ab}, L^{(1)}_{i_1 \ldots i_5}] = \varepsilon^{ab} L^{(1)}_{i_1 \ldots i_5} \quad , \qquad (6.49) \\
&[S^{ab}, L^{(1)}_{i_1 \ldots i_7}] = \varepsilon^{ab} L^{(1)}_{i_1 \ldots i_7} \quad , \quad [S_{ij}, L^{(1)}_{k}] = -L^{(1)}_{ijk} \quad , \\
&[S^{ij}, L^{(1)}_{k_1 k_2 k_3}] = -2L^{ij(1)}{}_{k_1 k_2 k_3} + 6L^{(1)}_{[k_1}\delta^{ij}_{k_2 k_3]} \quad , \\
&[S^{ij}, L^{(1)}_{k_1 \ldots k_5}] = 10L^{(1)}_{[k_1 k_2 k_3}\delta^{ij}_{k_4 k_5]} - L^{(1)ij}{}_{k_1 \ldots k_5} \quad , \\
&[S^{ij}, L^{(1)}_{k_1 \ldots k_7}] = 6 \cdot 7L^{(1)}_{[k_1 \ldots k_5}\delta^{ij}_{k_6 k_7]} \quad .
\end{aligned}$$

The commutators between $\mathcal{H}$ generators arising from IIA level $\pm 1$ generators are given by

$$\begin{aligned}
&[L^{(1)}_{i}, L^{(1)}_{j}] = 0 \quad , \quad [L^{(1)}_{i}, L^{(1)}_{j_1 j_2 j_3}] = 0 \quad , \quad [L^{(1)}_{i}, L^{(1)}_{j_1 \ldots j_5}] = -L^{(2)}_{ij_1 \ldots j_5} \quad , \\
&[L^{(1)}_{i}, L^{(1)}_{j_1 \ldots j_7}] = -L^{(2)}_{ij_1 \ldots j_7} + L^{(2)}_{j_1 \ldots j_7, i} \quad , \quad [L^{(1)}_{i_1 i_2 i_3}, L^{(1)}_{j_1 j_2 j_3}] = 2L^{(2)}_{i_1 i_2 i_3 j_1 j_2 j_3} \quad , \qquad (6.50) \\
&[L^{(1)}_{i_1 i_2 i_3}, L^{(1)}_{j_1 \ldots j_5}] = L^{(2)}_{i_1 i_2 i_3 j_1 \ldots j_5} - 5L^{(2)}_{i_1 i_2 i_3 [j_1 \ldots j_4, j_5]} \ ,
\end{aligned}$$

along with higher level relations we have not listed.

We now introduce the definition

$$L^{\alpha} = e^{-R} R^{\alpha} e^{R} + I_c(e^{-R} R^{\alpha} e^{-R}) \ , \quad R = \frac{1}{2}\varepsilon_{ab} R^{ab} \quad , \qquad (6.51)$$

One can verify that, apart from $J_{ab}$ and $S_{ab}$, all of the above generators take the form of $L^\alpha$. For example, we have

$$\begin{aligned} L_i^{(1)} &= e^{-R}R^i e^R + I_c(e^{-R}R^i e^R) = S_i + \frac{1}{2}\varepsilon^{a_1a_2}S_{a_1a_2i}\ , \\ L_{i_1i_2i_3}^{(1)} &= e^{-R}R^{i_1i_2i_3}e^R + I_c(e^{-R}R^{i_1i_2i_3}e^R) = S_{i_1i_2i_3} + \varepsilon^{a_1a_2}S_{a_1a_2i_1i_2i_3}\ . \end{aligned} \tag{6.52}$$

The content of [32] is devoted to showing this formula for $L_\alpha$ holds to all orders. The use of this little algebra in constructing the full irreducible representations of the IIA string is left to future work.

# Chapter 7

# Generalized Clifford Algebras and Fermions in E Theory

## 7.1 Generalized Clifford Algebra

Spinors have previously been introduced into E theory by hand [62] in that they do not follow naturally from the $E_{11}$ algebra but instead start from the familiar gravitino, with the requirement that it carry a representation of $I_c(E_{11})$, which was then constructed by hand. An alternative approach is to generalise the way spinors appear in the context of the Poincaré algebra. This is the approach we will consider in this chapter [31]. Recall that the Poincaré group is given by $\mathrm{SO}(1, D-1) \otimes_s T^D$, where $T^D$ is the vector representation of the Lorentz group $\mathrm{SO}(1, D-1)$, and it leads to a $D$ dimensional spacetime on which $\mathrm{SO}(1, D-1)$ acts. To each coordinate of the spacetime we introduce a $\gamma^a$ matrix which acts on a spinor. The existence of such $\gamma^a$ matrices is intrinsically tied to the existence of a Lorentz invariant metric

$$\eta_{ab} = \Lambda_a{}^c \eta_{cd} \Lambda^d{}_b \quad , \tag{7.1}$$

via the Clifford algebra relation

$$\{\gamma_a, \gamma_b\} = 2\eta_{ab} I \quad . \tag{7.2}$$

A very natural question is how this idea manifests in the generalized space-time of E theory [79], [84], [86], [57].

In the context of E theory we begin from the algebra $I_c(E_{11}) \otimes_s l_1$, which is also a semi-direct product, and indeed contains the Poincaré algebra as a subspace. Since the $l_1$ (vector) representation of $E_{11}$, $l_A \in \{P_a, Z^{a_1a_2}, ...\}$, leads to the generalized spacetime, we introduce operators $\Gamma^A$ which are in one to one correspondence with the elements $l_A$ of the vector representation. Thus, associated to the coordinates

$$x^A \in \{x^a, x^{a_1a_2}\ ,\ x^{a_1\ldots a_5}\ ,\ x^{a_1\ldots a_7,b},\ x^{a_1\ldots a_8}, \ldots\} \ , \tag{7.3}$$

in eleven dimensions, we introduce the following operators

$$\Gamma^A = \{\Gamma^a,\ \Gamma^{a_1a_2},\ \Gamma^{a_1\ldots a_5},\ \Gamma^{a_1\ldots a_7,b},\ \Gamma^{a_1\ldots a_8}, \ldots\} \ , \tag{7.4}$$

and require that these operators obey the 'generalized Clifford algebra' relations

$$\Gamma^A\Gamma^B + \Gamma^B\Gamma^A = 2K^{AB} \ . \tag{7.5}$$

Here $K^{AB}$ is the $I_c(E_{11})$-invariant tangent space metric [57]

$$K_{AB} = \begin{bmatrix} \eta_{ab} & 0 & 0 & \ldots \\ 0 & 2\delta^{a_1a_2,b_1b_2} & 0 & \ldots \\ 0 & 0 & 5!\delta^{a_1\ldots a_5,b_1\ldots b_5} & \ldots \\ 0 & 0 & 0 & \ddots \end{bmatrix} \ . \tag{7.6}$$

Unlike the finite-dimensional case, where an explicit matrix representation for the Clifford elements can be determined explicitly, e.g. using the representation theory of the finite Clifford group [85, Ch. 5], the infinite-dimensional nature of the 'Generalized Clifford group' associated to the $\Gamma^A$'s implies that the $\Gamma^A$ should simply be interpreted as abstract operators from the beginning.

## 7.2 Generalized Spinors and the Dirac Equation

Spinors in a Minkowski context can be interpreted as 'minimal left ideals' of a Clifford algebra [43, Sec. 5]. Thus, it is natural to think of generalized spinors as minimal left ideals of the above generalized Clifford algebra. Alternatively they can be introduced in a representation theory context. From this perspective, we can view a spinor $\Psi$ as carrying a representation of the $\Gamma^A$ operators, and transforming under $I_c(E_{11})$ as

$$\delta\Psi = U(S^\alpha)\Psi = -\mathcal{S}^\alpha\Psi \ , \tag{7.7}$$

where $U(S^\alpha)$ is the action of the generator $S^\alpha$ and the matrix $\mathcal{S}^\alpha$ its effect. We require that

$$[\mathcal{S}^\alpha, \Gamma^A] = \Gamma^B (\tilde{D}^\alpha)_B{}^A \quad , \tag{7.8}$$

where we used $S^\alpha := R^\alpha - R_\alpha$, $\tilde{D}^\alpha := D^\alpha - D_\alpha$, and the relation $[R^\alpha, l_A] = -(D^\alpha)_A{}^B l_B$. Thus the $\Gamma^A$'s transform under $I_c(E_{11})$ as the vector representation of $E_{11}$.

A generalised Dirac equation is given by

$$\Gamma^A \partial_A \Psi = 0 \quad . \tag{7.9}$$

This is invariant under $I_c(E_{11})$ transformations since under $h = I - \Lambda^{a_1 a_2 a_3} S_{a_1 a_2 a_3}$ the derivatives $\partial_A \equiv \frac{\partial}{\partial x^A}$ transform as $\partial'_A = D(h)_A{}^B \partial_B$, that is: $\delta(\partial_A) = -(\tilde{D}^\alpha)_A{}^B \partial_B$.

At level zero $I_c(E_{11})$ is just SO(1,10) and so at this level the above discussion just reduces to the standard discussion of the Dirac equation. The Gamma matrices corresponding to the higher level coordinates occur with derivatives with respect to these coordinates. Thus if we neglect the higher level derivatives this is just the familiar Dirac equation.

To account for the gravitino we can introduce the object $\Psi_A$ which is a spinor with the vector index $A$, and impose the on-shell conditions

$$\Gamma^B \partial_B \Psi_A = \Gamma^A \Psi_A = \partial^A \Psi_A = 0 \quad , \tag{7.10}$$

which reduce to the usual gravitino on-shell conditions [33, Eq. (4.37)] when restricted to the lowest levels $\Gamma^a$, $\partial_a$ and $\Psi_a$.

## 7.3 Generalised Supersymmetry

We can also try to introduce a generalised supersymmetry generator $Q^A$, which is a generalised spinor, into an $E_{11}$ context. In terms of the $I_c(E_{11})$ generators $S^\alpha$ it has the schematic relation

$$[S^\alpha, Q] = -\mathcal{S}^\alpha Q. \tag{7.11}$$

One may consider the possibility that an anti-commutator of the generic form exists

$$\{Q, Q\} = \Gamma^A l_A. \tag{7.12}$$

Indeed, thinking of supersymmetry as arising from seeking a ‘square root’ of the Dirac equation [52] similar to how the Dirac originally discovered the Dirac equation by seeking a ‘square root’ of the Klein-Gordon equation, the existence of a generalised supersymmetry associated to the generalised Dirac equation is a plausible generalization.

# Appendix A

# Lie Algebras

In this appendix we are going to introduce a brief introduction of Lie algebras and some associated concepts used in the text, largely based on [28], [36], [53], and [60].

## A.1 From Lie Groups to Lie Algebras

Consider a group $G$ whose elements $g$ are parametrized by a (potentially infinite) set of continuous 'parameters' $x^i$, $i = 1, 2, ...$ via

$$g = g(x^1, x^2, ...) = g(x^i) = g(x). \tag{A.1}$$

Let $g(x)$ depend smoothly on these parameters. We will refer to this continuous differentiable group as a Lie group. Let $g(0) = e$ be the identity element, and denote the inverse of $g(x)$ as

$$g(x)^{-1} = g(-x) = g(-x^1, -x^2, ...). \tag{A.2}$$

Group multiplication is now defined in terms of so-called 'structure functions' $\phi^i = \phi^i(x, y)$ defined via

$$g(x)g(y) = g[\phi(x, y)]. \tag{A.3}$$

The other properties of a group: existence of an identity, existence of inverses, and associativity, can be written in terms of structure functions as

$$\begin{aligned}
g(x) &= g(x)e = g(x)g(0) = g[\phi(x,0)] = g[\phi(0,x)] \\
&\rightarrow \quad \phi(x,0) = \phi(0,x) = x \quad \text{(identities of structure functions)}, \\
e &= g(x)g^{-1}(x) = g(x)g(-x) = g[\phi(x,-x)] = g[\phi(-x,x)] = g(0) \\
&\rightarrow \quad \phi(x,-x) = \phi(-x,x) = 0 \quad \text{(inverses of structure functions)}, \\
[g(x)g(y)]g(z) &= g(x)[g(y)g(z)] \quad \rightarrow \quad g\{\phi[\phi(x,y),z]\} = g\{\phi[x,\phi(y,z)]\} \\
&\rightarrow \quad \phi[\phi(x,y),z] = \phi[x,\phi(y,z)] \quad \text{(associativity of structure functions)}.
\end{aligned} \tag{A.4}$$

We now differentiate the associativity condition with respect to $z^k$

$$\frac{\partial \phi^i[\phi(x,y),z]}{\partial z^k} = \frac{\partial \phi^i[x,\phi(y,z)]}{\partial \phi^j(y,z)} \frac{\partial \phi^j(y,z)}{\partial z^k} \tag{A.5}$$

and then evaluate this at $z = 0$

$$\frac{\partial \phi^i[\phi(x,y),z]}{\partial z^k}|_{z=0} = \frac{\partial \phi^i(x,y)}{\partial y^j} \left( \frac{\partial \phi^j(y,z)}{\partial z^k}|_{z=0} \right). \tag{A.6}$$

Defining $\frac{\partial \phi^i(x,y)}{\partial y^j}|_{y=0} = u^i{}_j(x)$ this reads as

$$u^i{}_k[\phi(x,y)] = \frac{\partial \phi^i(x,y)}{\partial y^j} u^j{}_k(y). \tag{A.7}$$

We now isolate the partial derivative term by multiplying on the right by the inverse of $u$, denoted $v = u^{-1}$:

$$\frac{\partial \phi^i(x,y)}{\partial y^j} = u^i{}_m[\phi(x,y)] v^m{}_j(y) \quad , \quad (v = u^{-1}) \quad . \tag{A.8}$$

We can now require the following 'integrability conditions'

$$\begin{aligned}
\frac{\partial^2 \phi^i(x,y)}{\partial y^k \partial y^j} &= \frac{\partial^2 \phi^i(x,y)}{\partial y^j \partial y^k} \quad \rightarrow \\
\frac{\partial}{\partial y^k}\{u^i{}_m[\phi(x,y)] v^m{}_j(y)\} &= \frac{\partial}{\partial y^j}\{u^i{}_m[\phi(x,y)] v^m{}_k(y)\} \quad ,
\end{aligned} \tag{A.9}$$

which expand to give

$$\frac{\partial u^i{}_m(\phi)}{\partial \phi^p} \frac{\partial \phi^p}{\partial y^k} v^m{}_j(y) + u^i{}_m(\phi) \frac{\partial v^m{}_j(y)}{\partial y^k} = \frac{\partial u^i{}_m(\phi)}{\partial \phi^p} \frac{\partial \phi^p}{\partial y^j} v^m{}_k(y) + u^i{}_m(\phi) \frac{\partial v^m{}_k(y)}{\partial y^j} \quad , \tag{A.10}$$

and using (A.8) in this equation we find

$$\begin{aligned}&\frac{\partial u^i{}_m(\phi)}{\partial \phi^p}u^p{}_q(\phi)v^q{}_k(y)v^m{}_j(y)+u^i{}_m(\phi)\frac{\partial v^m{}_j(y)}{\partial y^k}\\&\qquad=\frac{\partial u^i{}_m(\phi)}{\partial \phi^p}u^p{}_q(\phi)v^q{}_j(y)v^m{}_k(y)+u^i{}_m(\phi)\frac{\partial v^m{}_k(y)}{\partial y^j}\quad .\end{aligned}\tag{A.11}$$

We can try to get the $y$ dependent terms on one side, and $\phi$ dependent terms on the other, by writing this as

$$\begin{aligned}&\left[\frac{\partial u^i{}_m(\phi)}{\partial \phi^p}u^p{}_q(\phi)-\frac{\partial u^i{}_q(\phi)}{\partial \phi^p}u^p{}_m(\phi)\right]v^q{}_k(y)v^m{}_j(y)=u^i{}_m(\phi)\left[\frac{\partial v^m{}_k(y)}{\partial y^j}-\frac{\partial v^m{}_j(y)}{\partial y^k}\right]\;\rightarrow\\&v^a{}_i(\phi)\left[\frac{\partial u^i{}_c(\phi)}{\partial \phi^p}u^p{}_b(\phi)-\frac{\partial u^i{}_b(\phi)}{\partial \phi^p}u^p{}_c(\phi)\right]=\left[\frac{\partial v^a{}_k(y)}{\partial y^j}-\frac{\partial v^a{}_j(y)}{\partial y^k}\right]u^k{}_b(y)u^j{}_c(y)\end{aligned}\tag{A.12}$$

On each side we have functions of different variables. Further, on the left hand side we have $\phi=\phi(x,y)$ where $x$ acts as a dummy variable. However if we choose it to be $-y$ we find $\phi(-y,y)=0$. Thus, both sides are constant, denoted $f_{bc}{}^a$, referred to as structure constants. We can thus analyse this relation in two separate ways.

Taking the right-hand side as being equal to $f_{bc}{}^a$ we can write this as

$$\frac{\partial v^a{}_k(y)}{\partial y^j}-\frac{\partial v^a{}_j(y)}{\partial y^k}=f^a{}_{bc}v^b{}_k(y)v^c{}_j(y)\quad .\tag{A.13}$$

We now need to find out what kind of $v^a{}_k(y)$ functions satisfy this relation, i.e. we need to establish 'integrability conditions' for this relation as well. Ignoring the $a$ index, the left-hand side is the same as the Maxwell field strength $F^{(a)}_{jk}=\partial_j v^{(a)}_k-\partial_k v^{(a)}_k=2\partial_{[j}v^{(a)}_{k]}$ (with $\partial_j=\partial/\partial y^j$). Thus we have

$$F^{(a)}_{jk}=f^a{}_{bc}v^b{}_k(y)v^c{}_j(y).\tag{A.14}$$

The integrability condition is then clearly just the Bianchi identity[1]

$$3\partial_{[i}F^{(a)}_{jk]}=0.\tag{A.15}$$

Applying this to the right-hand side gives

$$\begin{aligned}0&=3f^a{}_{bc}\partial_{[i}(v^c{}_j v^b{}_{k]})=3f^a{}_{bc}(\partial_{[i}v^c{}_j)v^b{}_{k]}+3f^a{}_{bc}v^c{}_{[j}(\partial_i v^b{}_{k]})\\&=3f^a{}_{bc}(2\partial_{[i}v^c{}_j)v^b{}_{k]}=3f_{bc}{}^a(f_{de}{}^c v^d{}_{[j}v^e{}_i)v^b{}_{k]}=3f_{[b|c|}{}^a(f_{de]}{}^c v^d{}_j v^e{}_i)v^b{}_k\\&=3f_{[ed}{}^c f_{|c|b]}{}^a v^e{}_i v^d{}_j v^b{}_k\quad .\end{aligned}\tag{A.16}$$

[1]The 3 is included because $\partial_{[i}F^{(a)}_{jk]}:=\delta^{pqr}_{ijk}\partial_p F^{(a)}_{qr}$ and we have taken the Generalized Kronecker-Delta $\delta^{k_1..k_n}_{i_1..i_n}$ to contain a $1/n$ term, i.e. $\delta^{pqr}_{ijk}=\frac{1}{3}(\delta^p_i\delta^{qr}_{jk}-\delta^q_i\delta^{pr}_{jk}+\delta^r_i\delta^{pq}_{jk})=\frac{1}{3}(\delta^p_i\delta^{qr}_{jk}+2\delta^{[q}_i\delta^{r]p}_{jk})$, see [85]. Thus we have $3\partial_{[i}F^{(a)}_{jk]}=\partial_i F^{(a)}_{jk}+2\partial_{[j}F^{(a)}_{k]i}=\partial_i F^{(a)}_{jk}+\partial_j F^{(a)}_{ki}-\partial_k F^{(a)}_{ji}=0$

Thus the integrability condition reduces to an algebraic constraint on the coefficients ${f_{bc}}^a$ known as the 'Jacobi identity',

$$3{f_{[ed}}^c {f_{|c|b]}}^a = 0. \tag{A.17}$$

No further integrability conditions will therefore arise.

Taking instead the left-hand side as being equal to ${f_{bc}}^a$ we can write this as

$$[{u^p}_b \partial_p {u^i}_c - {u^p}_c \partial_p {u^i}_b] = {f_{bc}}^a {u^i}_a \quad , \quad \partial_p = \tfrac{\partial}{\partial \phi^p} \quad , \tag{A.18}$$

Defining the operator

$$X_a = {u^i}_a \partial_i \ , \tag{A.19}$$

and the notation $[A, B] = AB - BA$, referred to as the commutator of $A$ and $B$, this relation then reads as

$$\begin{aligned} {f_{bc}}^a X_a &= {f_{bc}}^a {u^i}_a \partial_i = [{u^p}_b \partial_p {u^i}_c - {u^p}_c \partial_p {u^i}_b]\partial_i = [{u^p}_b \partial_p, {u^i}_c \partial_i] \\ &= [X_a, X_b] \ . \end{aligned} \tag{A.20}$$

The second integrability condition now arises in this language from the algebraic identity

$$3[[X_{[a}, X_b], X_{c]}] = 0 \ , \tag{A.21}$$

which can be verified directly, since when written in terms of the structure constants this reads as

$$0 = 3[[X_{[a}, X_b], X_{c]}] = 3{f_{[ab}}^d [[X_{|d|}, X_{c]}] = 3{f_{[ab}}^d {f_{|d|c]}}^e X_e \ . \tag{A.22}$$

The $X_a$ operators form a vector space. We will use these operators and the above properties to discuss an infinitesimal Lie group, and refer to the space they generate under the commutator, as a Lie algebra, summarized next.

## A.2 Lie Algebras

We define a Lie algebra as a vector space $\mathfrak{g}$ over a field $\mathbb{F}$ along with am operation $[\ ,\ ] : \mathfrak{g} \times \mathfrak{g} \to \mathfrak{g}$, called the commutator, which is bilinear, anti-commutative, and satisfies the

Jacobi identity:

$$\begin{aligned} [X, aY + bZ] = a[X, Y] + b[X, Z] \quad , \quad [aX + bY, Z] = a[X, Z] + b[Y, Z] \quad , \\ [X, Y] = -[Y, X] \quad , \\ 0 = [[X, Y], Z] + [[Y, Z], X] + [[Z, X], Y] \quad , a, b \in \mathbb{F} \ , \ X, Y, Z \in \mathfrak{g}. \end{aligned} \tag{A.23}$$

If we denote the basis of $\mathfrak{g}$ by $X_a$ then we can define the commutator between two basis vectors $X_a$ and $X_b$ as

$$[X_a, X_b] = f_{ab}{}^c X_c \quad , \tag{A.24}$$

where the $f_{ab}{}^c = (f_a)_b{}^c$ are called structure constants. The Jacobi identity in terms of the $X_a$ reads as

$$0 = 3[[X_{[a}, X_b], X_{c]}] = 3 f_{[ab}{}^d f_{|d|c]}{}^e X_e. \tag{A.25}$$

## A.3 Adjoint Representation and the Killing Metric

The Jacobi identity in terms of the structure constants can be written and simplified as

$$\begin{aligned} 0 = 3 f_{[ab}{}^d f_{|d|c]}{}^e &= (f_a)_b{}^d (f_d)_c{}^e + 2 (f_{[a})_{|c|}{}^d (f_{b]})_d{}^e \\ &= -(f_a)_b{}^d (-f_d)_c{}^e + [(-f_a), (-f_b)]_c{}^e \quad , \end{aligned} \tag{A.26}$$

so that, on defining matrices $T_a$ with matrix elements $(T_a)_c{}^e$ given as

$$(T_a)_c{}^e = (-f_a)_c{}^e, \tag{A.27}$$

we find

$$(f_a)_b{}^d (T_d)_c{}^e = [(T_a), (T_b)]_c{}^e \quad . \tag{A.28}$$

We can now directly interpret the Lie algebra in terms of matrices. This is the 'adjoint representation' of the Lie algebra $\mathfrak{g}$. In particular, in the case of finite-dimensional Lie algebras it is natural to take traces in the commutator relation and so use this to naturally define a metric on the adjoint of the Lie algebra via

$$(f_a)_b{}^d (T_d)_c{}^e (T_p)_e{}^c = [(T_a), (T_b)]_c{}^e (T_p)_e{}^c \quad \rightarrow \quad f_{abp} := (f_a)_b{}^d g_{dp} = \mathrm{tr}([T_a, T_b] T_p) \quad , \tag{A.29}$$

where we defined a metric, known as the 'Killing Metric', as

$$g_{ab} := (T_a)_c{}^e (T_b)_e{}^c = (-f_a)_c{}^e (-f_b)_e{}^c = (f_a)_c{}^e (f_b)_e{}^c \ . \tag{A.30}$$

Note $f_{abp}$ is completely anti-symmetric since (using $\text{tr}(T_a T_b T_c) = \text{tr}(T_b T_c T_a) = ...$) we can move $a$ forward twice:

$$\begin{aligned} f_{abp} &= \text{tr}([T_a, T_b]T_p) = \text{tr}(T_a T_b T_p - T_b T_a T_p) = \text{tr}(T_b T_p T_a - T_p T_b T_a) = \text{tr}([T_b, T_p]T_a) \\ &= f_{bpa}, \end{aligned} \tag{A.31}$$

so that, we find $f_{abc} = -f_{bac} = f_{cab} = -f_{acb} = f_{bca} = -f_{cba}$. This last result shows that

$$f_{abc} = \text{tr}([T_a, T_b]T_c) = \text{tr}(T_a[T_b, T_c]). \tag{A.32}$$

Treating the original $X_a$ as operators, we can simply define an inner product on the vector space of $X_a$ elements to be $(X_a, X_b) := g_{ab} = (f_a)_c{}^e (f_b)_e{}^c$ and so apply the same procedure to $[X_a, X_b] = (f_a)_b{}^c X_c$ to find $([X_a, X_b], X_c) = (f_a)_b{}^e g_{ec} = f_{abc}$. In particular, in infinite-dimensional Lie algebras, instead of using a trace we simply introduce an abstract (symmetric, bilinear) inner product ( , ) with $g_{ab} = (X_a, X_b)$ to find $([X_a, X_b], X_c) = (f_a)_b{}^e g_{ec} = f_{abc}$, and also require that it satisfy $g([T_a, T_b], T_c) = g(T_a, [T_b, T_c])$. These properties can be taken as abstract properties which define a Killing Metric on a Lie algebra.

## A.4 Semi-Simple Lie Algebras

Now that we have a metric $g_{ab} = (f_a)_c{}^e (f_b)_e{}^c$ on the Lie algebra $[T_a, T_b] = (f_a)_b{}^c T_c$, the most natural question to ask is whether we can use it to raise and lower indices on the generators via $T^a = g^{ab} T_b$ and $T_a = g_{ab} T^b$, so that we can define relations like $[T_a, T_b] = (f_a)_{bc} T^c$. Clearly the ability to raise and lower indices depends on whether the metric/matrix $g_{ab}$ is invertible or not. The metric $g_{ab}$ is invertible if the determinant of $g_{ab}$ is non-zero, $\det(g_{ab}) \neq 0$. In this case we say $g_{ab}$ is a non-degenerate metric. Since the determinant of a matrix is the product of its eigenvalues $\lambda_c$, $\det(g_{ab}) = \Pi_c \lambda_c$, the question reduces to whether $g_{ab}$ has any zero eigenvalues, i.e. whether an eigenvector $v^b$ exists such that $g_{ab} v^b = 0$ holds, in which case we would say that $g_{ab}$ is a degenerate metric. Since $g_{ab} = (f_a)_c{}^e (f_b)_e{}^c$, we see such a vector $v^b$ would exist if $(f_a)_c{}^e = 0$ held for all values of $c$ and $e$, since then regardless of the value of $v^b$ we would have $g_{ab} v^b = (f_a)_c{}^e (f_b)_e{}^c v^b = 0$. This would be the case if $[T_a, T_c] = (f_a)_c{}^e T_e = 0$ held, i.e. $[T_a, T_c] = 0$, i.e. $T_a$ is a generator which commutes with all elements of the Lie algebra.

An invariant subalgebra $I$ of a Lie algebra $\mathfrak{g}$ (also called an 'ideal') is a subset which satisfies $[\mathfrak{g}, I] \subset I$. In terms of the basis generators $T_a$ of $\mathfrak{g}$, if we denote the generators of $I$ by $T_{a'}$, then $[T_a, T_{b'}] = (f_a)_{b'}{}^{c'} T_{c'}$ holds in an ideal. If the generators of $I$ are commutative we say that $I$ is an abelian invariant subalgebra. Clearly if $[T_{a'}, T_b] = 0$ holds then $T_{a'}$ generates an abelian invariant subalgebra. We now define a '*semi-simple Lie algebra*' to be a Lie algebra possessing no abelian invariant subalgebras.

## A.5 Serre Presentation of a Semi-Simple Lie Algebra

Summarizing [85, Sec. 16.1], in a semi-simple Lie algebra $\mathfrak{g}$, a maximal set of simultaneously commuting generators $H_i$, $i = 1, ..., r$, $[H_i, H_j] = 0$, generating the 'Cartan subalgebra' of $\mathfrak{g}$, can always be found, for which the operators $[H_i, \ ]$ are also diagonalizable. We can then find a basis for the rest of the Lie algebra in which $[H_i, E_\alpha] = \alpha_i E_\alpha$ holds. The r-component vectors $\alpha = (\alpha_1, \ldots, \alpha_r)$ are referred to as 'root vectors' of the semi-simple Lie algebra $\mathfrak{g}$. It can be shown that if $\alpha$ is a root then $-\alpha$ is also a root, that $[E_\alpha, E_\beta]$ is non-zero if $\alpha + \beta$ is a root, and that a Euclidean inner product $(\alpha_a, \alpha_b)$ exists on the space of roots. In the space of roots, a 'positive root' is then defined (with respect to a given ordered basis) as a root vector whose first non-zero component is positive, and a 'simple root' is a positive root which cannot be written as a sum of two positive roots. It can be shown that there are $r$ simple roots $\alpha_a$, $a = 1, ..., r$, and the generators associated to these simple roots can be denoted $E_{\alpha_a}$. Given the simple roots we can define the 'Cartan matrix' $A_{ab} = 2\frac{(\alpha_a, \alpha_b)}{(\alpha_a, \alpha_a)}$. This matrix must satisfy $A_{aa} = 2$ by construction, it can be shown that this matrix must satisfy [87]

$$\begin{aligned} A_{ab} \leq 0 \quad , \quad (a \neq b) \quad , \quad \text{if} A_{ab} = 0 \ \text{ then } \ A_{ba} = 0 \quad (a \neq b) \quad , \\ v^a A_{ab} v^b \geq 0 \quad , \quad \text{for any real vector } v^a \quad . \end{aligned} \tag{A.33}$$

The positive-definite nature of $A_{ab}$, due to the last condition, then implies that $A_{ab}$, for $a \neq b$, can only the take on the values 0, $-2$, or $-3$, in which case $A_{ba}$ can then only take on the corresponding value 0, 1, or $-1$, or $A_{ab}$ can take on the value $-1$, in which case $A_{ba}$ can then take on any of the values $-1, -2, -3$. We may now define 'Chevalley generators' $E_a = \sqrt{\frac{2}{(\alpha_a,\alpha_a)}} E_{\alpha_a}$, $F_a = \sqrt{\frac{2}{(\alpha_a,\alpha_a)}} E_{-\alpha_a}$, $H_a = \frac{2\alpha_a^i H_i}{(\alpha_a,\alpha_a)}$, which satisfy the relations

$$[H_a, H_b] = 0 \quad , \quad [H_a, E_b] = A_{ab} E_b \quad , \quad [H_a, F_b] = -A_{ab} F_b \quad , \quad [E_a, F_b] = \delta_{a,b} H_a. \tag{A.34}$$

Defining (ad $X$)($Y$) = $[X, Y]$, Serre proved that, given a Cartan matrix $A_{ab}$ as defined above, a semi-simple Lie algebra can be reconstructed by generators $E_a, F_a, H_a$ satisfying the relations of equation (A.34) along with the conditions

$$(\text{ad } E_a)^{1-A_{ab}}(E_b) = 0 \quad , \quad (\text{ad } F_a)^{1-A_{ab}}(F_b) = 0 \quad . \tag{A.35}$$

Here we see the conditions of equation (A.35) are completely determined by the Cartan matrix $A_{ab}$. The theory of Kac-Moody algebras arises by relaxing the requirement that $A_{ab}$ be positive-definite.

## A.6 Fundamental Representations of $A_{D-1} = \mathrm{SL}(D)$

The Lie algebra of $A_{D-1} = \mathrm{SL}(D)$ is given by generators $\hat{K}^a{}_b$, with $a, b = 1, \ldots, n$, satisfying

$$\begin{aligned} [\hat{K}^a{}_b, \hat{K}^c{}_d] &= \delta_b{}^c \hat{K}^a{}_d - \delta^a{}_d \hat{K}^b{}_c \quad , \\ \sum_c \hat{K}^c{}_c &= 0. \end{aligned} \tag{A.36}$$

Defining fermionic oscillators

$$\hat{b}^{\dagger a} \quad , \quad \hat{b}_a \quad , \tag{A.37}$$

these satisfy, using the notation $\{A, B\} = AB + BA$, the anti-commutation relations

$$\{\hat{b}^{\dagger a}, \hat{b}_b\} = \delta^a{}_b \quad , \quad \{\hat{b}^{\dagger a}, \hat{b}^{\dagger b}\} = \{\hat{b}_a, \hat{b}_b\} = 0 \ . \tag{A.38}$$

These $A_{D-1}$ generators can now be written as

$$\hat{K}^a{}_b = \hat{b}^{\dagger a} \hat{b}_b - \frac{1}{D} \delta^a{}_b \sum_c \hat{b}^{\dagger c} \hat{b}_c \tag{A.39}$$

and satisfy the above commutation relations. Introducing a 'vacuum' satisfying $\hat{b}_a|0\rangle = 0$, the anti-symmetric tensors

$$T^{a_1 \ldots a_n} = \hat{b}^{\dagger a_1} \ldots \hat{b}^{\dagger a_n} |0\rangle \tag{A.40}$$

satisfy

$$\hat{K}^a{}_b T^{c_1\ldots c_n} = n\delta^{[c_1}{}_b T^{|a|c_2\ldots c_n]} - \frac{n}{D}\delta^a{}_b T^{c_1\ldots c_n} \quad , \tag{A.41}$$

and so provide irreducible representations of SL($D$) for each $n = 1, ..., D$. For $n = 1$ we find the vector representation, for $n = 2$ we find the rank-two anti-symmetric tensor representation, and so on. We refer to the irreducible representations

$$T^a, T^{a_1a_2}, \ldots \tag{A.42}$$

as the first, second, etc... 'fundamental representations' of SL($D$). One may also introduce symmetric tensor representations using bose oscillators, and introduce tensors with mixed symmetries (for example, $R^{a_1\ldots a_8,b}$ of SL(11) satisfying $R^{[a_1\ldots a_8,b]} = 0$) using the oscillator approach [18, Ch. 3].

## A.7 Level Decomposition of $A_4$ Over $A_3$

We now provide some motivation for the notion of a level decomposition used in the text. To do this we consider the Lie algebra $A_4$

$$\bullet - \bullet - \bullet - \bullet$$

and temporarily ignore its fourth node, 'deleting it', resulting in $A_3$, which we denote as

$$\bullet - \bullet - \bullet - \otimes$$

We know that we can describe $A_4$ by

$$\hat{K}^a{}_b, \hat{K}^a{}_5, \hat{K}^5{}_b, \hat{K}^5{}_5. \tag{A.43}$$

Here $\hat{K}^a{}_b$ is the adjoint representation of $A_3$, associated to the root

$$\alpha_{ab} = \mathbf{e}_a - \mathbf{e}_b \tag{A.44}$$

of $A_4$ which is also a root of $A_3$. The generator $\hat{K}^a{}_5$ associated to a positive root $\alpha_{a5} = \mathbf{e}_a - \mathbf{e}_5$ of $A_4$ transforms as a vector under $A_3$, i.e. it transforms as the first fundamental representation of $A_3$. The generator $\hat{K}^5{}_b$ associated to the negative root $-\alpha_{a5} = \mathbf{e}_5 - \mathbf{e}_a$ also transforms as a vector representation of $A_3$, and $\hat{K}^5{}_5$ transforms as a scalar representation of $A_3$, which we can interpret as being associated to a U(1) subgroup, i.e. $\hat{K}^a{}_b$ and $\hat{K}^5{}_5$ generate a $A_3 \times \mathrm{U}(1)$ subgroup of $A_4$. Thus we have completely described the Lie algebra $A_4$ in terms of representations of its $A_3 \times \mathrm{U}(1)$ subalgebra by deleting the fourth node of $A_4$.

The simple roots of $A_4$ are $\alpha_i = \mathbf{e}_i - \mathbf{e}_{i+1}$ for $i = 1, \ldots, 4$ are associated to the four nodes of the Dynkin diagram. Note $\alpha_{a5} = \alpha_{a4} + \alpha_4$. We can refer to the 'level' of a generator of $A_4$ as the number of times the simple root $\alpha_4$ of $A_4$ appears, positive level if $+\alpha_4$ appears, negative level if $-\alpha_4$ appears. Thus $\hat{K}^a{}_b$ and $\hat{K}^5{}_5$ are of level zero, while $\hat{K}^a{}_5$ is of level one, and $\hat{K}^5{}_a$ is of level minus one. Using the program [49] we find which agrees with this analysis. Recall that $[R^\alpha, R^\beta]$ is non-zero if $\alpha + \beta$ is a root [85]. Thus in $[\hat{K}^a{}_b, \hat{K}^c{}_d]$ we see from $\mathbf{e}_a - \mathbf{e}_b + \mathbf{e}_c - \mathbf{e}_d$ that we will find $\hat{K}^a{}_d$ if $b = c$ and $-K^c{}_b$ if $a = d$ in order to end up with a root consistent with the index symmetries. Thus, the commutator of a level zero generator with a level one generator cannot affect the level since it involves adding roots, so one we can always expect to find a representation of the level zero algebra at level one, and similarly in general the levels of roots add. These properties generalize to Kac-Moody algebras [85, Sec. 16.6] and we will assume these properties.

Table A.1: $A_3$ Representations in $A_4$

| $l$ | $A_3$ Irrep | $A_4$ Root | $\alpha^2$ | $d_r$ | $\mu$ | $R^\alpha$ |
|---|---|---|---|---|---|---|
| -1 | 0 0 1 | 0 0 0 -1 | 2 | 4 | 1 | $\hat{K}^5{}_b$ |
| 0 | 0 0 0 | 0 0 0 0 | 0 | 1 | 1 | $\hat{K}^5{}_5$ |
| 0 | 1 0 1 | 1 1 1 0 | 2 | 15 | 1 | $\hat{K}^a{}_b$ |
| 1 | 1 0 0 | 1 1 1 1 | 2 | 4 | 1 | $\hat{K}^a{}_5$ |

If we had instead deleted the first node of $A_4$, we would expect the exact same analysis (just replacing 4 by 1 above). Technically, since we originally attached the $A_4$ node to the third node of $A_3$, we would actually have expected to find the rank-three anti-symmetric tensor representation of $A_3$ at level one, given by $T^{abc}$, just as we found the vector representation by attaching a node to the first node of $A_3$ (a property used when first motivating the $l_1$ representation in Section 3.1). However the third fundamental representation is dual to the vector representation since $T_a = \varepsilon_{abcd}T^{bcd}$ thus at level one the above representation is indeed the vector representation of $A_3$ as expected.

## A.8 Constructing New Lie Algebras From Old Ones

Semi-simple Lie algebras are the Lie algebras where an intrinsic invertible metric $g_{ab}$ can be introduced so as to allow for an unambiguous 'group-intrinsic' notion of raising and lowering indices on the generators of the group. Starting from the four families of so-called 'classical semi-simple Lie algebras', one can interpret the search for further, less obvious, semi-simple Lie algebras as amounting to the following. Given the classical Lie algebras, we can construct vector, tensor, or spinor, fundamental representations of these Lie algebras. The question is now whether we can construct a new semi-simple Lie algebra simply by starting from that given classical Lie algebra, and adding extra generators which arose as fundamental representations of that Lie algebra. This fixes the commutation relations between the semi-simple Lie algebra generators and the generators of the fundamental representation. The only question now is whether one can propose commutation relations, involving those generators associated to the fundamental representations, such that the result is a semi-simple Lie algebra.

For example, given the semi-simple Lie algebra $\mathrm{SO}(1,3)$ generated by $M_{ab}$, we can try to form a new Lie algebra by the addition of the first fundamental representation $P_c$ to these. The commutation relations $[M^{ab}, M_{cd}] = -4\delta^{[a}{}_{[c}M^{b]}{}_{d]}$ and $[M_{ab}, P_c] = 2\eta_{c[a}P_{b]}$ are fixed by assumption. The only question is as to how $[P_a, P_b]$ is defined[2]. We now simply state that if one takes $[P_a, P_b] = 0$, the result, referred to as the 'Poincaré algebra', is *not* a semi-simple Lie algebra. For example, if we use the differential operator representation $P_a = \partial_a$ and $M_{ab} = 2x_{[a}\partial_{b]}$, this is completely natural. However, if one takes a clifford algebra representation $P_a = \gamma_a/R$, where $\gamma_a$ is a clifford algebra element satisfying $\{\gamma_a, \gamma_b\} = 2\eta_{ab}I$, and $R$ is a constant, and $M_{ab} = \frac{1}{4}[\gamma_a, \gamma_b]$, then $[P_a, P_b] = \frac{4}{R^2}M_{ab}$, and so one instead finds the *semi-simple* 'De-Sitter Group' $\mathrm{SO}(1,4)$ [39, Ch. 14]. Incidentally, taking the $R \to \infty$ limit is known as the 'Wigner-Inönü contraction' and results in the non-semi-simple Poincaré

[2]One can rewrite these as $[M^{ab}, M^{cd}] = f^{ab,cd}{}_{ef}M^{ef} + f^{ab,cd}{}_{e}P^e$ with $f^{ab,cd}{}_{e} = 0$ and $[M^{ab}, P^c] = f^{ab,c}{}_{d}P^d + f^{ab,c}{}_{ef}M^{ef}$ with $f^{ab,c}{}_{ef} = 0$, and so seek to analyse $[P^a, P^b] = f^{a,b}{}_{c}P^c + f^{a,b}{}_{cd}M^{cd}$.

algebra.

Similarly, the semi-simple Lie algebra $G_2$ arises in trying to construct a new semi-simple Lie algebra from SU(3) generated by $\hat{K}^a{}_b$, along with its vector representation $T^a$ and a dual vector $T_a = \frac{1}{2}\varepsilon_{abc}T^{bc}$, where $T^a$ is just the first fundamental representation of SU(3), and $T^{ab}$ is the second fundamental representation of SU(3). Thus $T^a$ and $T_a$ transform as a vector and dual vector respectively under $\hat{K}^a{}_b$. The only question now is whether the $[T^a, T^b]$, $[T_a, T_b]$ and $[T^a, T_b]$ commutators can be defined such that the overall result is a semi-simple Lie algebra. The result is that the Lie algebra of $G_2$ is [59], [18, Sec. 11.6]:

$$\begin{aligned} [\hat{K}^a{}_b, \hat{K}^c{}_d] &= \delta_b{}^c \hat{K}^a{}_d - \delta_d{}^a \hat{K}^c{}_b \quad , \\ [\hat{K}^a{}_b, T^c] &= \delta^c{}_b T^a - \frac{1}{3}\delta^a{}_b T^c \quad , \quad [\hat{T}^a{}_b, T_c] = -\delta^a{}_c T_b + \frac{1}{3}\delta^a{}_b T_c \quad , \\ [T^a, T^b] &= 2\varepsilon^{abc}A_c \quad , \quad [T_a, T_b] = -2\varepsilon_{abc}T^c \quad , \quad [T^a, T_b] = 3\hat{K}^a{}_b \ . \end{aligned} \tag{A.45}$$

Here the 2 coefficient is an arbitrary normalisation, whereas the 3 is fixed by the Jacobi identities.

As a simpler example, consider SU(2), $\hat{K}^a{}_b$ (where $\hat{K}^c{}_c = 0$), along with the **2**, $T_a$, of SU(2) and try to treat $\{\hat{K}^a_b, T_a\}$ as a *simple* Lie algebra. The commutators $[\hat{K}^a_b, \hat{K}^c_d]$ are $[\hat{K}^a{}_b, \hat{K}^c{}_d] = \delta^c{}_b\hat{K}^a{}_d - \delta^a{}_d\hat{K}^c{}_b$ and $[\hat{K}^a{}_b, T_c] = \delta^a{}_c T_b - \frac{1}{2}\delta^a{}_b T_c$. The only question is as to what $[T_a, T_b]$ must be. Since it comes from a tensor of the form $T_{ab} = T_{(ab)} + T_{[ab]}$ i.e. $\mathbf{2}\otimes\mathbf{2} = \mathbf{1}\oplus\mathbf{3}$, and the only possibility could be

$$[T_a, T_b] = A_1\epsilon_{ab}\hat{K}^c_{\ c} + 2A_2\epsilon_{[a|c}\hat{K}^c_{\ |b]}. \tag{A.46}$$

The first term is zero since the trace $\hat{K}^c_{\ c}$ is zero. The second term reduces to the trace since for example

$$[T_1, T_2] = 2\epsilon_{[1|c}\hat{K}^c_{\ |2]} = \epsilon_{1c}\hat{K}^c_{\ 2} - \epsilon_{2c}\hat{K}^c_{\ 1} = \epsilon_{12}\hat{K}^2_{\ 2} - \epsilon_{21}\hat{K}^1_{\ 1} = \hat{K}^2_{\ 2} + \hat{K}^1_{\ 1} = 0 \quad , \tag{A.47}$$

thus we must have $[T_a, T_b] = 0$. Therefore $\{T_a\}$ is an Abelian invariant subalgebra and so this new Lie algebra is not semi-simple.

Every non-semi-simple Lie algebra we discuss in this thesis is constructed by starting from a semi-simple Lie algebra, along with the addition of extra generators, which are simply representations of that semi-simple Lie algebra. For example, in Section 4.1 the generators $R^{a_1a_2a_3}$ and $R^{a_1\cdots a_6}$ in $G_{11}$ (and $E_{11}$) transform as representations of $\hat{K}^a{}_b$.

Thus, given a specific set of generators in advance, the above principles of ensuring the commutators have the correct index symmetries and are consistent with the Jacobi identities are the basis for constructing new algebras. When applied to Kac-Moody algebra, in principle one must also verify that the Serre relations discussed in Appendix B are also verified. The above level decomposition discussed in the last section, now applied to a Kac-Moody algebra decomposed with respect to a semi-simple Lie algebra, is what leads these kinds of set of generators in the first place, to which we can then apply the above principles.

# Appendix B

# Kac-Moody Algebras

We now briefly introduce some of the basic concepts of Kac-Moody algebras. For a more detailed introduction, please refer to [85, Ch. 16].

A Kac-Moody algebra is defined in terms of a generalized Cartan matrix - we focus on the symmetric case as these are the cases we will consider and the definitions simplify [25]. A symmetric generalized $r \times r$ Cartan matrix $A_{ab}, a, b = 1, ..., r$ is a matrix such that

- $A_{aa} = 2$,
- $A_{ab}$ is a negative integer or zero for $a \neq b$.

We also consider the case where the matrix $A_{ab}$ cannot be reduced to block-diagonal form by swapping rows and columns, so that the associated algebra will not consist of commuting subalgebras. If the Cartan matrix is positive-definite, the associated Lie algebra can be shown to be a finite-dimensional Lie algebra, in which case the Cartan matrix is invertible. If it is positive semi-definite, the associated Lie algebra can be shown to be an affine Lie algebra, in which case the Cartan matrix is not invertible.

The Kac-Moody algebra associated to the above symmetric generalised Cartan matrix is the Lie algebra generated by the elements $H_a, E_a, F_a$, $a = 1, ..., r$, satisfying the following commutation relations

$$\begin{aligned} [H_a, H_b] = 0 \ , \qquad & [H_a, E_b] = A_{ab} E_b, \\ [E_a, F_b] = \delta_{ab} H_b \ , \qquad & [H_a, F_b] = -A_{ab} F_b, \\ [E_a, ...[E_a, E_b]...] = 0 \ , \qquad & [F_a, ...[F_a, F_b]...] = 0, \end{aligned} \tag{B.1}$$

where in the last line there are $1 - A_{ab}$ nested commutators involving $E_a$, and similarly for $F_a$. These are the Chevalley-Serre relations of a finite-dimensional simple Lie algebra when applied to a symmetric Cartan matrix. We refer to the subalgebra generated by the $H_a$ generators as the Cartan subalgebra, and a given $H_a$ commuted with the $r$ $E_{b_j}$ generators, $j = 1, ..., r$, generates an $r$-component vector (depending on $H_a$) $\alpha = \alpha(H_a) = (A_{ab_1}, ..., A_{ab_r})$ referred to as a weight/root vector, and a notion of positive roots can be constructed, with the $E_a$ associated to positive roots $\alpha$ and the $F_a$ associated to the negative roots $-\alpha$. Similarly a notion of simple roots $\alpha_a$, $a = 1, ..., r$, can be constructed. The Chevalley-Serre relations are invariant under the map $E_a \to -F_a$, $F_a \to -E_a$, $H_a \to -H_a$, referred to as the Cartan involution, and so the generators $E_a - F_a$, referred to as even, produce a subalgebra referred to as the Cartan-involution invariant subalgebra, while the combinations $E_a + F_a$, referred to as odd, are sent to their negatives.

Using a symmetric generalized Cartan matrix $A_{ab}$ a bilinear form $B$ can be defined in the associated Kac-Moody algebra by setting $B(E_a, F_b) = \delta_{a,b}$, $B(H_a, H_b) = A_{ab}$, and zero otherwise, where $B$ is symmetric $B(X, Y) = B(Y, X)$ and satisfies $B([X, Y], Z) = B(X, [Y, Z])$, the same properties that the Killing metric of a finite-dimensional semi-simple Lie algebra satisfies for a symmetric Cartan matrix. Using the bilinear form, the Cartan matrix can be shown to be expressed in terms of the simple roots as

$$A_{ab} = \frac{2(\alpha_a, \alpha_b)}{(\alpha_a, \alpha_a)}, \tag{B.2}$$

and since $A_{ab}$ is symmetric the simple roots $\alpha_a$ must all have the same length, which we fix as $(\alpha_a, \alpha_a) = 2$, $a = 1, ..., r$.

A symmetric $r \times r$ Cartan matrix can be represented by a diagram consisting of $r$ nodes labelled from 1 to $r$ which is such that between the nodes $a$ and $b$ there are $-A_{ab}$ lines, referred to as its Dynkin diagram. The Dynkin diagram of a finite-dimensional Lie algebra is said to be of finite type, and that of an affine Lie algebra said to be of affine type. We refer to a Kac-Moody algebra as a Lorentzian Kac-Moody algebra if its Dynkin diagram is a connected diagram possessing at least one node, referred to as a central node, whose deletion yields a diagram whose connected components are of finite type except for at most one of affine type. We will focus on the cases where at least one node exists such that its deletion yields a diagram whose connected components are all of finite type.

Recall [26, Sec.8.14] that for a finite-dimensional simple Lie algebra of rank $m$, an irreducible representation is fixed by specifying a highest weight $\mu$, that the highest weight satisfies $\frac{2\alpha^i\cdot\mu}{(\alpha^i,\alpha^i)} = l^i$ for non-negative integers $l^i$, leading to the expansion of the highest weight $\mu$ in terms of fundamental weights $\lambda_i$, $\mu = \sum_i l^i\lambda_i$, where the fundamental weights satisfy $\delta_{ik} = \frac{2(\alpha_i,\lambda_k)}{(\alpha_i,\alpha_i)}$. We thus define fundamental weights for a Kac-Moody algebra to be weights $\lambda_j$ defined by the condition

$$\delta_{ij} = \frac{2(\alpha_i, \lambda_j)}{(\alpha_i, \alpha_i)}. \tag{B.3}$$

The fundamental weights can be expanded in a basis of the simple roots as

$$\lambda_j = \sum_k \tfrac{(\alpha_j,\alpha_j)}{(\alpha_k,\alpha_k)}(A^{-1})_{jk}\alpha_k \quad , \tag{B.4}$$

and the inverse Cartan matrix can be expressed in terms of the highest weights via

$$(A^{-1})_{ij} = \frac{2(\lambda_i, \lambda_j)}{(\alpha_i, \alpha_i)} \quad . \tag{B.5}$$

Thus, assuming the roots are normalized to 2 we see from $\lambda_j = \sum_k (A^{-1})_{jk}\alpha_k$ that the simple roots $\alpha_i$ can be expressed in terms of the fundamental weights $\lambda_j$ via

$$\alpha_i = \sum_j A_{ij}\lambda_j. \tag{B.6}$$

Labelling the simple roots of the reduced diagram as $\alpha_1, ..., \alpha_{r-1}$ and the root of the cenral node as $\alpha_c$, the Cartan matrix between the central root and any of the finite type roots is $A_{ci} = (\alpha_c, \alpha_i)$. On expressing $\alpha_c$ in terms of the simple root basis $\alpha_i$ and a root $x$ orthogonal to the space of simple roots of finite type, we can expand $\alpha_c$ in terms of the fundamental weights of the reduced diagram and $x$

$$\alpha_c = -\nu + x \quad \text{with} \quad \nu = -\sum_{i=1}^{r-1} A_{ci}\lambda_i \quad , \tag{B.7}$$

since this directly satisfies $A_{ci} = (\alpha_c, \alpha_i)$, leaving only the condition

$$A_{cc} = 2 = \nu^2 + x^2. \tag{B.8}$$

Using the simple roots we can build up the positive roots (and thus the entire algebra) as in the case of finite-dimensional simple Lie algebras. However, as in the finite type case, the possible roots must satisfy certain conditions which we describe next. An arbitrary positive root of the Kac-Moody algebra can be expanded in terms of the simple roots using non-negative integer linear combinations

$$\alpha = l\alpha_c + \sum_{i=1}^{r-1} m_i \alpha_i \quad , \tag{B.9}$$

where we refer to the (non-negative) integer $l$ as the level of the (positive) root $\alpha$ (with similar statements for negative roots), and note that $l$ is the number of times $E_{\alpha_c}$ in a multiple commutator which gives $E_\alpha$. The usefulness of our previous results involving the fundamental weights now arises as we can express this root in terms of the fundamental weights of the reduced diagram, referring to the Cartan matrix of the reduced (finite type) Dynkin diagram as $A^f$, via

$$\begin{aligned} \alpha &= l\alpha_c + \sum_{i=1}^{r-1} m_i\alpha_i = l(-\nu + x) + \sum_{i=1}^{r-1} m_i \sum_s A^f_{is}\lambda_s \\ &= lx - \Lambda \quad , \end{aligned} \tag{B.10}$$

where we labelled the part of the root involving the reduced diagram fundamental weights as

$$\Lambda = -l\sum_{i=1}^{r-1} A_{ci}\lambda_i - \sum_{i,j=1}^{r-1} m_i A^f_{ij}\lambda_j. \tag{B.11}$$

The root $\alpha$, at each given level $l$, can only contain a highest weight representation of the reduced Dynkin diagram if it actually contains a highest weight of that reduced Dynkin diagram. Since $\Lambda$ fully characterizes the reduced Dynkin diagram, we need only have that $\Lambda$ satify the condition that it be a highest weight for the reduced diagram, meaning [26, (8.73)] it must satisfy $\frac{2(\alpha_i,\Lambda)}{(\alpha_i,\alpha_i)} = p^i$ for $p^i$ a non-negative integer (we invoke the normalization $\alpha^2 = 2$ below). Thus the first condition that a root $\Lambda$ must satisfy at a given non-negative level $l$ is that $\Lambda$ expands as

$$\Lambda = \sum_{k=1}^{r-1} p_k\lambda_k \quad , \tag{B.12}$$

where $p_k$ is a positive integer (a zero integer means the representation trivially doesn't arise), where $p_k$ is defined by

$$p_k = (\Lambda, \alpha_k) = -lA_{ck} - \sum_i^{r-1} m_i A^f_{ik} \quad . \tag{B.13}$$

As we expect that the non-negative $m_i$ integers are fixed by the analysis, we isolate them by multiplying this expression on the right by the inverse reduced Cartan matrix

$$\sum_{k=1}^{r-1}(A^f)^{-1}_{kj}p_k = -l\sum_k A_{ck}(A^f)^{-1}_{kj} - m_j. \tag{B.14}$$

The roots are constrained by a further requirement - the roots of a Kac-Moody algebra with a symmetric Cartan matrix are such that any given root $\alpha$ has its squared length bounded by $2, 0, -2, ...$ [80] and so $\alpha = lx - \Lambda = lx - \sum_{i=1}^{r-1} p_i\lambda_i$ must satisfy

$$\begin{aligned}\alpha^2 &= l^2x^2 + \sum_{i,j=1}^{r-1} p_i(\lambda_i, \lambda_j)p_j \\ &= l^2x^2 + \sum_{i,j=1}^{r-1} p_i(A^f)^{-1}_{ij}p_j \leq 2, 0, -2, ...\end{aligned} \tag{B.15}$$

The value of $x^2$ can be expressed in terms of the determinant of the Cartan matrix $A_C$ and the reduced Cartan matrix $A_{C_R}$

$$x^2 = \frac{\det A_C}{\det A_{C_R}} \quad . \tag{B.16}$$

This can be seen by noting that, given $(A_C)_{ab} = (\alpha_a, \alpha_b)$, if we now expand the roots in an orthonormal basis $(A_C)_{ab} = \alpha^i_a\alpha^j_b\delta_{ij}$ (or $G_{ij}$ more generally) we see $\det(A_C) = (\det\alpha^i_a)(\det\alpha^j_b)\det(\delta_{ij}) = (\det\alpha^i_a)^2$. Since the $x$ component is orthogonal to all other components, and the other components of the matrix $\alpha^i_a$ have no components in the $x$ direction, the determinant should factor out to $x$ times the square root of the determinant of $C_R$, so that $\det A_C = x^2 \det A_{C_R}$, giving the above result.

We thus find it is necessary for $(p_1, ..., p_{r-1})$ to satisfy the two conditions

$$\begin{aligned}\sum_{k=1}^{r-1}(A^f)^{-1}_{kj}p_k &= -l\sum_k A_{ck}(A^f)^{-1}_{kj} - m_j \quad , \\ \alpha^2 = l^2\frac{\det A_C}{\det A_{C_R}} &+ \sum_{i,j=1}^{r-1} p_i(A^f)^{-1}_{ij}p_j \leq 2, 0, -2, ... \quad ,\end{aligned} \tag{B.17}$$

at a given positive level $l$ in order for the root $\alpha$ to be a root containing the highest weight of an irreducible representation of the reduced Dynkin diagram. These conditions, however, are not sufficient to guarantee that a given root $\alpha$ is actually a root of the Kac-Moody algebra. This is because roots of the Kac-Moody algebra are generated by the simple roots subject to the Chevalley-Serre relations, which restricts the possible roots, so the above conditions will have to at least contain the roots of the Kac-Moody algebra, but they may allow for more solutions that are not in the Kac-Moody algebra.

A general method for determining whether a highest weight representation actually occurs is to compute the 'outer multiplicity' of that representation [50], denoted $\mu$, which is the number of times that representation actually occurs. However, one can also work directly by working out the algebra of the possible generators up to a given level $l$ and enforcing consistency checks such as the Jacobi identity between all the generators up to that level, and verifying the Chevalley-Serre relations between the generators that remain.

Furthermore, if a representation of the reduced diagram occurs at a given level $l$, its roots will not affect the level, so that if a representation occurs it must be included entirely in, say, the positive root space, with a copy in the negative root space as $-\alpha$ is a root when $\alpha$ is a root.

Next we note that the fundamental weights of the reduced Dynkin diagrams are associated to tensors and spinors [60, Ch. 8]. In the case where the reduced Dynkin diagram is the simple Lie algebra $A_{r-1}$, the case we will consider, the fundamental weight $\lambda_{r-k}$ for $k = 1, ..., r-1$ can be associated to a rank $k$ anti-symmetric tensor $A^{i_1 \cdots i_k}$ with raised indices, and we can define lowered tensors as $A_{i_1..i_m} = \varepsilon_{i_1..i_{r-1}} A^{i_{m+1} \cdots i_{r-1}}$ [26, Ch. 13]. We thus see at a given level $l$ the admissible $p_k$ values should produce tensors with blocks of anti-symmetric indices. These tensors will satisfy constraints that can be read off from the associated Dynkin diagrams.

We thus see that once we are given the Dynkin diagram of a Lorentzian Kac-Moody algebra, whose connected components are of finite type, and we specify the central node, we must compute its Cartan matrix and its determinant, as well as the inverse Cartan matrix and determinant of the Cartan matrix of the reduced Dynkin diagram. We then need to solve the equations (B.17) to determine the admissible $(p_1, ..., p_{r-1})$ values at each level $l$, which we then relate to tensors with blocks of anti-symmetric indices. Combining these results with the method of nonlinear realizations will then produce field content for a theory possessing the symmetries of the associated Kac-Moody algebra.

This process of computing these quantities has been automated for low levels in the Nutma computer program SimpLie [49]. In this program one may simply draw the Dynkin diagram of a Lorentzian Kac-Moody algebra, specify the central node, and the program produces the $(p_1, ..., p_{r-1})$ vectors up to a selected level, along with the dimension $d_r$ of that representation, and the outer multiplicity $\mu$.

We now specialise our considerations to Kac-Moody algebras possessing the algebra $A_{r-1}$ when we delete the central node, as it is known the gravity sector of a nonlinear realization must involve $A_{r-1}$ [6], [41]. We will find that the generators found by the above methods at low levels will not produce a generator that could be interpreted as a translation generator. In the case of the Poincaré algebra, it is known that the translation generator arises from an abelian representation of the Lorentz algebra, and so we seek a representation containing the fundamental weight $\lambda_{r-1}$, or equivalently $\lambda_1$, of the reduced diagram $A_{r-1}$ which can naturally be associated to translations.

To find such a representation we note that, although we have so far only considered the fundamental representations $\lambda_i$, $i = 1, ..., r-1$, of the reduced Dynkin diagram, we can also define fundamental weights for a Lorentzian Kac-Moody algebra. These can be shown to take the form [25]

$$l_i = \lambda_i + \frac{(\lambda_i, \nu)}{(x,x)} x \quad , \quad i = 1, ..., r-1, \quad , \quad l_c = \frac{1}{(x,x)} x. \tag{B.18}$$

These fundamental weights provide irreducible representations for the whole Kac-Moody algebra. Note $l_1 = \lambda_1 + \frac{(\lambda_1,\nu)}{(x,x)} x$ contains the first fundamental weight $\lambda_1 = \lambda_{r-(r-1)}$ of the reduced diagram, which is the vector representation of the reduced diagram. In the case that the reduced diagram is $A_{r-1}$, this fundamental weight $\lambda_{r-(r-1)}$ is associated to a rank $(r-1)$ tensor, or equivalently a rank 1 vector. For example, in the case of $A_1^{+++}$ decomposed with respect to the reduced diagram $A_{4-1}$, we have seen in Chapter 3 that the rank $(4-1)$ tensor $P^{bcd}$ is an element of the vector representation, and its rank-1 dual is $P_a = \frac{1}{3!}\varepsilon_{abcd}P^{bcd}$. We refer to the vector representation of a Lorentzian Kac-Moody algebra as the $l_1$ representation. This representation can also be classified in terms of levels, and SimpLie produces the tensor content of the $l_1$ representation, as we discuss in Chapters 3 and 4. It is consistent to treat the generators in the $l_1$ representation as commutative [84]. With this background, one could in principle compute the roots of $E_{11}$ and its $l_1$ representation directly, rather than relying on SimpLie. A review is given in [85, Sec. 16.7], and [63].

# Appendix C

# IIA Commutators

In this Appendix we give the IIA algebra to level one, including only those generators involving the underlying $E_{11}$ generators to $E_{11}$ level three, that is needed in Section 6.2. The algebra of the generators at IIA level zero is [82]

$$\begin{gathered}[K^{\underline{a}}{}_{\underline{b}}, K^{\underline{c}}{}_{\underline{d}}] = \delta^{\underline{c}}{}_{\underline{b}} K^{\underline{a}}{}_{\underline{d}} - \delta^{\underline{a}}{}_{\underline{d}} K^{\underline{c}}{}_{\underline{b}} \ , \ [K^{\underline{a}}{}_{\underline{b}}, \tilde{R}] = 0 \ , \ [K^{\underline{a}}{}_{\underline{b}}, R^{\underline{cd}}] = 2\delta^{[\underline{c}}{}_{\underline{b}} R^{|\underline{a}|\underline{d}]} \ , \\ [K^{\underline{a}}{}_{\underline{b}}, R_{\underline{cd}}] = -2\delta^{\underline{a}}{}_{[\underline{c}} R_{|\underline{b}|\underline{d}]} \ , \ [\tilde{R}, \tilde{R}] = 0 \ , \ [\tilde{R}, R^{\underline{ab}}] = 0 \ , \ [\tilde{R}, R_{\underline{ab}}] = 0 \ , \\ [R^{\underline{ab}}, R^{\underline{cd}}] = 0 \ , \ [R^{\underline{ab}}, R_{\underline{cd}}] = 4\delta^{[\underline{a}}{}_{[\underline{c}} K^{\underline{b}]}{}_{\underline{d}]} \ , \ [R_{\underline{ab}}, R_{\underline{cd}}] = 0 \ .\end{gathered} \tag{C.1}$$

We note that these generators satisfy the algebra of $D_{10} \otimes GL(1) = SO(10, 10) \otimes GL(1)$.

The commutators between IIA level one generators are given by

$$\begin{gathered}[R^{\underline{a}}, R^{\underline{b}}] = 0 \ , \ [R^{\underline{a}}, R^{\underline{b}_1\underline{b}_2\underline{b}_3}] = 0 \ , \ [R^{\underline{a}}, R^{\underline{b}_1\cdots\underline{b}_5}] = -R^{\underline{a}\,\underline{b}_1\cdots\underline{b}_5} \ , \\ [R^{\underline{a}}, R^{\underline{b}_1\cdots\underline{b}_7}] = -2R^{\underline{a}\,\underline{b}_1\cdots\underline{b}_7} - 7R^{\underline{a}[\underline{b}_1\cdots\underline{b}_6,\underline{b}_7]} \ , \ [R^{\underline{a}_1\underline{a}_2\underline{a}_3}, R^{\underline{b}_1\underline{b}_2\underline{b}_3}] = 2R^{\underline{a}_1\underline{a}_2\underline{a}_3\underline{b}_1\underline{b}_2\underline{b}_3} \ , \\ [R^{\underline{a}_1\underline{a}_2\underline{a}_3}, R^{\underline{b}_1\cdots\underline{b}_5}] = -3R^{\underline{b}_1\cdots\underline{b}_5[\underline{a}_1\underline{a}_2,\underline{a}_3]} \ , \ [R^{\underline{a}_1\underline{a}_2\underline{a}_3}, R^{\underline{b}_1\cdots\underline{b}_7}] = 0 \ , \\ [R^{\underline{a}_1\cdots\underline{a}_5}, R^{\underline{b}_1\cdots\underline{b}_5}] = 0 \ , \ [R^{\underline{a}_1\cdots\underline{a}_5}, R^{\underline{b}_1\cdots\underline{b}_7}] = 0 \ , \ [R^{\underline{a}_1\cdots\underline{a}_7}, R^{\underline{b}_1\cdots\underline{b}_7}] = 0 \ ,\end{gathered} \tag{C.2}$$

and for those with IIA level minus one by

$$\begin{gathered}[R_{\underline{a}}, R_{\underline{b}}] = 0 \ , \ [R_{\underline{a}}, R_{\underline{b}_1\underline{b}_2\underline{b}_3}] = 0 \ , \ [R_{\underline{a}}, R_{\underline{b}_1\cdots\underline{b}_5}] = +R_{\underline{a}\,\underline{b}_1\cdots\underline{b}_5} \ , \\ [R_{\underline{a}}, R_{\underline{b}_1\cdots\underline{b}_7}] = +2R_{\underline{a}\,\underline{b}_1\cdots\underline{b}_7} + 7R_{\underline{a}[\underline{b}_1\cdots\underline{b}_6,\underline{b}_7]} \ , \ [R_{\underline{a}_1\underline{a}_2\underline{a}_3}, R_{\underline{b}_1\underline{b}_2\underline{b}_3}] = 2R_{\underline{a}_1\underline{a}_2\underline{a}_3\underline{b}_1\underline{b}_2\underline{b}_3} \ , \\ [R_{\underline{a}_1\underline{a}_2\underline{a}_3}, R_{\underline{b}_1\cdots\underline{b}_5}] = -3R_{\underline{b}_1\cdots\underline{b}_5[\underline{a}_1\underline{a}_2,\underline{a}_3]} \ , \ [R_{\underline{a}_1\underline{a}_2\underline{a}_3}, R_{\underline{b}_1\cdots\underline{b}_7}] = 0 \ , \\ [R_{\underline{a}_1\cdots\underline{a}_5}, R_{\underline{b}_1\cdots\underline{b}_5}] = 0 \ , \ [R_{\underline{a}_1\cdots\underline{a}_5}, R_{\underline{b}_1\cdots\underline{b}_7}] = 0 \ , \ [R_{\underline{a}_1\cdots\underline{a}_7}, R_{\underline{b}_1\cdots\underline{b}_7}] = 0 \ .\end{gathered} \tag{C.3}$$

The algebra of IIA level zero generators with IIA level one generators are given by

$$\begin{aligned}
&[K^{\underline{a}}{}_{\underline{b}}, R^{\underline{c}}] = R^{\underline{a}}\delta^{\underline{c}}{}_{\underline{b}} - \frac{1}{2}\delta^{\underline{a}}{}_{\underline{b}}R^{\underline{c}} \ , \ [K^{\underline{a}}{}_{\underline{b}}, R^{\underline{c}_1\underline{c}_2\underline{c}_3}] = 3\delta^{[\underline{c}_1}{}_{\underline{b}}R^{|\underline{a}|\underline{c}_2\underline{c}_3]} - \frac{1}{2}\delta^{\underline{a}}{}_{\underline{b}}R^{\underline{c}_1\underline{c}_2\underline{c}_3} \ ,\\
&\qquad [K^{\underline{a}}{}_{\underline{b}}, R^{\underline{c}_1\cdots\underline{c}_5}] = 5\delta^{[\underline{c}_1}{}_{\underline{b}}R^{|\underline{a}|\underline{c}_2\cdots\underline{c}_5]} - \frac{1}{2}\delta^{\underline{a}}{}_{\underline{b}}R^{\underline{c}_1\cdots\underline{c}_5} \ ,\\
&\qquad [K^{\underline{a}}{}_{\underline{b}}, R^{\underline{c}_1\cdots\underline{c}_7}] = 7\delta^{[\underline{c}_1}{}_{\underline{b}}R^{|\underline{a}|\underline{c}_2\cdots\underline{c}_7]} - \frac{1}{2}\delta^{\underline{a}}{}_{\underline{b}}R^{\underline{c}_1\cdots\underline{c}_7} \ ,\\
&[\tilde{R}, R^{\underline{a}}] = -3R^{\underline{a}} \ , \ [\tilde{R}, R^{\underline{a}_1\underline{a}_2\underline{a}_3}] = -3R^{\underline{a}_1\underline{a}_2\underline{a}_3} \ , \ [\tilde{R}, R^{\underline{a}_1\cdots\underline{a}_5}] = -3R^{\underline{a}_1\cdots\underline{a}_5} \ ,\\
&\qquad [\tilde{R}, R^{\underline{a}_1\cdots\underline{a}_7}] = -3R^{\underline{a}_1\cdots\underline{a}_7} \ , \ [R^{\underline{a}_1\underline{a}_2}, R^{\underline{b}}] = -R^{\underline{a}_1\underline{a}_2\underline{b}} \ ,\\
&[R^{\underline{a}_1\underline{a}_2}, R^{\underline{b}_1\underline{b}_2\underline{b}_3}] = -2R^{\underline{a}_1\underline{a}_2\underline{b}_1\underline{b}_2\underline{b}_3} \ , \ [R^{\underline{a}_1\underline{a}_2}, R^{\underline{b}_1..\underline{b}_5}] = -R^{\underline{a}_1\underline{a}_2\underline{b}_1..\underline{b}_5} \ ,\\
&[R^{\underline{a}_1\underline{a}_2}, R^{\underline{b}_1..\underline{b}_7}] = 0 \ , \ [R_{\underline{a}_1\underline{a}_2}, R^{\underline{b}}] = 0 \ , \ [R_{\underline{a}_1\underline{a}_2}, R^{\underline{b}_1\underline{b}_2\underline{b}_3}] = -6\delta^{[\underline{b}_1\underline{b}_2}_{\underline{a}_1\underline{a}_2}R^{\underline{b}_3]} \ ,\\
&[R_{\underline{a}_1\underline{a}_2}, R^{\underline{b}_1..\underline{b}_5}] = -10R^{[\underline{b}_1\underline{b}_2\underline{b}_3}\delta^{\underline{b}_4\underline{b}_5]}_{\underline{a}_1\underline{a}_2} \ , \ [R_{\underline{a}_1\underline{a}_2}, R^{\underline{b}_1..\underline{b}_7}] = -42R^{[\underline{b}_1..\underline{b}_5}\delta^{\underline{b}_6\underline{b}_7]}_{\underline{a}_1\underline{a}_2} \ .
\end{aligned} \tag{C.4}$$

The commutators of the IIA level zero with the IIA level minus one are given by

$$\begin{aligned}
&[K^{\underline{a}}{}_{\underline{b}}, R_{\underline{c}}] = -R_{\underline{b}}\delta^{\underline{a}}{}_{\underline{c}} + \frac{1}{2}\delta^{\underline{a}}{}_{\underline{b}}R_{\underline{c}} \ , \ [K^{\underline{a}}{}_{\underline{b}}, R_{\underline{c}_1\underline{c}_2\underline{c}_3}] = -3\delta^{[\underline{a}}{}_{\underline{c}_1}R_{|\underline{b}|\underline{c}_2\underline{c}_3]} + \frac{1}{2}\delta^{\underline{a}}{}_{\underline{b}}R^{\underline{c}_1\underline{c}_2\underline{c}_3} \ ,\\
&\qquad [K^{\underline{a}}{}_{\underline{b}}, R_{\underline{c}_1\cdots\underline{c}_5}] = -5\delta^{[\underline{a}}{}_{\underline{c}_1}R_{|\underline{b}|\underline{c}_2\cdots\underline{c}_5]} + \frac{1}{2}\delta^{\underline{a}}{}_{\underline{b}}R_{\underline{c}_1\cdots\underline{c}_5} \ ,\\
&\qquad [K^{\underline{a}}{}_{\underline{b}}, R_{\underline{c}_1..\underline{c}_7}] = -7\delta^{\underline{b}}{}_{[\underline{c}_1}R_{|\underline{b}|\underline{c}_2..\underline{c}_7]} + \frac{1}{2}\delta^{\underline{a}}{}_{\underline{b}}R_{\underline{c}_1..\underline{c}_7},\\
&[\tilde{R}, R_{\underline{a}}] = 3R_{\underline{a}} \ , \ [\tilde{R}, R_{\underline{a}_1\underline{a}_2\underline{a}_3}] = 3R_{\underline{a}_1\underline{a}_2\underline{a}_3} \ , \ [\tilde{R}, R_{\underline{a}_1\cdots\underline{a}_5}] = 3R_{\underline{a}_1\cdots\underline{a}_5} \ ,\\
&[\tilde{R}, R_{\underline{a}_1\cdots\underline{a}_7}] = 3R_{\underline{a}_1\cdots\underline{a}_7} \ , \ [R^{\underline{a}_1\underline{a}_2}, R_{\underline{b}}] = 0 \ , \ [R^{\underline{a}_1\underline{a}_2}, R_{\underline{b}_1\underline{b}_2\underline{b}_3}] = 6\delta^{\underline{a}_1\underline{a}_2}_{[\underline{b}_1\underline{b}_2}R_{\underline{b}_3]} \ ,\\
&[R^{\underline{a}_1\underline{a}_2}, R_{\underline{b}_1..\underline{b}_5}] = -10R_{[\underline{b}_1\underline{b}_2\underline{b}_3}\delta^{\underline{a}_1\underline{a}_2}_{\underline{b}_4\underline{b}_5]} \ , \ [R^{\underline{a}_1\underline{a}_2}, R_{\underline{b}_1..\underline{b}_7}] = -42R_{[\underline{b}_1..\underline{b}_5}\delta^{\underline{a}_1\underline{a}_2}_{\underline{b}_6\underline{b}_7]} \ ,\\
&\qquad [R_{\underline{a}_1\underline{a}_2}, R_{\underline{b}}] = R_{\underline{a}_1\underline{a}_2\underline{b}} \ , \ [R_{\underline{a}_1\underline{a}_2}, R_{\underline{b}_1\underline{b}_2\underline{b}_3}] = -2R_{\underline{a}_1\underline{a}_2\underline{b}_1\underline{b}_2\underline{b}_3} \ ,\\
&\qquad [R_{\underline{a}_1\underline{a}_2}, R_{\underline{b}_1..\underline{b}_5}] = -2R_{\underline{a}_1\underline{a}_2\underline{b}_1..\underline{b}_5} \ , \ [R_{\underline{a}_1\underline{a}_2}, R_{\underline{b}_1..\underline{b}_7}] = 0 \ .
\end{aligned} \tag{C.5}$$

The IIA level one generators with IIA level minus generators are given by

$$
\begin{aligned}
&[R^{\underline{a}}, R_{\underline{b}}] = K^{\underline{a}}{}_{\underline{b}} - \frac{1}{6}\delta^{\underline{a}}{}_{\underline{b}}(3D - \tilde{R}) \ \ , \ \ [R^{\underline{a}}, R_{\underline{b}_1\underline{b}_2\underline{b}_3}] = -3\delta^{\underline{a}}{}_{[\underline{b}_1} R_{\underline{b}_2\underline{b}_3]} \ , \\
&[R^{\underline{a}}, R_{\underline{b}_1\ldots\underline{b}_5}] = 0 \ \ , \ \ [R^{\underline{a}}, R_{\underline{b}_1\ldots\underline{b}_7}] = 0 \ , \ [R^{\underline{a}_1\underline{a}_2\underline{a}_3}, R_{\underline{b}}] = -3\delta^{[\underline{a}_1}{}_{\underline{b}} R^{\underline{a}_2\underline{a}_3]} \ \ , \\
&[R^{\underline{a}_1\underline{a}_2\underline{a}_3}, R_{\underline{b}_1\underline{b}_2\underline{b}_3}] = 18\delta^{[\underline{a}_1\underline{a}_2}_{[\underline{b}_1\underline{b}_2} K^{\underline{a}_3]}{}_{\underline{b}_3]} - \delta^{\underline{a}_1\underline{a}_2\underline{a}_3}_{\underline{b}_1\underline{b}_2\underline{b}_3}(3D - \tilde{R}) \ \ , \\
&[R^{\underline{a}_1\underline{a}_2\underline{a}_3}, R_{\underline{b}_1..\underline{b}_5}] = 30\delta^{\underline{a}_1\underline{a}_2\underline{a}_3}_{[\underline{b}_1\underline{b}_2\underline{b}_3} R_{\underline{b}_4\underline{b}_5]} \ , \ [R^{\underline{a}_1\underline{a}_2\underline{a}_3}, R_{\underline{b}_1..\underline{b}_7}] = 0 \ , \\
&[R^{\underline{a}_1..\underline{a}_5}, R_{\underline{b}}] = 0 \ , \ [R^{\underline{a}_1..\underline{a}_5}, R_{\underline{b}_1\underline{b}_2\underline{b}_3}] = -30\delta^{[\underline{a}_1\underline{a}_2\underline{a}_3}_{[\underline{b}_1\underline{b}_2\underline{b}_3} R^{\underline{a}_4\underline{a}_5]} \ \ , \\
&[R^{\underline{a}_1..\underline{a}_5}, R_{\underline{b}_1\ldots\underline{b}_5}] = -5\cdot 5\cdot 6\delta^{[\underline{a}_1..\underline{a}_4}_{[\underline{b}_1\ldots\underline{b}_4} K^{\underline{a}_5]}{}_{\underline{b}_5]} + 5\delta^{\underline{a}_1..\underline{a}_5}_{\underline{b}_1\ldots\underline{b}_5}(3D - \tilde{R}) \ \ , \\
&[R^{\underline{a}_1..\underline{a}_5}, R_{\underline{b}_1\ldots\underline{b}_7}] = -9\cdot 70\delta^{\underline{a}_1..\underline{a}_5}_{[\underline{b}_1\ldots\underline{b}_5} R_{\underline{b}_6\underline{b}_7]} \ , \ [R^{\underline{a}_1..\underline{a}_7}, R_{\underline{b}}] = 0 \ , \\
&[R^{\underline{a}_1..\underline{a}_7}, R_{\underline{b}_1\underline{b}_2\underline{b}_3}] = 0 \ , \ [R^{\underline{a}_1..\underline{a}_7}, R_{\underline{b}_1\ldots\underline{b}_5}] = 9\cdot 70 R^{[\underline{a}_1\underline{a}_2}\delta^{\underline{a}_3\ldots\underline{a}_7]}_{\underline{b}_1\ldots\underline{b}_5} \ \ , \\
&[R^{\underline{a}_1..\underline{a}_7}, R_{\underline{b}_1..\underline{b}_7}] = 7\cdot 7\cdot 180\delta^{[\underline{a}_1\ldots\underline{a}_6}_{[\underline{b}_1\ldots\underline{b}_6} K^{\underline{a}_7]}{}_{\underline{b}_7]} - 7\cdot 6\cdot 5\delta^{\underline{a}_1..\underline{a}_7}_{\underline{b}_1..\underline{b}_7}(3D - \tilde{R}) \ \ ,
\end{aligned}
\tag{C.6}
$$

where we have set

$$
D = \sum_{\underline{a}=0}^{9} K^{\underline{a}}{}_{\underline{a}}. \tag{C.7}
$$

We now give the commutators of the IIA generators with the $l_1$ generators. Again these results can be derived directly or equivalently from the commutators of the $E_{11} \otimes_s l_1$ algebra in its eleven dimensional formulation [20]. Here we consider the algebra to level one in the IIA and $l_1$ representations and again only to $E_{11}$ level three. The IIA level zero generators with the IIA level zero $l_1$ generators have the commutators

$$
\begin{aligned}
&[K^{\underline{a}}{}_{\underline{b}}, P_{\underline{c}}] = -\delta^{\underline{a}}{}_{\underline{c}} P_{\underline{b}} \ \ , \ \ [\tilde{R}, P_{\underline{c}}] = -3P_{\underline{c}} \ \ , \\
&[R^{\underline{a}_1\underline{a}_2}, P_{\underline{b}}] = -2\delta^{[\underline{a}_1}{}_{\underline{b}} Q^{\underline{a}_2]} \ \ , \ \ [R_{\underline{a}_1\underline{a}_2}, P_{\underline{b}}] = 0 \ \ , \\
&[K^{\underline{a}}{}_{\underline{b}}, Q^{\underline{c}}] = +\delta^{\underline{c}}{}_{\underline{b}} Q^{\underline{a}} \ \ , \ \ [\tilde{R}, Q^{\underline{c}}] = -3Q^{\underline{c}} \ \ , \\
&[R^{\underline{a}_1\underline{a}_2}, Q^{\underline{c}}] = 0 \ \ , \ \ [R_{\underline{a}_1\underline{a}_2}, Q^{\underline{c}}] = +2\delta^{\underline{c}}{}_{[\underline{a}_1} P_{\underline{a}_2]} \ \ .
\end{aligned}
\tag{C.8}
$$

The IIA level zero $IIA$ generators with the IIA level one $l_1$ generators have the commu-

tators

$$[K^{\underline{a}}{}_{\underline{b}}, Z] = -\frac{1}{2}\delta^{\underline{a}}{}_{\underline{b}} Z \ , \ [K^{\underline{a}}{}_{\underline{b}}, Z^{\underline{c}_1\underline{c}_2}] = 2\delta^{[\underline{c}_1}{}_{\underline{b}} Z^{|\underline{a}|\underline{c}_2]} - \frac{1}{2}\delta^{\underline{a}}{}_{\underline{b}} Z^{\underline{c}_1\underline{c}_2} \ ,$$
$$[K^{\underline{a}}{}_{\underline{b}}, Z^{\underline{c}_1\cdots\underline{c}_4}] = 4\delta^{[\underline{c}_1}{}_{\underline{b}} Z^{|\underline{a}|\underline{c}_2\underline{c}_3\underline{c}_4]} - \frac{1}{2}\delta^{\underline{a}}{}_{\underline{b}} Z^{\underline{c}_1\cdots\underline{c}_4} \ ,$$
$$[K^{\underline{a}}{}_{\underline{b}}, Z^{\underline{c}_1\cdots\underline{c}_6}] = 6\delta^{[\underline{c}_1}{}_{\underline{b}} Z^{|\underline{a}|\underline{c}_2\cdots\underline{c}_6]} - \frac{1}{2}\delta^{\underline{a}}{}_{\underline{b}} Z^{\underline{c}_1\cdots\underline{c}_6} \ ,$$
$$[\tilde{R}, Z] = -6Z \ , \ [\tilde{R}, Z^{\underline{a}_1\underline{a}_2}] = -6Z^{\underline{a}_1\underline{a}_2} \ , \ [\tilde{R}, Z^{\underline{a}_1\cdots\underline{a}_4}] = -6Z^{\underline{a}_1\cdots\underline{a}_4} \ , \tag{C.9}$$
$$[\tilde{R}, Z^{\underline{a}_1\cdots\underline{a}_6}] = -6Z^{\underline{a}_1\cdots\underline{a}_6} \ , \ [R^{\underline{a}_1\underline{a}_2}, Z] = Z^{\underline{a}_1\underline{a}_2} \ , \ [R_{\underline{a}_1\underline{a}_2}, Z] = 0 \ ,$$
$$[R^{\underline{a}_1\underline{a}_2}, Z^{\underline{b}_1\underline{b}_2}] = Z^{\underline{a}_1\underline{a}_2\underline{b}_1\underline{b}_2} \ , \ [R_{\underline{a}_1\underline{a}_2}, Z^{\underline{b}_1\underline{b}_2}] = 2\delta^{\underline{a}_1\underline{a}_2}_{\underline{b}_1\underline{b}_2} Z \ ,$$
$$[R^{\underline{a}_1\underline{a}_2}, Z^{\underline{b}_1\cdot\cdot\underline{b}_4}] = \frac{1}{3} Z^{\underline{a}_1\underline{a}_2\underline{b}_1\cdot\cdot\underline{b}_4} \ , \ [R_{\underline{a}_1\underline{a}_2}, Z^{\underline{b}_1\cdot\cdot\underline{b}_4}] = 12\delta^{[\underline{b}_1\underline{b}_1}_{\underline{a}_1\underline{a}_2} Z^{\underline{b}_3\underline{b}_4]} \ ,$$
$$[R^{\underline{a}_1\underline{a}_2}, Z^{\underline{b}_1\cdot\cdot\underline{b}_6}] = 0 \ , \ [R^{\underline{a}_1\underline{a}_2}, Z^{\underline{b}_1\cdot\cdot\underline{b}_6}] = 2\cdot 45\delta^{[\underline{b}_1\underline{b}_2}_{\underline{a}_1\underline{a}_2} Z^{\underline{b}_3\cdot\cdot\underline{b}_6]} \ .$$

The commutators of the IIA level one $IIA$ generators with level zero $l_1$ generators are given by

$$[R^{\underline{a}}, P_{\underline{b}}] = -\delta^{\underline{a}}{}_{\underline{b}} Z \ , \ [R^{\underline{a}_1\underline{a}_2\underline{a}_3}, P_{\underline{b}}] = 3\delta^{[\underline{a}_1}{}_{\underline{b}} Z^{\underline{a}_2\underline{a}_3]} \ ,$$
$$[R^{\underline{a}_1\cdots\underline{a}_5}, P_{\underline{b}}] = -\frac{5}{2}\delta^{[\underline{a}_1}{}_{\underline{b}} Z^{\underline{a}_2\cdots\underline{a}_5]} \ , \ [R^{\underline{a}_1\cdots\underline{a}_7}, P_{\underline{b}}] = \frac{7}{6}\delta^{[\underline{a}_1}{}_{\underline{b}} R^{\underline{a}_2\cdots\underline{a}_7]} \ , \tag{C.10}$$
$$[R^{\underline{a}}, Q^{\underline{b}}] = Z^{\underline{a}\underline{b}} \ , \ [R^{\underline{a}_1\underline{a}_2\underline{a}_3}, Q^{\underline{b}}] = -Z^{\underline{a}_1\underline{a}_2\underline{a}_3\underline{b}} \ ,$$
$$[R^{\underline{a}_1\cdots\underline{a}_5}, Q^{\underline{b}}] = \frac{1}{6} Z^{\underline{a}_1\cdots\underline{a}_5\underline{b}} \ , \ [R^{\underline{a}_1\cdots\underline{a}_7}, Q^{\underline{b}}] = 0 \ ,$$

and the level minus one $IIA$ with level zero $l_1$ commutators are given by

$$[R_{\underline{a}}, P_{\underline{b}}] = 0 \ , \ [R_{\underline{a}_1\underline{a}_2\underline{a}_3}, P_{\underline{b}}] = 0 \ , \ [R_{\underline{a}_1\cdots\underline{a}_5}, P_{\underline{b}}] = 0 \ , [R_{\underline{a}_1\cdots\underline{a}_7}, P_{\underline{b}}] = 0 \ , \tag{C.11}$$
$$[R_{\underline{a}}, Q^{\underline{b}}] = 0 \ , \ [R_{\underline{a}_1\underline{a}_2\underline{a}_3}, Q^{\underline{b}}] = 0 \ , \ [R_{\underline{a}_1\cdots\underline{a}_5}, Q^{\underline{b}}] = 0 \ , \ [R_{\underline{a}_1\cdots\underline{a}_7}, Q^{\underline{b}}] = 0 \ .$$

The commutators of the level one $IIA$ generators with the level one $l_1$ commutators are given by

$$[R^{\underline{a}}, Z] = 0 \ , \ [R^{\underline{a}}, Z^{\underline{b}_1\underline{b}_2}] = 0 \ , \ [R^{\underline{a}}, Z^{\underline{b}_1\cdot\cdot\underline{b}_4}] = Z^{\underline{a}\underline{b}_1\cdot\cdot\underline{b}_4} \ , \ [R^{\underline{a}_1\underline{a}_2\underline{a}_3}, Z] = 0 \ ,$$
$$[R^{\underline{a}}, Z^{\underline{b}_1\cdot\cdot\underline{b}_6}] = Z^{\underline{a}\underline{b}_1\cdot\cdot\underline{b}_6} + Z^{\underline{b}_1\cdot\cdot\underline{b}_6,\underline{a}} \ , \ [R^{\underline{a}_1\underline{a}_2\underline{a}_3}, Z^{\underline{b}_1\underline{b}_2}] = Z^{\underline{a}_1\underline{a}_2\underline{a}_3\underline{b}_1\underline{b}_2}$$
$$[R^{\underline{a}_1\underline{a}_2\underline{a}_3}, Z^{\underline{b}_1\cdot\cdot\underline{b}_4}] = Z^{\underline{b}_1\cdot\cdot\underline{b}_4[\underline{a}_1\underline{a}_2,\underline{a}_3]} - \overline{Z}^{\underline{a}_1\underline{a}_2\underline{a}_3\underline{b}_1\cdot\cdot\underline{b}_4} \ , \ [R^{\underline{a}_1\underline{a}_2\underline{a}_3}, Z^{\underline{b}_1\cdot\cdot\underline{b}_6}] = 0 \ ,$$
$$[R^{\underline{a}_1\cdot\cdot\underline{a}_5}, Z] = \frac{1}{2} R^{\underline{a}_1\cdot\cdot\underline{a}_5} \ , \ [R^{\underline{a}_1\cdot\cdot\underline{a}_5}, Z^{\underline{b}_1\underline{b}_2}] = -\overline{Z}^{\underline{a}_1\cdot\cdot\underline{a}_5\underline{b}_1\underline{b}_2} - \frac{1}{3} Z^{\underline{a}_1\cdot\cdot\underline{a}_5[\underline{b}_1,\underline{b}_2]} \ , \tag{C.12}$$
$$[R^{\underline{a}_1\cdot\cdot\underline{a}_5}, Z^{\underline{b}_1\cdot\cdot\underline{b}_4}] = 0 \ , \ [R^{\underline{a}_1\cdot\cdot\underline{a}_5}, Z^{\underline{b}_1\cdot\cdot\underline{b}_6}] = 0 \ ,$$
$$[R^{\underline{a}_1\cdot\cdot\underline{a}_7}, Z] = -\frac{3}{2}\overline{Z}^{\underline{a}_1\cdot\cdot\underline{a}_7} - \frac{1}{6} Z^{\underline{a}_1\cdot\cdot\underline{a}_7} \ , \ [R^{\underline{a}_1\cdot\cdot\underline{a}_7}, Z^{\underline{b}_1\underline{b}_2}] = 0 \ ,$$
$$[R^{\underline{a}_1\cdot\cdot\underline{a}_7}, Z^{\underline{b}_1\cdot\cdot\underline{b}_4}] = 0 \ , \ [R^{\underline{a}_1\cdot\cdot\underline{a}_7}, Z^{\underline{b}_1\cdot\cdot\underline{b}_6}] = 0 \ ,$$

and level minus one $IIA$ with the level one $l_1$ commutators are given by

$$
\begin{aligned}
&[R_{\underline{a}}, Z] = -P_{\underline{a}} \ , \ [R_{\underline{a}}, Z^{\underline{b}_1\underline{b}_2}] = 2\delta^{[\underline{b}_1}{}_{\underline{a}} Q^{\underline{b}_2]} \ , \ [R_{\underline{a}}, Z^{\underline{b}_1..\underline{b}_4}] = 0 \ , \ [R_{\underline{a}}, Z^{\underline{b}_1..\underline{b}_6}] = 0 \ , \\
&[R_{\underline{a}_1\underline{a}_2\underline{a}_3}, Z] = 0 \ , \ [R_{\underline{a}_1\underline{a}_2\underline{a}_3}, Z^{\underline{b}_1\underline{b}_2}] = 6\delta^{\underline{b}_1\underline{b}_2}_{[\underline{a}_1\underline{a}_2} P_{\underline{a}_3]} \ , \\
&[R_{\underline{a}_1\underline{a}_2\underline{a}_3}, Z^{\underline{b}_1..\underline{b}_4}] = -24\delta^{[\underline{b}_1\underline{b}_2\underline{b}_3}_{\underline{a}_1\underline{a}_2\underline{a}_3} Q^{\underline{b}_4]} \ , \ [R_{\underline{a}_1\underline{a}_2\underline{a}_3}, Z^{\underline{b}_1..\underline{b}_6}] = 0 \ , \\
&[R_{\underline{a}_1..\underline{a}_5}, Z] = 0, \ [R_{\underline{a}_1..\underline{a}_5}, Z^{\underline{b}_1\underline{b}_2}] = 0 \ , \ [R_{\underline{a}_1..\underline{a}_5}, Z^{\underline{b}_1..\underline{b}_4}] = 72\delta^{\underline{b}_1..\underline{b}_4}_{[\underline{a}_1..\underline{a}_4} P_{\underline{a}_5]} \ , \\
&[R_{\underline{a}_1..\underline{a}_5}, Z^{\underline{b}_1..\underline{b}_6}] = -135\cdot 8\delta^{[\underline{b}_1..\underline{b}_5}_{\underline{a}_1..\underline{a}_5} Q^{\underline{b}_6]} \ , \ [R_{\underline{a}_1..\underline{a}_7}, Z] = 0 \ , \ [R_{\underline{a}_1..\underline{a}_7}, Z^{\underline{b}_1\underline{b}_2}] = 0 \ , \\
&[R_{\underline{a}_1..\underline{a}_7}, Z^{\underline{b}_1..\underline{b}_4}] = 0 \ , \ [R_{\underline{a}_1..\underline{a}_7}, Z^{\underline{b}_1..\underline{b}_6}] = 8\cdot 7\cdot 135 P_{[\underline{a}_1} \delta^{\underline{b}_1..\underline{b}_6}_{\underline{a}_1..\underline{a}_7]} \ .
\end{aligned} \tag{C.13}
$$

# Appendix D

# Author Publication List

The following is a list of all papers co-written by the author that contributed to this thesis:

4 **The String Little Algebra.** [32]

K. Glennon and P. West. IJMPA **37** 10, 2250051 (2022). arXiv: 2202.01106 [hep-th].

3 **The Massless Irreducible Representation in E Theory and How bosons** [31] **Can Appear as Spinors.**

K. Glennon and P. West, IJMPA **36** 16, 2150096 (2021). arXiv: 2102.02152 [hep-th].

2 **The Non-Linear Dual Gravity Equation of Motion in Eleven Dimensions** [30]

K. Glennon and P. West, Phys.Lett.B 809 (2020) 135714. arXiv: 2006.02383 [hep-th].

1 **Gravity, Dual Gravity and $A_1^{+++}$** [29]

K. Glennon and P. West, IJMPA **35** 14, 2050068 (2020). arXiv: 2004.03363 [hep-th].